\documentclass[useAMS,usenatbib,onecolumn,showpacs,superscriptaddress,floatfix,nofootinbib,noeprint,prd]{revtex4-2}

\usepackage{graphicx}
\usepackage{epsf}
\usepackage{epsfig}
\usepackage{amssymb,amsmath}
\usepackage[usenames]{color}
\usepackage{amssymb}
\usepackage{times}
\usepackage{mathrsfs}
\usepackage{hyperref}
\hypersetup{
  colorlinks=true,        
  linkcolor=blue,         
  citecolor=cyan,         
}
\usepackage{float}

\newcommand{\ie}{{i.e.,}~}

\renewcommand{\u}{\boldsymbol{u}}
\newcommand{\halb}{\frac{1}{2}}

\newcommand{\x}{\boldsymbol{x}}
\newcommand{\n}{\boldsymbol{n}}

\renewcommand{\u}{\boldsymbol{u}}

\newcommand{\q}{\boldsymbol{q}}
\newcommand{\F}{\boldsymbol{F}}
\renewcommand{\S}{\boldsymbol{S}}
\newcommand{\f}{\boldsymbol{f}}

\newcommand{\A}{\boldsymbol{A}}
\newcommand{\B}{\boldsymbol{B}}
\newcommand{\emm}{m}

\allowdisplaybreaks

\begin{document}

\title[High order DG schemes for FO-BSSNOK system]{High-order discontinuous Galerkin schemes with subcell finite volume
	limiter and adaptive mesh refinement for a monolithic first-order BSSNOK
	formulation of the Einstein-Euler equations}

\date{\today}
\label{firstpage}

\author{Michael Dumbser}
\affiliation{
Laboratory of Applied Mathematics, University of Trento,
Via Mesiano 77, 38123 Trento, Italy
}

\author{Olindo Zanotti}
\affiliation{
Laboratory of Applied Mathematics, University of Trento,
Via Mesiano 77, 38123 Trento, Italy
}

\author{Ilya Peshkov}
\affiliation{
	Laboratory of Applied Mathematics, University of Trento,
	Via Mesiano 77, 38123 Trento, Italy
}

\begin{abstract}
We propose a high order discontinuous Galerkin (DG) scheme with subcell finite volume (FV) limiter to solve a monolithic first--order hyperbolic BSSNOK formulation of the coupled Einstein--Euler equations. The numerical scheme runs with adaptive mesh refinement (AMR) in three space dimensions, is endowed with time-accurate local time stepping (LTS) and is able to deal with both conservative and non-conservative hyperbolic systems.  The system of governing partial differential equations was shown to be strongly hyperbolic and is solved in a monolithic fashion with one numerical framework that can be simultaneously applied to both the conservative matter subsystem as well as the non-conservative subsystem for the spacetime.  
Since high order unlimited DG schemes are well-known to produce spurious oscillations in the presence of discontinuities and singularities, our subcell finite volume limiter is crucial for the robust discretization of shock waves arising in the matter as well as for the stable treatment of puncture black holes.
We test the new method on a set of classical test problems of numerical general relativity, showing good agreement with available exact or numerical reference solutions. In particular, we perform the first long term evolution of the inspiralling merger of two puncture black holes with a high order ADER-DG scheme.
\end{abstract}

\maketitle


\section{Introduction}


The advent of a third generation of 
gravitational wave (GW) detectors~\citep{Luck2022}, specifically 
the Einstein Telescope\footnote{\url{https://www.et-gw.eu/}} in Europe and the Cosmic Explorer\footnote{\url{https://cosmicexplorer.org/}} in the US,  will provide large enough signal-to-noise ratios to allow for a high precision
GW astronomy. 
The planned  sensitivity $h$ for these instruments
will be in the range $10^{-25}\div10^{-24}\,\rm{Hz}^{-1/2}$, 
thus representing an invaluable tool in fundamental physics research,
black hole physics and high energy astrophysics.
Indeed, the actual detection of GWs since September 14 2015 was preceded and continuously accompanied  by 
more and more accurate numerical simulations, which are now crucial for interpreting
the signals that we receive and for  understanding the complex physics behind the high energy processes  involved on the cosmic scale. Hence, numerical relativity (NR) is by now a mature field of research, having obtained
a large amount of scientific results which are virtually impossible to document properly in a few lines
(see, among the others, the monographs by \cite{Thornburg2007LRR,Alcubierre:2008,Font08,Grandclement2009,Bona-and-Palenzuela-Luque-2005:numrel-book, Baumgarte2010,Gourgoulhon2012,Faber2012LRR,Rezzolla_book:2013,Bishop2016LRR,Kyutoku2021LRR}).

In view of numerical applications, 
the Einstein equations can be cast in many different forms, some of which have not
been fully explored yet. 
The first numerical simulation of a binary 
system \cite{Pretorius2005}
has been obtained within the generalized harmonic (GH) formulation,
which was later cast as a first-order system by \cite{Lindblom2006},
allowing for important scientific results (see \cite{Tichy2023} and references therein).
However, the most widely adopted formulations of the Einstein equations 
are represented by the BSSNOK/Z4/CCZ4/Z4c family, which are based on the
famous BSSNOK system \cite{Nakamura87,Shibata95,Baumgarte1998}.
The BSSNOK formulation arises naturally as a mixed first and second order (in space derivatives) PDE system, and as such, 
central finite difference numerical schemes have always been  
the most natural choice. This is in fact  the case for
a wide class of numerical codes like
Einstein--Toolkit~\citep{Loffler2012}, LazEv~\citep{Zlochower2005,Lousto2023}, BAM~\citep{Bruegmann2008,Thierfelder2011},
McLachlan~\citep{McLachlan:web}, GRChombo~\citep{Clough_2015},
AMReX~\citep{Peterson_2023}, SACRA~\citep{Yamamoto2009,Kiuchi2017}, MHDueT~\cite{Palenzuela2018},
ExaGRyPE~\cite{Zhang2024}.

However, in the context of numerical analysis for hyperbolic conservation laws rather advanced numerical schemes have been developed over the last decades in alternative to classical finite differencing, that could be rather beneficial to NR. A special case in this respect is represented by high order Discontinuous Galerkin (DG) schemes
\cite{cbs1,cbs2,cbs3,cbs4}, that come with  a virtually negligible numerical dissipation, 
allow for an optimal parallelization scalability and  
can reach at least in principle arbitrary high order of  accuracy, though not requiring
any spatial reconstruction. 
The application of DG schemes to the relativistic context has been initially obtained discarding the
Einstein equations but rather focusing on the evolution of the field content, i.e. solving the equations
$\nabla_\mu T^{\mu\nu}=0$ in a stationary  spacetime (either flat or curved), see e.g. the seminal works \cite{DumbserZanotti} and \cite{Radice2011}. 
When doing so, one faces a well--known problem of DG schemes, namely that of producing strong oscillations at
discontinuities, unless special interventions are performed. This fostered the research for a proper \emph{limiting} of DG schemes, with quite interesting results obtained by various groups
proposing a variety of different approaches including artificial viscosity 
\cite{Hartman_02,Persson_06,Feistauer4}, or filtering \cite{Radice2011}, or
\textit{a priori} limiting of  the troubled cells
\cite{cbs4,QiuShu1,Qiu_2004,balsara2007,Zhu_2008,Zhu_13,Luo_2007,Krivodonova_2007}, 
and, finally, \textit{a posteriori} limiting \cite{Dumbser2014,Boscheri2017}.
In this way, a number of relevant results in the field of relativistic hydrodynamics and 
magnetohydrodynamics have been achieved, see e.g.  \cite{Radice2011,DumbserZanotti,Zanotti2011,Zanotti2015d,Teukolsky2015,Bugner2016,Zhao2017,ADERGRMHD,Hebert2018,
	Junming2020,Deppe2022,Deepak2023}. 

We should also mention that DG schemes come in two flavors as far as the time discretization is concerned.
From one side there are the classical Runge--Kutta DG schemes of Cockburn and Shu \cite{cbs4,CBS-convection-dominated,CBS-convection-diffusion} for first and second order systems, which appeared in the general relativistic context for the first time in \cite{Radice2011}. On the other side there are ADER-DG schemes~\cite{dumbser_jsc,QiuDumbserShu}, which are fully-discrete one-step schemes that are uniformly high order accurate in both space and time and which were first applied to relativistic hydrodynamics and magnetohydrodynamics in \cite{dumbser2008unified,DumbserZanotti}.  

When one moves to the numerical solution of the Einstein equations, however, the situation complicates considerably, worsened by the fact that DG schemes can be naturally applied to first--order systems, 
while they pose a number of hard challenges when applied to second--order systems. Of course there 
have been various attempts with rather promising results: for instance, 
\cite{Miller2017} successfully 
solved  a number of fundamental tests of NR with a DG scheme within the second--order BSSNOK formulation;
\cite{Tichy2023} could evolve a single black hole using DG methods in combination with
the first--order generalized harmonic formulation by \cite{Lindblom2006}.
Quite recently, and within the same formulation,
\cite{Deppe2024} have presented important results
on the binary merger of neutron stars using a hybrid DG-finite difference method
with the SpEC code~\cite{Kidder2000,Duez2008,Haas2016}.
However, to the best of our knowledge, it has not been possible yet to successfully evolve binary systems of black holes with high order DG schemes using a second--order version of either the BSSNOK or the GH formulation. 

Therefore, a general consensus seems to be forming in the last decade or so that a strongly hyperbolic first--order formulation of the Einstein equations is required if one really wants to use DG schemes in NR calculations.  
Possibly under these motivations, \cite{Brown2012} proposed a first--order version
of BSSNOK which they implemented both with finite difference and Runge--Kutta DG schemes. However, 
in spite of the promises of that approach, there are no evidences that the project has
been pushed further on. 
Within our group, we have initially explored the capabilities of a first--order version of the Z4 formulation, proposed originally by \cite{Bona:2003fj,Bona:2004} with the aim of treating
properly the Einstein constraints, to perform binary merger simulations.
Hence, in \cite{Dumbser2017strongly,Dumbser2020GLM} we applied high order ADER--DG methods  
to a new first--order version (FO-CCZ4) of the conformal and covariant Z4 formulation of the Einstein equations of \cite{Alic:2011a}, showing also first preliminary results for a head-on collision of two puncture black holes. Based on some preliminary studies on well-balanced schemes for the Einstein equations in one space dimension \cite{Gaburro2021WBGR1D}, more recently in \cite{DumbserZanottiGaburroPeshkov2023} we have presented a new first--order Z4 formulation (FO-Z4) solved with high order exactly well--balanced ADER--DG methods,
which can be particularly suitable to perform numerical perturbative analysis of compact objects. 
However, in spite of some progress made, none of these approaches was robust enough to
perform a stable long term evolution of binary systems with high order DG schemes.
 
In this work we finally reach a major step forward 
by solving a new strongly hyperbolic first--order non--conservative version of BSSNOK, denoted by FO-BSSNOK in the following, with high order ADER--DG schemes. 
Our new FO-BSSNOK formulation of the Einstein equations 
has been recently presented in \cite{CWENOBSSNOK} in combination with an entirely different class of numerical schemes, namely central WENO (CWENO) finite difference schemes \cite{Levy2000,Levy2001}. High order WENO finite difference schemes were for the first time applied to the Einstein field equations in \cite{Balsara2024c}, where a first order Z4 formulation has been discretized. 

Since DG schemes are so little diffusive, they become extremely sensitive to any, even tiny, deviation from strong hyperbolicity. Hence, our guiding principle for designing FO-BSSNOK in \cite{CWENOBSSNOK} has been to prove strong hyperbolicity of the corresponding PDE system 
by effectively showing a complete set of real eigenvalues and associated eigenvectors. 
In this paper we show that the FO-BSSNOK system proposed in \cite{CWENOBSSNOK} can successfully evolve binary systems of black holes also over longer timescales using high order ADER--DG schemes with suitable subcell finite volume limiting, in principle to any desired order of accuracy in space and time.  
A key feature of our new approach is that \emph{the same numerical scheme} can be applied to the matter and to the Einstein sectors of the PDE system, with the only caveat of taking into account non--conservative terms in the spacetime evolution. This represents an enormous advantage with respect to traditional approaches where totally different numerical schemes must be used for the two parts of the PDE system (matter sector and spacetime sector).


The structure of the paper is as follows. In Section~\ref{sec:second} we present the new first order version of the BSSNOK formulation of the coupled Einstein-Euler equations.
In Section~\ref{sec.ader} we describe the new numerical
scheme that can be used in a monolithic way for the full system of the Einstein--Euler PDEs. 
In Section~\ref{sec:tests} we show a number of benchmark results
to demonstrate the correctness of both the formulation and the numerical
solver. Finally, the conclusions are summarized in Section
\ref{sec:conclusions}.

We work in a geometrized set of units, in which the speed of light and
the gravitational constant are set to unity, \ie $c=G=1$. Greek indices
run from $0$ to $3$, Latin indices run from $1$ to $3$ and we use the
Einstein summation convention of repeated indices.

\section{The first--order monolithic BSSNOK formulation of the coupled Einstein--Euler system}
\label{sec:second}

In \cite{CWENOBSSNOK} we have presented the FO-BSSNOK system, a new strongly hyperbolic first--order reformulation of the original 
second--order BSSNOK system~\cite{Alcubierre:2008,Baumgarte2010,Gourgoulhon2012,Rezzolla_book:2013}), which was directly coupled with the general relativistic Euler equations of hydrodynamics in a monolithic manner, solving both subsystems for matter and spacetime with one and the same numerical method in one single set of first order hyperbolic equations. The coupling between the matter and spacetime subsystems occurs merely via algebraic source terms on the right hand side of the system. In the following we provide a concise description of our approach, which inherits several definitions and conventions 
quite common in numerical relativity. As customary in the 3+1 formalism, the line element of the metric is split as
\begin{equation}
	ds^2 =-(\alpha^2- \beta_i \beta^i)dt^2 + 2\beta_i\,dx^i \,dt + \gamma_{ij}dx^i\,dx^j\,,
\end{equation}
where $\alpha$ is the lapse, $\beta^i$ is the shift and $\gamma_{ij}$ is the purely spatial metric tensor. 
A key feature of the BSSNOK approach is to factorize  $\gamma_{ij}$ in terms of a unit-determinant auxiliary metric tensor 
$\tilde\gamma_{ij}$, i.e.
\begin{equation}
\gamma_{ij}=\psi^{4}\tilde\gamma_{ij}=e^{4\phi}\tilde\gamma_{ij}\,,
\end{equation}
with $\psi=\gamma^{1/12}$ being the conformal factor, and $\gamma = \det(\gamma_{ij})$.
Following the same logic, the symmetric extrinsic curvature $K_{ij}$, is replaced by a rescaled trace-free tensor
\begin{equation}
\tilde{A}_{ij}=\psi^{-4}A_{ij}=e^{-4\phi}\left(K_{ij}-\frac{1}{3}\gamma_{ij}K\right)\,,
\end{equation}
where $K = \gamma^{ij} K_{ij}$ is the trace of $K_{ij}$.
The price to pay for getting a first--order formulation of the Einstein equations is to enlarge the set of 
evolved quantities, which must now involve the space derivatives of the metric terms. Hence, just like 
in \cite{Dumbser2017strongly,CWENOBSSNOK}, 
we introduce the following 30 auxiliary variables 
\begin{align}
\label{eq:Auxiliary}
A_k := \partial_k\ln\alpha = \frac{\partial_k \alpha }{\alpha}\,, \qquad
B_k^{\,\,i} := \partial_k\beta^i\,, \qquad
D_{kij} := \frac{1}{2}\partial_k\tilde\gamma_{ij}\,. \qquad
P_k := \frac{\partial_k \psi}{\psi} = \partial_k \phi = \frac{1}{12}\partial_k\ln \gamma\,.
\end{align}
In general, we will also be interested in the evolution of an external field, described by an
energy--momentum tensor $T^{\mu\nu}$ of an arbitrarily complex physical system. 
In this paper, though, for the sake of simplicity, we will limit
our attention to a perfect (non--dissipative) fluid,  for which we
have
\begin{equation}
	T^{\mu\nu}=p\,g^{\mu\nu} + (e+p) u^\mu u^\nu\,.
\end{equation}
Here, $p$ is the gas pressure, $e$ is the energy density and $u^\mu$ is the four velocity of the fluid.
The equation of state is that of an ideal gas, with $p=\rho\epsilon(\gamma-1)$, where $\rho$ the rest mass density, $\epsilon$ is the specific internal energy, while $\gamma$ is the adiabatic index. 
In the reference frame of the so--called Eulerian observer defined by $n^\mu=(1/\alpha,-\beta^i/\alpha)$, the relativistic mass density, the momentum density, the energy density  and the spatial part of the energy momentum tensor are given, respectively, by~\cite{Rezzolla_book:2013}
\begin{eqnarray}
	D   &=& \rho W\,,\\
	S_i &=& \rho h W^2 v_i\,,\\ 
	E   &=& \rho h W^2 - p\,,\\
	S_{ij} &=& \rho h W^2 v_i v_j + p\, \gamma_{ij} \,,
\end{eqnarray}
where $h=1 + \epsilon + p/\rho$ is the specific enthalpy, $v^i$ is the fluid velocity and $W=1/\sqrt{1-v^i v_i}$ is the Lorentz factor of the fluid relative to $n^\mu$.

With all these definitions in mind,
and using the traditional metric variables $\tilde\gamma_{ij}$, $\phi$, $\tilde A_{ij}$, $K$ of the second--order BSSNOK
formulation, as well as
of the additional variables given by \eqref{eq:Auxiliary},
the new first--order BSSNOK formulation of the Einstein-Euler equations leads to the following system of PDEs
(see \cite{CWENOBSSNOK})
\begin{align}
\label{eqn.rho}
&\partial_t (\sqrt{\gamma}D)+\partial_i\left[\sqrt{\gamma}(\alpha v^i D - \beta^i D)\right]=0\,,\\ 
\label{eqn.S}
&\partial_t (\sqrt{\gamma}S_j)+\partial_i\left[\sqrt{\gamma}(\alpha S^i_{\,\,j} - \beta^i S_j)\right]=\sqrt{\gamma}\left[\alpha S^{ik}D_{jik}  +
S_i  B_j^{\,\,i} - \alpha E A_j   \right]\,,\\ 
\label{eqn.E}
&\partial_t (\sqrt{\gamma}E)+\partial_i\left[\sqrt{\gamma}(\alpha S^i - \beta^i E)\right]=\sqrt{\gamma}\left[\alpha S^{ij}e^{4\phi}\left(\tilde A_{ij} + \frac{1}{3}\tilde\gamma_{ij}K\right) - \alpha S^j A_j  \right]\,,\\ 
\label{eqn.gamma}
&\partial_t\tilde\gamma_{ij} - \beta^k\partial_k\tilde\gamma_{ij}=\tilde\gamma_{ik} B_{j}^{\,\,k}  + \tilde\gamma_{kj} B_{i}^{\,\,k} -\frac{2}{3}\tilde\gamma_{ij}
B_{k}^{\,\,k} - 2\alpha \tilde A_{ij}\,,\\
\label{eqn.phi}
&\partial_t \phi  - \beta^k\partial_k \phi= \frac{1}{6}B_{k}^{\,\,k}  -\frac{1}{6}\alpha K  \,,\\
\label{eqn.Kij}
&\partial_t \tilde A_{ij} - \beta^k \partial_k \tilde A_{ij}  + \alpha e^{-4\phi} \left( \partial_{(i} {A}_{j)} -\frac{1}{3}\tilde\gamma_{ij}\tilde\gamma^{mn}\partial_{(m} {A}_{n)}\right)
-\alpha e^{-4\phi} \bigg[ (R_{ij})^{TF}_{\textrm{ncp}}     \bigg] = \nonumber \\
&  \qquad \qquad 
\tilde A_{ik} B_j^{\,\,k} + \tilde A_{kj} B_i^{\,\,k} -\frac{2}{3}\tilde A_{ij}B_k^{\,\,k} 
- \alpha e^{-4\phi} \left[ A_i A_j -  \Gamma^k_{ij} A_k -\frac{1}{3}\tilde\gamma_{ij}\tilde\gamma^{mn}(A_m A_n -  \Gamma^k_{mn} A_k)  \right] +\nonumber\\
&  \qquad \qquad 
+\alpha e^{-4\phi} \bigg[ (R_{ij})^{TF}_{\textrm{src}}      \bigg] 
- 8\pi\alpha e^{-4\phi} \left(S_{ij}-\frac{1}{3}e^{4\phi}\tilde\gamma_{ij}\,S\right)  
 +  \alpha ( K \tilde A_{ij} - 2  \tilde A_{il} \tilde\gamma^{lm} \tilde A_{mj} ) \,, \\
\label{eqn.K}
&\partial_t K - \beta^k \partial_k K + \alpha e^{-4\phi}\tilde\gamma^{ij}\partial_{(i} A_{j)} 
= -\alpha e^{-4\phi}\tilde\gamma^{ij}  \left( A_i A_j  -\Gamma^k_{ij} A_k \right) + \alpha \left(\tilde A_{ij}\tilde A^{ij} + \frac{1}{3}K^2 + 4\pi(E+S)  \right)\,,\\
\label{eqn.slicing}
&\partial_t \ln\alpha - \beta^k{\partial_k\ln\alpha} = -g(\alpha)\alpha(K-K_0)\,,\\
\label{g-driver1}
&\partial_t \beta^i - b\, s\,\beta^k\partial_k \beta^i  = \frac{3}{4}\,s\,b^i, \\
\label{g-driver2}
&\partial_t b^i - b\, s\,(\beta^k\partial_k b^i - \beta^k\partial_k \tilde\Gamma^i) = s(\partial_t \tilde\Gamma^i - \eta b^i)\,,\\
\label{eqn.Gammai}
&  \partial_t\tilde\Gamma^i -s\,\bigg[ \beta^k\partial_k \tilde\Gamma^i
+\tilde\gamma^{jk}\partial_{(j} B_{k)}^{\,\,i}
 + \frac{1}{3}\tilde\gamma^{ij}\partial_{(j} B_{k)}^{\,\,k}  
-\frac{4}{3}\alpha\tilde\gamma^{ij}\partial_j K 
 \bigg]
 =\nonumber\\ 
& \qquad \qquad s\,\bigg[\frac{2}{3}\tilde\Gamma^i B_k^{\,\,k} - \tilde\Gamma^k B_k^{\,\,i}
	-2 \alpha \tilde A^{ij} A_j   + 
	2\alpha\left(\tilde\Gamma^i_{jk}\tilde A^{jk} + 6\tilde A^{ij}P_j  -8\pi e^{4\phi} M^i \right)\bigg]\,,\\
\label{eqn.A}
& \partial_t A_{i} - {\beta^k \partial_k A_i} + \alpha g(\alpha) \left( \partial_i K - \partial_i K_0  \right)
=-\alpha A_i \left( K - K_0  \right) \left( g(\alpha) + \alpha g^\prime(\alpha)  \right)
  +   B_i^{\,\,k} ~A_{k} 
\\
\label{eqn.B}
& \partial_t B_k^{\,\,i}  - s\left[  \frac{3}{4} \partial_k b^i  
+ b\left(\beta^m\partial_m B_k^{\,\,i}  \right)
- 
\mu\, \alpha^2  \gamma^{ij} \gamma^{nl} \left( \partial_k D_{ljn} - \partial_l D_{kjn} \right)  
 \right]
=		s\,b\, B_m^{\,\,i} B_k^{\,\,m} \,,
\\
\label{eqn.D}
&
\partial_t D_{kij} - {\beta^m \partial_m D_{kij}} 
         - \frac{1}{2} {\tilde\gamma}_{mi} \partial_{(k} {B}_{j)}^{\,\,m}
         - \frac{1}{2} {\tilde\gamma}_{mj} \partial_{(k} {B}_{i)}^{\,\,m}
         + \frac{1}{3} {\tilde\gamma}_{ij} \partial_{(k} {B}_{m)}^{\,\,m}
				 +  \alpha \partial_k {\tilde A}_{ij} = B_k^{\,\,m} D_{mij} + B_j^{\,\,m} D_{kmi} + B_i^{\,\,m} D_{kmj}  \nonumber \\ 
&	\qquad \qquad 	-\frac{2}{3}B_m^{\,\,m} D_{kij} - \alpha A_k  {\tilde A}_{ij}
+\frac{1}{3}\alpha \tilde\gamma_{ij}\left[ \tilde\gamma^{nm} \partial_k \tilde A_{nm} + \tilde A_{nm} \partial_k \tilde\gamma^{nm} \right]
\,,\\
\label{eqn.P}
&
\partial_t P_i - \beta^k\partial_k P_i+\frac{1}{6}\alpha\partial_i K-\frac{1}{6}\partial_{(i} B_{k)}^{\,\,k}=P_k B_i^{\,\,k} - \frac{1}{6}\alpha K A_i \, ,
\end{align}
where the superscript $TF$ denotes the trace-free part of the corresponding tensor, $S = \gamma^{ij}S_{ij}$, while
the Hamiltonian and the momentum  constraints are given by
\begin{eqnarray}
	\label{eqn.adm1}
	H &=& R - \tilde A_{ij} \tilde A^{ij} + \frac{2}{3}K^2 - 16\pi E = 0\,, \\
	\label{eqn.adm2}
	M^i &=& \partial_j\tilde A^{ij}+\tilde\Gamma^i_{jk}\tilde A^{jk}+6\tilde A^{ij}\partial_j\phi-\frac{2}{3}\tilde\gamma^{ij}\partial_j K  - 8 \pi S^i = 0\,.
\end{eqnarray}
In addition, we stress the following facts about the above system:  
\begin{itemize}
	\item Eq.~\eqref{eqn.rho}--\eqref{eqn.P} form a strongly hyperbolic PDE system in which the matter and the spacetime sectors are coupled only through the source terms.
	The full sets of real eigenvalues and corresponding eigenvectors for the case $s=0$ are reported in the Appendix of \cite{CWENOBSSNOK}.
	\item The insertion of the auxiliary variables in Eq.~\eqref{eq:Auxiliary}, which are all
	gradients of primary evolution quantities, generates a set of 
	second-order ordering constraints, namely:
	\begin{equation}
		\label{eq:curlfree}
		\mathcal{A}_{lk}   = \partial_l A_k - \partial_k A_l = 0, \quad
		\mathcal{B}_{lk}^i = \partial_l B_k^{\,\,i} - \partial_k B_l^{\,\,i} = 0, \quad
		\mathcal{D}_{lkij} = \partial_l D_{kij} - \partial_k D_{lij} = 0, \quad
		\mathcal{P}_{lk}   = \partial_l P_k - \partial_k P_l = 0\,,
	\end{equation}
	the third of which can be recognized in Eq.~\eqref{eqn.B}, multiplied by a variable factor $\mu$.
	An additional differential constraint comes from the trace-free character of $\tilde{A}_{ij}$, which gives
	\begin{equation}
		\mathcal{C}_k =  
		\tilde{A}_{ij} \partial_k \tilde{\gamma}^{ij}   + 
		\tilde{\gamma}^{ij} \partial_k \tilde{A}_{ij} = 0\,,
	\end{equation}
	which is used, for instance, in Eq.~\eqref{eqn.D} for the time evolution of $D_{kij}$.
	\item Eqs.~\eqref{eqn.slicing}--\eqref{eqn.Gammai} are the so-called {\emph {gauge conditions}} for the lapse and for the shift. In particular, we adopt the canonical {\emph{gamma--driver}} to treat
	moving punctures, thus promoting the contracted Christoffel symbol ${\tilde\Gamma}^i  =   {\tilde\gamma}^{jk}\, {\tilde\Gamma}^i_{jk}$ to a primary variable~\cite{Alcubierre2004}. Hence,
	$s$ is a binary parameter, either to 1 or 0, which is used to switch the {\emph{ Gamma--driver}} on or off in Eqs.~\eqref{g-driver1}--\eqref{eqn.Gammai},
	depending on the test being considered.
	\item The terms $(R_{ij})^{TF}_{ncp}$ and $(R_{ij})^{TF}_{src}$ in Eq.~\eqref{eqn.Kij} separate
	the non--conservative products from the purely algebraic factors, that are included in the Ricci tensor.
	\item In those tests that involve fluids, we will adopt an efficient and robust conversion from the conservative to the primitive variables introduced originally in \cite{DumbserZanottiGaburroPeshkov2023}, which has proven to be valid even in the presence of zero density
	atmospheres around neutron stars.
\end{itemize}
As a mixed conservative--non conservative first--order system, Eq.~\eqref{eqn.rho}--\eqref{eqn.P} can be discretized via ADER--DG schemes in the way we describe below.

\section{ADER Discontinuous Galerkin schemes for the FO-BSSNOK non conservative system}
\label{sec.ader}
\subsection{The unlimited ADER-DG scheme}

Discontinuous Galerkin schemes are usually proposed for hyperbolic systems in conservation form.
However, they can be as well extended to non conservative (but still hyperbolic) systems and the first
analysis of this kind, although not in the relativistic
context, can be found in \cite{Rhebergen2008,Dumbser2009a,Dumbser2010}. The first application
 to the general relativistic context
was presented in \cite{Dumbser2017strongly} for the FO-CCZ4 equations.
The
numerical approach that we propose here, \emph{mutatis mutandis},  is the same one considered in that work.
Hence, we first write  the system \eqref{eqn.rho}--\eqref{eqn.P} as
\begin{equation}
	\label{eqn.pde.mat.preview}
	\frac{\partial \u }{\partial t} 
	+ \frac{\partial \f^i}{\partial x_i} +
	\B^i(\u) \frac{\partial \u}{\partial x_i} 
	= \S(\u),
	\qquad 
	\textnormal{or, equivalently,}
	\qquad 
	\frac{\partial \u }{\partial t} + \nabla \cdot \F(\u)
	+ \B(\u) \cdot \nabla \u  
	= \S(\u)\,,
\end{equation}
where $\u$ is the state vector of 63 dynamical variables, namely 
\begin{equation}
\u=\left( D, S_i, E, \tilde\gamma_{ij}, \ln{\alpha}, \beta^i, \ln{\phi}, \tilde
A_{ij}, K, \tilde\Gamma^i, b^i, A_k, B^i_k, D_{kij}, P_k, K_0 \right)\,.
\end{equation}
According to the DG approach, 
at time $t^n$ in each
element  $\Omega_{i} = [x_i - \halb \Delta x_i,
x_i + \halb \Delta x_i] \times [y_i - \halb \Delta y_i, y_i + \halb
\Delta y_i] \times [z_i - \halb \Delta z_i, z_i + \halb \Delta z_i] $
the discrete solution is written in terms of 
spatial basis functions $\Phi_l(\x)$ as 
\begin{equation}
	\u_h(\x,t^n) = \sum \limits_l \hat{\u}_{i,l}
	\Phi_l(\x) := \hat{\u}_{i,l}^n \Phi_l(\x)\,,
	\label{eqn.ansatz.uh}
\end{equation}
where the coefficients $\hat{\u}_{i,l}^n$ 
are the so-called \emph{degrees of freedom}, while 
$l=(l_1,l_2,l_3)$ is a multi-index. 
Concerning the choice of the spatial basis functions, we first  define
one-dimensional basis functions $\varphi_{k}(\xi)$ on the reference
element $[0,1]$. These are given  
by the Lagrange interpolation polynomials, all of degree $N$, which pass through the $(N + 1)$
Gauss–Legendre quadrature nodes $\xi_j$. Then,
the spatial basis functions $\Phi_l(\x)$ can be obtained via the tensor product of
the one-dimensional basis functions $\varphi_{k}(\xi)$, i.e.
$\Phi_l(\x) = \varphi_{l_1}(\xi) \varphi_{l_2}(\eta)
\varphi_{l_3}(\zeta)$. We use a  simple linear map 
from the physical coordinates $\x \in
\Omega_i$ to reference coordinates $\boldsymbol{\xi}=\left(
\xi,\eta,\zeta \right) \in [0,1]^3$, namely $\x =
\x_i - \halb \Delta \x_i + (\xi \Delta x_i, \eta \Delta
y_i, \zeta \Delta z_i)^T$. 

As customary for the DG discretization, we 
multiply the governing equations
\eqref{eqn.pde.mat.preview} by a test function $\Phi_k \in \mathcal{U}_h$, for us the same one
of Eq.~\eqref{eqn.ansatz.uh}, and then we integrate over
the space-time control volume $\Omega_i \times
[t^n;t^{n+1}]$, to get
\begin{equation}
	\label{eqn.pde.nc.gw1}
	\int \limits_{t^n}^{t^{n+1}} \int \limits_{\Omega_i}
	\Phi_k \frac{\partial \u}{\partial t} d\x\, d t
	+\int \limits_{t^n}^{t^{n+1}} \int \limits_{\Omega_i}
	\Phi_k \left( \A(\u) \cdot \nabla \u  \right) d\x\, d t
	= \int \limits_{t^n}^{t^{n+1}} \int \limits_{\Omega_i}   \Phi_k
	\S(\u) d\x\, d t\,,
\end{equation}
where $\A(\u) = \partial \f / \partial \u + \B(\u)$.
In the ADER-DG framework, the time integration of Eq.~\eqref{eqn.pde.nc.gw1} 
is not performed through traditional Runge--Kutta schemes, but rather
after resorting to an element-local
space-time predictor, denoted by $\q_h(\x,t)$,
available at any intermediate time between $t^n$ and $t^{n+1}$ (see Sect.~\ref{sec.ADER}), in such a way that
the final time update can be obtained through a single time-step numerical scheme.
When this approach is applied to non--conservative systems, further complications arise since
the non conservative terms are in principle vulnerable to non-smooth solutions.
These can, however, be treated via the so--called \emph{path-conservative} schemes, which over the years
have been developed both for finite volume and DG schemes~\cite{Castro2006,Pares2006,Castro2006,Rhebergen2008,Dumbser2009a,Dumbser2010},
following the pioneering theoretical investigations of \cite{DalMaso1995}.
Hence, assuming that 
the space-time predictor $\q_h(\x,t)$ is available, the master DG
equation resulting from~\eqref{eqn.pde.nc.gw1}, after \eqref{eqn.ansatz.uh} is adopted,  is
\begin{equation}
	\label{eqn.pde.nc.gw2}
	\left( \int \limits_{\Omega_i}  \Phi_k \Phi_l d\x \right)
	\left( \hat{\u}^{n+1}_{i,l} - \hat{\u}^{n}_{i,l}  \right)
	+ \int \limits_{t^n}^{t^{n+1}} \! \! \int \limits_{\Omega_i^\circ}
	\Phi_k \left( \A(\q_h) \cdot \nabla \q_h  \right)  d\x d t
	+ \int \limits_{t^n}^{t^{n+1}} \! \! \int \limits_{\partial \Omega_i}
	\Phi_k \mathcal{D}^-\left( \q_h^-, \q_h^+ \right) \cdot \n \,d S d t
	= \int \limits_{t^n}^{t^{n+1}} \! \! \int \limits_{\Omega_i}
	\Phi_k \S(\q_h)  d\x\, d t\,.
\end{equation}
Eq.~\eqref{eqn.pde.nc.gw2} is the basis for constructing our DG numerical scheme, provided we
have the predictor solution $\q_h$ at disposal, as described in Sec.~\ref{sec.ADER}.
The third integral on the left hand side of Eq.~\eqref{eqn.pde.nc.gw2} is precisely 
responsible for the jumps $\q_h^+ - \q_h^-$  across the element interfaces and 
it can be computed as follows:

\begin{enumerate}
\item 
Firstly, a simple segment path is introduced, which connects the states $\q_h^-$
	and $\q_h^+$:
	\begin{equation}
		\label{line-path}
		\boldsymbol{\psi} = \boldsymbol{\psi}(\q_h^-, \q_h^+, s) =
		\q_h^- + s \left( \q_h^+ - \q_h^- \right),
		\qquad 0 \leq s \leq 1\,.
	\end{equation}
	%
\item 
Secondly,  the jump terms are computed through a
path-integral in phase space along the segment path \eqref{line-path}
\begin{equation}
	\label{eqn.pc.scheme}
	\mathcal{D}^-\left( \q_h^-, \q_h^+ \right) \cdot \n =
	\frac{1}{2} \left( \, \int \limits_{0}^1 \A(\boldsymbol{\psi})
	\cdot \n \, d s \right) \left( \q_h^+ - \q_h^- \right)  -
	\frac{1}{2} \boldsymbol{\Theta} \left( \q_h^+ - \q_h^- \right),
\end{equation}
where $\A \cdot \n = \A_1 n_1 + \A_2 n_2 +
\A_3 n_3$ is the system matrix in normal direction
and where the path integral is  computed numerically via a
Gauss-Legendre quadrature formula of sufficient order.
%
\item 
%
The extra term $\boldsymbol{\Theta} > 0$ in Eq.~\eqref{eqn.pc.scheme} plays the role
of a numerical viscosity matrix. Because of the complexity of the Einstein equations, we have adopted a simple Rusanov-type viscosity matrix~\cite{Dumbser2009a}, namely
\begin{equation}
	\label{eqn.rusanov}
	\boldsymbol{\Theta}_{\textnormal{Rus}} = s_{\max} \boldsymbol{I},
	\qquad \textnormal{with} \qquad s_{\max} = \max \left( \left| \Lambda(\q_h^-)
	\right|, \left| \Lambda(\q_h^+) \right| \right)\,,
\end{equation}
and where $s_{\max}$ denotes the maximum wave speed found at the
interface and $\boldsymbol{I} = \delta_{ij}$ is the 
identity matrix. 
\end{enumerate}
%
In alternative to the simple Rusanov method proposed above also more sophisticated HLL-type  Riemann solvers may be used, see e.g. \cite{toro-book} for an overview of modern state-of-the-art approximate Riemann solvers. 
Next, we will briefly
describe the computation of the local space-time predictor solution
$\q_h$ needed in Eq. \eqref{eqn.pde.nc.gw2} and \eqref{eqn.pc.scheme}.
\subsection{ADER space-time predictor}
\label{sec.ADER}
In the modern version of ADER schemes proposed in \cite{DumbserEnauxToro,dumbser2008unified,Dumbser2009a}, 
which differs
from the original one of Titarev and Toro \cite{Titarev2002,Titarev2005,Toro2006}, 
the element-local space-time predictor solution $\q_h(\x,t)$ is
computed from the known discrete solution $\u_h(\x,t^n)$ at time
$t^n$ using  the DG approach, which operates locally for each cell.
Over the years this strategy has been applied to more and more complex systems of PDEs,
ranging from relativistic ideal magnetohydrodynamics~\cite{Zanotti2015d}, to
resistive--relativistic magnetohydrodynamics~\cite{Dumbser2011}, and finally to full
general relativity~\cite{Dumbser2017strongly,Dumbser2020GLM}.
According to this approach,
the solution $\q_h$ is expanded into a local space-time basis
\begin{equation}
	\label{eqn.spacetime}
	\q_h(\x,t) = \sum \limits_l \theta_l(\x,t)
	\hat{\q}_{i,l} := \theta_l(\x,t) \hat{\q}_{i,l}\,,
\end{equation}
with the multi-index $l=(l_0,l_1,l_2,l_3)$ and where the space-time basis
functions are constructed in quite a similar way as done for 
$\Phi_l(\x)$ in Eq.~\eqref{eqn.ansatz.uh}, but with a time dependence included,
i.e.
$\theta_l(\x,t) = \varphi_{l_0}(\tau)
\varphi_{l_1}(\xi) \varphi_{l_2}(\eta) \varphi_{l_3}(\zeta) $, where  
$\tau \in [0,1]$ such that $t = t^n + \tau \Delta t$. 
We now multiply \eqref{eqn.pde.mat.preview}
by $\theta_l(\x,t)$ and we
integrate over the space–time control volume, to obtain
\begin{equation}
	\label{eqn.pde.st1}
	\int \limits_{t^n}^{t^{n+1}} \! \! \int \limits_{\Omega_i}
	\theta_k(\x,t) \frac{\partial \q_h}{\partial t} d\x\, d t
	+\int \limits_{t^n}^{t^{n+1}} \! \! \int \limits_{\Omega_i}
	\theta_k(\x,t) \left( \A(\q_h) \cdot \nabla \q_h  \right) d\x\, d t
	= \int \limits_{t^n}^{t^{n+1}} \! \! \int \limits_{\Omega_i}
	\theta_k(\x,t) \S(\q_h) d\x\, d t\,.
\end{equation}
Since the vector $\q_h$ remains local in space, 
the numerical integration of \eqref{eqn.pde.st1}
does not require the solution of any Riemann problem, nor it needs to
account for the jumps in $\q_h$ across the element interfaces, which are 
instead invoked in the global scheme \eqref{eqn.pde.nc.gw2}.
Hence, integration of \eqref{eqn.pde.st1} by parts in time provides
\begin{eqnarray}
	\label{eqn.pde.st2}
	\int \limits_{\Omega_i}   \theta_k(\x,t^{n+1}) \q_h(\x,t^{n+1}) d\x  -
	\int \limits_{\Omega_i}   \theta_k(\x,t^{n}) \u_h(\x,t^{n}) d\x
	- \int \limits_{t^n}^{t^{n+1}} \! \! \int \limits_{\Omega_i}
	\frac{\partial }{\partial t} \theta_k(\x,t) \q_h(\x,t)
	d\x\, d t = && \nonumber \\
	\int \limits_{t^n}^{t^{n+1}} \! \! \int \limits_{\Omega_i}  \theta_k(\x,t)\,
	\S(\q_h)  d\x\, d t
	- \int \limits_{t^n}^{t^{n+1}} \! \! \int \limits_{\Omega_i} \theta_k(\x,t)
	\left( \A(\q_h) \cdot \nabla \q_h  \right)  d\x\, d t\,. &&
\end{eqnarray}
Eq.~\eqref{eqn.pde.st2} forms a non-linear system to be solved in the unknown expansion coefficients 
 $\hat{\q}_{i,l}$ and it can be conveniently  
 tackled through a simple and fast converging fixed point iteration method, the convergence of which was proven in \cite{BCDGP20}.

\subsection{ADER-WENO finite volume subcell limiter}
\label{sec:limiter}
Even if the fields arising from the Einstein equations are linearly degenerate and therefore no shocks can form in the metric, strong discontinuities can still form, in particular when a
spacetime singularity is produced due to the merger of already existing black holes. If matter is also present,
the problem is then exacerbated, for instance during the gravitational collapse of a star,
 because strong gas shocks are easily produced. From the numerical viewpoint
this is a challenge for DG methods, which are linear in the sense of Godunov's theorem \cite{godunov}, and they are hence
exposed to the appearance of strong oscillations (Gibbs phenomenon), unless specific procedures are set forth.
The strategies that are usually adopted to cope with this problem can be roughly classified as
\begin{itemize}
	\item \emph{Filtering}. This is a rather natural approach to remove unphysical oscillations from the solution, and in the relativistic context it has been successfully adopted in \cite{Radice2011}.
	\item \emph{A priori slope and moment limiting}. This kind of approach is perhaps the most widely adopted among existing DG codes, proposed in quite different versions~\cite{cbs4,QiuShu1,Qiu_2004,balsara2007,Zhu_2008,Zhu_13,Luo_2007,Krivodonova_2007,Kidder2017,Deppe2022b}. In the special relativistic context we also mention the interesting family of positivity preserving and entropy stable limiters for DG schemes~\cite{Kailiang2017,Deepak2023}. 
	\item \emph{A posteriori subcell FV limiting}. In an attempt of preserving as much as possible the accuracy of the DG schemes, the detection of troubled numerical cells can be performed after updating the solution with an unlimited DG scheme. This line of research has been developed within our group~\cite{Dumbser2014,Zanotti2015c,Zanotti2015d} and in \cite{Sonntag2} and it has been shown to produce quite reliable results even in full general relativity \cite{Dumbser2017strongly,Deppe2022}. 
\end{itemize}
Following therefore the third of the above approaches, we use
a third order ADER-WENO finite volume scheme at the subcell level as a remedy to the Gibbs phenomenon.
The whole scheme was already in place for our previous investigations within the CCZ4 formulation of the
Einstein equations~\cite{Dumbser2017strongly}. Hence, we just provide here an essential description of the idea, addressing the reader to the above-mentioned literature for further details. 
\begin{enumerate}
	\item We first apply the pure (unlimited) DG scheme of Eq.~\eqref{eqn.pde.nc.gw2} to evolve the solution from time $t^n$ to $t^{n+1}$. This produces a candidate solution $\u_h^*(\x,t^{n+1})$ inside each cell $\Omega_i$. 
	\item The polynomial representing the candidate solution is then checked to verify that 
	it  lies between the minimum and the maximum 
	of all the polynomials representing the solution at the previous time step and belonging to the set ${\cal{V}}_i$. If this is not the case, 
	the cell is flagged as \emph{troubled}. 
	The set ${\cal{V}}_i$ contains the cell $\Omega_i$ and all its neighbor cells sharing a node with $\Omega_i$. When matter is present, this numerical admissibility criterion can be augmented by a physical admissibility one, requiring, for instance, that the velocity of the gas does not become superluminal.
	\item All the troubled cells are then covered by a local sub-grid
formed by $N_s= 2N+1$ cells per space dimension, denoted as $\Omega_{i,s}$, such that 
$\Omega_i = \bigcup_s \Omega_{i,s}$. Each sub-cell $\Omega_{i,s}$ is endowed with a cell average
using the un-corrupted DG polynomial at the \textit{previous time level} $t^n$, i.e.
\begin{equation}
	\label{eqn.subcellaverage}
	\bar{\u}^n_{i,s} = \frac{1}{|\Omega_{i,s}|} \int \limits_{\Omega_{i,s}}
	\u(\x,t^n) d\x\,.
\end{equation}
\item  The new data provided by Eq.~\eqref{eqn.subcellaverage} are now evolved in time using a finite volume scheme on the sub-grid. In this work we have essentially used a pragmatic 
piecewise quadratic reconstruction, leading to a nominally third order accurate scheme on the subgrid. However, 
we stress that WENO schemes of arbitrary order can be applied as well.
\item 
As a final step, the new solution at time $t^{n+1}$ over the sub-grid is projected  back to the main grid, via a classical constrained least squares reconstruction~\cite{Titarev2002,Titarev2004}, to find
a new limited DG polynomial $\u'_h(\x,t^{n+1})$ as
\begin{equation}
	\frac{1}{|\Omega_{i,s}|} \int \limits_{\Omega_{i,s}} \u'_h(\x,t^{n+1}) d\x
	= \bar{\u}^{n+1}_{i,s} \qquad \forall \Omega_{i,s} \in \Omega_i,
	\qquad \textnormal{ and } \qquad \int \limits_{\Omega_{i}} \u'_h(\x,t^{n+1})
	d\x = \sum \limits_{\Omega_{i,s} \in \Omega_i} |\Omega_{i,s}| \bar{\u}^{n+1}
	{i,s}\,.
\end{equation}
\end{enumerate}
We note, for instance,  that the successful simulation of the inspiralling merger of two black holes described in Sect.\ref{sec:BBHs} was only possible thanks to the activation of the limiter just described. To simplify the computational setup, it is of course also possible to activate the subcell FV limiter always in a certain area of the computational domain, without making use of the dynamic data-dependent detection criteria outlined above.  

\subsection{Well-balancing and adaptive mesh refinement}
Our numerical scheme is enriched by two extra features, that make it particularly appealing for NR applications.

\subsubsection{Well-balancing}
\label{sec:wb}
A numerical method is called \emph{well-balanced} if 
it is designed to solve exactly, or at least with machine precision  accuracy, certain relevant stationary solutions of the governing PDE system to which it is applied. Such a property does not come
for free, and, if not present, an equilibrium solution $\u_e$, which satisfies  $\partial_t\u_e=0$ on a piece of paper, will not always remain stationary in the code.
Well--balanced numerical schemes have been proposed using a lot of different  techniques, 
sometimes limited to certain PDEs systems \cite{Bermudez1994,leveque1998balancing,gosse2001well,audusse2004fast,BottaKlein,Castro2017Book,castro2020well,Gaburro2021WBGR1D,xu2024}.
Since the Einstein equations are so complex, as proposed by \cite{DumbserZanottiGaburroPeshkov2023,PareschiRey,berberich2021high,Balsara2024c}, we have chosen the rather pragmatic approach of
subtracting at each timestep the \textit{a priori} known equilibrium solution all together from the governing PDE. This means that the following augmented system must be solved
\begin{align}
	{{\partial }_{t}}\u+{{\partial }_{x}}\mathbf{F}\left( \u \right)-{{\partial }_{x}}\mathbf{F}\left( \u_e \right)
	+\A\left( \u \right){{\partial }_{x}}\u
	-\A\left( \u_e \right){{\partial }_{x}}\u_e=\S(\u)-\S(\u_e); \qquad {{\partial }_{t}}\u_e = 0. 
	\label{eq:afdnc_pde}
\end{align}
As a result, the discretization errors inherent to the numerical scheme are polished 
and stationarity can be preserved \textit{exactly} at the discrete level for any \textit{a priori} known equilibrium $\u_e$. We refer to \cite{DumbserZanottiGaburroPeshkov2023} for  
further details on the way how this can be practically implemented in an existing general purpose solver for hyperbolic PDE. Here we just recall that the 
vector of evolved quantities must be necessarily doubled, becoming $\tilde{\u}=[\u,\u_e]$. In fact, even if the equilibrium sector $\u_e$ is effectively excluded from the time evolution, it must 
nevertheless be included in  all the spatial discretization procedures of the numerical scheme.	
Hence, our well--balanced DG scheme is essentially a factor 2 slower than the non well-balanced one.	
	
\subsubsection{Adaptive Mesh Refinement}
\label{sec:amr}
	
Following our previous works documented in 
\cite{AMR3DCL,Zanotti2015c,Zanotti2015,ExaHype2020}, our DG scheme has been combined with Adaptive Mesh Refinement (AMR) and time-accurate local timestepping (LTS) according to a ''cell-by-cell'' approach~\cite{Khokhlov1998}. This means that  every cell $\Omega_i$ 
is individually refined, avoiding grid patches. 
A suitable refinement criterion involves the computation of the scalar 
\begin{equation} 
	\label{eq:chi_m}
	\chi_\emm (\Phi) =  \sqrt{ \frac{\sum_{k,l}{\left(\left. \partial ^2 \Phi \middle/ \partial x_k \partial x_l \right. \right)^2}}{ \sum_{k,l}{ \left[ \left. \Big( \left| \left. \partial \Phi \middle/ \partial x_k\right.\right|_{i+1} + \left| \left.\partial \Phi \middle/ \partial x_k \right. \right|_i \Big) \middle/ \Delta x_l \right. + \epsilon \left|\frac{\partial^2 }{\partial x_k\partial x_l} \right|\left| \Phi \right|    \right]^2}}  }\,.
\end{equation}
where the \emph{indicator function} $\Phi$ is usually set equal to the conformal factor $\psi$. 
When a cell $\Omega_i$ is flagged for refinement according to $\chi_i > \chi_\text{ref}$, 
it generates $\mathfrak{r}^3$ \emph{children cells}, with $\mathfrak{r}=3$. We also set
a maximum level of refinement $\ell_\text{max}$, such that $0\le\ell\le\ell_\text{max}$.
At any level of refinement $\ell$ each cell has a specific status, namely it can be 
either an \textit{active cell}, updated through the standard ADER-DG scheme;
or a \textit{virtual child cell}, updated using 
$L_2$ projection of the high order polynomial of the mother cell at the $(\ell-1)$-th level; or, finally,
a \textit{virtual mother cell}, updated by averaging over all children cells from higher refinement levels.  
If the subcell limiter of Sect.~\ref{sec:limiter} is also active, the whole AMR strategy becomes more involved and the following caveats come in place
\begin{itemize}
	\item Virtual children cells inherit the limiter status of their active mother cell.
	\item If even a single active child requires the limiter, then the (virtual) mother will also call for the limiter.
	\item Cells which need the subcell limiter cannot be recoarsened.
\end{itemize}
Further details about the typical  AMR operations of  projection and averaging at the sub-grid level involving different levels of refinement can be found in \cite{AMR3DCL,Zanotti2015c,ExaHype2020}. Of course, our AMR facility can also be used in a fixed refinement fashion if physical and computational conditions suggest to do so.

\section{Numerical tests}
\label{sec:tests}

In this Section we show the numerical results obtained for several classical test problems of numerical general relativity, obtained with our new high order ADER-DG schemes applied to the first order BSSNOK formulation of the coupled Einstein-Euler system. The classical standard tests for vacuum spacetimes are taken from \cite{Alcubierre2004,Babiuc_2008}.

\subsection{Linearized gravitational wave }
\label{sec:linearized-grav-wave}

We start with the classical linear gravitational wave test case proposed in \cite{Alcubierre2004}, which consists in a very small perturbation of flat Minkowski spacetime. The metric is given by 
\begin{equation} 
	\label{eqn.lw.metric}
	d s^2 = - d t^2 + d x^2 + (1+b)\,d y^2 + (1-b)\,d z^2, \quad \textnormal{with} \quad b = \epsilon \sin \left( 2 \pi (x-t) \right), 
\end{equation} 
hence all metric terms can be directly deduced from~\eqref{eqn.lw.metric}, such as the extrinsic curvature, which is 
$K_{ij} = \partial_t  \gamma_{ij} /(2\alpha)$. For small values of $\epsilon$ the overall dynamics can be considered to be linear and thus the terms depending on $\epsilon^2$ can be neglected. Here, we choose $\epsilon = 10^{-8}$. Furthermore, we use a harmonic lapse and the \emph{gamma--driver} can be deactivated, i.e. $s=0$. 
\begin{figure}[!htbp]
	\includegraphics[width=0.45\textwidth]{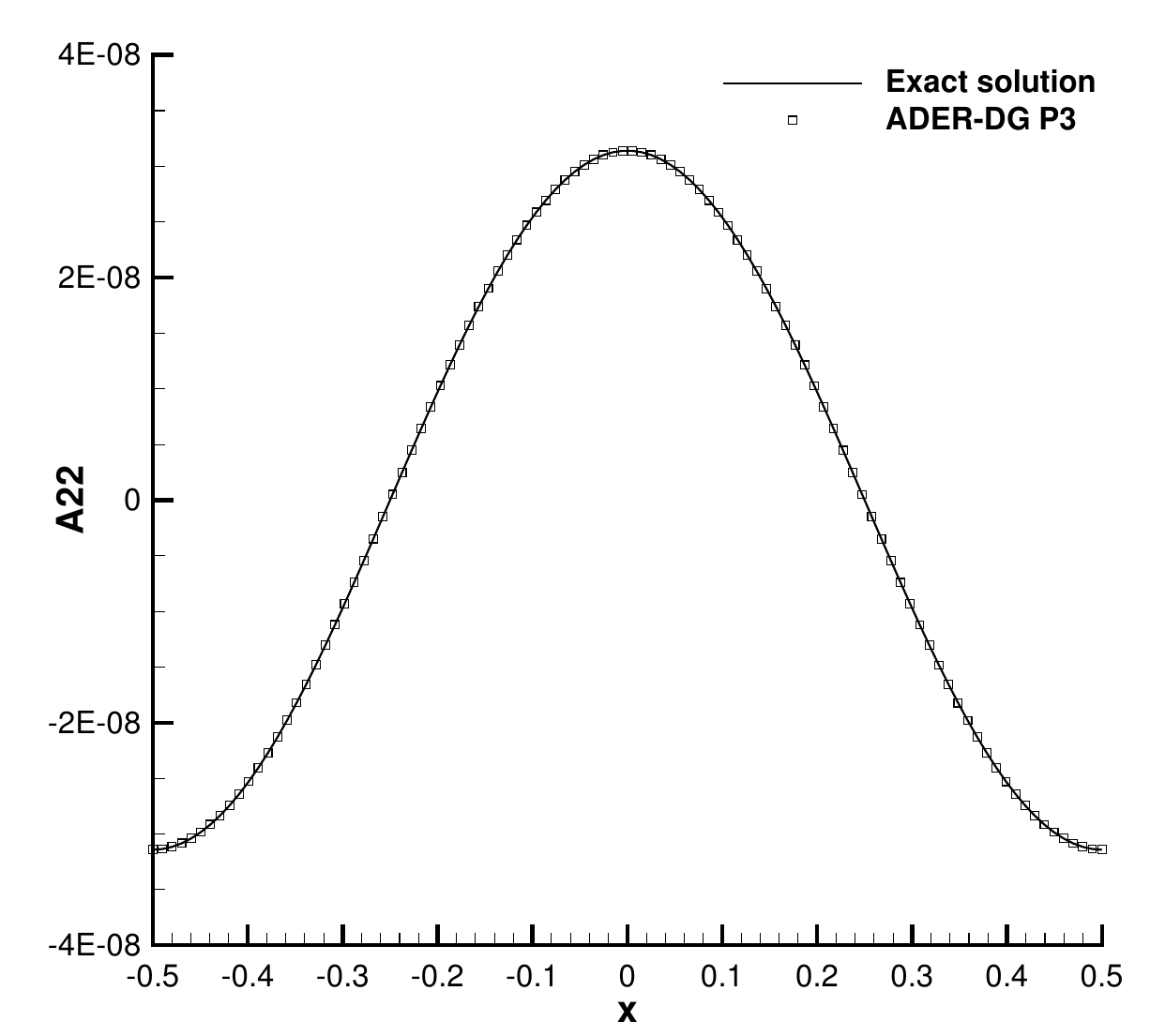}
	\includegraphics[width=0.45\textwidth]{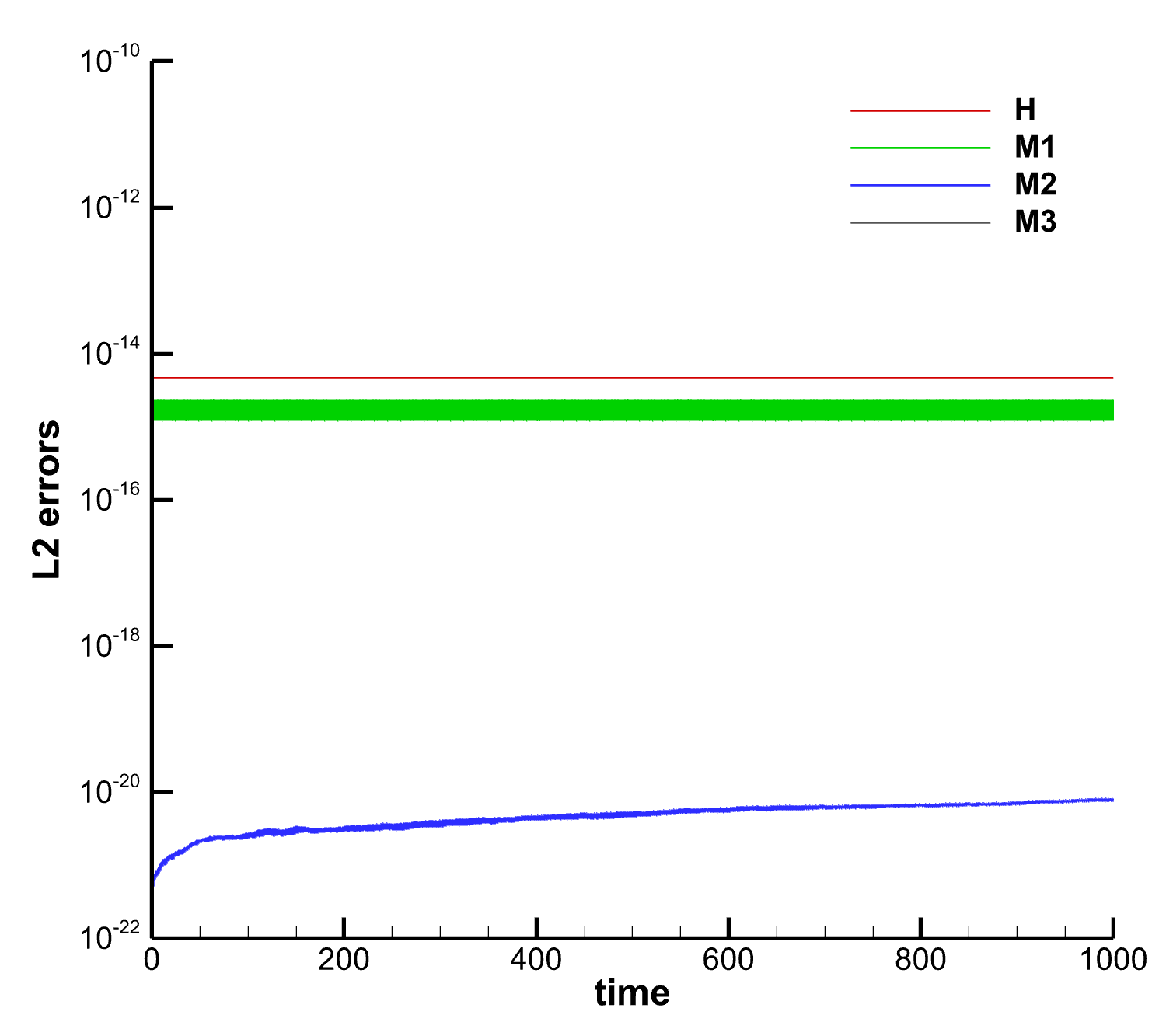}
		\caption{Linearized gravitational wave test case solved with a fourth order ADER-DG scheme up to $t=1000$. 
		Left panel: $\tilde A_{22}$ component of the conformal extrinsic curvature at the final time $t=1000$, compared to the exact solution. Right panel: time evolution of the Einstein constraints.}
		\label{fig.linearized-grav-wave}
\end{figure}
There is no matter present in this test problem. The computational domain $\Omega = [-0.5,0.5]\times[-0.05,0.05]$ is discretized using  $32 \times 4$ DG elements of polynomial approximation degree $N=3$. We employ periodic boundary conditions in all  coordinate directions.

In Figure~\ref{fig.linearized-grav-wave} we depict the computational results of the ADER-DG scheme up to a final time of $t=1000$. In the left panel the profiles of $\tilde A_{22}$ are compared to the exact solution at the final simulation time of $t=1000$.  
One can observe an excellent agreement between the numerical results obtained with the fourth order DG scheme and the exact solution. 
In the right panel of the same figure we show the time evolution of the Einstein constraints, which remain close to machine precision during the entire simulation.

\subsection{The robust stability test}
\label{sec:stability-test}

\begin{figure}[!htbp]
	\begin{center}
		\begin{tabular}{cc}
			\includegraphics[trim=10 10 10 10,clip,width=0.45\textwidth]{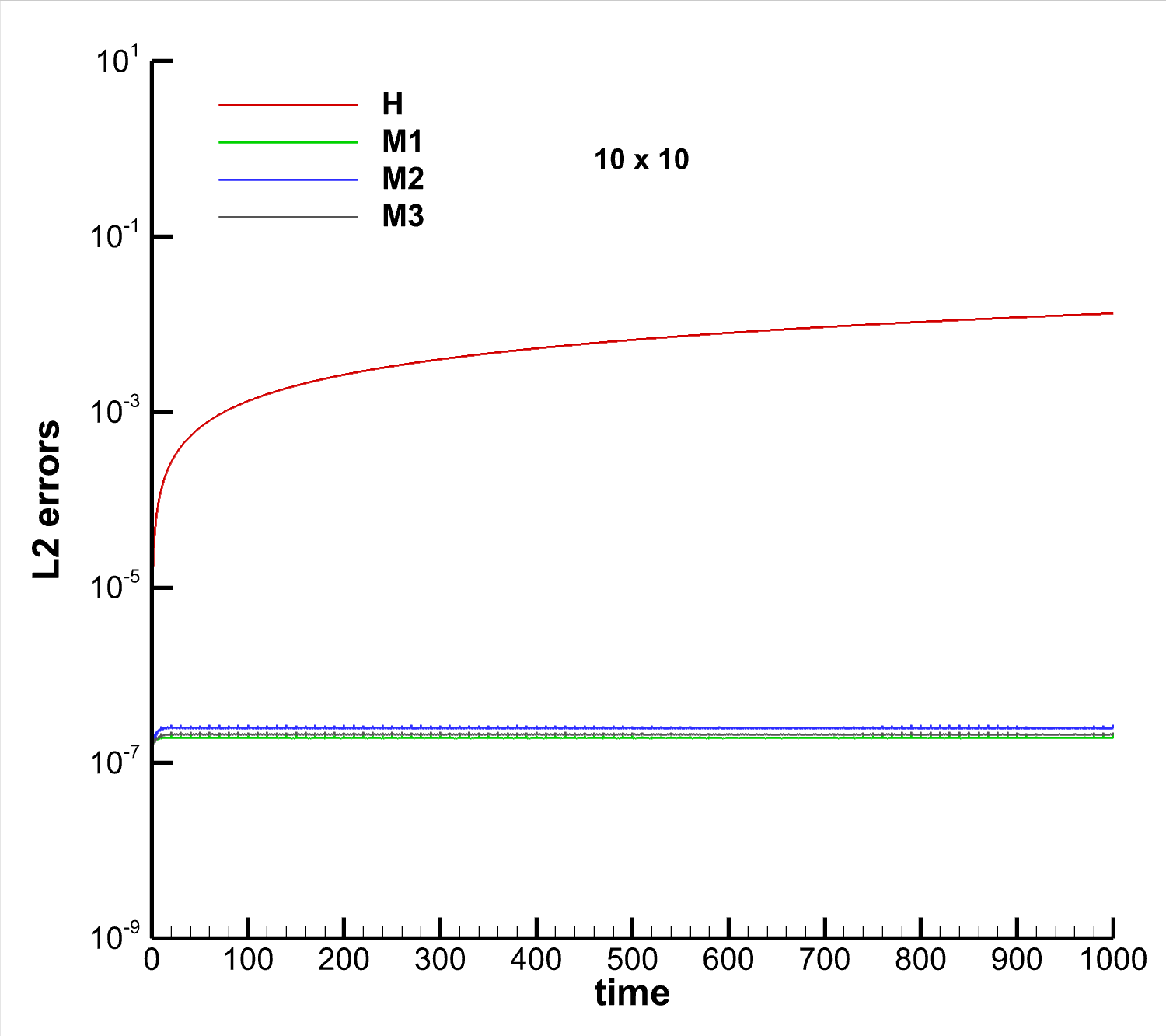} &
			\includegraphics[trim=10 10 10 10,clip,width=0.45\textwidth]{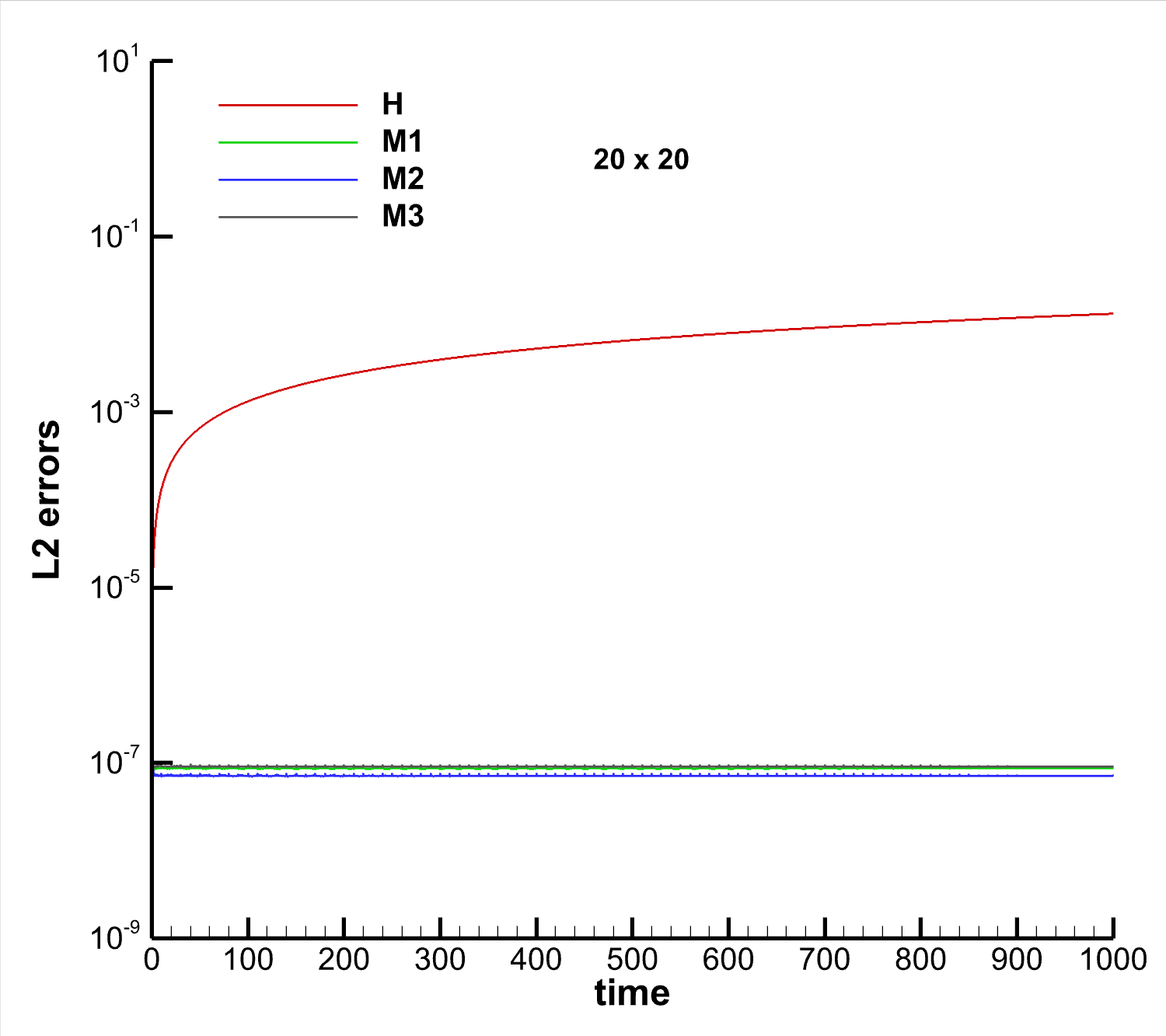} \\
			\includegraphics[trim=10 10 10 10,clip,width=0.45\textwidth]{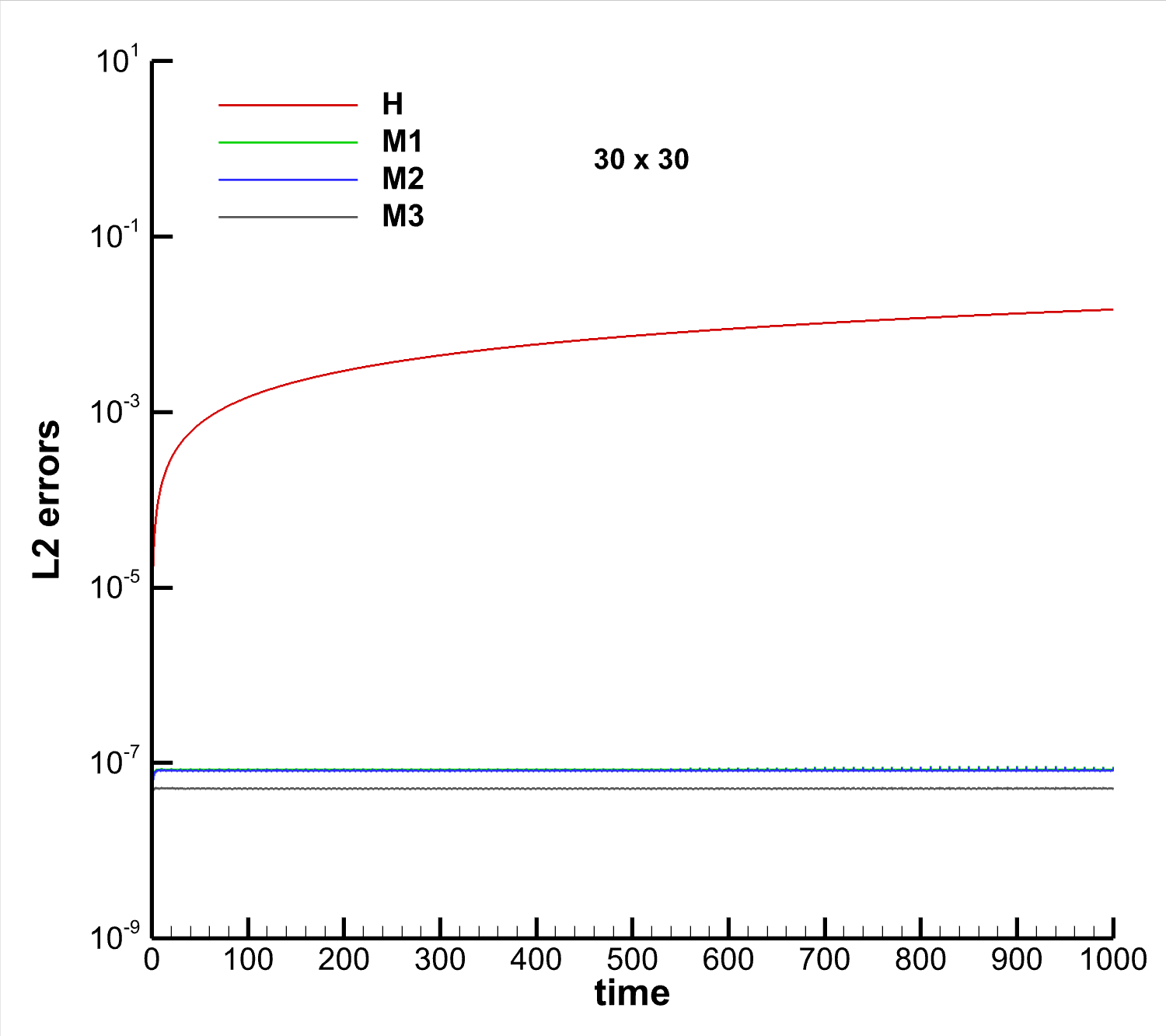} &
			\includegraphics[trim=10 10 10 10,clip,width=0.45\textwidth]{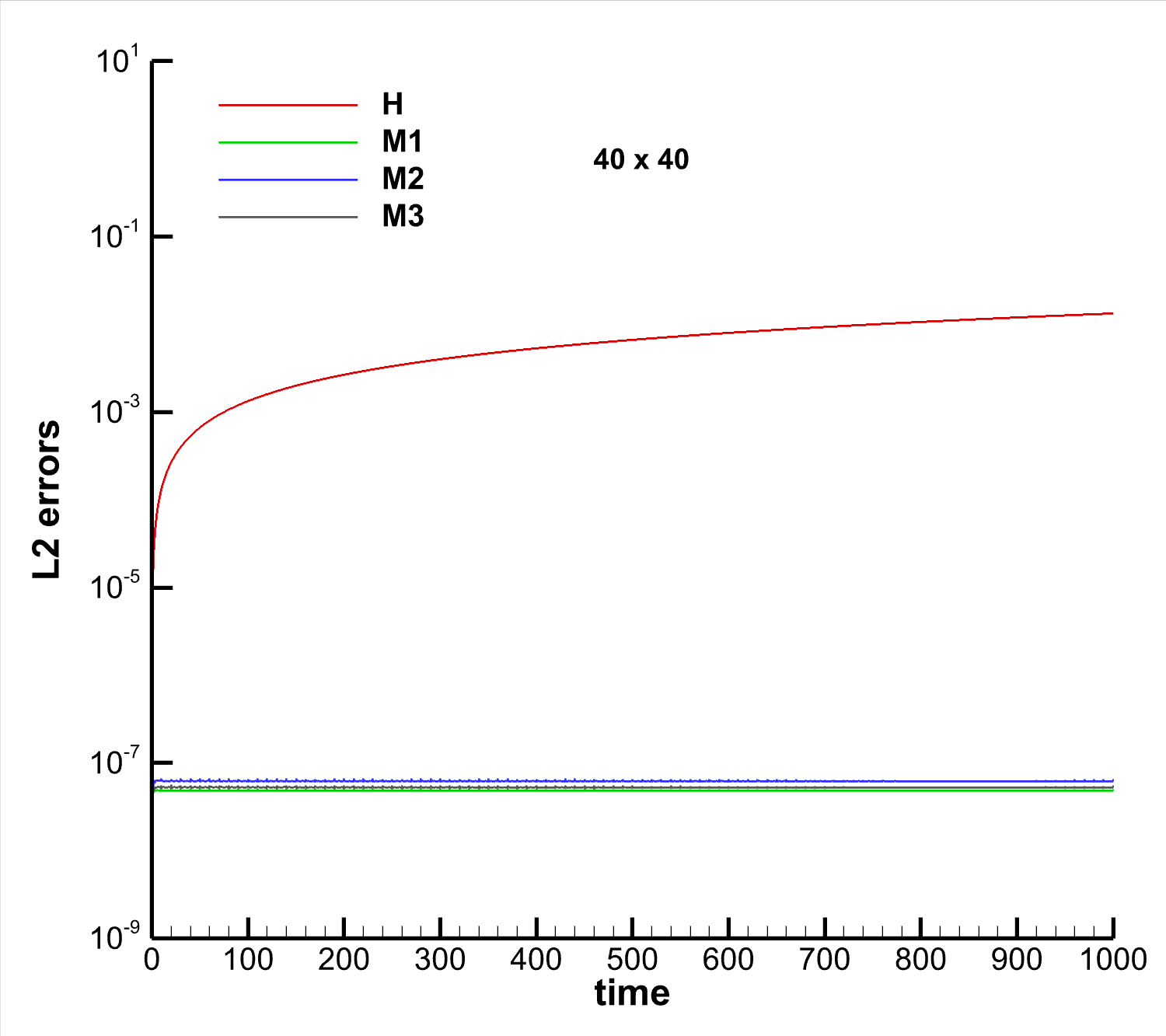}  
		\end{tabular}
		\caption{Robust stability test for the FO-BSSNOK formulation with {\emph {gamma--driver}} shift condition and $1+\log$ slicing. A random initial perturbation of amplitude
			$10^{-8}/\varrho^2$ in all quantities has been applied on a sequence of successively
			refined meshes on the unit square in 2D. 
			The simulation has been carried out with a fourth order ADER-DG scheme ($N=3$). 
			Top left: $10\times10$ elements ($\varrho=1$). Top right: $20\times20$ elements ($\varrho=2$). 
			Bottom left: $30\times30$ elements ($\varrho=3$). Bottom right: $40\times40$ elements ($\varrho=4$). }
		\label{fig.robstab}
	\end{center}
\end{figure}

The {\emph {robust stability test}} was proposed in \cite{Alcubierre2004} and is typically used as an empirical hyperbolicity 
check of the chosen formulation of the field equations. It can discover potentially unstable and therefore exponentially growing modes in the numerical solution. We solve this test on the 2D domain $\Omega = [-0.5;0.5]\times[-0.5;0.5]$, 
perturbing a flat Minkowski spacetime with a uniformly distributed random perturbation with amplitude $\pm 10^{-8}/\varrho^2$ in all variables. 
As usual in this test, the parameter $\varrho$ is used to scale the perturbation with the mesh resolution, see \cite{Alcubierre2004}, where the computational mesh is composed of $10 \varrho \times 10 \varrho$ finite elements. 
The {\emph {gamma--driver}} shift condition is activated with $\mu=0$ and a $1+\log$ slicing is used for the lapse, i.e. $g=2/\alpha$.  
Note that our initial perturbation is two orders of magnitude larger than the one suggested in \cite{Alcubierre2004}.

Fig.~\ref{fig.robstab} reports the time evolution of the Einstein constraints obtained with a fourth order ADER-DG scheme ($N=3$) on a sequence of successively refined meshes with $\varrho \in \left\{ 1, 2, 3, 4 \right\}$. We observe at most linear growth in the constraint errors, which indicates a stable long-time evolution, as expected, since our FO-BSSNOK formulation has been proven to be strongly hyperbolic for general spacetimes, see \cite{CWENOBSSNOK}. 
The linear growth of the Hamiltonian constraint reflects the fact that inside the BSSNOK formulation no special care is taken of the Hamiltonian constraint. To control this constraint explicitly, another formulation of the PDE would be necessary, such as the Z4 or the CCZ4 system, where the errors in the Hamiltonian constraint are explicitly propagated away via a dedicated cleaning scalar $\Theta$.  For high order ADER-DG results obtained for this testcase using the Z4 and the CCZ4 formulation, see \cite{DumbserZanottiGaburroPeshkov2023,Dumbser2017strongly}. 
%

\subsection{The gauge wave}
\label{sec:gauge-wave}

\begin{figure}[!htbp]
	\includegraphics[trim=10 10 10 10,clip,width=0.45\textwidth]{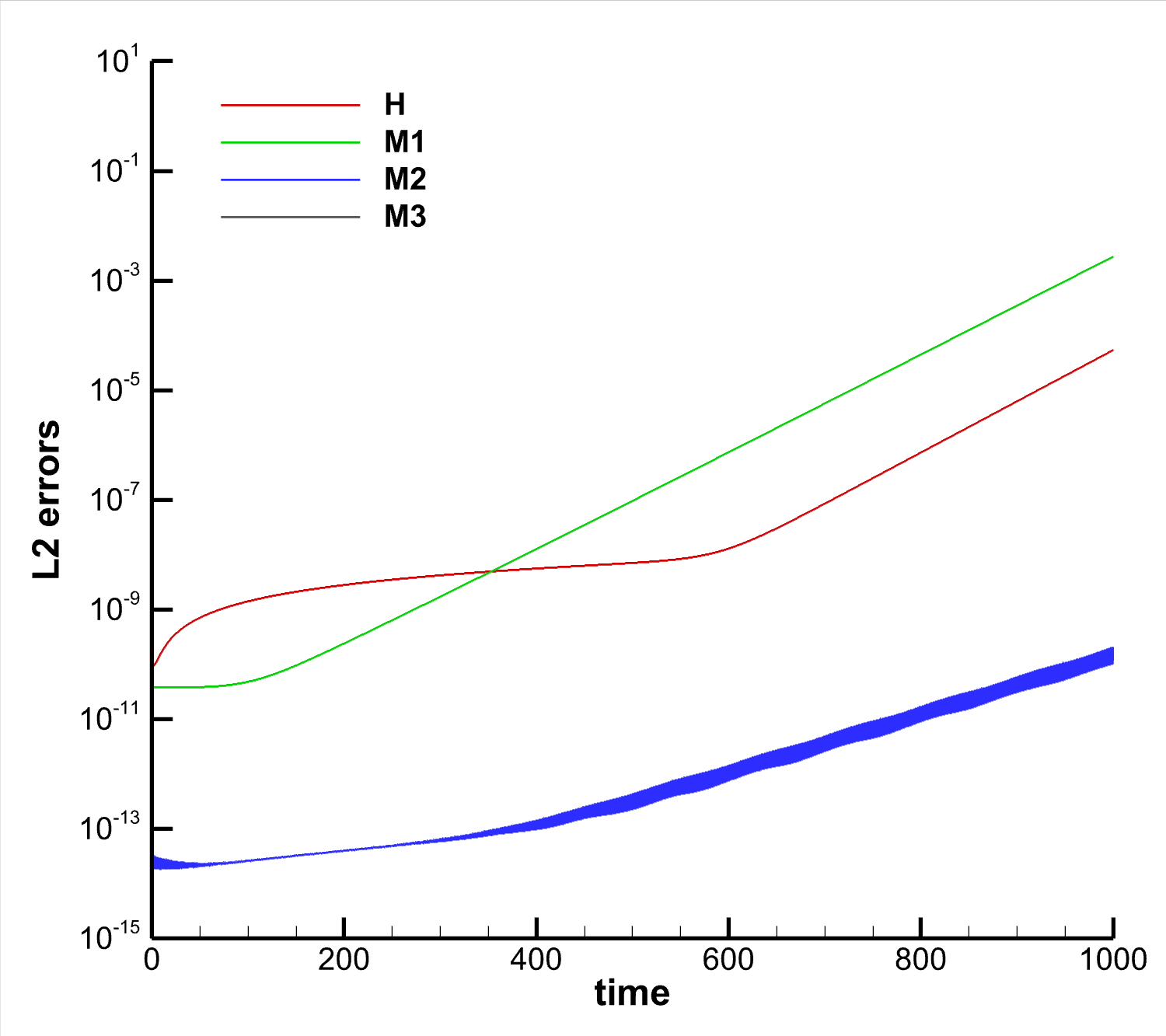}
	\includegraphics[width=0.45\textwidth]{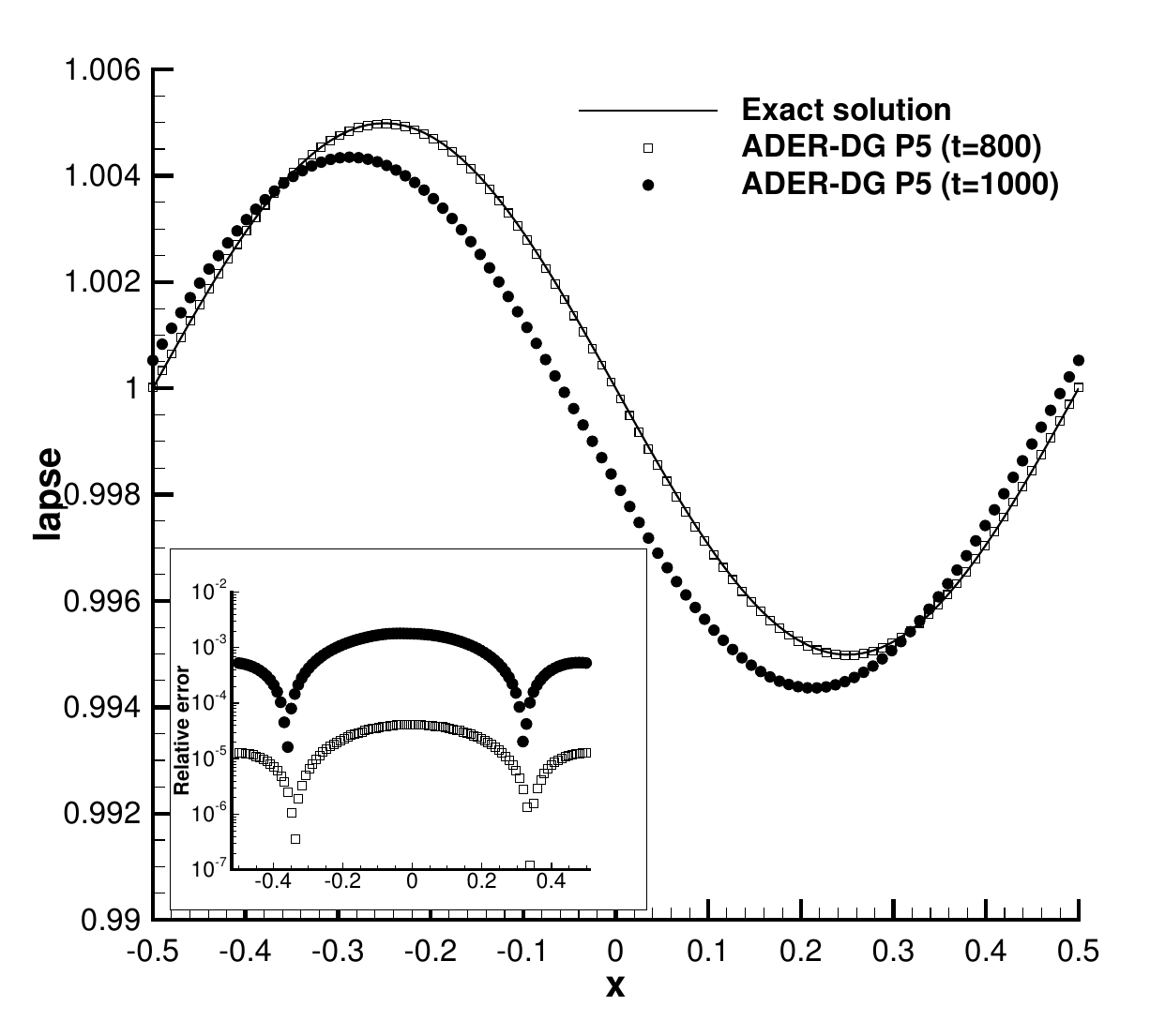}
	\caption{Solution of the gauge wave test with $A=0.01$ using a sixth order ADER-DG scheme ($N=5$). 
		Left panel: Temporal evolution of the Einstein constraints. One observes the usual exponential growth of the constraints
		that is typical for the gauge wave test when applied to BSSNOK formulations of the Einstein equations. 
		Right panel: profiles of the lapse $\alpha$ at $t=800$ and at $t=1000$ compared to the exact solution. 	  
	}
	\label{fig.gaugewavetest-smallA}
\end{figure}

The next classical test case, also taken from~\cite{Alcubierre2004}, is the so-called \textit{gauge wave test}. It is very well-known to be particularly difficult for BSSNOK formulations of the Einstein field equations
(see~\cite{Alic:2011a,Brown2012}).   
With our numerical experiments we can confirm that this is still true even for a high order DG discretiztion of 
our first order BSSNOK formulation. 
Instead, excellent results were reported in \cite{Dumbser2017strongly} for a first--order reformulation of the CCZ4 system
and, more recently, for a first order reformulation of the original Z4 system, see \cite{DumbserZanottiGaburroPeshkov2023,Balsara2024c}. 
In the gauge wave test the metric is obtained via a nonlinear coordinate transformation applied to the Minkowski spacetime. It reads 
\begin{equation}
	\label{eqn.gw.metric}
	d s^2 = - H(x,t) \, d t^2 + H(x,t) \, d x^2 + d y^2 + d z^2, \quad \text{where} \quad H(x,t) = 
	1-A\,\sin \left( 2\pi(x-t) \right).
\end{equation}
%
%
A \emph{harmonic gauge condition} is imposed for the lapse, while the shift is switched off, i.e. $s=0$. We impose periodic boundary conditions everywhere.

We first run the test with a small amplitude of $A=0.01$ on the rectangular computational domain $\Omega = [-0.5,0.5] \times [-0.5, 0.5]$ using a sixth order accurate ADER-DG scheme ($N=5$) on a computational grid composed of $N_x = 40$ and $N_y = 4$ equidistant finite elements. The final time is set to $t=1000$. As expected, the gauge wave test shows several pathologies when evolved 
over a long timescale using a BSSNOK formulation of the Einstein field equations, and these well-known problems do neither disappear with a first order reformulation of the BSSNOK system, nor with the use of high order discontinuous Galerkin schemes (see also \cite{Alcubierre2004,Tichy2009,Brown2012,CWENOBSSNOK}).
Instead, we can confirm the usual behavior of the BSSNOK formulation for the Gauge wave test case, which is an exponential growth of the Einstein constraints that starts after a finite time and which is shown in the left panel 
of Figure~\ref{fig.gaugewavetest-smallA}. In the right panel we show the numerical solution for the lapse $\alpha$ at 
$t=800$ and at the final time $t=1000$, when the discrepancy with respect to the exact solution is appreciable. In addition, the inset shows the relative error, again with respect to the exact solution. 

In spite of these known problems of the BSSNOK formulation of the field equations for the gauge wave test case, it is still possible to perform a numerical convergence analysis for this test in the case the final time is short enough. In this case 
the above-mentioned pathologies are not able to spoil the properties of the numerical scheme. In Table \ref{tab.conv1} we present the results of a numerical convergence analysis for a gauge wave with amplitude $A=0.1$ at a final time of $t=1$. One can clearly observe that the high order ADER-DG schemes properly achieve their designed order of accuracy of $N+1$. 

Limited to this test, we have also performed a careful comparison, in terms of CPU time and accuracy, with the results obtained through an entirely different numerical scheme, namely a central WENO finite difference scheme (FD-CWENO), that we have recently proposed and analyzed in \cite{CWENOBSSNOK}. We stress that the formulation of the Einstein equations is exactly the same in the two approaches. We also stress that the time integration philosophy in the two schemes is instead quite different, being Runge--Kutta for FD-CWENO and ADER for DG.

Under these circumstances, we have set up two different kinds of comparison in 3D, using a computational domain $\Omega = [-0.5,0.5]^3$, and adopting periodic boundary conditions along each spatial direction,
\begin{itemize}
\item In a first battery of runs, with results shown in Tab.~\ref{tab.cpu-nelements},
we keep the total number of \emph{elements constant}, while comparing the two numerical schemes at two different orders, i.e. orders 3 and 5.  
\item In a second battery of runs, with results shown in Tab.~\ref{tab.cpu-ndof},
we keep the total number of \emph{degrees of freedom constant}, while comparing the two numerical schemes at two different orders, i.e. orders 3 and 5.  
\end{itemize}
For each simulation we report the total Wall--Clock--Time (WCT), the time per element update (TEU),
and the time per degree of freedom update (TDU), each of them normalized to the FD-CWENO values, which are considered as reference for the computational cost. We recall that in the DG framework each element contains $(N+1)^3$ degrees of freedom, corresponding to the number of  Gaussian quadrature points, while in the finite different method each grid point is one degree of freedom on its own.
In addition, we also show the L2-error of one representative quantity, in this case the lapse. 
The conclusions that can be drawn from these numbers are twofold. From one side (see Tab.~\ref{tab.cpu-nelements}), if one uses the same number of elements in the two numerical schemes, ADER-DG pays a huge price in terms of WCT, which is typically hundreds of times larger than that of FD-CWENO. Even in these un-favorable conditions, however, the TDU of the ADER-DG scheme is smaller than that of FD-CWENO. 
This can be explained by recalling that ADER-DG requires only
6 computations of the non-conservative product per \textit{element} and \textit{time step}. On the contrary, FD-CWENO needs 6 computations of the non-conservative product per \textit{degree of freedom} and \textit{Runge--Kutta stage}. 
Moreover, if we look at the L2-errors of the lapse, we find that 
the gain in accuracy of DG vs FD is rather significant, beyond 4 orders of magnitude at order 5 (see the bottom right panel of Tab.~\ref{tab.cpu-nelements}).

From another side (see Tab.~\ref{tab.cpu-ndof}), if one uses the same number of degrees of freedom, ADER-DG becomes extremely competitive, both in terms of CPU time and in accuracy. As an example, with $216000$ degrees of freedom, which for ADER-DG P4 (order 5) implies a grid as coarse as $12^3$, compared to the $60^3$ grid of FD--CWENO, the ADER-DG simulations is only a factor 1.01 slower that FD-CWENO, but it has an L2-error which is still one order of magnitude smaller. 
We believe that in realistic applications to astrophysical contexts, a good compromise can be reached between more time consuming numerical simulations and a substantial gain in  accuracy.

\begin{table}
\caption{Numerical convergence results for the gauge wave
    with $A=0.1$ at a final time of $t=1$ for ADER-DG schemes with polynomial approximation degree
    $N \in \left\{ 2, 3, 4, 5 \right\}$.  The $L_2$
    errors and the corresponding observed convergence order are reported for
    the variables $K$, $\psi$ and $\alpha$.}
\begin{center}
\begin{tabular}{ccccccc}
\hline
  $N_x$ & ${L_2}$ error $K$ & $\mathcal{O}(K)$  & ${L_2}$ error $\psi$ & $\mathcal{O}(\psi)$  & ${L_2}$ error $\alpha$ & $\mathcal{O}(\alpha)$  \\
\hline
  \multicolumn{7}{c}{ADER-DG $N=2$}   \\
\hline
$ 16$  & 8.5750E-05 &     & 2.2393E-06 &      & 1.3295E-05 &      \\
$ 32$  & 1.0324E-05 & 3.1 & 2.8161E-07 & 3.0  & 1.6711E-06 & 3.0  \\
$ 64$  & 1.2824E-06 & 3.0 & 3.5259E-08 & 3.0  & 2.0917E-07 & 3.0  \\
$128$  & 1.5878E-07 & 3.0 & 4.4090E-09 & 3.0  & 2.6151E-08 & 3.0  \\
\hline
  \multicolumn{7}{c}{ADER-DG $N=3$}   \\
\hline
$ 16$  & 2.2685E-06 &     & 2.8875E-08 &      & 1.6887E-07 &      \\
$ 32$  & 1.4077E-07 & 4.0 & 1.7569E-09 & 4.0  & 1.0270E-08 & 4.0  \\
$ 64$  & 8.8059E-09 & 4.0 & 1.0896E-10 & 4.0  & 6.3690E-10 & 4.0  \\
$128$  & 5.4125E-10 & 4.0 & 6.7922E-12 & 4.0  & 3.9705E-11 & 4.0  \\
\hline
  \multicolumn{7}{c}{ADER-DG $N=4$}   \\
\hline
$16$   & 7.1006E-08  &     & 1.0155E-09 &      & 5.9480E-09 &      \\
$24$   & 9.2795E-09  & 5.0 & 1.3617E-10 & 5.0  & 7.9688E-10 & 5.0  \\
$32$   & 2.2321E-09  & 5.0 & 3.2513E-11 & 5.0  & 1.9020E-10 & 5.0  \\
$64$   & 6.9099E-11  & 5.0 & 1.0226E-12 & 5.0  & 5.9794E-12 & 5.0  \\
\hline
  \multicolumn{7}{c}{ADER-DG $N=5$}   \\
\hline
$ 8$  & 1.3232E-07 &      & 1.2354E-09 &       & 7.2339E-09 &       \\
$16$  & 2.0489E-09 & 6.0  & 1.7720E-11 &  6.1  & 1.0343E-10 &  6.1  \\
$24$  & 1.8116E-10 & 6.0  & 1.5182E-12 &  6.1  & 8.8570E-12 &  6.1  \\
$32$  & 3.2365E-11 & 6.0  & 2.7118E-13 &  6.0  & 1.5821E-12 &  6.0  \\
\hline
\end{tabular}
\end{center}
\label{tab.conv1}
\end{table}

\begin{table}
	\caption{Comparison among FD-CWENO and ADER-DG schemes for the gauge wave when the number of total elements is kept constant. Note that $\rm{TDU}=\rm{TEU}/(N+1)^3$.}
	\begin{center}
		\begin{tabular}{c|cc|cc}
			\hline
			\hline
			& Elements = $15^3=3375$  &  & Elements = $20^3=8000$\\
			\hline
			\bf{WCT (normalized)} & FD-CWENO & ADER-DG  & FD-CWENO  &  ADER-DG \\
			\hline
			Order 3  & 1 & 64.25      & 1 & 81.81  \\
			Order 5  & 1 & 1243.47   & 1 &  723.32\\
			\hline
			\bf{TEU (normalized)} & FD-CWENO & ADER-DG & FD-CWENO & ADER-DG \\
			\hline
			Order 3  & 1 & 10.95       & 1 & 14.02   \\
			Order 5  & 1 & 87.55       & 1 &  50.29\\
			\hline
			\bf{TDU (normalized)} & FD-CWENO & ADER-DG & FD-CWENO&  ADER-DG \\
			\hline
			Order 3  & 1 &  0.41       &  1& 0.52 \\
			Order 5  & 1 & 0.70        &  1& 0.40 \\
			\hline
			\bf{L2-error (lapse)} & FD-CWENO & ADER-DG &FD-CWENO &  ADER-DG \\
			\hline
			Order 3  & 3.85E-03 & 1.42E-05  & 1.64E-03 & 6.05E-06  \\
			Order 5  & 9.44E-05 & 7.01E-09  & 2.43E-05 & 1.70E-09\\			
			\hline
			\hline
		\end{tabular}
	\end{center}
	\label{tab.cpu-nelements}
\end{table}

\begin{table}
	\caption{Comparison among FD-CWENO and ADER-DG schemes for the gauge wave when the number of total degrees of freedom is kept constant. Note that $\rm{TDU}=\rm{TEU}/(N+1)^3$.}
	\begin{center}
		\begin{tabular}{c|cc|cc}
			\hline
			\hline
			& DoFs = 91125  &  & DoFs = 216000 \\
			\hline
			\bf{WCT (normalized)} & FD-CWENO & ADER-DG  & FD-CWENO  &  ADER-DG \\
			\hline
			Order 3  & 1 & 0.94      &  1   & 0.87  \\
			Order 5  & 1 & 1.52   &  1   & 1.09\\
			\hline
			\bf{TEU (normalized)} & FD-CWENO & ADER-DG & FD-CWENO &ADER-DG  \\
			\hline
			Order 3  & 1 & 12.91       & 1 & 11.89  \\
			Order 5  & 1 & 65.29    & 1 & 46.50\\
			\hline
			\bf{TDU (normalized)} & FD-CWENO & ADER-DG & FD-CWENO &  ADER-DG  \\
			\hline
			Order 3  & 1 & 0.47     & 1 & 0.44 \\
			Order 5  & 1 & 0.52     & 1 & 0.37\\
			\hline
			\bf{L2-error (lapse)} & FD-CWENO & ADER-DG &FD-CWENO &  ADER-DG \\
			\hline
			Order 3  & 1.15E-04 & 1.42E-05     &4.97E-05 & 6.05E-06 \\
			Order 5  & 4.53E-07 & 8.40E-08     &1.08E-07 & 2.09E-08\\			
			\hline
			\hline
		\end{tabular}
	\end{center}
	\label{tab.cpu-ndof}
\end{table}

\subsection{Special relativistic Riemann Problems in the Cowling approximation }
\label{sec:SR}

\begin{figure}
{\includegraphics[angle=0,width=5.5cm,height=5.0cm]{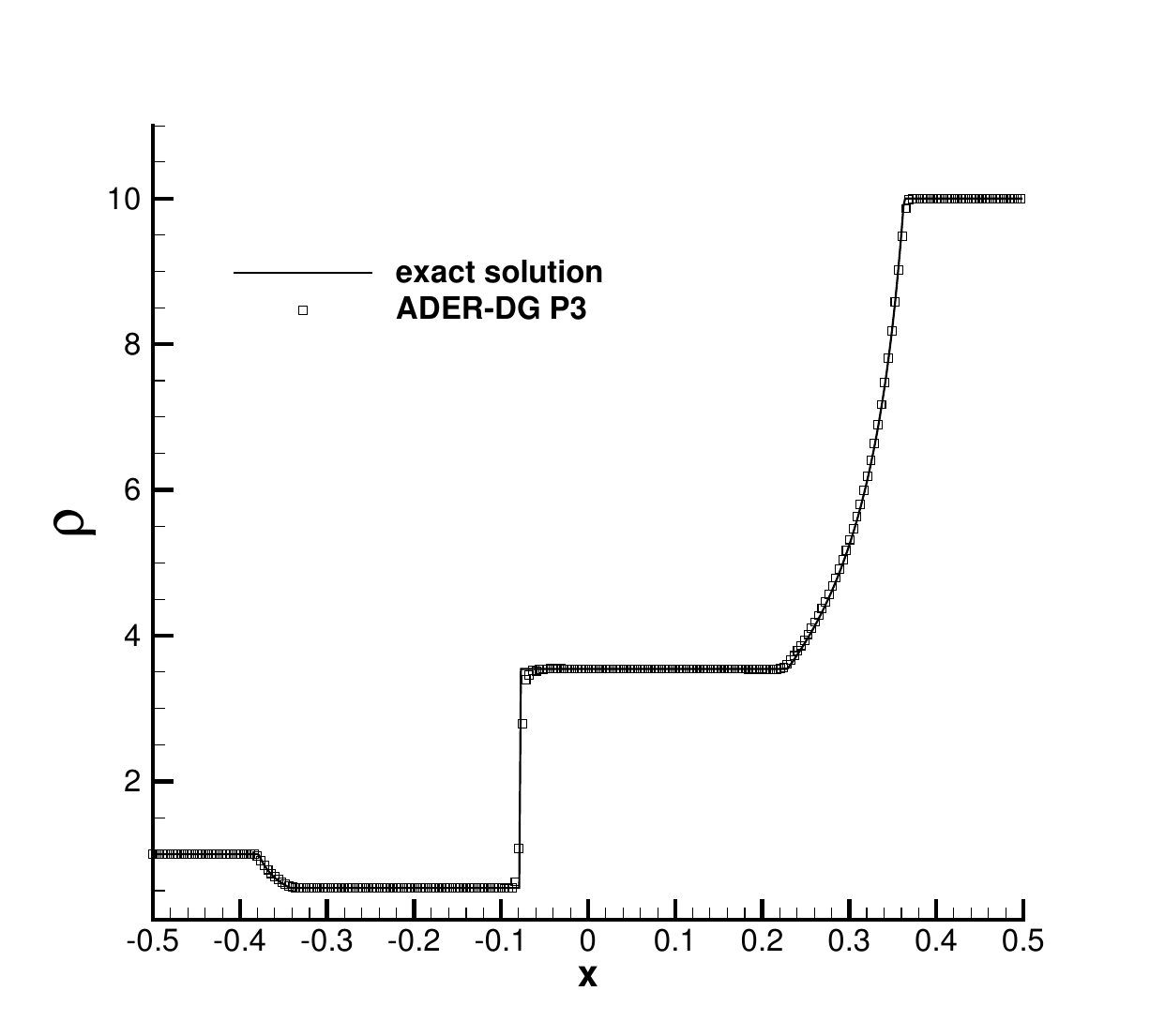}}
{\includegraphics[angle=0,width=5.5cm,height=5.0cm]{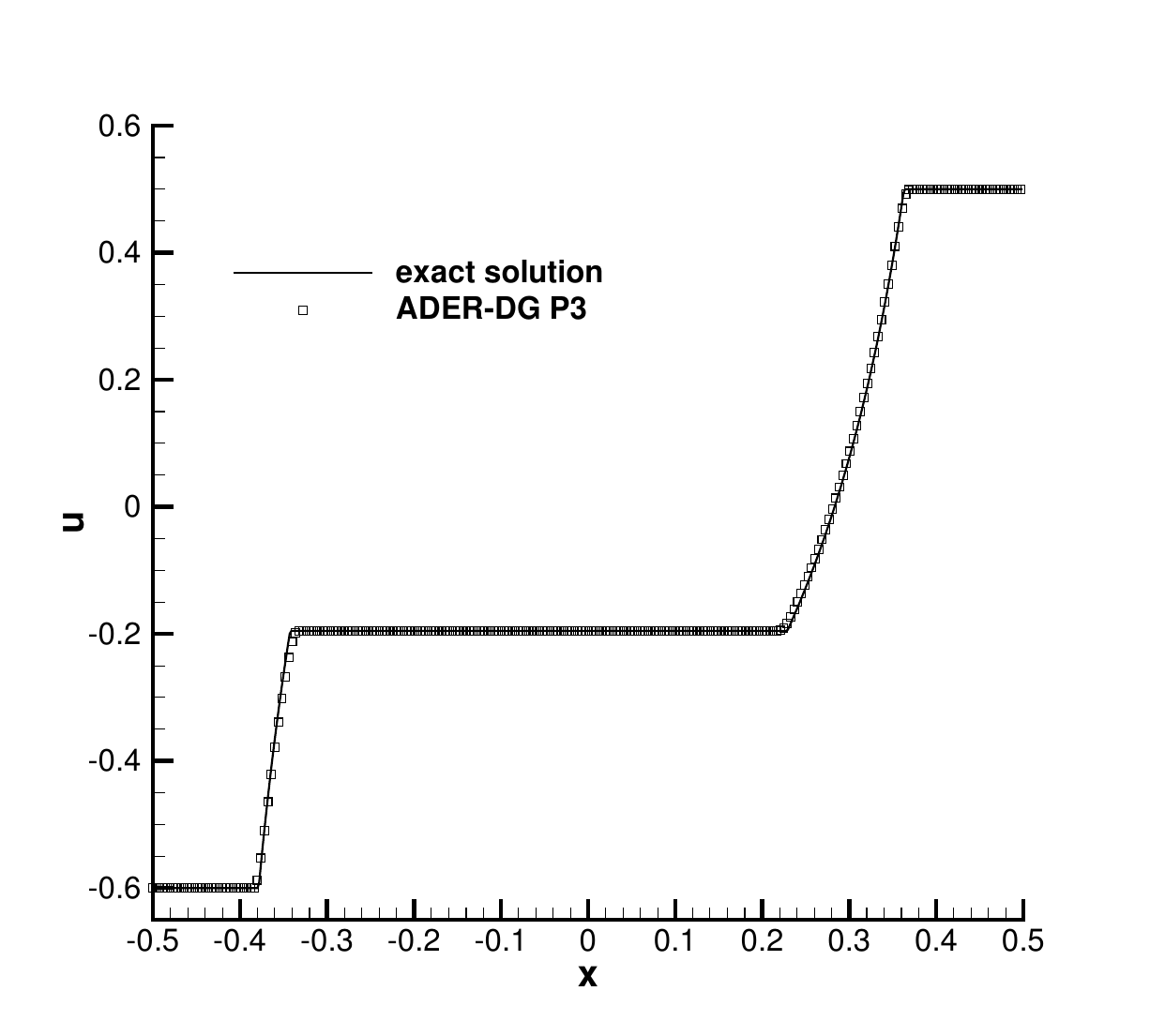}}
{\includegraphics[angle=0,width=5.5cm,height=5.0cm]{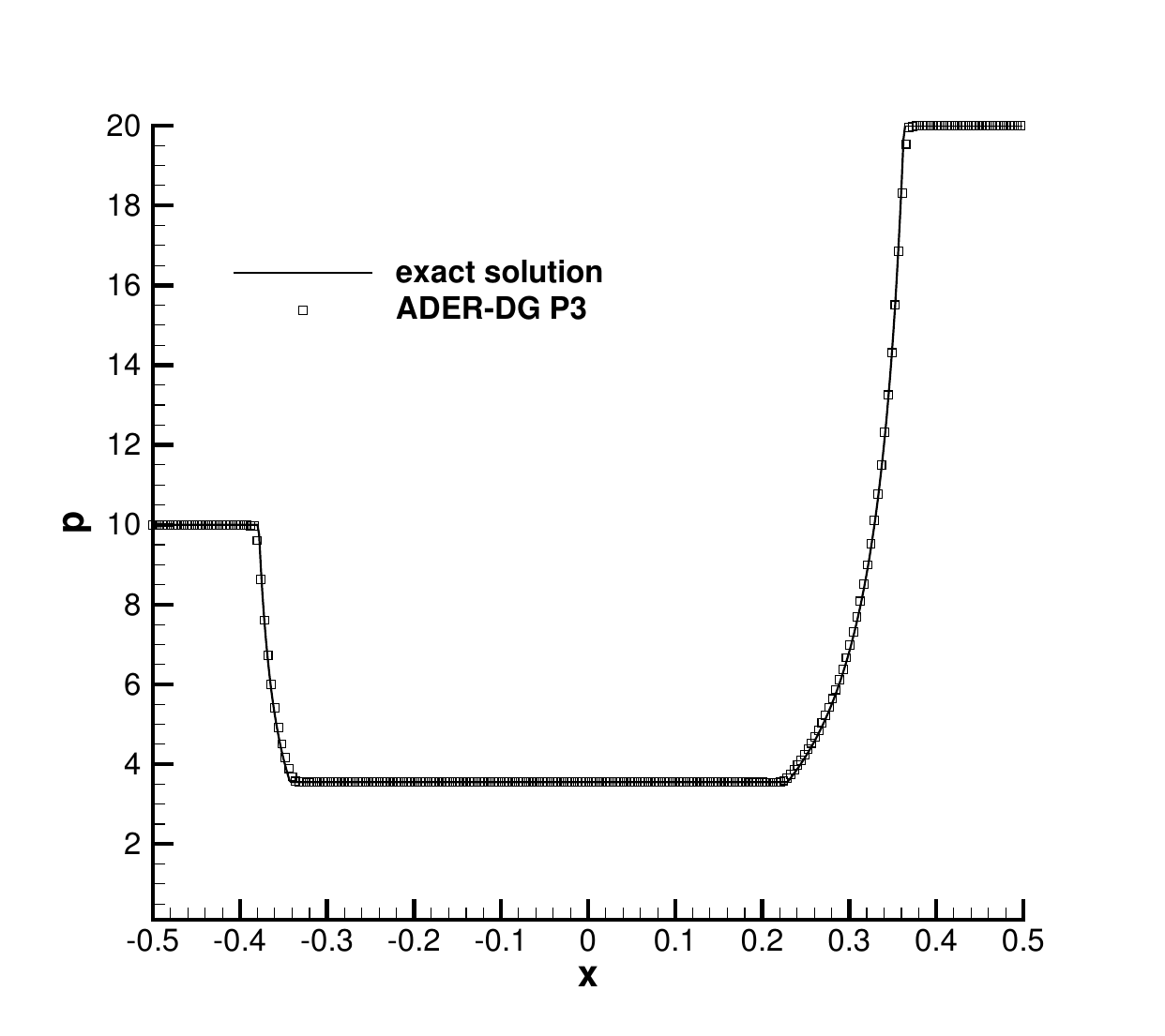}}
\caption{Solution of Riemann Problem 1  at time $t=0.4$.}
\label{fig:shock-tube-2R}
\end{figure}

\begin{figure}
{\includegraphics[angle=0,width=5.5cm,height=5.0cm]{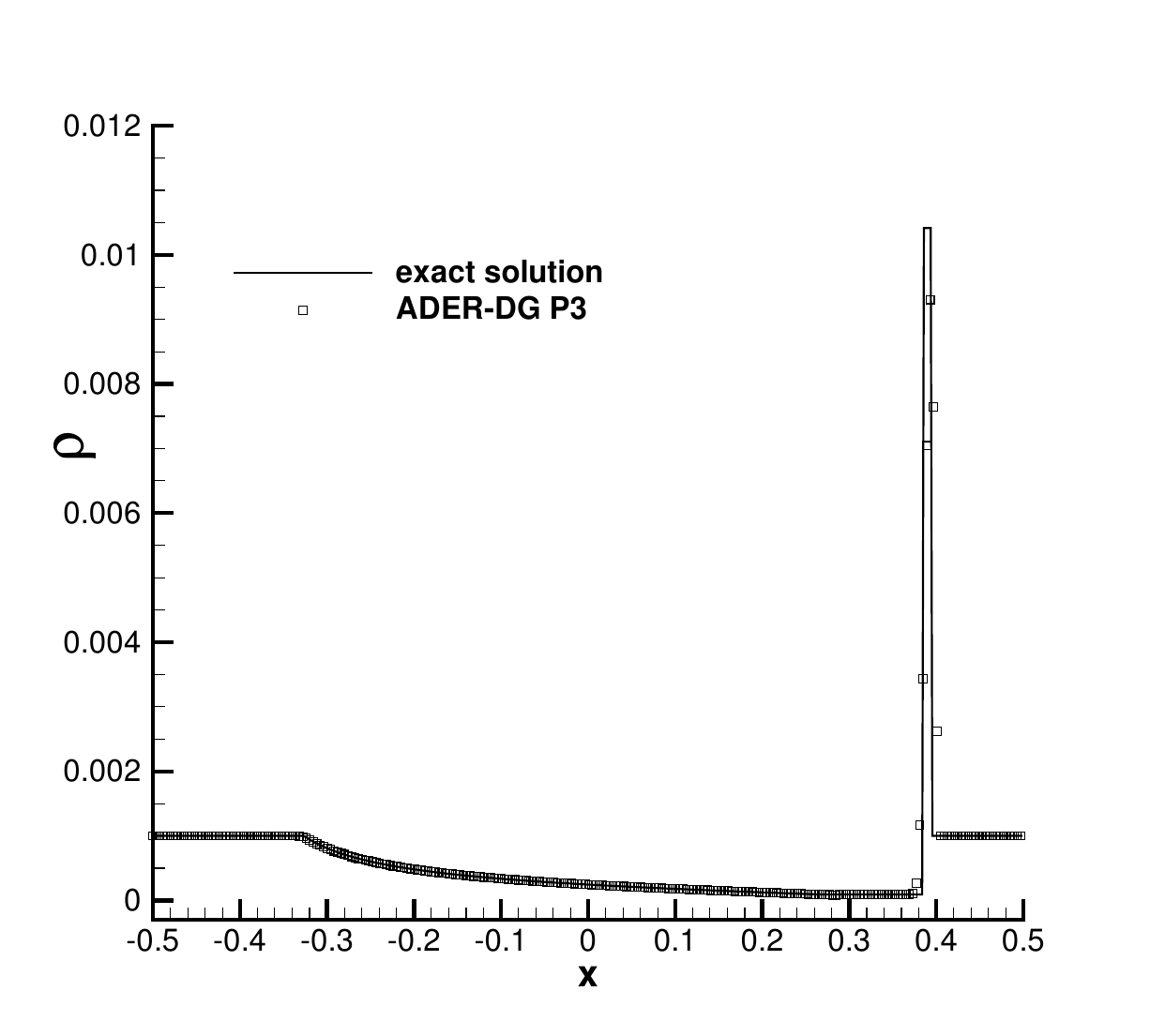}}
{\includegraphics[angle=0,width=5.5cm,height=5.0cm]{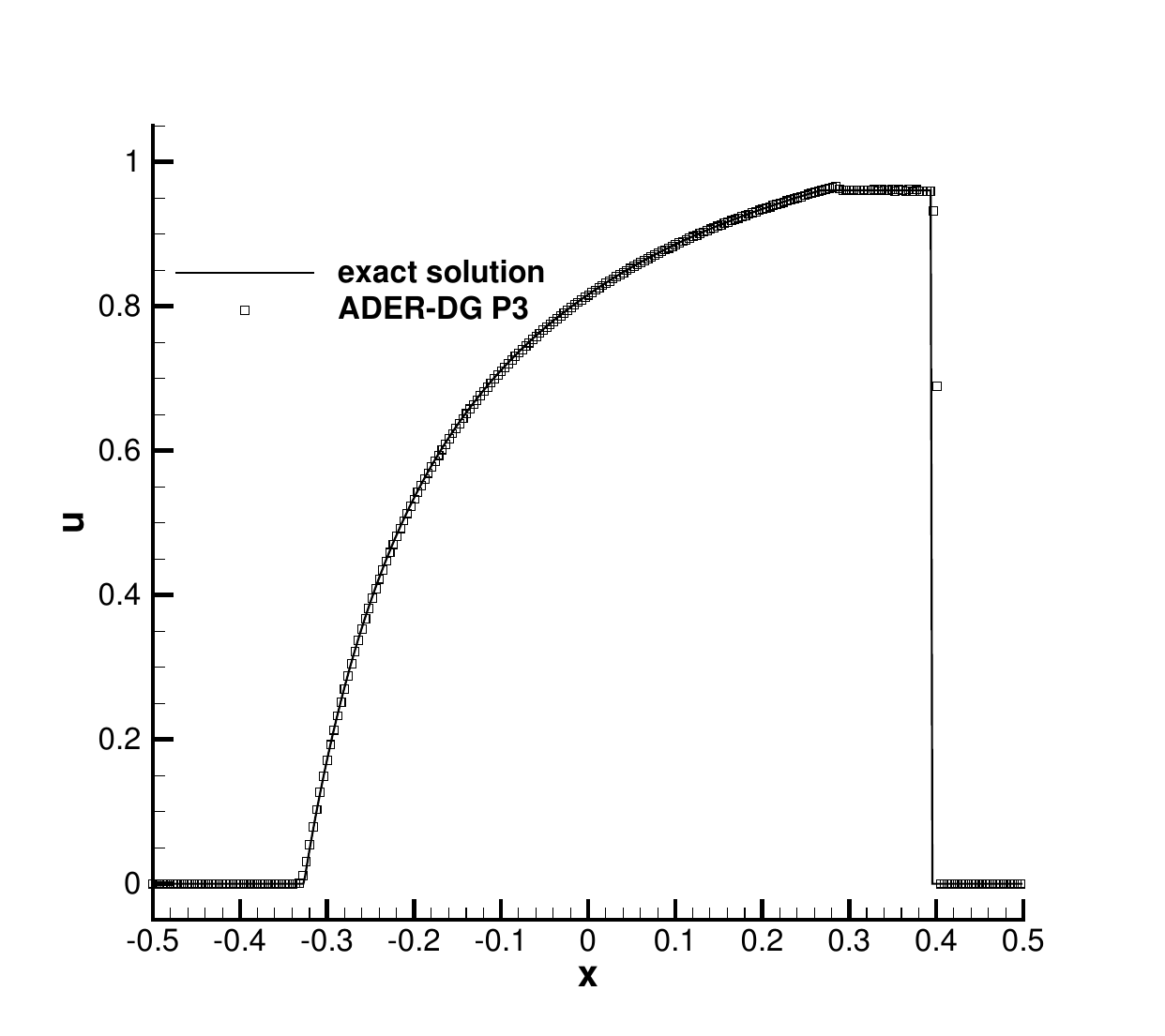}}
{\includegraphics[angle=0,width=5.5cm,height=5.0cm]{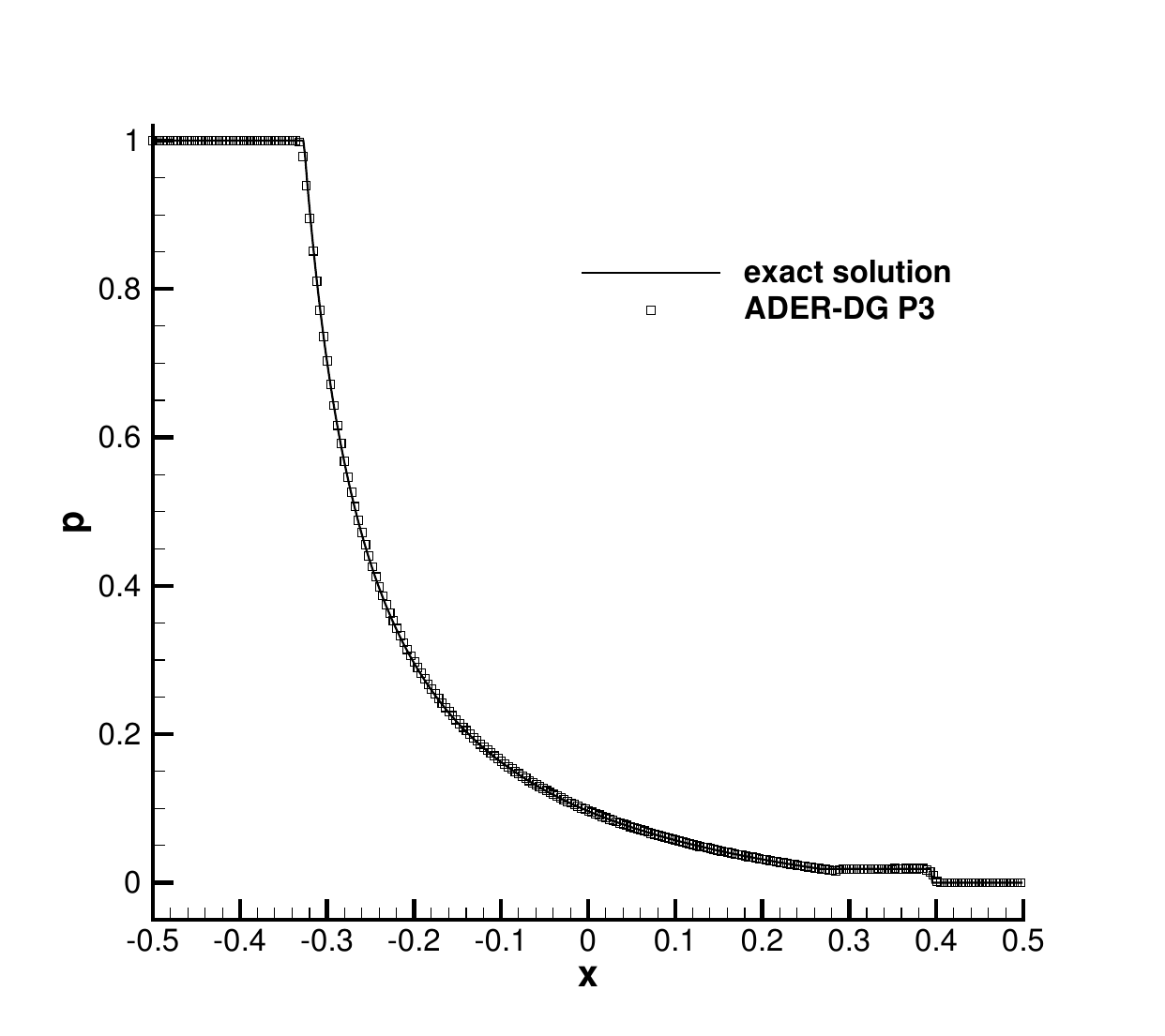}}
\caption{Solution of  Riemann Problem 2 at time $t=0.4$.}
\label{fig:shock-tube-RS}
\end{figure}

\begin{figure}
{\includegraphics[angle=0,width=5.5cm,height=5.0cm]{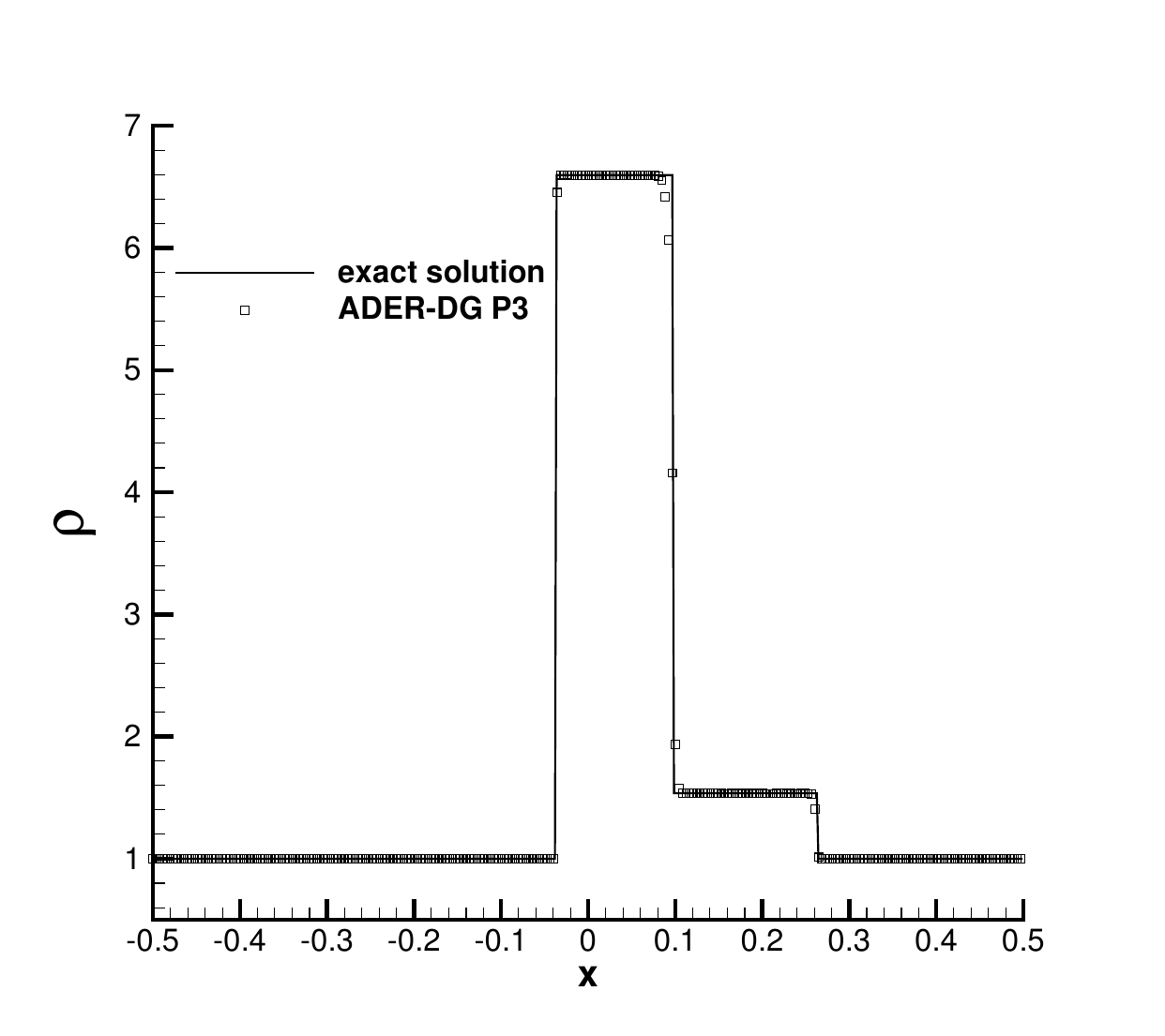}}
{\includegraphics[angle=0,width=5.5cm,height=5.0cm]{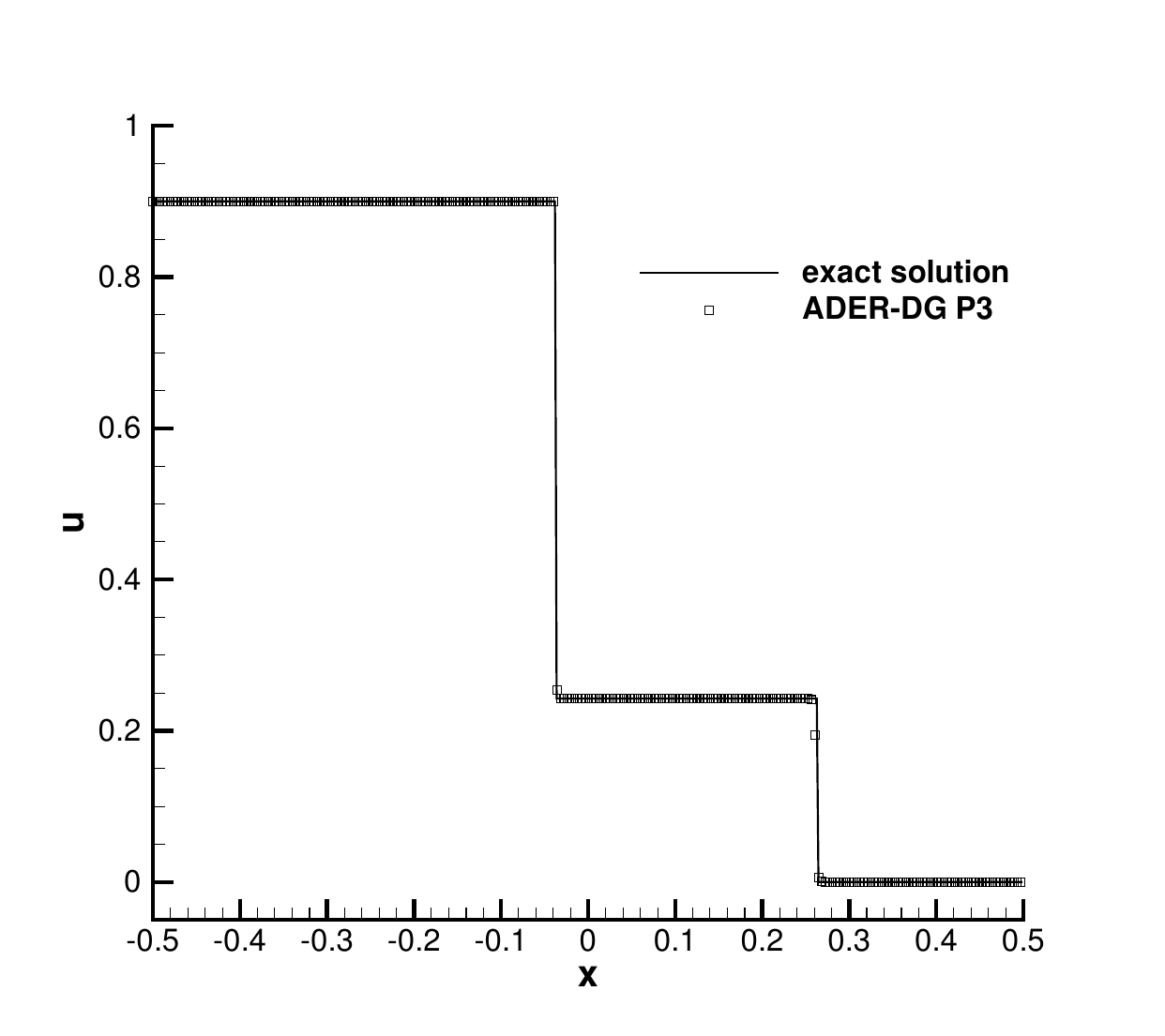}}
{\includegraphics[angle=0,width=5.5cm,height=5.0cm]{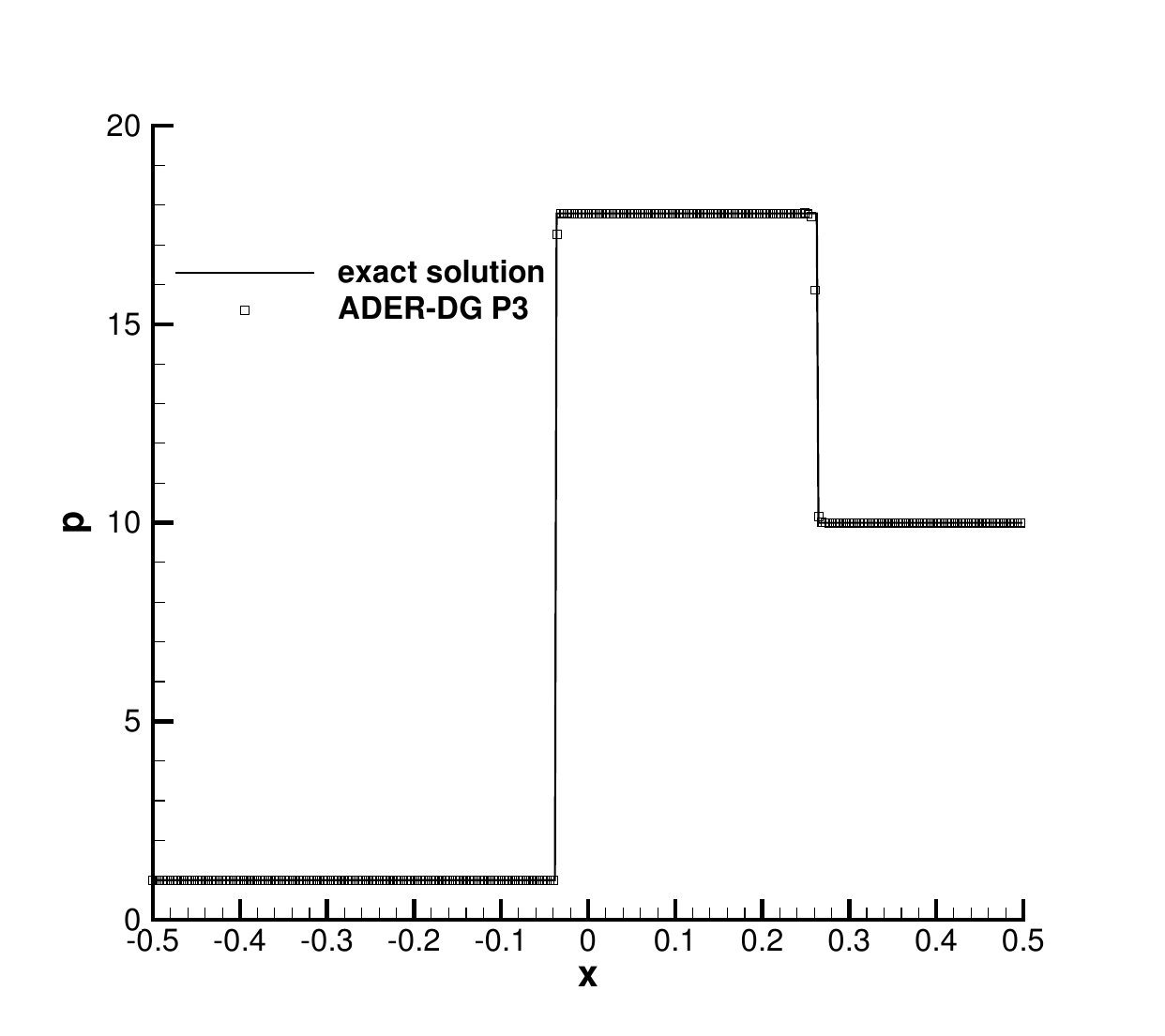}}
\caption{Solution of  Riemann Problem 3 at time $t=0.4$.}
\label{fig:shock-tube-2S}
\end{figure}

If we neglect the evolution of the spacetime, namely we freeze all the spacetime variables and we evolve only
the matter part, then the numerical code can be used to solve the relativistic hydrodynamics equations with a prescribed fixed background metric. After setting to the flat Minkowski spacetime, we have first 
considered three Riemann problems 
with initial conditions chosen as follows:
\begin{enumerate}
	\item $(\rho,v,p)_L=(1,-0.6,10)$ and $(\rho,v,p)_R=(10,0.5,20)$ with adiabatic index $\gamma=5/3$. This configuration, already considered by \cite{Mignone2005}, produces a two--rarefactions wave pattern, that
	we show  in Fig.~\ref{fig:shock-tube-2R} at $t=0.4$.
    \item $(\rho,v,p)_L=(10^{-3},0,1)$ and $(\rho,v,p)_R=(10^{-3},0,10^{-5})$ with adiabatic index $\gamma=5/3$. This configuration, already considered by \cite{Radice2012a}, produces a one-rarefaction--one-shock wave pattern, that
    we show  in Fig.~\ref{fig:shock-tube-RS} at $t=0.4$.
	\item $(\rho,v,p)_L=(1,0.9,1)$ and $(\rho,v,p)_R=(1,0,10)$ with adiabatic index $\gamma=4/3$. This configuration, already considered by \cite{Zhang2006}, produces a two--shocks wave pattern, that
	we show  in Fig.~\ref{fig:shock-tube-2S} at $t=0.4$.
\end{enumerate}
We have solved each of these Riemann problems over a
computational grid composed of $256\times 4$ equidistant finite elements.
Because of the presence of various discontinuities in the solution, for these tests it is mandatory to activate
the subcell FV limiter. Hence we have used an ADER-DG P3 version of our scheme in combination with a second order TVD sub-cell limiter. As apparent from Figs.~\ref{fig:shock-tube-2R}--\ref{fig:shock-tube-2S}, all the tests considered 
reproduce the exact solution, available from \cite{Marti94,Rezzolla2001}, in a rather satisfactory way.

%
%
%
\subsection{Spherical accretion }
\label{sec:Michel}

As a first test involving matter  in a truly curved but otherwise stationary spacetime, we study the spherical  accretion
 of a  gas onto a non rotating black hole~\cite{Rezzolla_book:2013,Anton06}. This test is a classic one,
 and we have solved it in two space dimensions $(r,\theta)$
after adopting spherical Kerr--Schild coordinates. The Lorentz factor of the fluid with respect to the Eulerian observer
 is given by
\begin{equation}
W = \alpha\frac{-z u^r - \sqrt{(u^r)^2 - z + 1}}{z - 1}\hspace{1cm}{\textrm{with}}\,\,\,\,\,u^r<0\,,\,\,\,\,\,z=2/r\,.
\end{equation}
The computational domain is $ [0.5;10]\times[0+\epsilon;\pi-\epsilon]$, with $\epsilon=0.05$, covered by a $64\times32$ uniform grid. What makes this test challenging are the parameters chosen for the hydrodynamics.
In fact, placing the critical radius at
$r_c=5$ with a critical density  $\rho_c=1.006\times10^{-7}$, makes the overall solution rather rarefied,
with consequent difficult conditions for the recovering of the primitive variables from the conserved ones.
Indeed, these numbers imply
a mass accretion rate (computed as $4\pi r_c^2 \rho_c u_c^r$) as small as  
$-1.0\times10^{-5}$, meaning that the total mass accreted onto the black hole from $t=0$ to $t=1000\,M$ is just $1/100$ of  the total mass $M$ of the central black hole. Hence, at least over the timescale considered, the physical assumption of a stationary spacetime  is justified, namely the Cowling approximation with Einstein equations frozen.
Moreover, we have 
perturbed the  rest mass density using a Gaussian profile peaked at the critical radius, with a perturbation amplitude $\delta\rho = 10^{-2}\rho_c$.  The adiabatic  index of the gas is $\gamma=5/3$.

\begin{figure}[!htbp]
\begin{center}
\begin{tabular}{cc} 
{\includegraphics[angle=0,width=7.3cm,height=7.3cm]{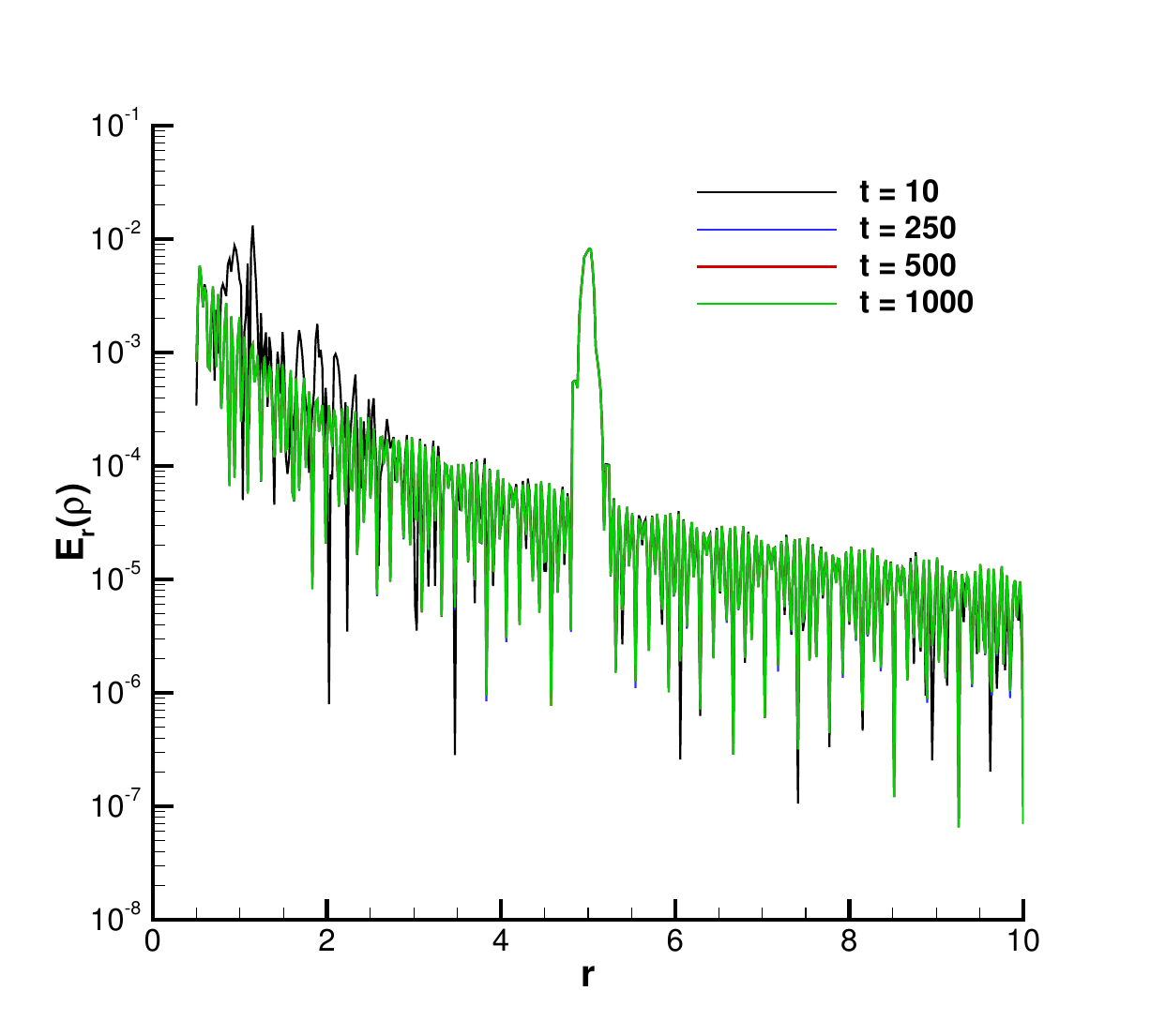}} &
{\includegraphics[angle=0,width=7.3cm,height=7.3cm]{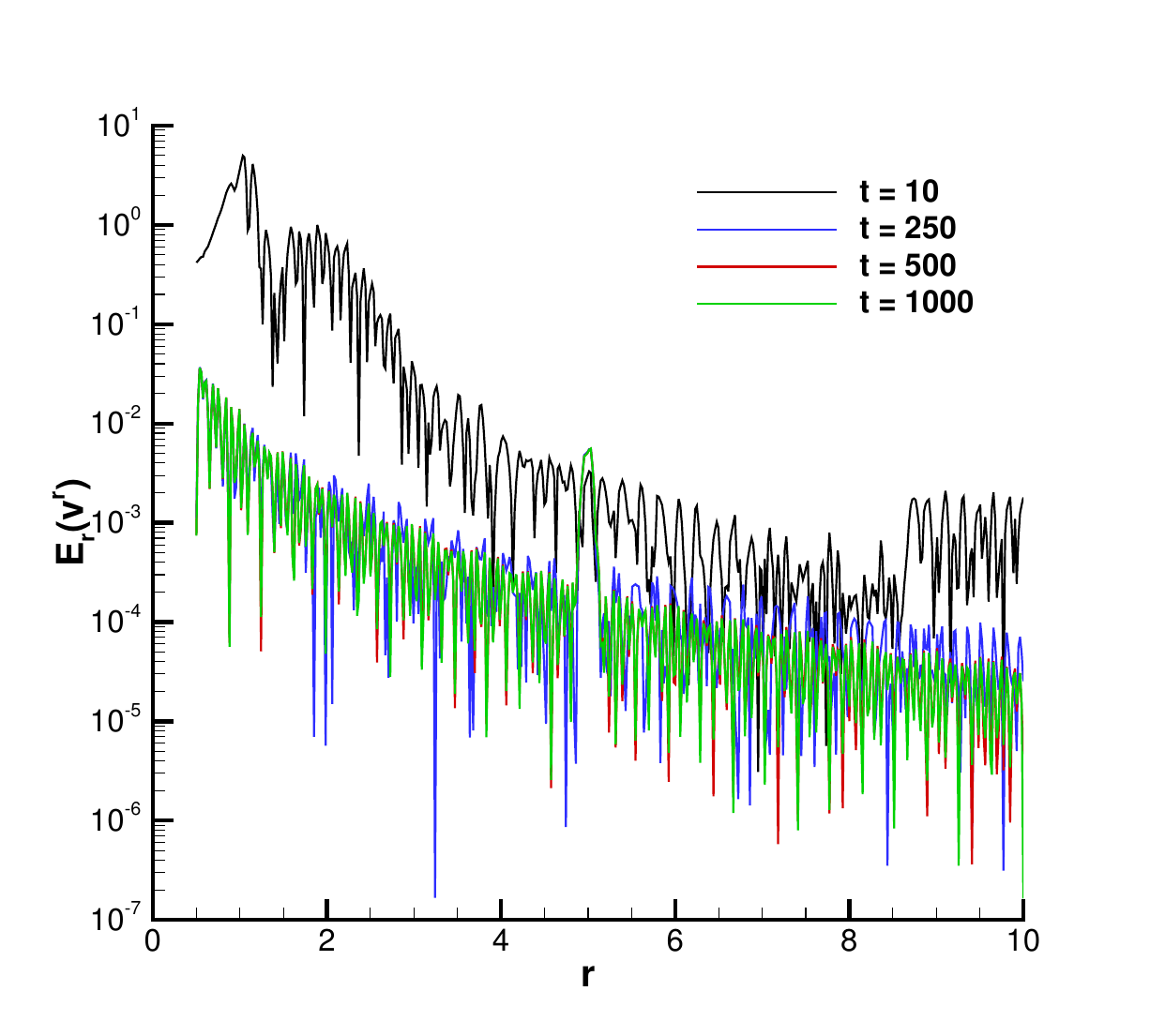}} \\
\end{tabular} 
\caption{Spherical accretion of matter onto a non-rotating black hole using Kerr-Schild coordinates. 
The profiles of the relative errors
are reported, both for the rest mass density (left panel) and for the radial velocity (right panel). 
An ADER-DG P3 scheme with Well Balancing is adopted.} 
\label{fig:SA-1D}
\end{center}
\end{figure}

We could only solve such a test, under these specific conditions, after activating the 
well--balancing property of the numerical scheme,  described in Sect.~\ref{sec:wb}. 
Also in this case we have chosen the ADER-DG P3 version of the scheme, in combination with a
simple Rusanov Riemann solver. Being a completely smooth solution, there was no need to activate the limiter. 
Fig.~\ref{fig:SA-1D} shows the relative errors of the 
mass density and of the radial velocity, with respect to the exact solution, at few representative times.

\subsection{The equilibrium TOV star }
\label{sec:TOV}

\begin{figure}[!htbp]
	\begin{center}
		\begin{tabular}{cc} 
			{\includegraphics[width=0.49\textwidth]{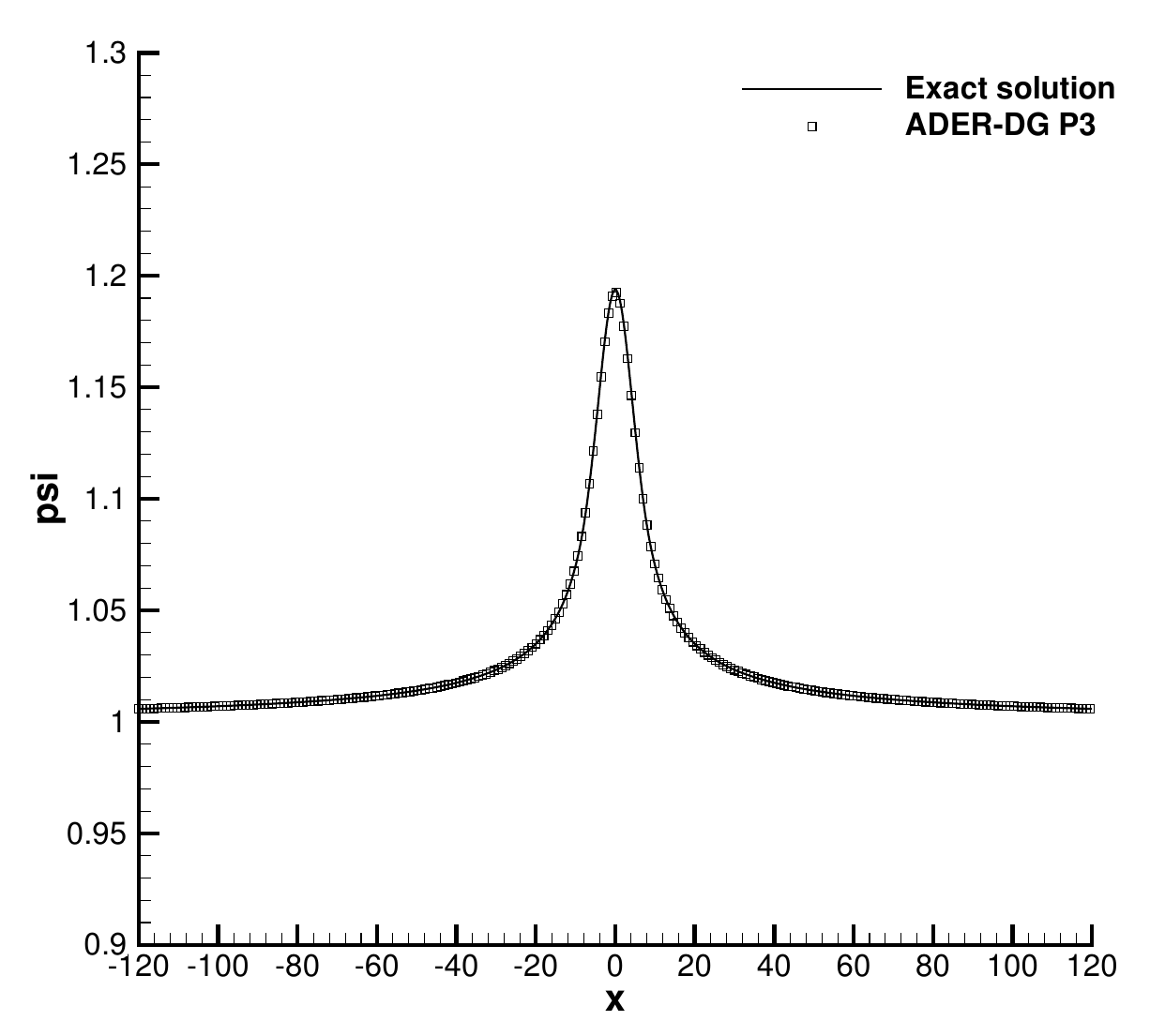}} &
			{\includegraphics[width=0.49\textwidth]{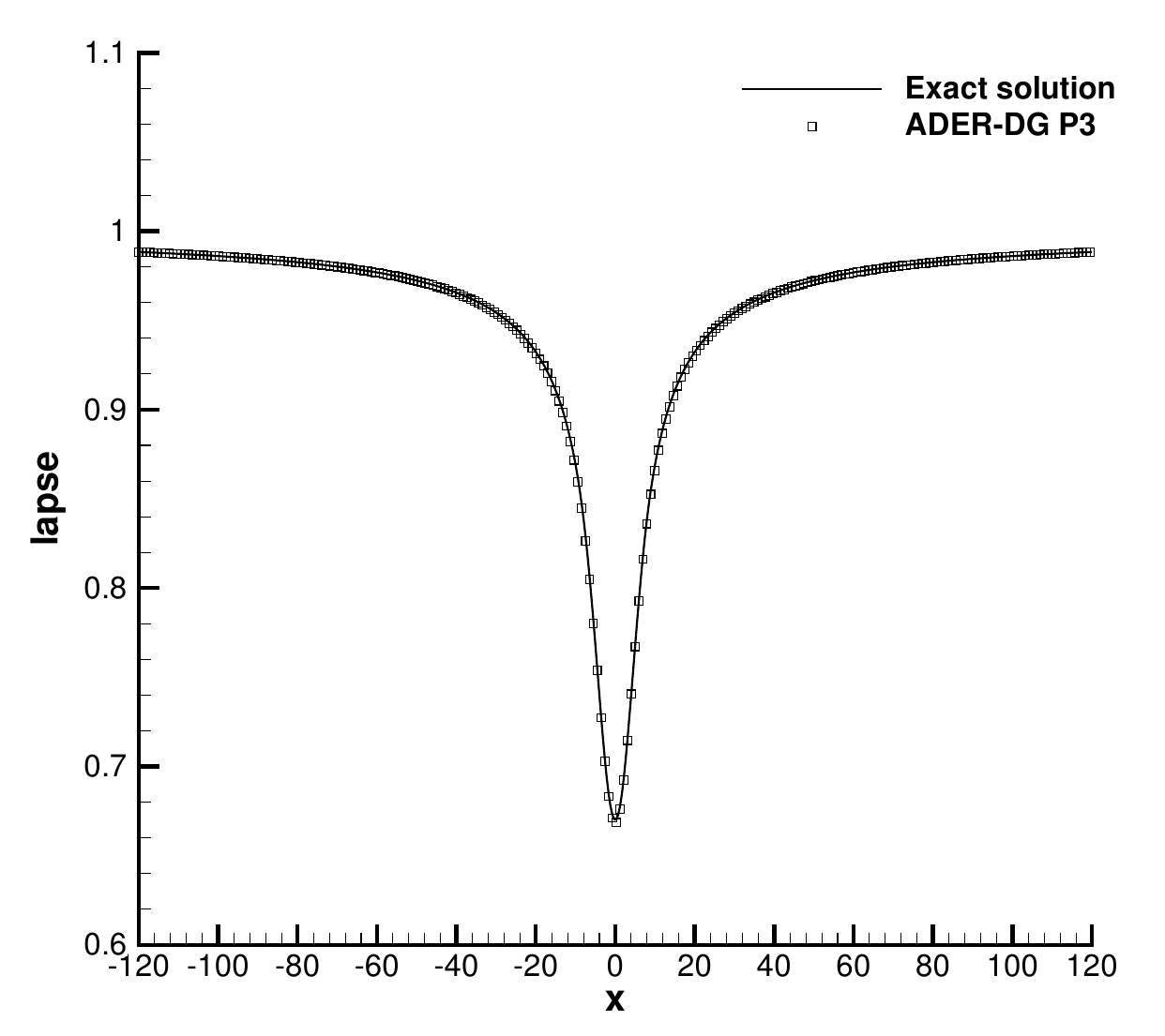}}\\
			{\includegraphics[width=0.49\textwidth]{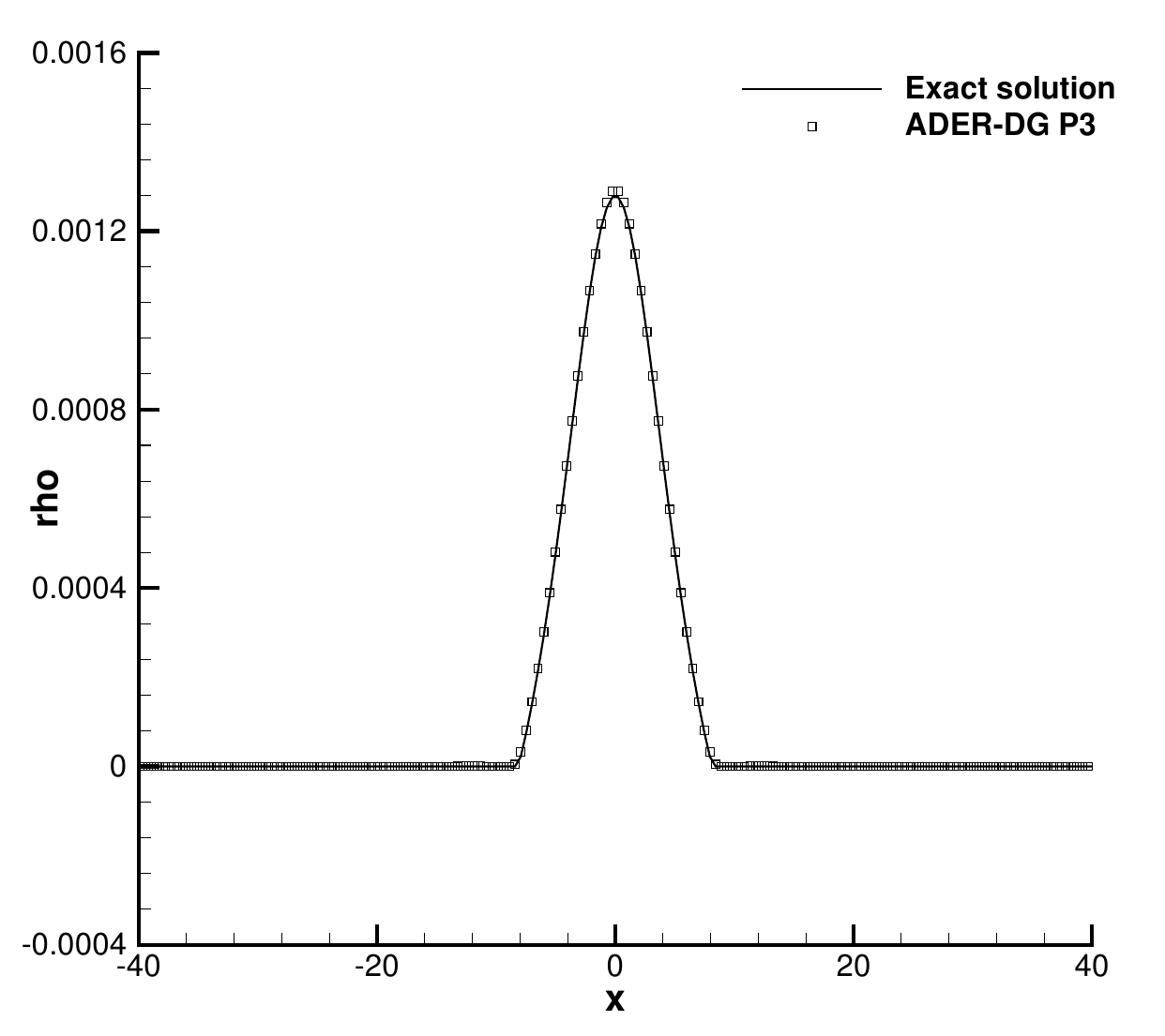}} &
			{\includegraphics[width=0.49\textwidth]{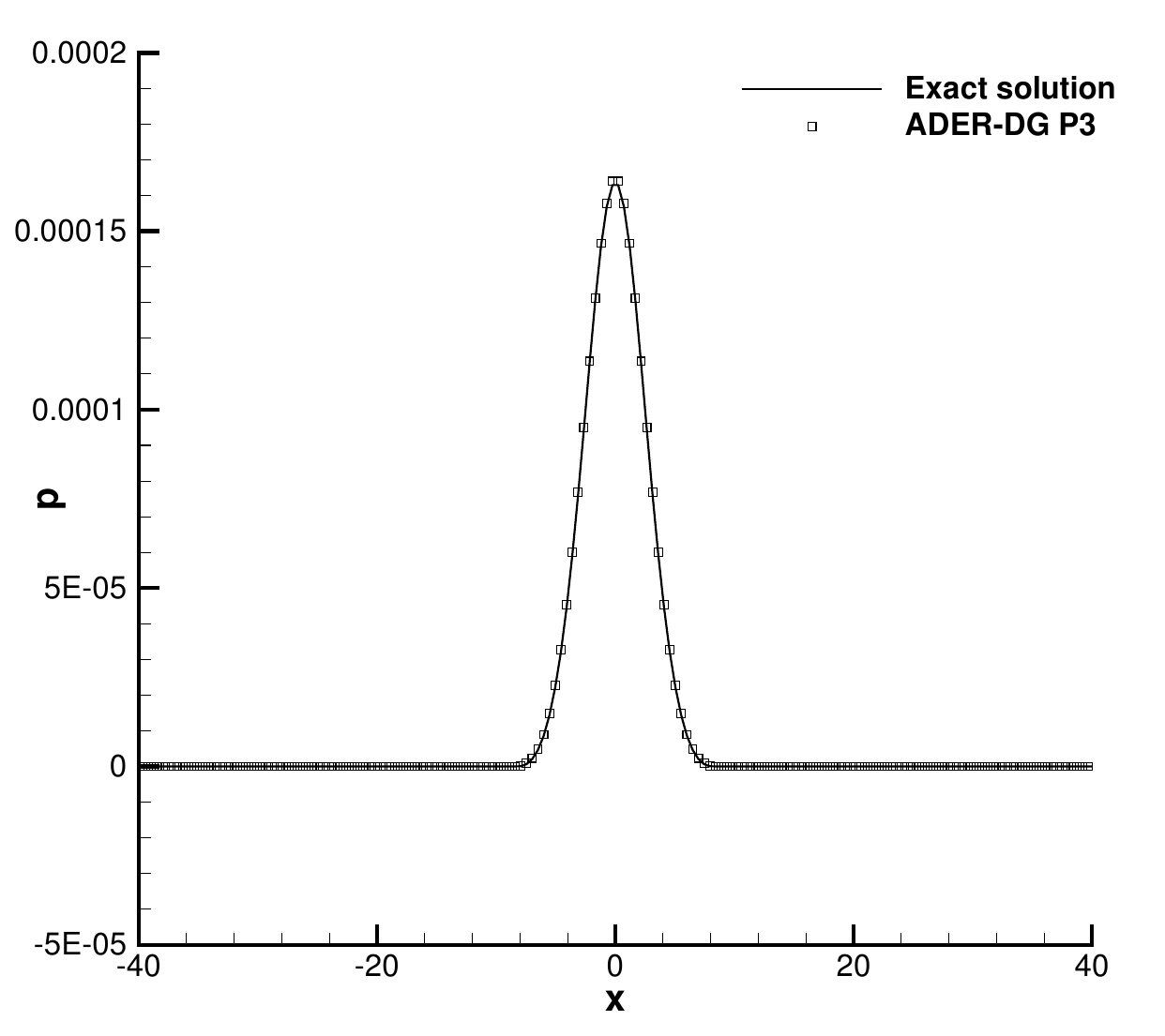}}\\
		\end{tabular} 
		\caption{Solution for the stable TOV star obtained with a fourth order ADER-DG scheme ($N=3$) at time $t=100$. Conformal factor (top left),
			lapse (top right), rest mass density (bottom left) and pressure (bottom right). } 
		\label{fig:TOV-profiles}
	\end{center}
\end{figure}

\begin{figure}[!htbp]
	\begin{center}
		\begin{tabular}{cc} 
			{\includegraphics[width=0.49\textwidth]{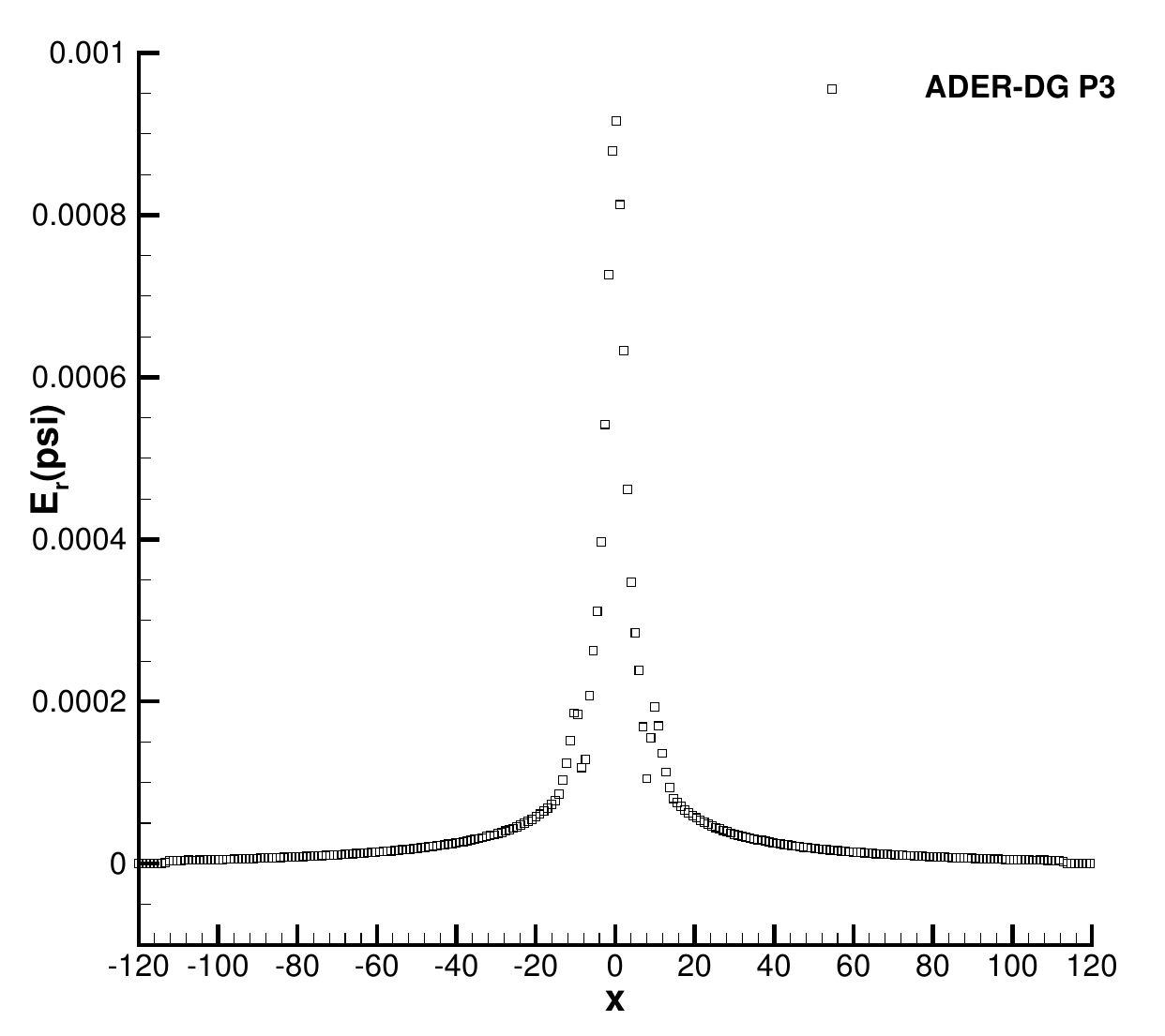}} &
			{\includegraphics[width=0.49\textwidth]{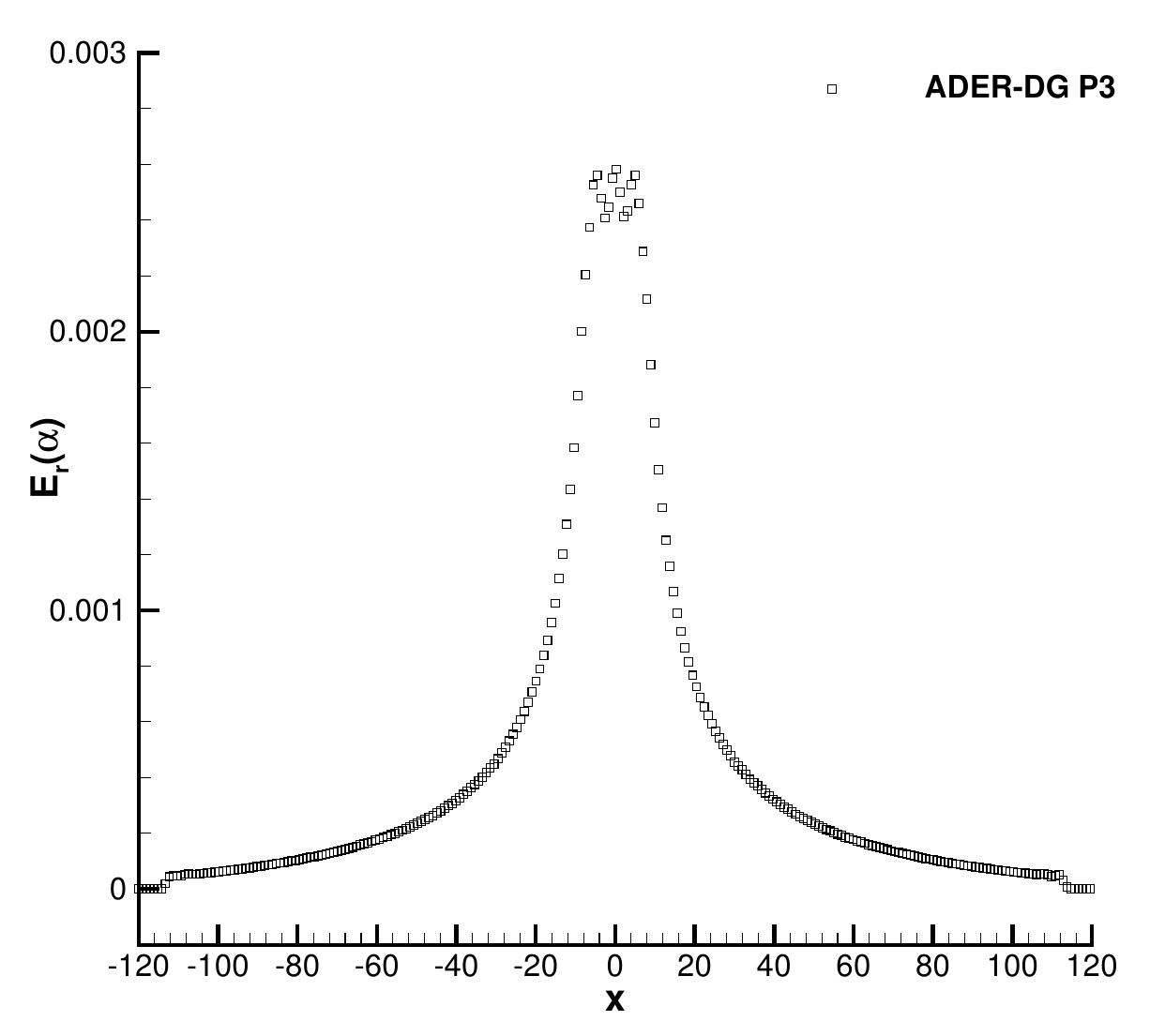}}\\
			{\includegraphics[width=0.49\textwidth]{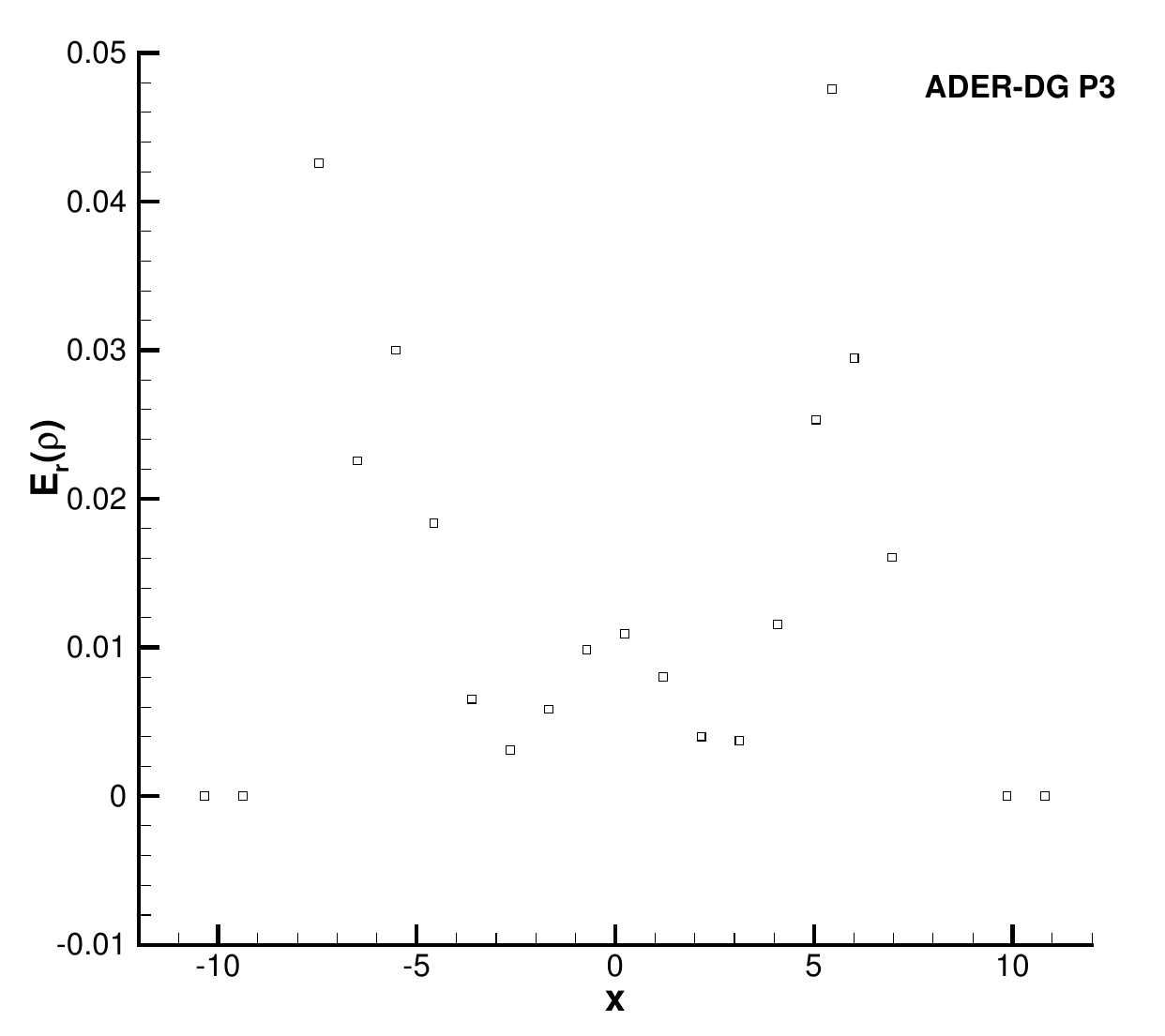}} &
			{\includegraphics[width=0.49\textwidth]{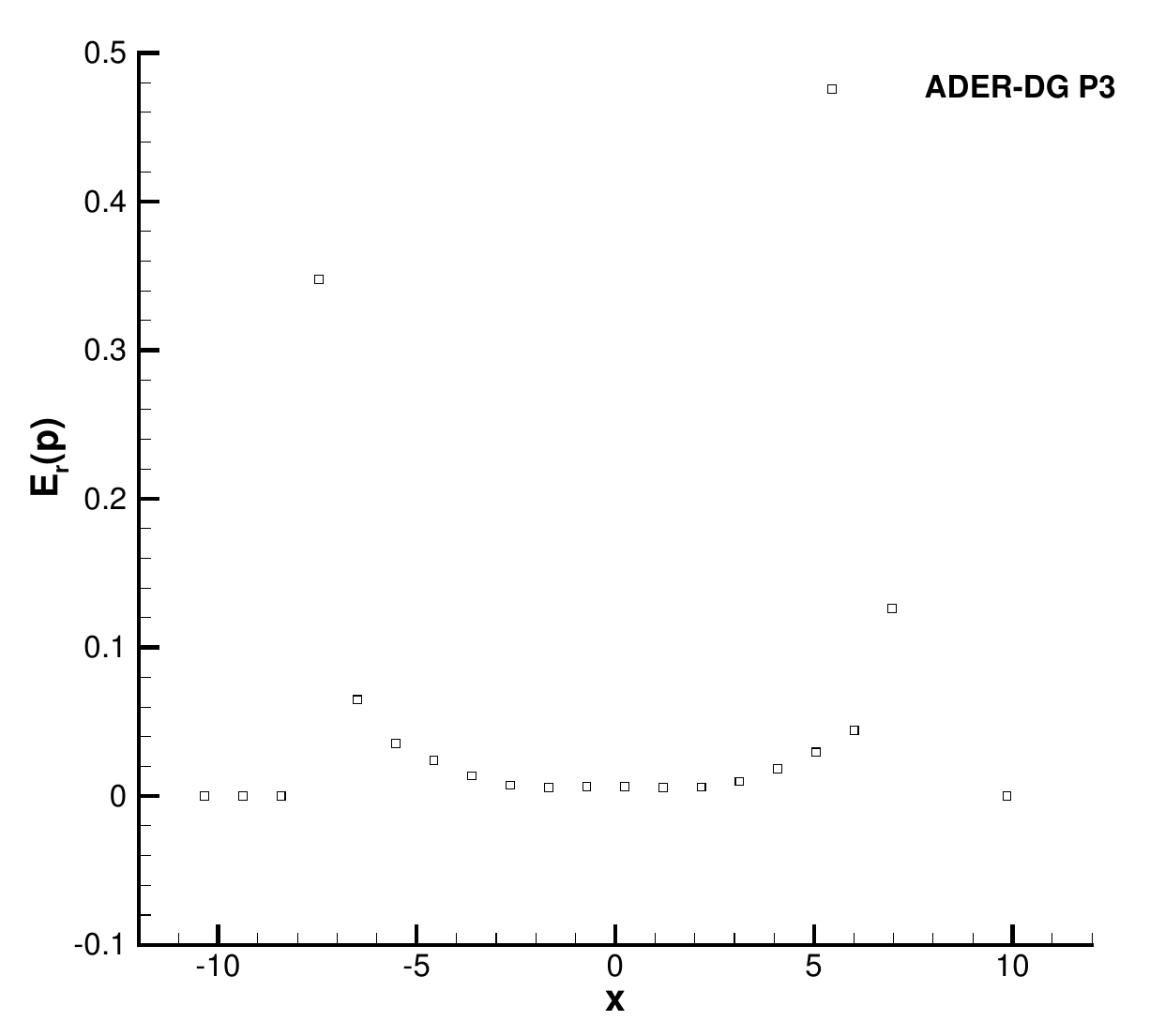}}\\
		\end{tabular} 
		\caption{Solution for the stable TOV star obtained with a fourth order ADER-DG scheme ($N=3$) at time $t=100$. Relative errors (with respect to the exact solution) are reported for the conformal factor (top left), the
				lapse (top right), the rest mass density (bottom left) and the pressure (bottom right). } 
		\label{fig:TOV-profiles.err}
	\end{center}
\end{figure}

\begin{figure}[!hpb]
	\begin{center}
			{\includegraphics[angle=0,width=0.6\textwidth]{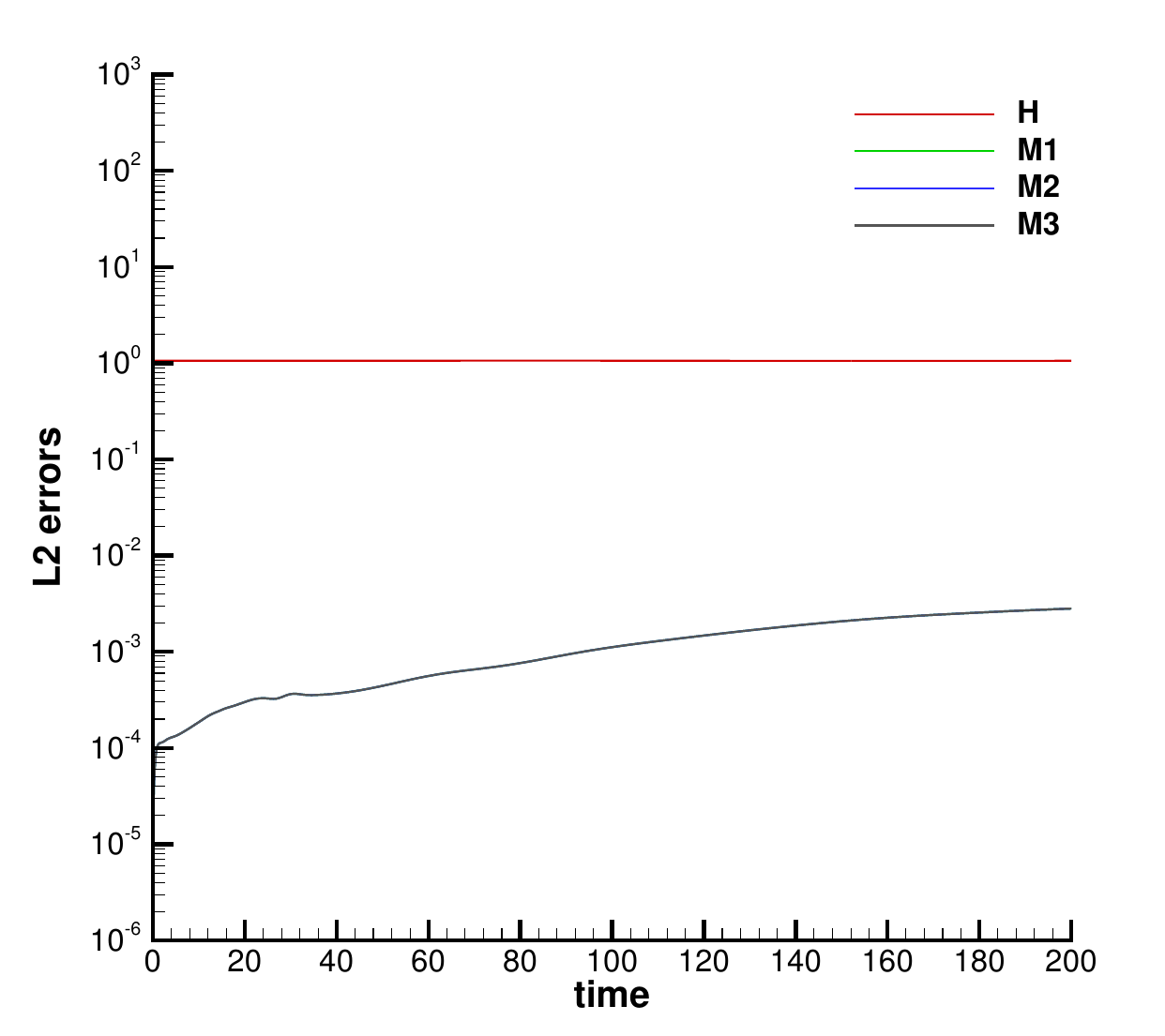}} 
		\caption{Time evolution of the Einstein constraints for the 3D TOV star, simulated with a fourth order ADER-DG scheme until $t=200$.} 
		\label{fig:TOV-timeseries}
	\end{center}
\end{figure}

In this section we consider the fully coupled Einstein-Euler system by simulating a stable neutron star. 
We therefore consider the classical Tolman--Oppenheimer--Volkoff (TOV) system~\cite{Tolman,Oppenheimer39b,Rezzolla_book:2013,Camenzind2007} 
for a polytropic gas with $p= K\,\rho^\gamma$. 
The parameters are chosen as in~\cite{Font2002}, i.e. a central rest mass density of $\rho_c=1.28\times 10^{-3}$ with the constants chosen as $K=100$ and $\gamma=2$. We then integrate the ODE system that governs the TOV star via a high order Discontinuous Galerkin solver 
for ODE \cite{ADERNSE}, obtaining a total mass $M=1.4\,M_{\odot}$ and a radius $R=14.15\,{\textrm {km}}$.  
The FO-BSSNOK system is solved with a fourth order ADER-DG scheme ($N=3$) on the domain $\Omega=[-120,+120]^3$ using an adaptive computational mesh  
with one layer of static refinement in the box $\Omega=[-30,+30]^3$. The mesh spacing on the coarsest mesh is $\Delta x^0 = 6$, in the inner box it is $\Delta x^1 = 2$ and in the innermost area $\Omega=[-12,+12]^3$ we use a third order ADER-WENO subcell finite volume limiter with mesh spacing $\Delta x = 2/7$. In this test we also employ the new ``filtering'' technique proposed by \cite{DumbserZanottiGaburroPeshkov2023}, which allows to perform the primitive variable recovery even in a $\rho=0$ atmosphere outside the compact object. 
Fig.~\ref{fig:TOV-profiles} reports the results of this set--up by showing 
a few representative quantities at time $t=100$, computed along one dimensional x-cuts, 
while in Fig.~\ref{fig:TOV-profiles.err} the corresponding relative errors with respect to the exact solution at time $t=0$ are presented. 
We notice that, while the errors are very small for the metric quantities $\psi$ and $\alpha$,
they are substantially larger for the density and the pressure, especially at the star surface.
Fig.~\ref{fig:TOV-timeseries} contains the temporal evolution of the Einstein constraints, which present only a moderate linear growth in time for the momentum constraints, while the Hamiltonian appears to be perfectly stable. 

%

\subsection{Single puncture black hole }
\label{sec:Trumpet}
\begin{figure}[!htbp]
	\begin{center}
		\begin{tabular}{cc}
			\includegraphics[trim=10 10 10 10,clip,width=0.48\textwidth]{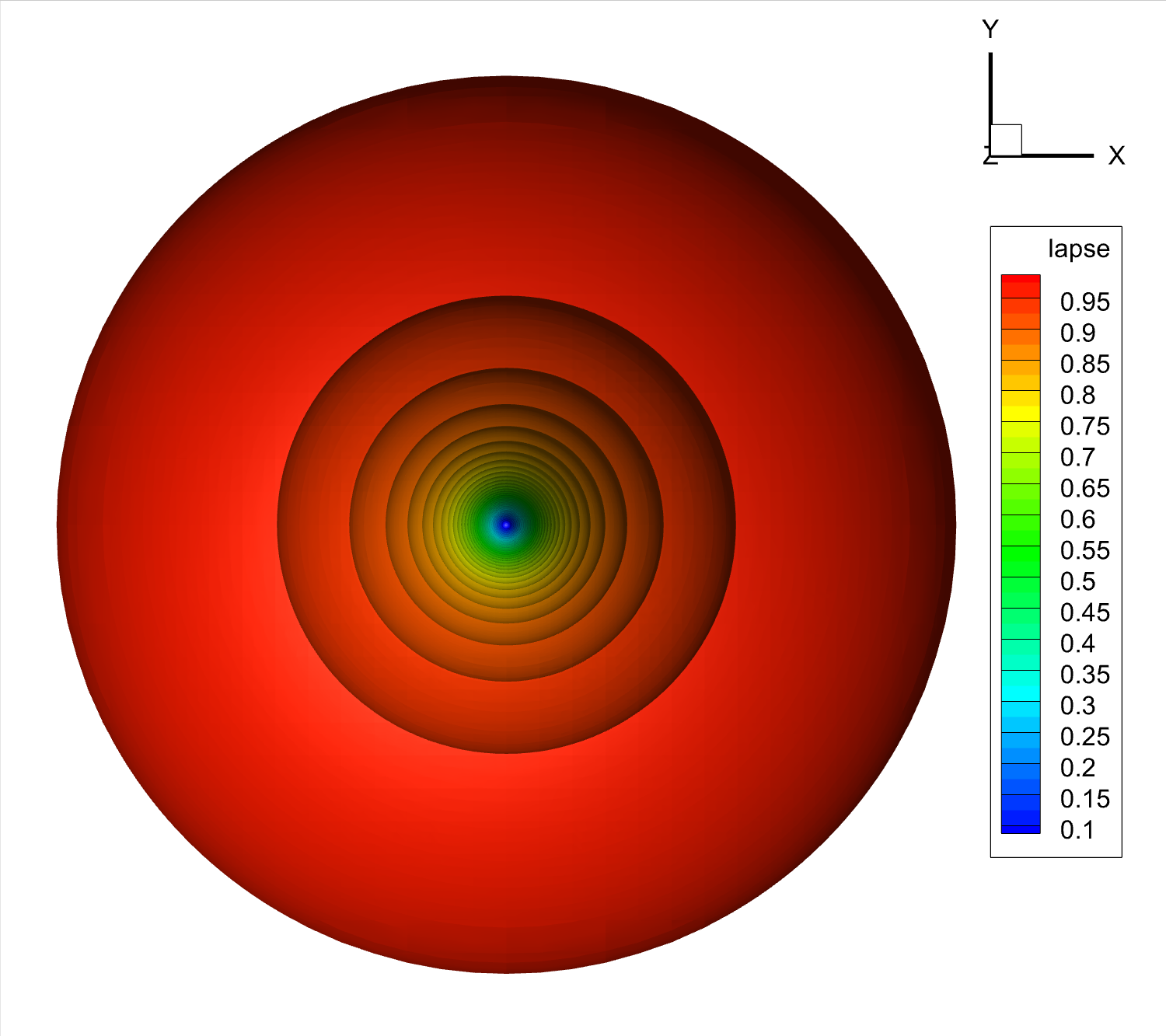} & 
			\includegraphics[width=0.48\textwidth]{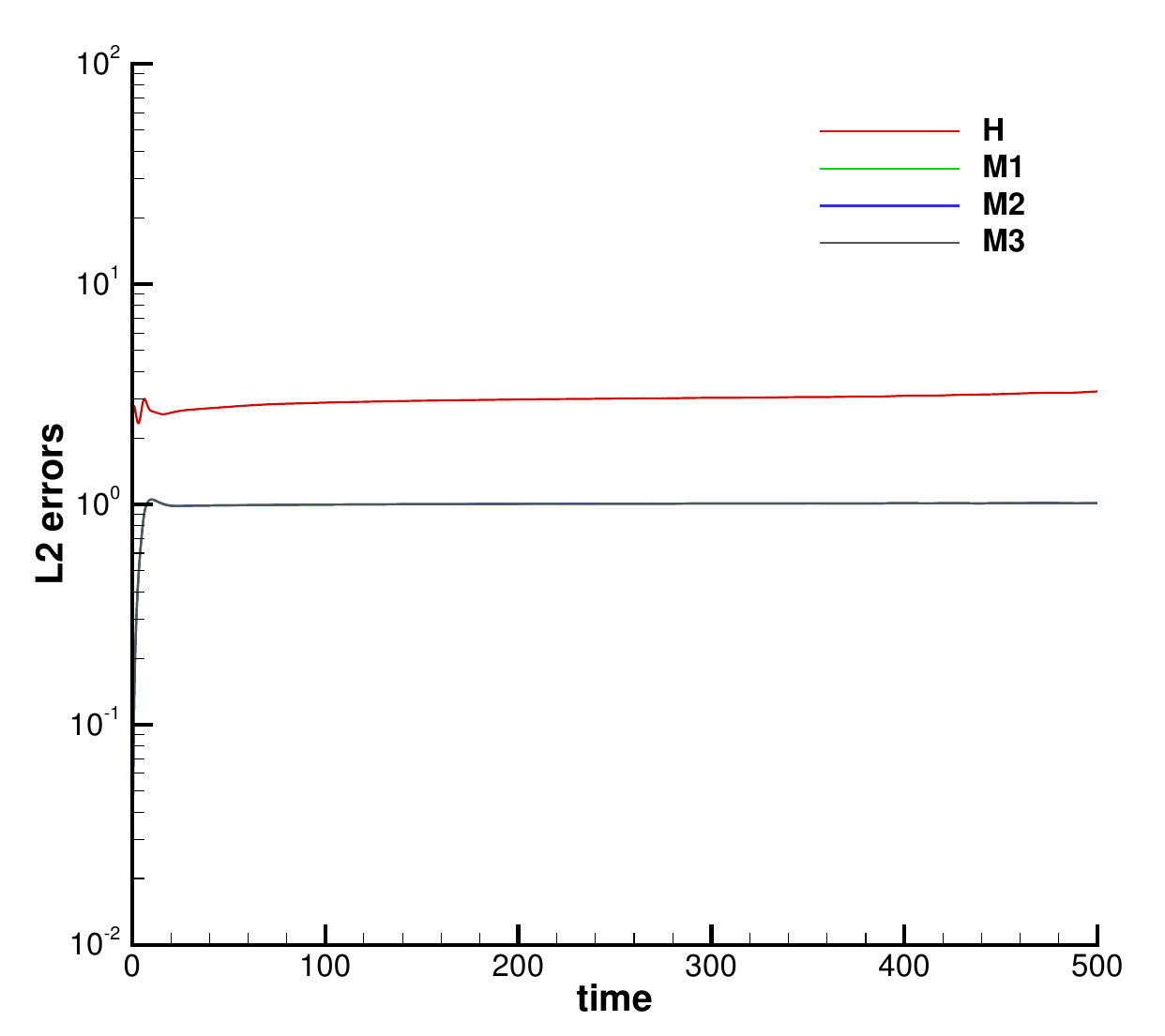}  \\ 
			\multicolumn{2}{c}{
						 \includegraphics[width=0.6\textwidth]{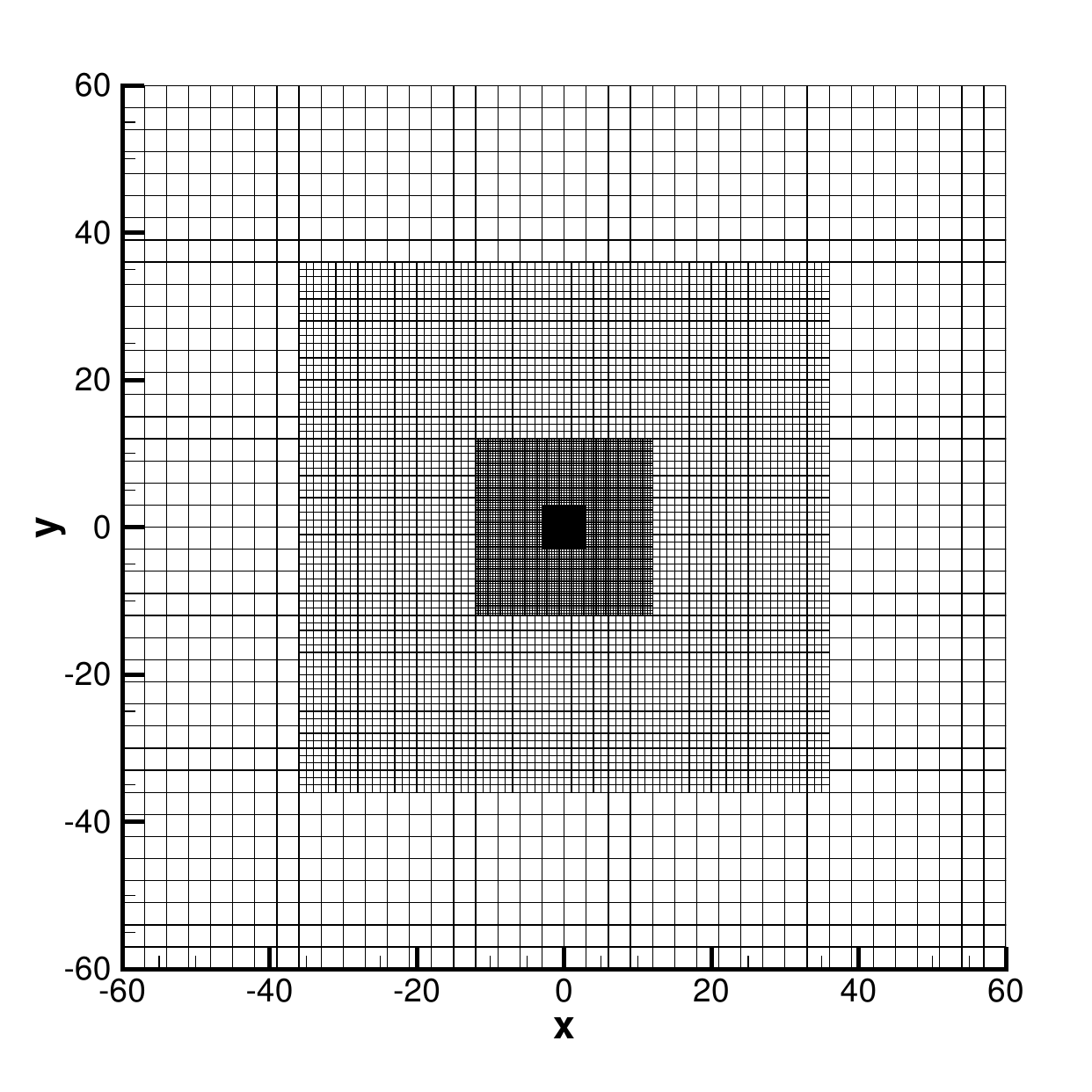}  
 			}
		\end{tabular}
		\caption{Simulation of a single puncture black hole until $t=500$ using a fourth order ADER-DG scheme ($N=3$). 
			Iso-surface contours of the lapse at the final time (top left), time-evolution of the Einstein constraints (top right) and mesh setup in the $z=0$ plane, with two levels of adaptive mesh refinement (AMR) for the DG scheme and the innermost mesh corresponding to the subcell FV limiter (bottom center). 		}
		\label{fig.OnePunctureSetup}
	\end{center}
\end{figure}

\begin{figure}[!htbp]
	\begin{center}
		\begin{tabular}{cc}
			\includegraphics[width=0.45\textwidth]{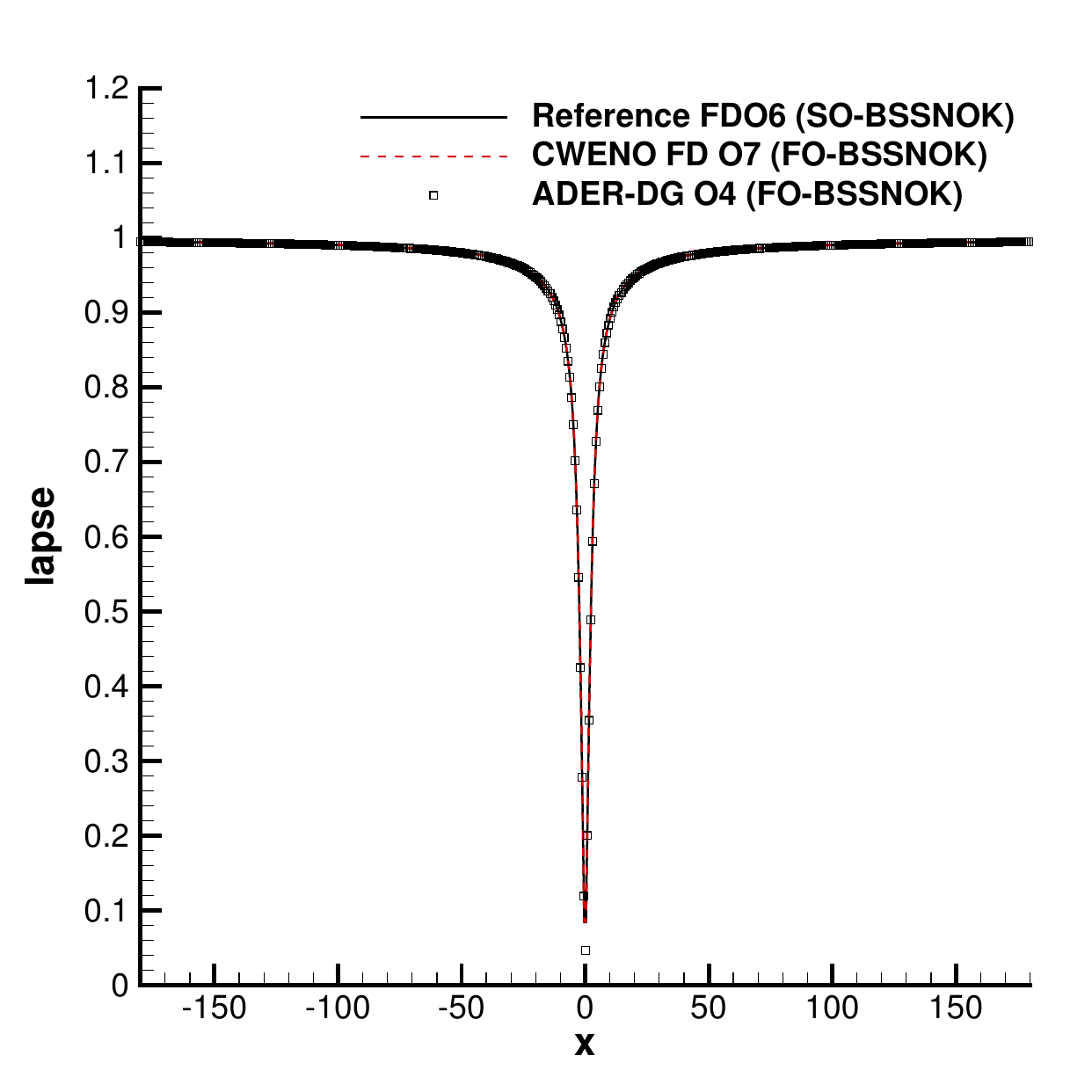} &
			\includegraphics[width=0.45\textwidth]{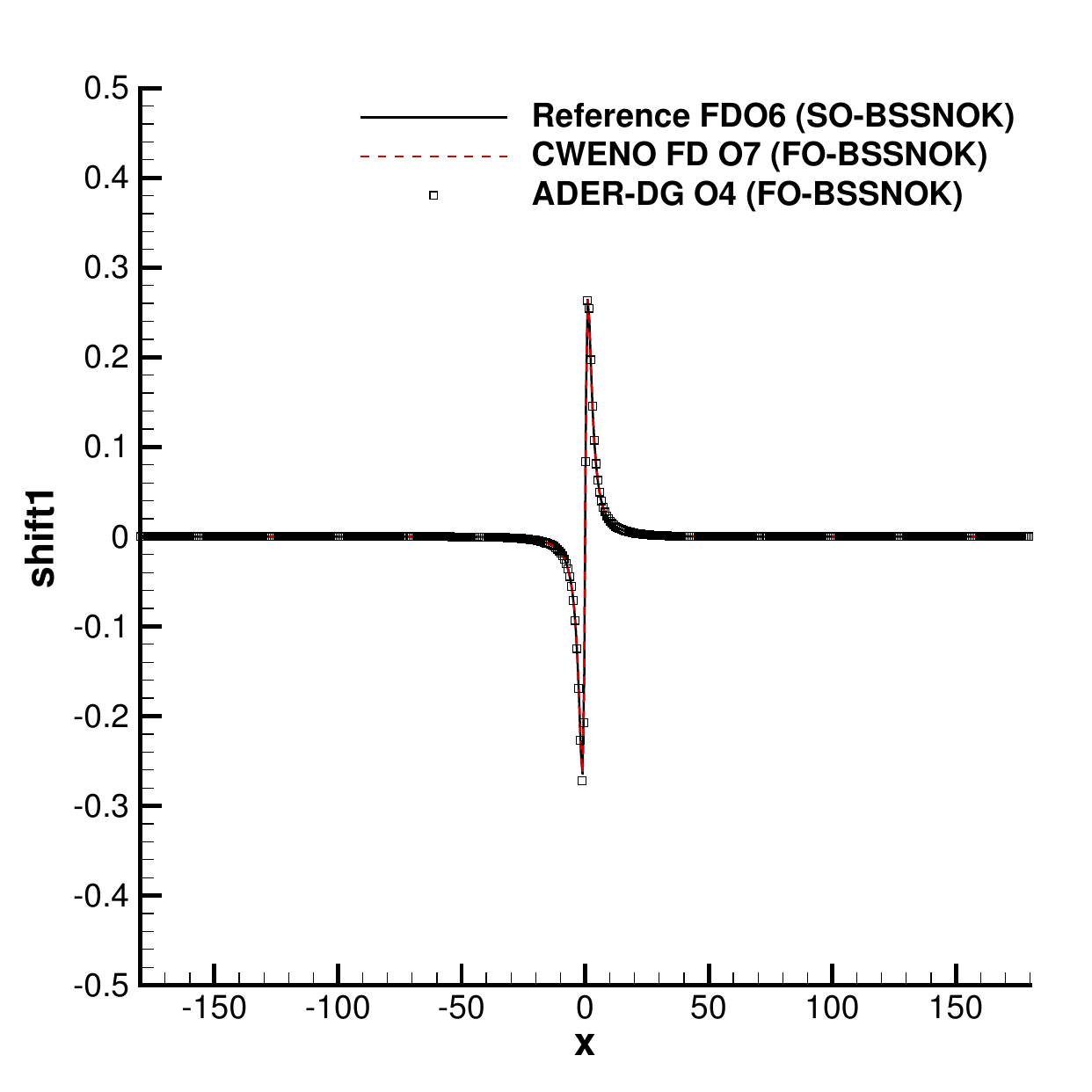} \\
			\includegraphics[width=0.45\textwidth]{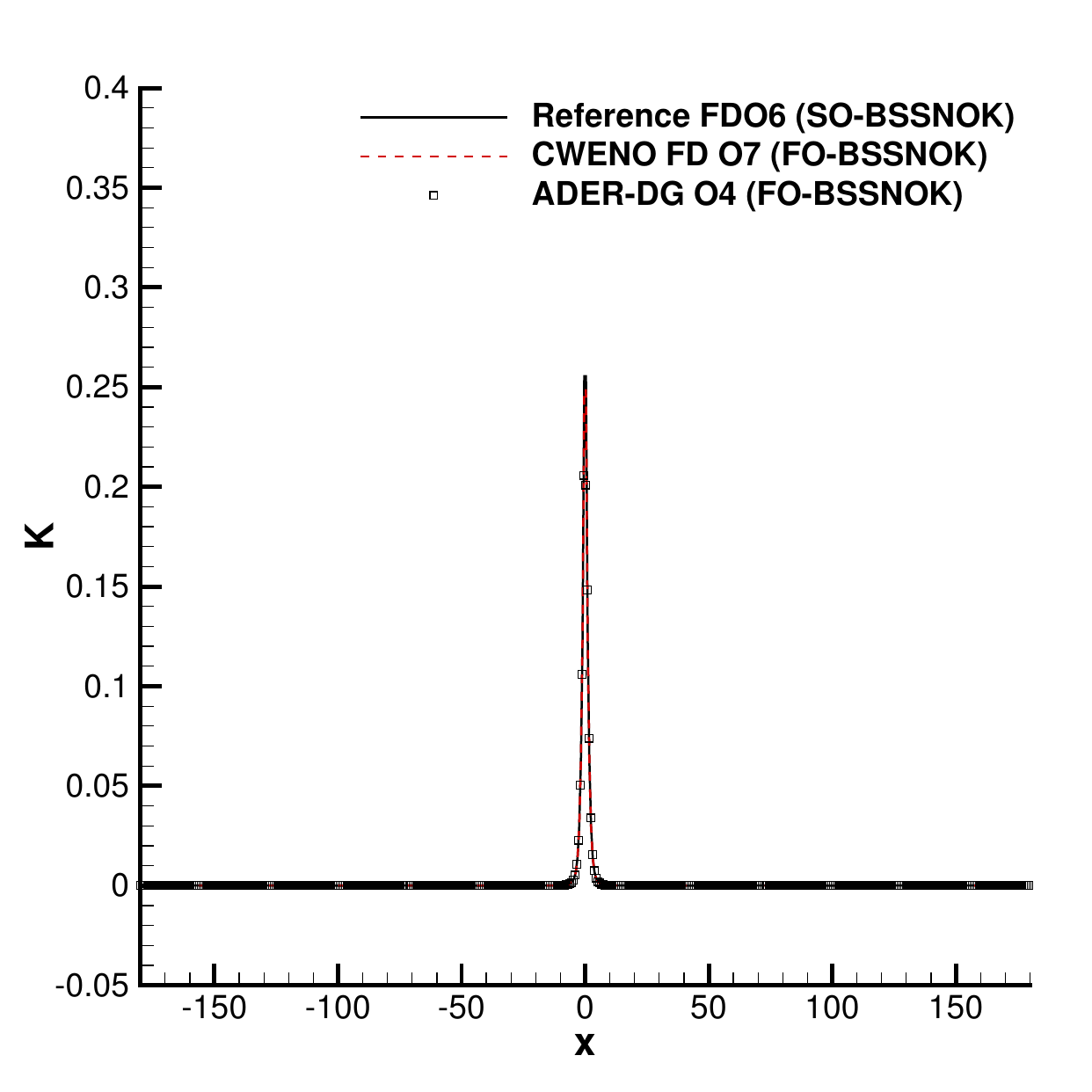} &
			\includegraphics[width=0.45\textwidth]{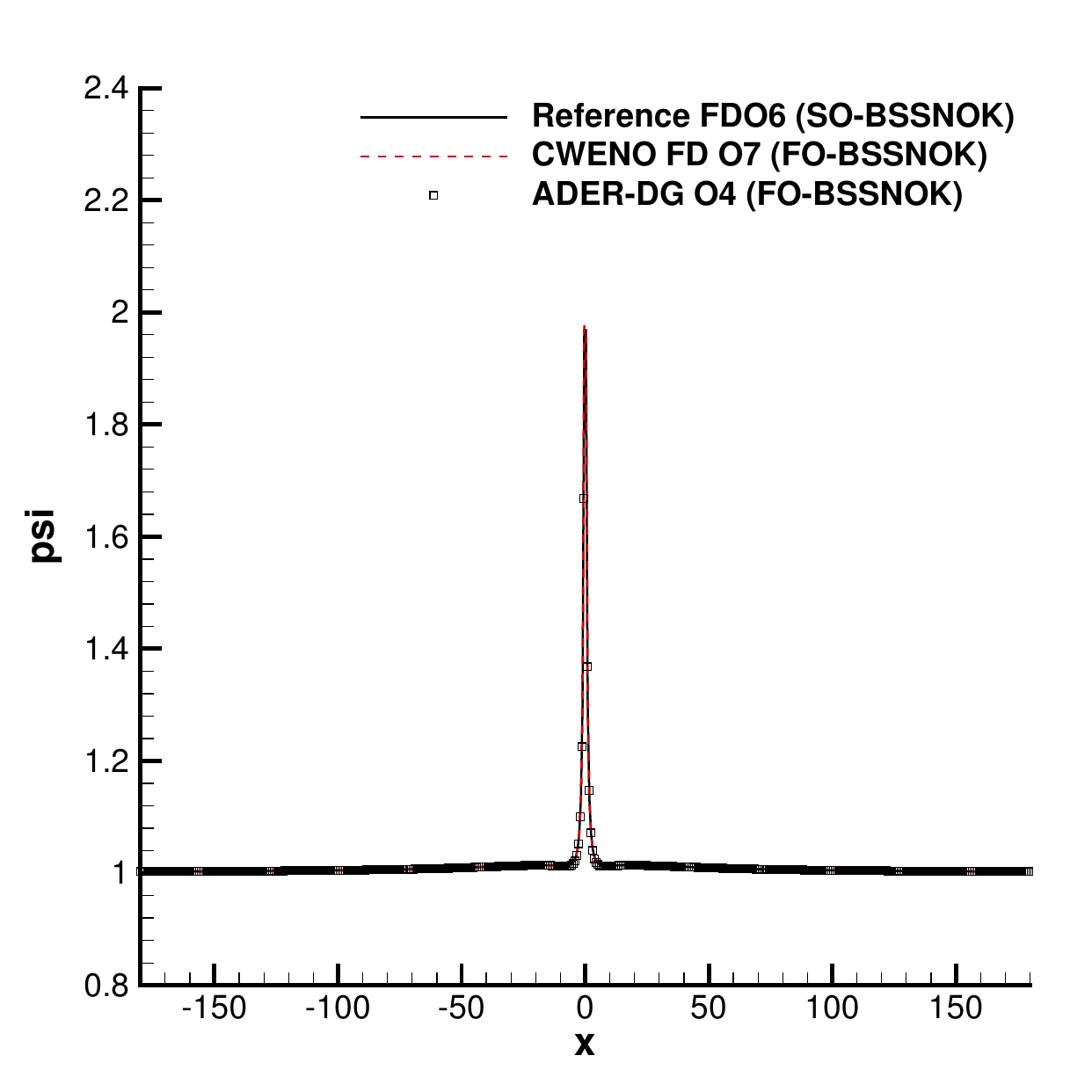} 
		\end{tabular}
		\caption{Numerical solution at time $t=500$ for a single puncture black hole obtained with the FO-BSSNOK system using a fourth order ADER-DG scheme ($N=3$) with AMR and subcell ADER-WENO FV limiter. Comparison of the DG scheme with classical sixth order central finite difference schemes applied to the standard second order formulation of BSSNOK (SO-BSSNOK) with conservative seventh order CWENO finite difference schemes \cite{CWENOBSSNOK} applied to the first order FO-BSSNOK system.  
			1D cuts along the $x_1$ axis are shown: lapse $\alpha$ (top left), shift vector component $\beta^1$ (top right), trace of the extrinsic curvature $K$ (bottom left) and conformal factor $\psi$ (bottom right). 		}
		\label{fig.OnePuncture}
	\end{center}
\end{figure}

\begin{figure}[!htbp]
	\begin{center}
		\begin{tabular}{cc}
			\includegraphics[width=0.45\textwidth]{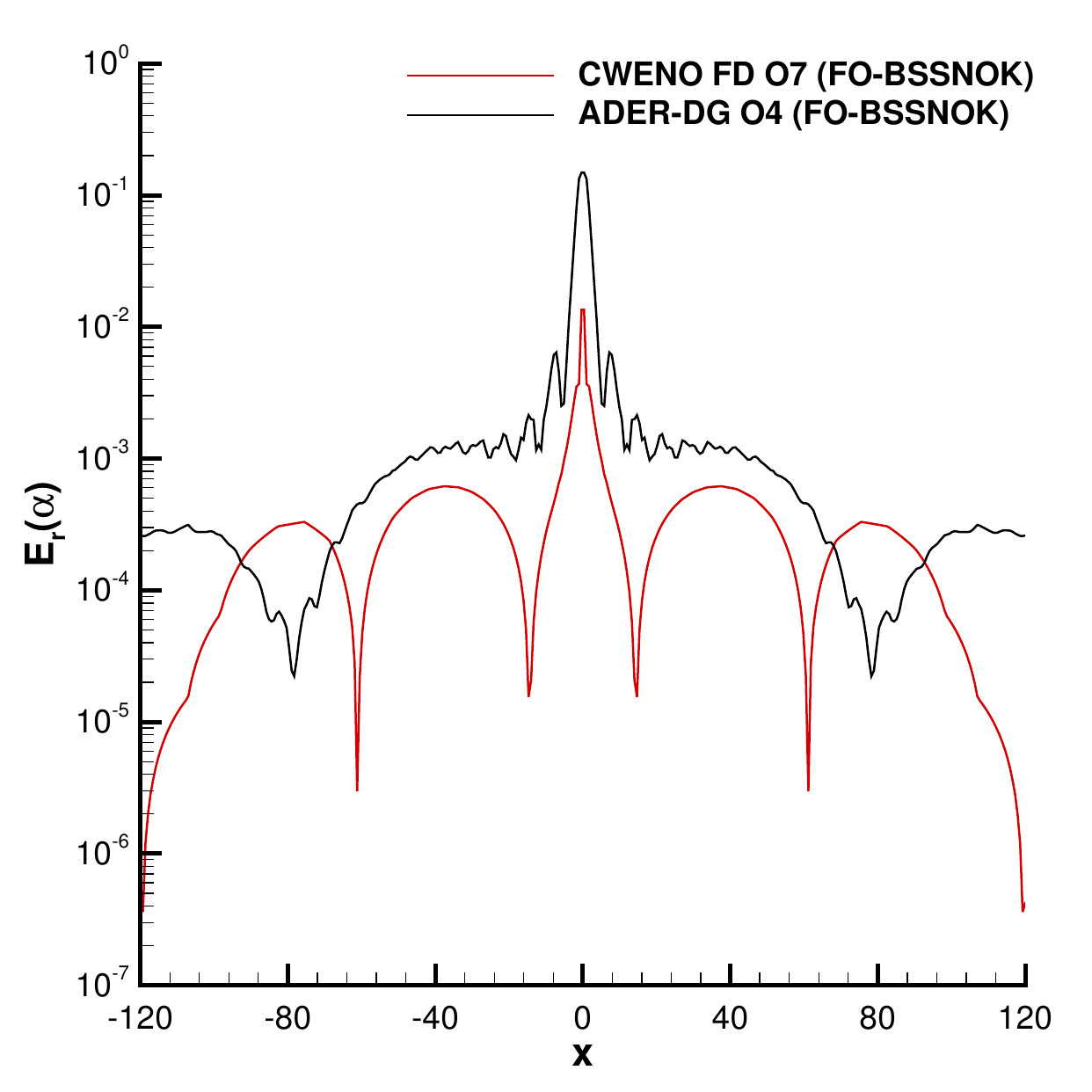} &
			\includegraphics[width=0.45\textwidth]{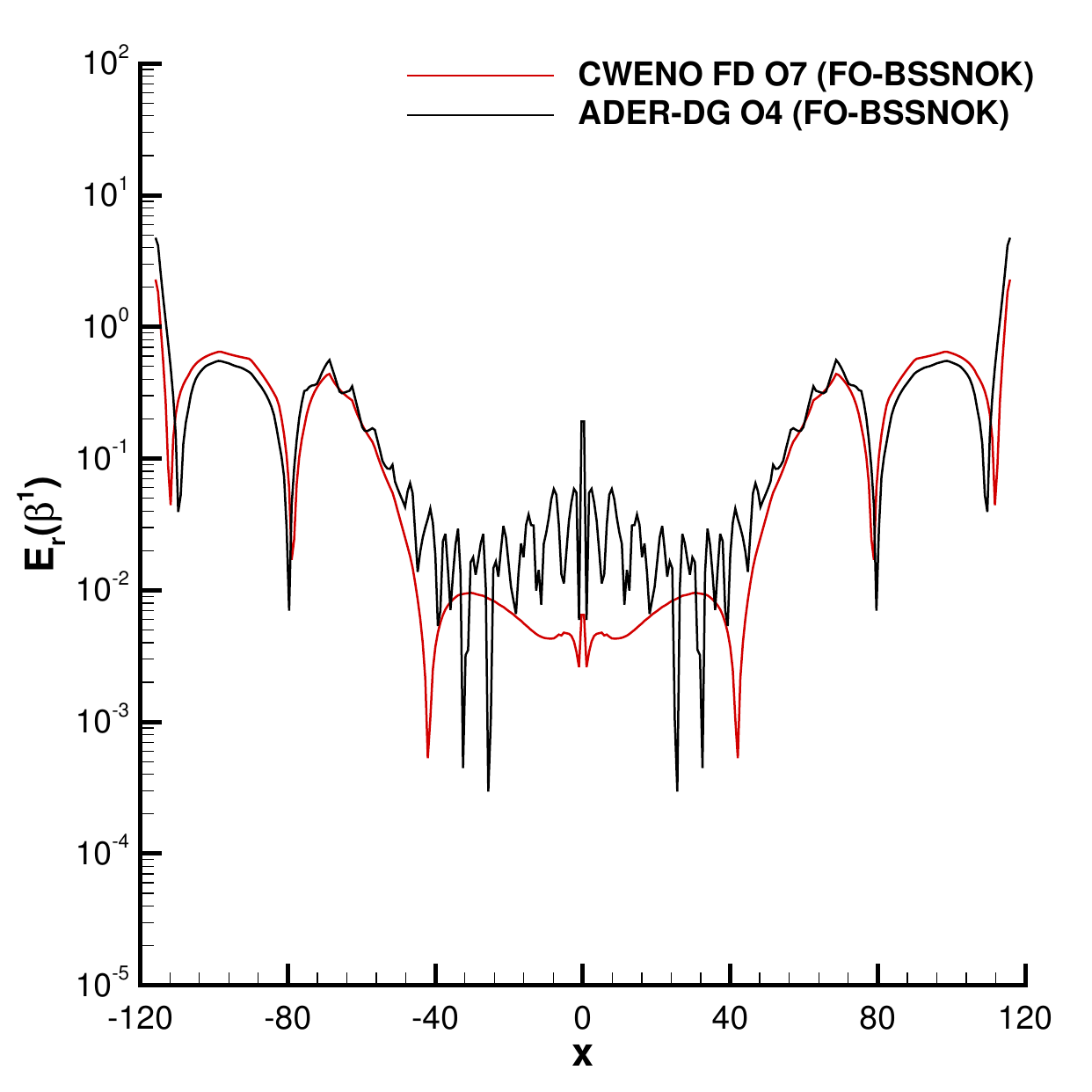} \\
			\includegraphics[width=0.45\textwidth]{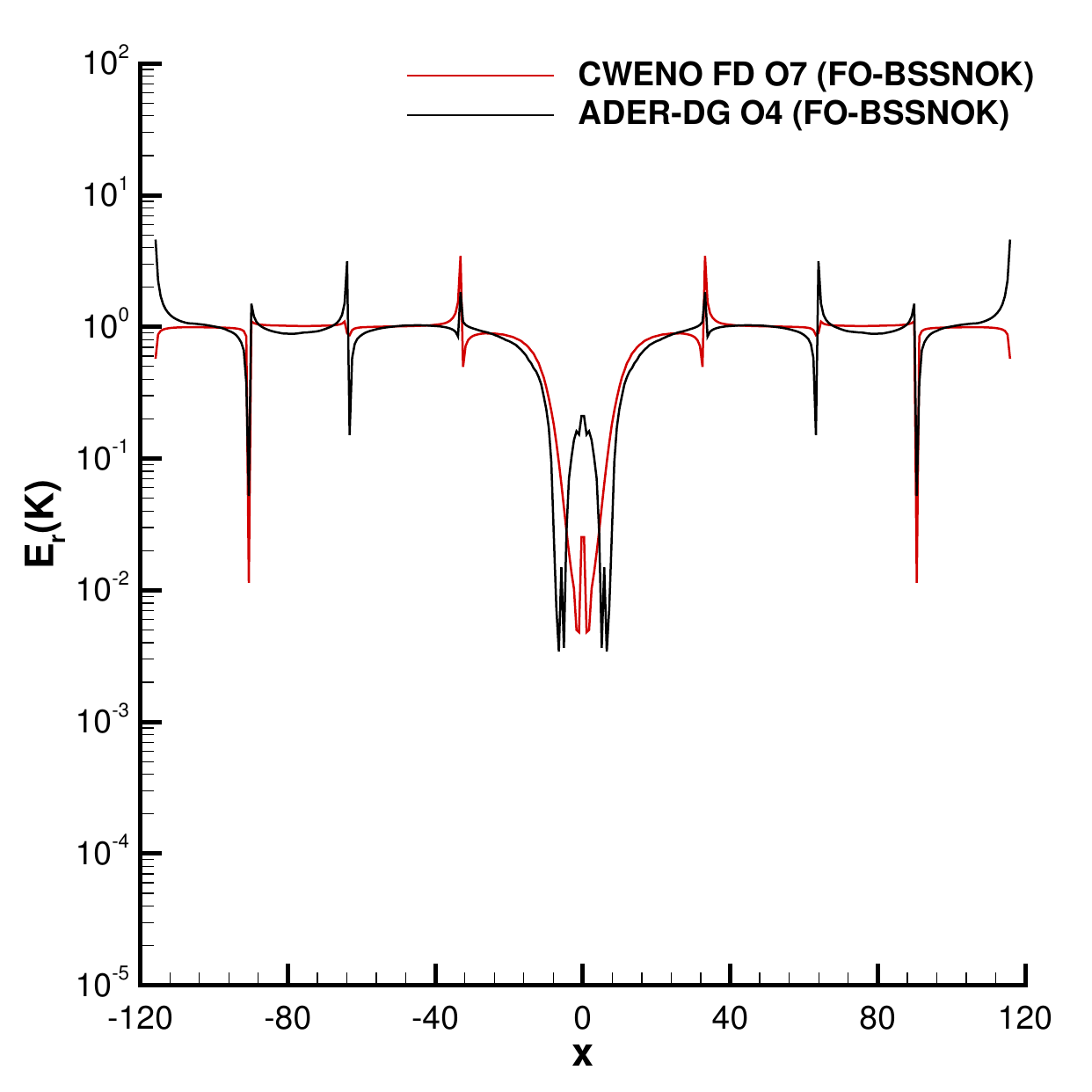} &
			\includegraphics[width=0.45\textwidth]{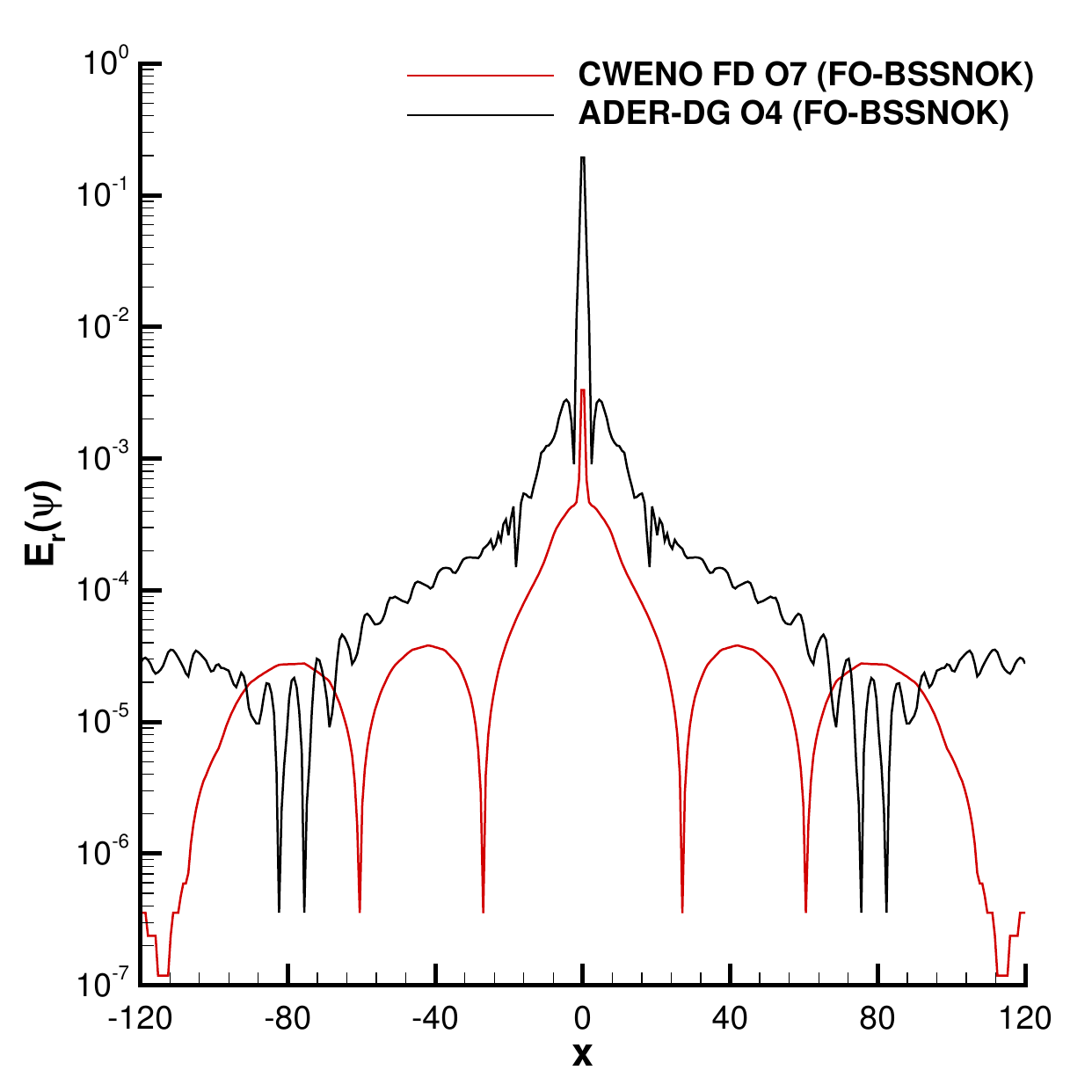} 
		\end{tabular}
		\caption{Numerical solution at time $t=500$ for a single puncture black hole obtained with the FO-BSSNOK system using a fourth order ADER-DG scheme ($N=3$) with AMR and subcell ADER-WENO FV limiter. Relative errors with respect to the SO-BSSNOK solution: lapse $\alpha$ (top left), shift vector component $\beta^1$ (top right), trace of the extrinsic curvature $K$ (bottom left) and conformal factor $\psi$ (bottom right). 		}
		\label{fig.OnePuncture.rel.err}
	\end{center}
\end{figure}

In this section we consider a single puncture black hole \cite{Brandt1997,Alcubierre2003,Alcubierre:2008} centered in the origin $\x_c=(0,0,0)$. The initial conformal metric is set to the identity matrix, i.e. $\tilde{\gamma}_{ij} = \boldsymbol{I}$, the initial conformal factor is $\psi=1+M/(2r)$, with $r = \left| \x-\x_c \right |$ and the lapse is initialized with $\alpha = \psi^{-2}$. The black hole has a unit mass of $M=1$ and zero spin. Initially the shift and the extrinsic curvature are zero. From there the auxiliary quantities of the FO-BSSNOK system $A_k$, $P_k$, $D_{kij}$ and $B_k^i$ can be easily obtained. 
The three dimensional computational domain is given by $\Omega = [-180, +180]^3$, which is direcretized with an adaptive Cartesian mesh with two levels of adaptive mesh refinement (AMR). The mesh on the coarsest level $\ell=0$ has $40^3$ elements, i.e. a uniform mesh spacing of $\Delta x^0_i = 9$. We use a refinement factor between two successive levels of $\mathfrak{r}=3$, see also \cite{AMR3DCL,ExaHype2020}.  
The mesh on the first refinement level, which is applied in the static box $[-36,+36]^3$ has a uniform mesh spacing of $\Delta x^1_i = 3$, while the inner box $[-12,+12]^3$ has a mesh spacing of $\Delta x^2_i = 1$. The third order ADER-WENO subcell finite volume limiter with $2N+1$ subcells per space dimension is applied in the innermost static area $[-3,+3]^3$, leading to a spatial resolution of $\Delta x^{FV}_i = 1/7$ around the black hole. The \emph{gamma--driver} is turned on, i.e. $s=1$, with parameters $\eta=2$ and $\mu=0$. We solved this test problem with a fourth order ADER-DG scheme ($N=3$) up to a final time of $t=500$.  
In Fig.~\ref{fig.OnePunctureSetup} we present a sketch of the AMR mesh setup and the time evolution of the ADM constraints.
Simulations are stable up to the final simulation time since the Einstein constraints show only a very small linear increase in time, which is acceptable over the timescales considered. 
In Fig. \ref{fig.OnePuncture} we present 1D cuts at time $t=500$ along the $x_1$ axis of some fundamental physical quantities, such as the lapse, the shift vector component $\beta^1$, the trace of the extrinsic curvature $K$, and the conformal factor $\psi$. One can clearly see that our high order ADER-DG scheme with subcell ADER-WENO FV limiter reaches the classical trumpet solution that is typical for a single puncture black hole. To verify our numerical results we also compare with a classical second order SO-BSSNOK solution obtained with a simple central finite difference scheme, as well as with the FO-BSSNOK solution obtained via the CWENO finite difference approach presented in \cite{CWENOBSSNOK} and \cite{Balsara2024c}. The corresponding relative errors with respect to the SO-BSSNOK solution are reported in Fig. \ref{fig.OnePuncture.rel.err}. We emphasize that also the SO-BSSNOK solution is only a numerical solution of the problem, and not an exact one. 
%
%
\subsection{Head on collision of two puncture black holes }
\label{sec:Head}
After the stable long-time evolution of a single puncture black hole shown in the previous section we now consider the head-on collision of two non--spinning puncture black holes. The two black holes have equal masses $m_1 = m_2 = 1$, are initially at rest with respect to each other and are placed at an initial distance of $d=2\,M$ in positions $\x_1 = (-1,0,0)$ and $\x_2 = (+1,0,0)$.   
The initial metric is set to $\tilde{\gamma}_{ij} = \boldsymbol{I}$ and the initial conformal factor is given by $\psi = 1 + \frac{m_1}{2 r_1} + \frac{m_2}{2 r_2}$, 
with the radial distances $r_i = \left\| \x - \x_i \right\|$, see also \cite{Alcubierre:2008}. Initially, the lapse is again simply set to $\alpha = \psi^{-2}$. 
The initial extrinsic curvature and the initial shift are set to zero, i.e. $K_{ij}=0$ and $\beta^i=0$. From these definitions the initial conditions for the auxiliary quantities are readily obtained.  

The three dimensional computational domain $\Omega = [-180, +180]^3$ is direcretized with a fourth order ADER-DG scheme ($N=3$) with third order ADER-WENO subcell FV limiter. We use an adaptive Cartesian mesh with two levels of adaptive mesh refinement (AMR). The mesh on the coarsest level $\ell=0$ has $40^3$ elements, i.e. a uniform mesh spacing of $\Delta x^0_i = 9$. We use a refinement factor between two successive levels of $\mathfrak{r}=3$, see also \cite{AMR3DCL,ExaHype2020}.  
The mesh on the first refinement level, which is applied in the static box $[-36,+36]^3$ has a uniform mesh spacing of $\Delta x^1_i = 3$, while the inner box $[-12,+12]^3$ has a mesh spacing of $\Delta x^2_i = 1$. The third order ADER-WENO subcell finite volume limiter with $2N+1$ subcells per space dimension is applied in the box $[-5,+5]^3$, leading to a spatial resolution of $\Delta x^{FV}_i = 1/7$ around the black holes.

The \emph{gamma--driver} is turned on ($s=1$) with $\eta = 2$ and $\mu=0$. Simulations are run until a final time of $t=200$, when the two black holes have already merged and converged to a final configuration consisting in a single black hole. 

\begin{figure}[!htbp]
\begin{center}
    \begin{tabular}{cc}
  \includegraphics[trim=10 10 10 10,clip,width=0.45\textwidth]{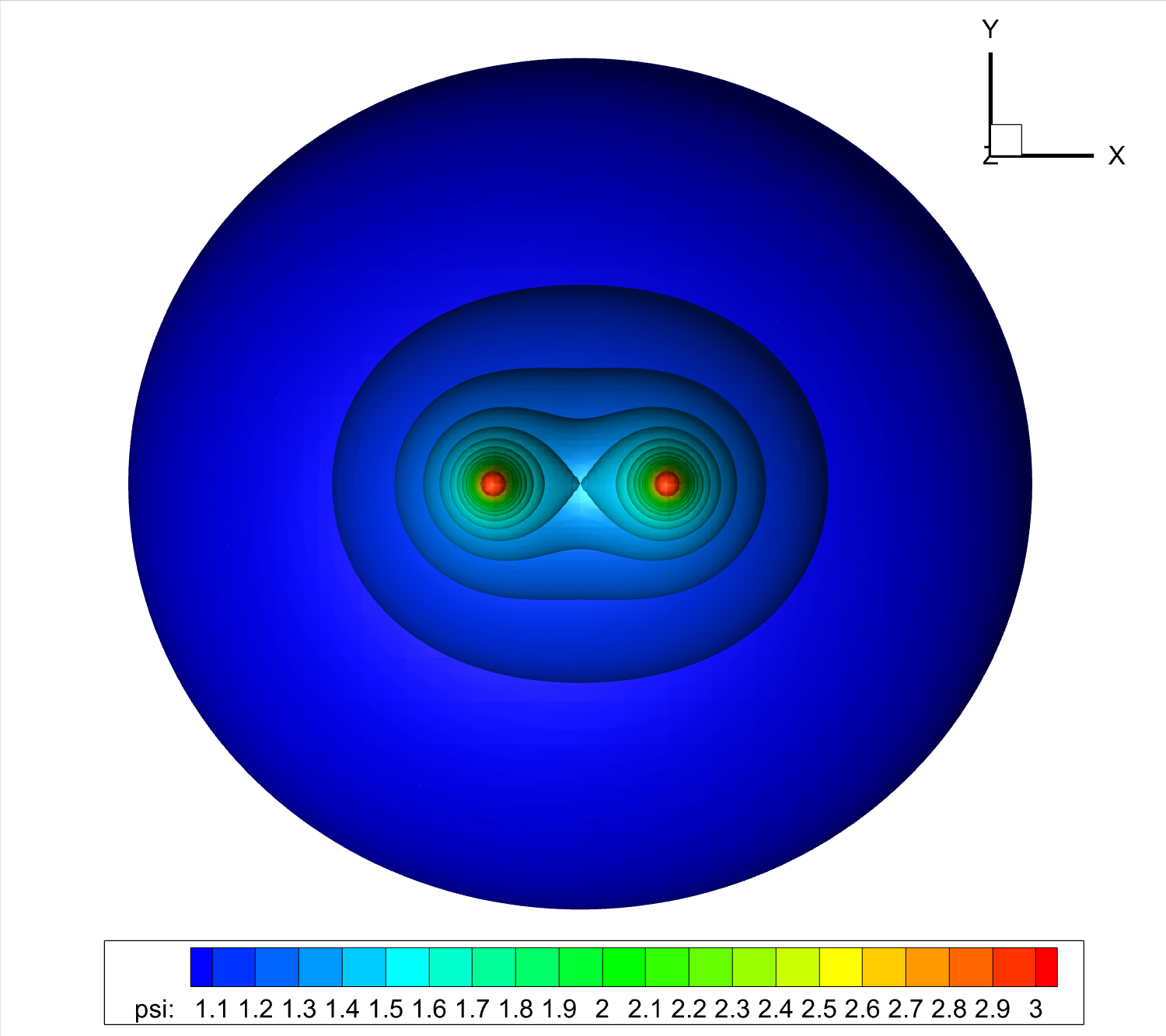} &
  \includegraphics[trim=10 10 10 10,clip,width=0.45\textwidth]{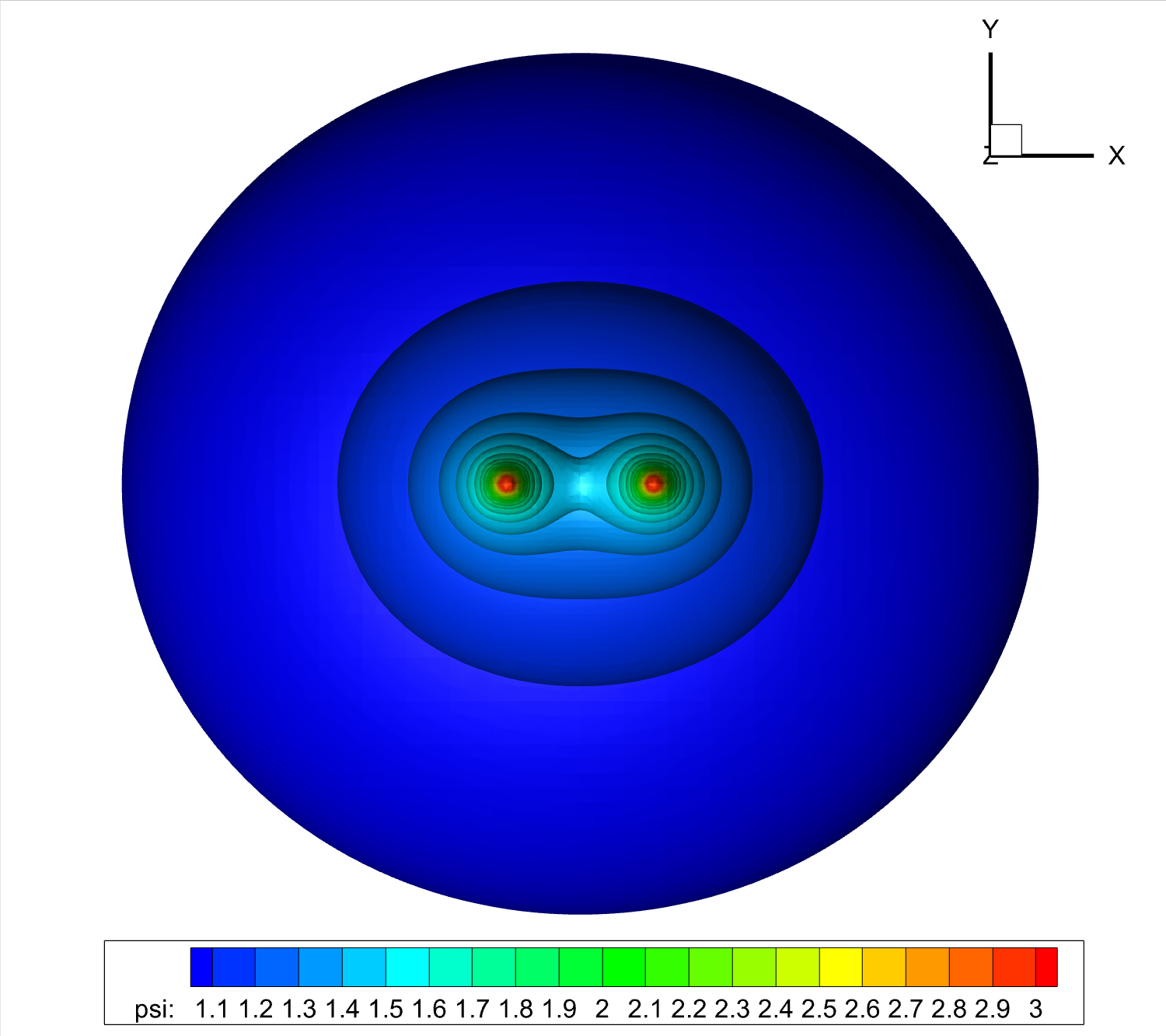} \\
  \includegraphics[trim=10 10 10 10,clip,width=0.45\textwidth]{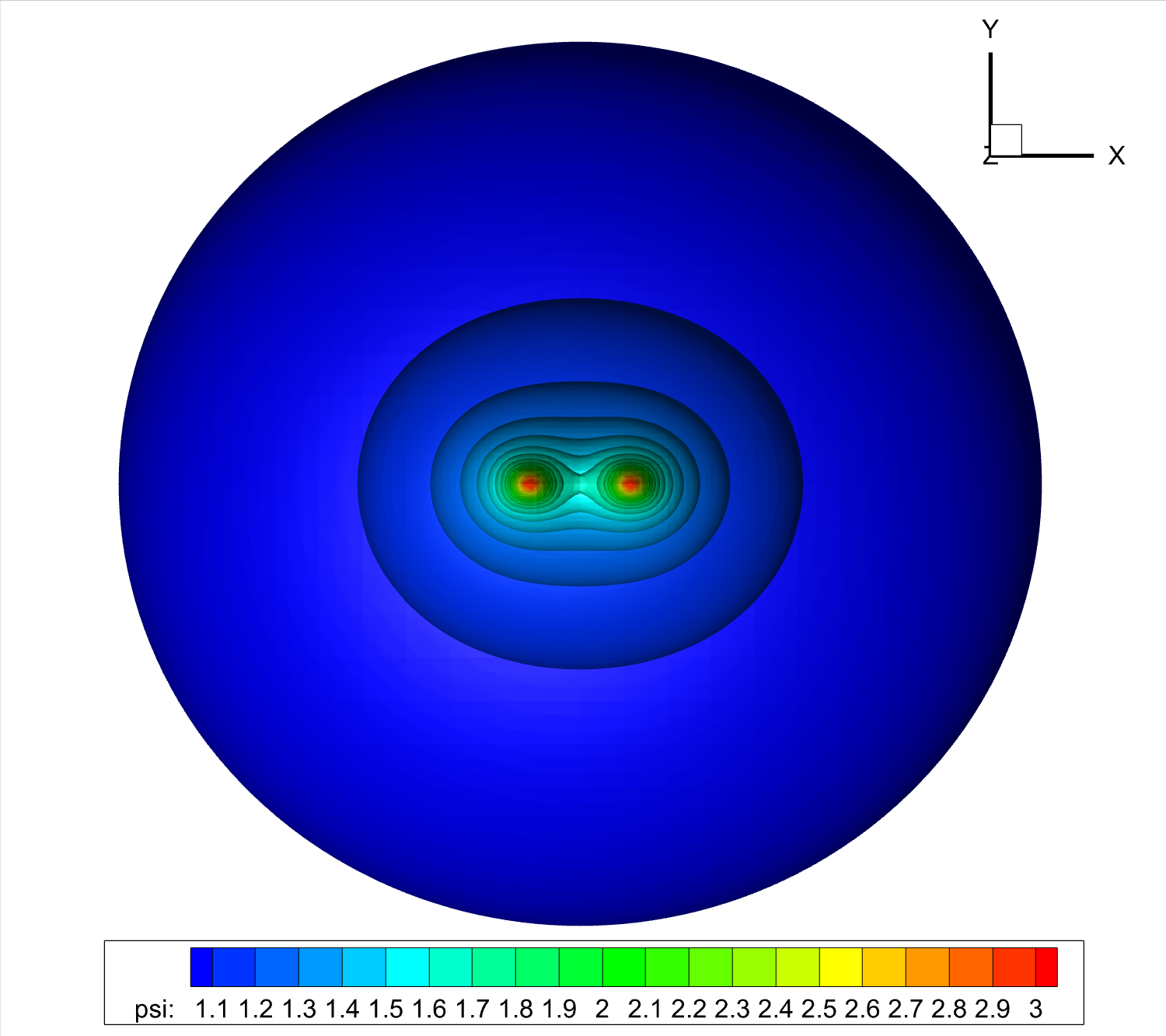} &
  \includegraphics[trim=10 10 10 10,clip,width=0.45\textwidth]{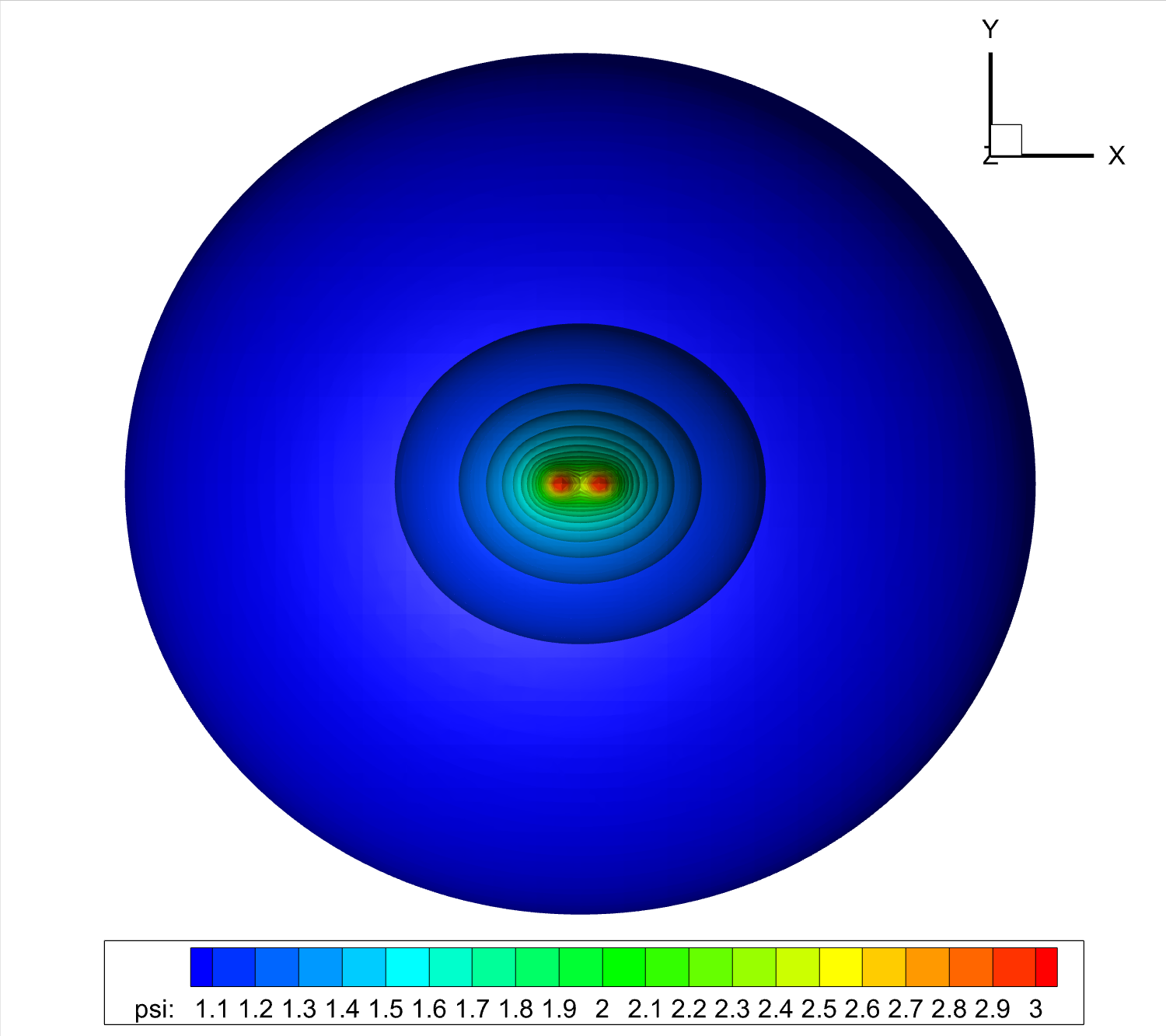} \\
  \includegraphics[trim=10 10 10 10,clip,width=0.45\textwidth]{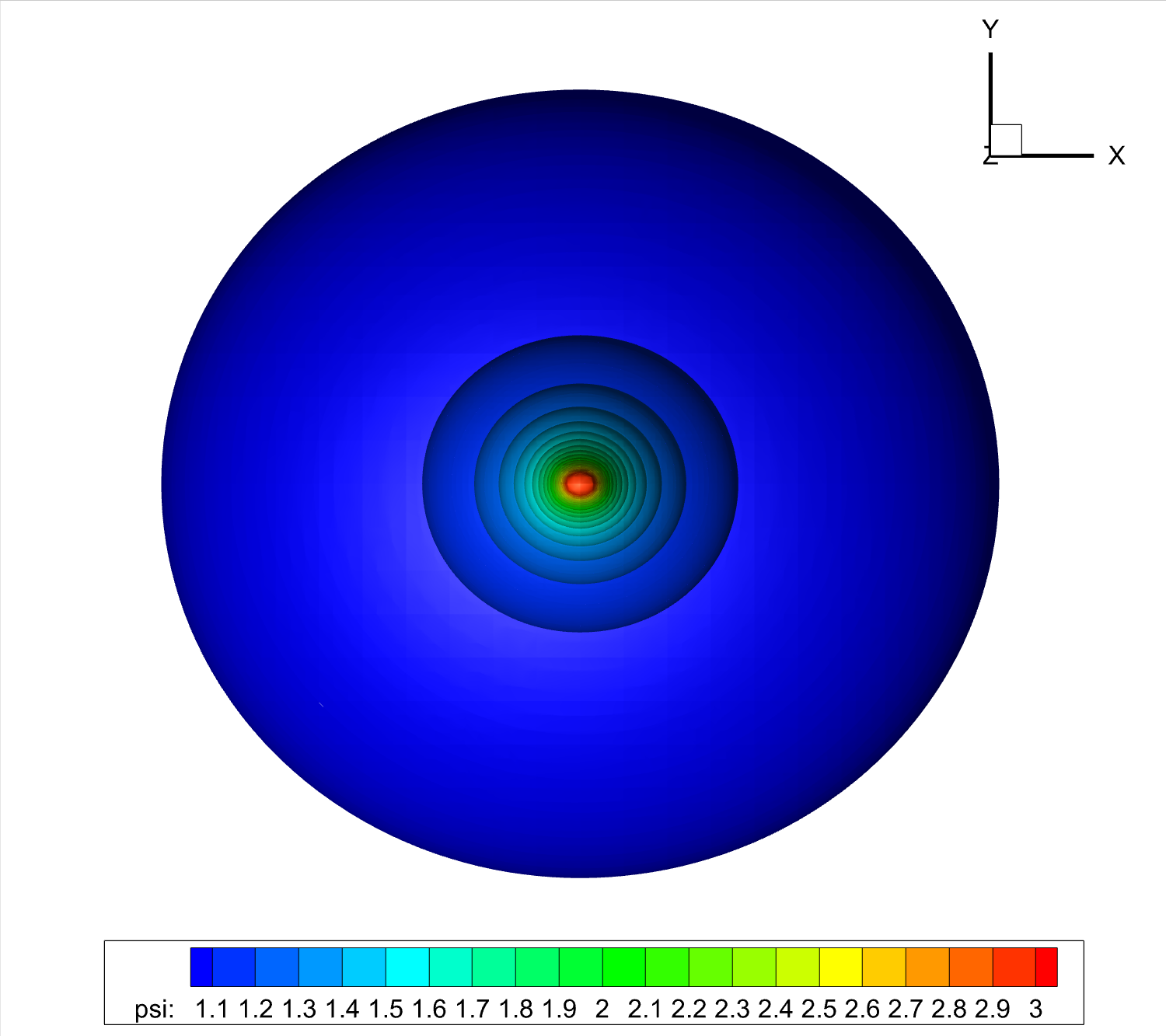} &
  \includegraphics[trim=10 10 10 10,clip,width=0.45\textwidth]{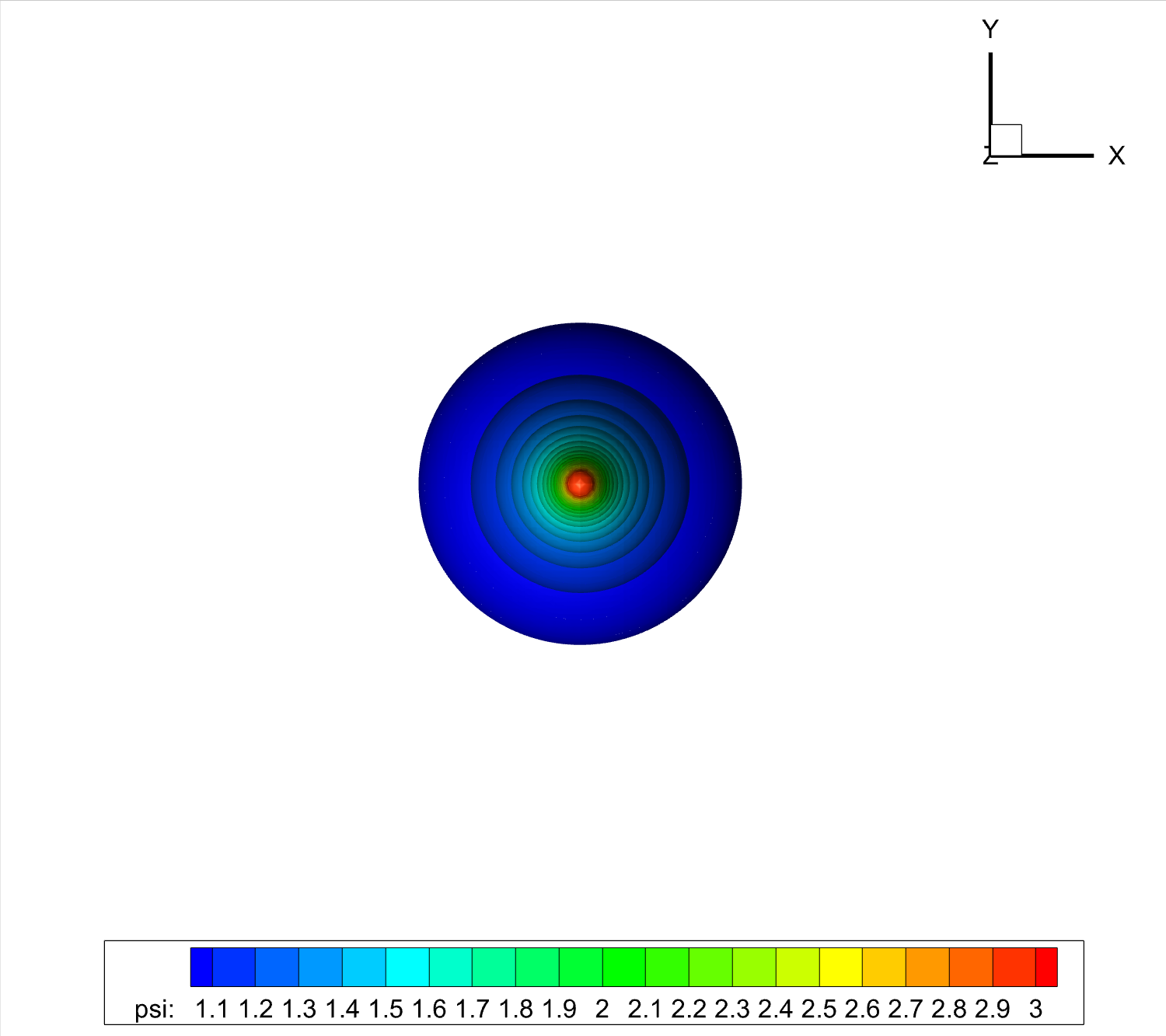} \\
    \end{tabular}
    \caption{Head on collision of two puncture black holes obtained with a fourth order ADER-DG scheme ($N=3$) and third order ADER-WENO subcell finite volume limiter. From top left to bottom right: $t=0$, $t=10$, $t=15$, $t=20$, $t=25$ and $t=100$. }
    \label{fig.HeadOn}
\end{center}
\end{figure}

In Fig.~\ref{fig.HeadOn} we show the contour plots of the scalar $\psi$  over the $z=0$ plane. The two black holes merge as expected and the simulation remains stable also after the merger, which can be seen via the corresponding temporal evolution of the Einstein constraints that is shown in the left panel of Fig.~\ref{fig.binaryConstraints}.  

\subsection{Inspiralling merger of two black holes}
\label{sec:BBHs}

In this last test case we show the potential of our new high order ADER-DG schemes to solve also  rotating binary systems of black holes. After the breakthrough calculations of  \cite{Pretorius2005,Campanelli2006,Baker2006}, 
numerical investigation of binary black hole mergers are routinely performed by several research groups,
having by now reached a high level of sophistication in terms of physical information that can be extracted
(see, among the other, \cite{Babiuc_2008,Reisswig2009,Hannam2010}).
In this Section we employ the procedure initially proposed by  \cite{Brandt1997,Ansorg:2004ds} to create the initial conditions for two moving puncture black holes without excision.  
Hence, using the TWOPUNCTURES library \cite{Ansorg:2004ds}, we place the two black holes at a distance of $d=5\,M$ among each other, with zero individual spins and opposite linear momenta along the $y$ direction, \ie $p_1=(0,0.5,0)$, $p_2=(0,-0.5,0)$. The masses of the two black holes are $m_1=m_2=1$. The initial shift is set to zero. The auxiliary quantities are computed via simple second order finite differences applied to the initial data for the metric, the conformal factor and the lapse. 
The \emph{gamma--driver} is again turned on, i.e. $s=1$, with damping parameter $\eta=2$ and $\mu=0$. For our simulations we employ a fourth order ADER-DG scheme ($N=3$). 
The computational domain is $\Omega=[-180,+180]^3$, which is direcretized with an adaptive Cartesian mesh (AMR) with two levels of refinement. The chosen refinement factor between two levels is $\mathfrak{r}=3$, see also \cite{AMR3DCL,ExaHype2020}. 
The mesh on the coarsest level $\ell=0$ has a uniform mesh spacing of $\Delta x^0_i = 9$, the mesh on the first refinement level, which is applied in the static box $[-36,+36]^3$ has a uniform mesh spacing of $\Delta x^1_i = 3$, while the inner box $[-12,+12]^3$ has a mesh spacing of $\Delta x^2_i = 1$. The third order ADER-WENO subcell finite volume limiter with $2N+1$ subcells per space dimension is applied in the static area $[-5,+5]^3$, leading to a spatial resolution of $\Delta x^{FV}_i = 1/7$ around the black holes.   

\begin{figure}[!htbp]
\begin{center}
    \begin{tabular}{cc}
	\includegraphics[trim=10 10 10 10,clip,width=0.4\textwidth]{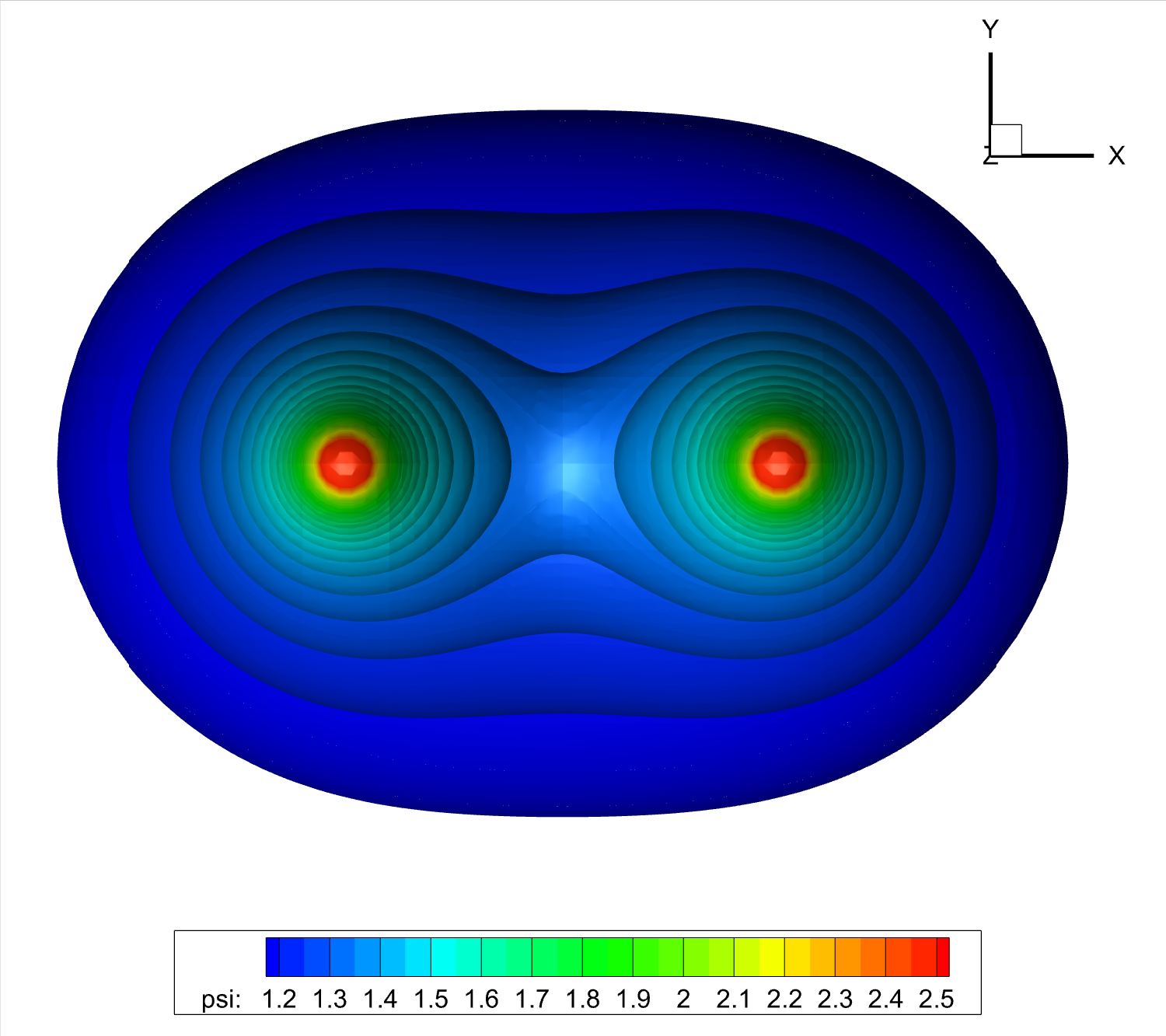} &
	\includegraphics[trim=10 10 10 10,clip,width=0.4\textwidth]{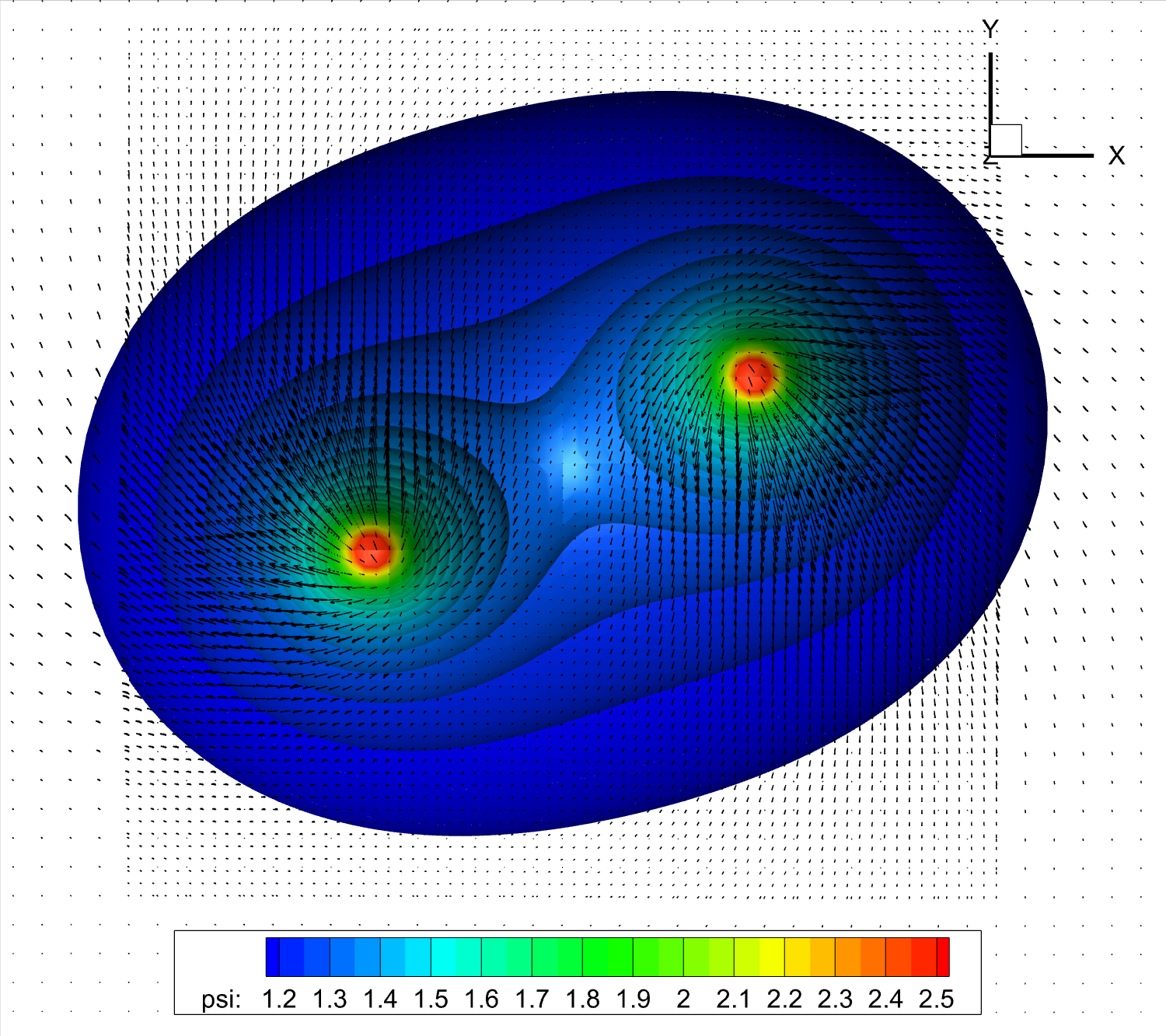} \\
	\includegraphics[trim=10 10 10 10,clip,width=0.4\textwidth]{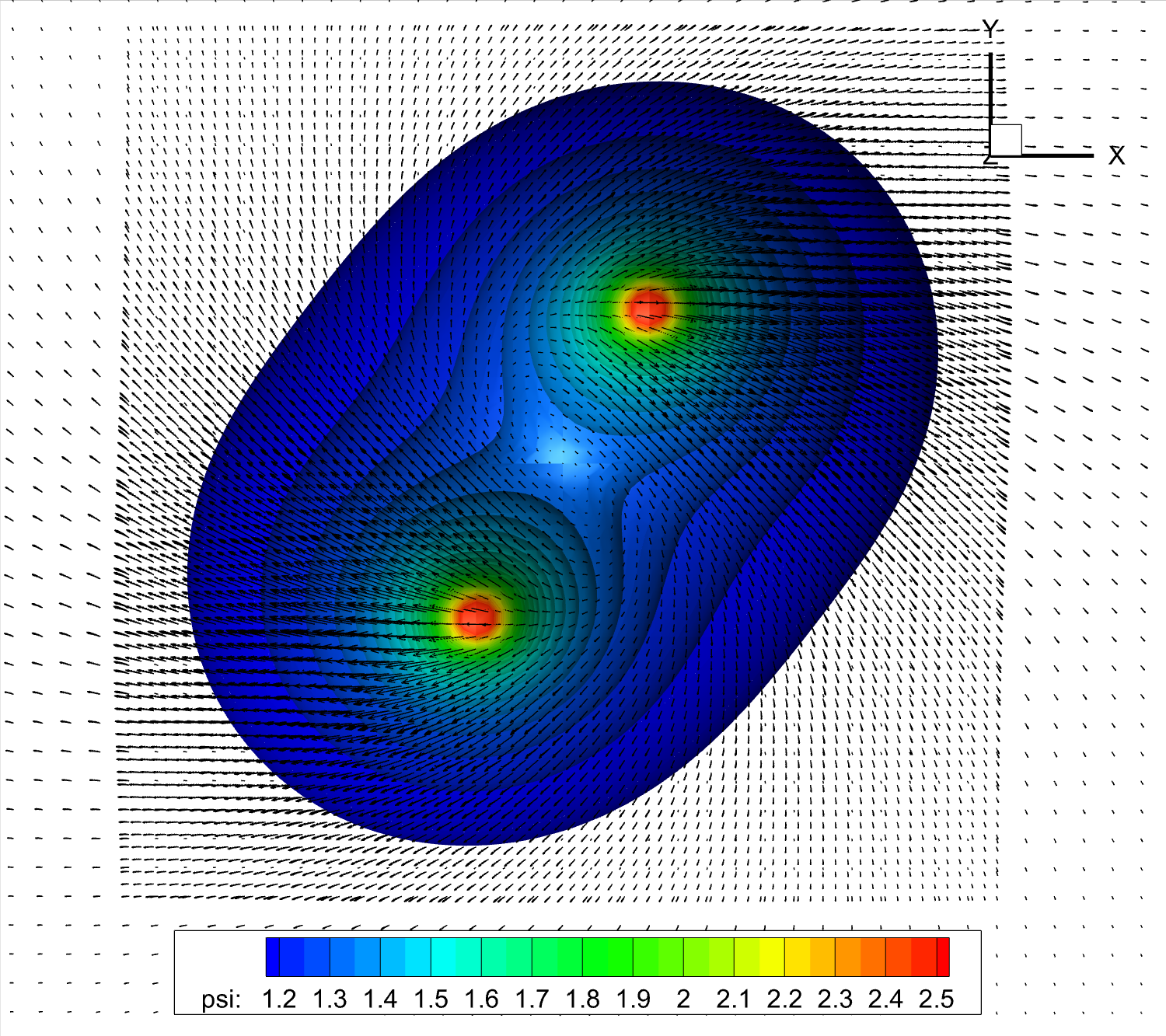} &
	\includegraphics[trim=10 10 10 10,clip,width=0.4\textwidth]{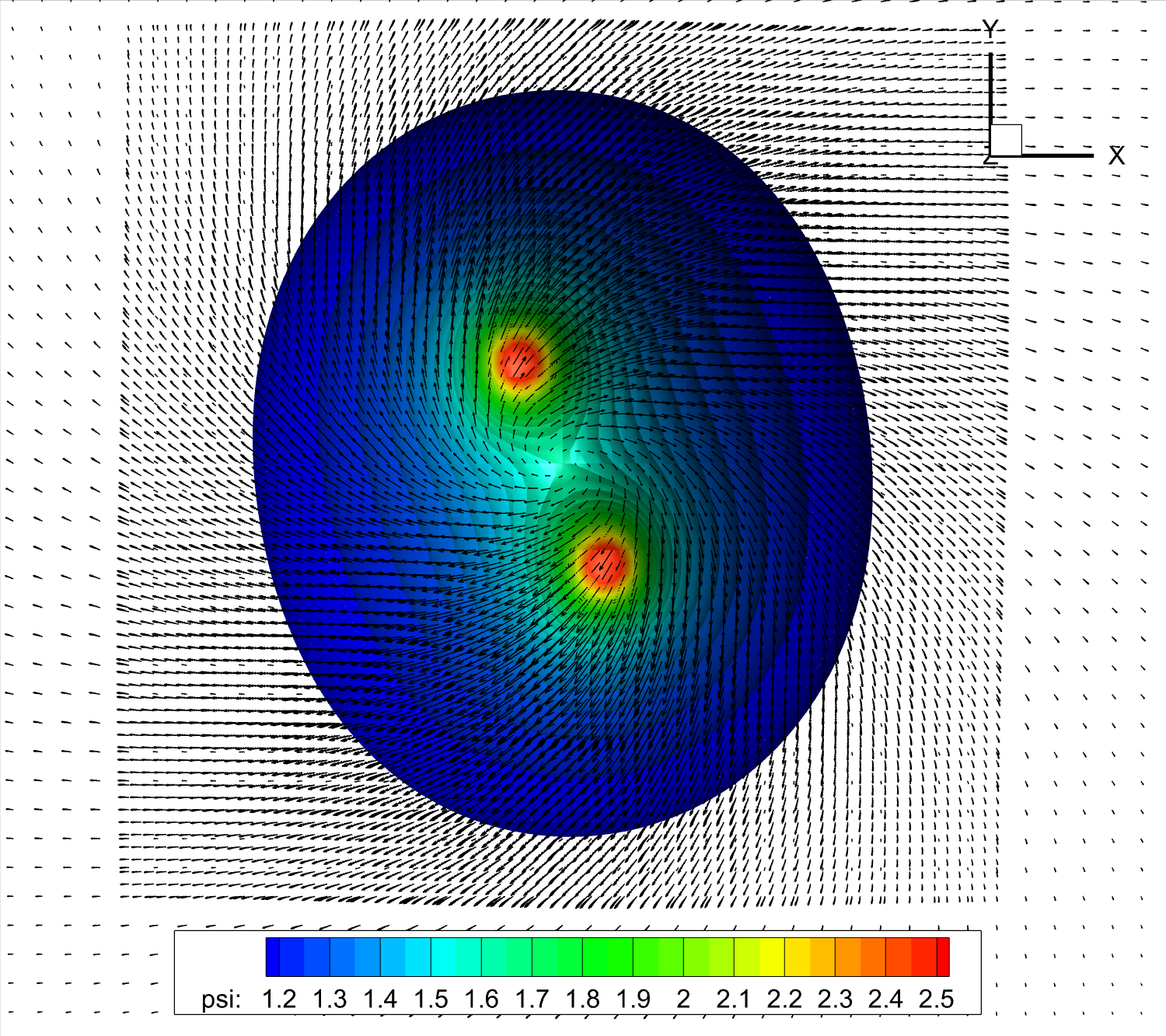} \\
	\includegraphics[trim=10 10 10 10,clip,width=0.4\textwidth]{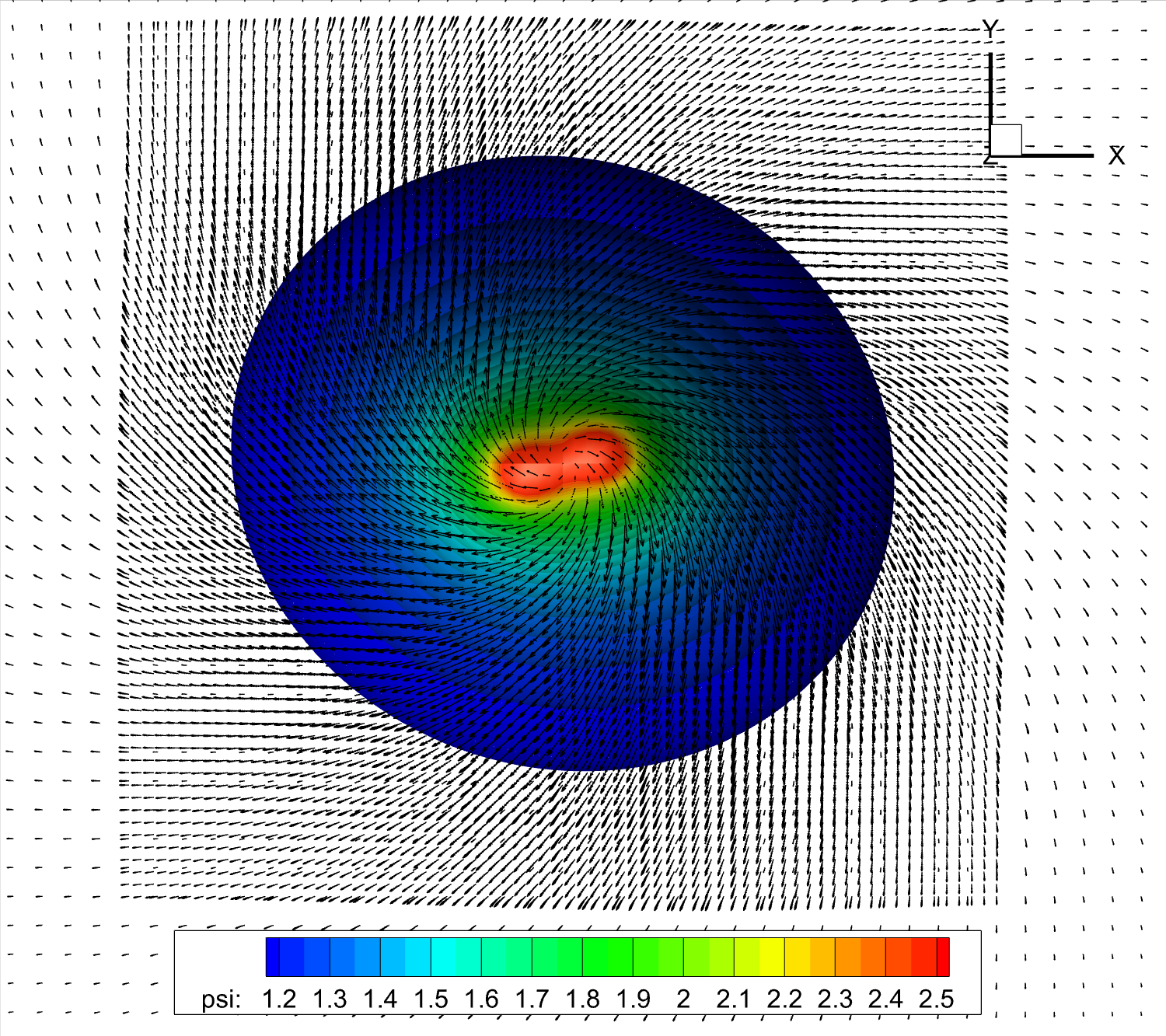} &
	\includegraphics[trim=10 10 10 10,clip,width=0.4\textwidth]{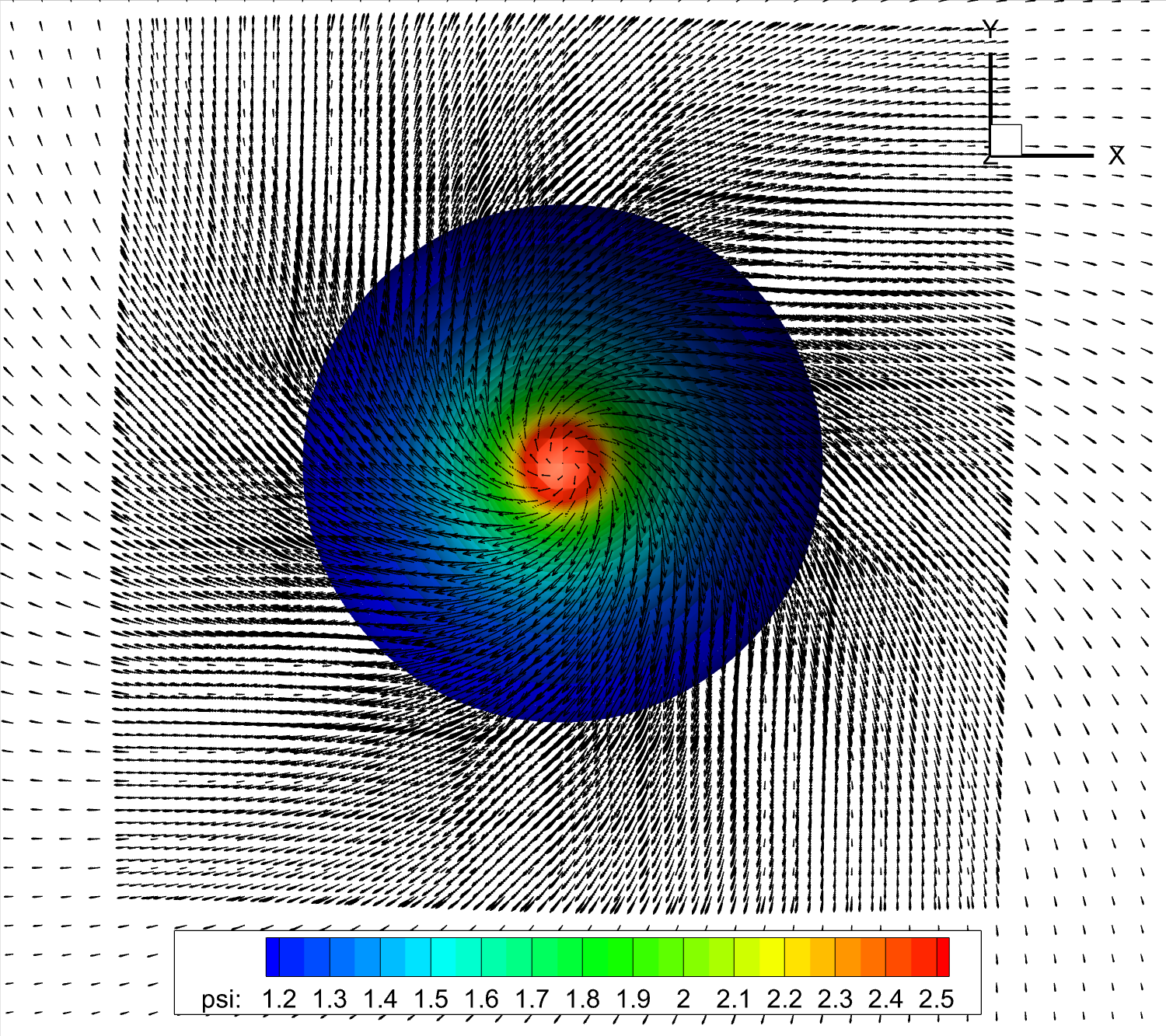} 
    \end{tabular}
    \caption{Inspiralling merger of two black holes obtained with a fourth order ADER-DG scheme ($N=3$) with third order ADER-WENO-FV subcell limiter on an adaptive mesh with two levels of refinement. Time evolution of the iso-surfaces of the conformal factor $\psi$ and of the shift vector. From top left to bottom right: $t=0$, $t=10$, $t=20$, $t=30$, $t=40$ and $t=100$. }
    \label{fig.BinaryMerger}
\end{center}
\end{figure}

\begin{figure}[!htbp]
\begin{center}
    \begin{tabular}{cc}
    \includegraphics[width=0.45\textwidth]{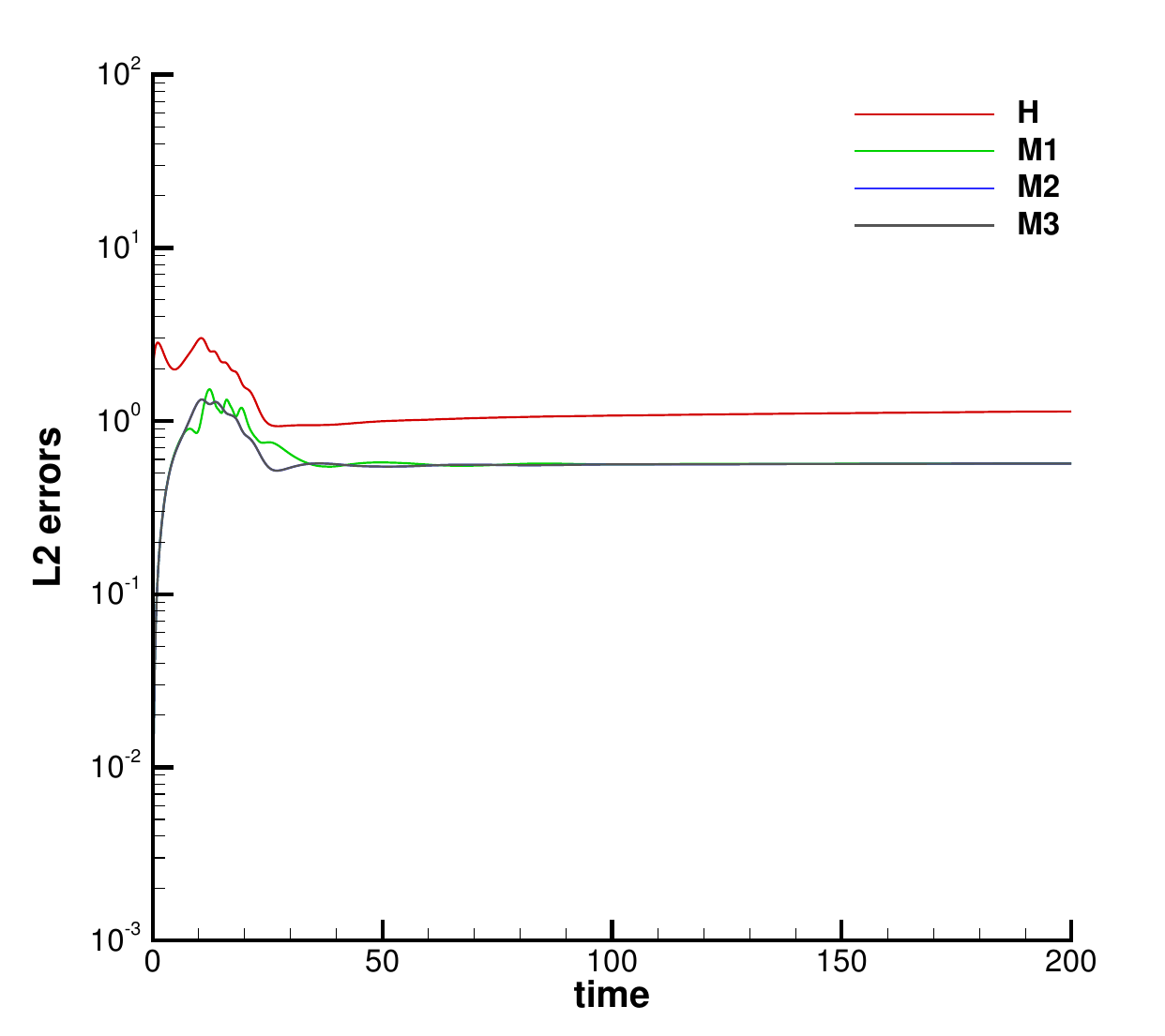} &
    \includegraphics[width=0.45\textwidth]{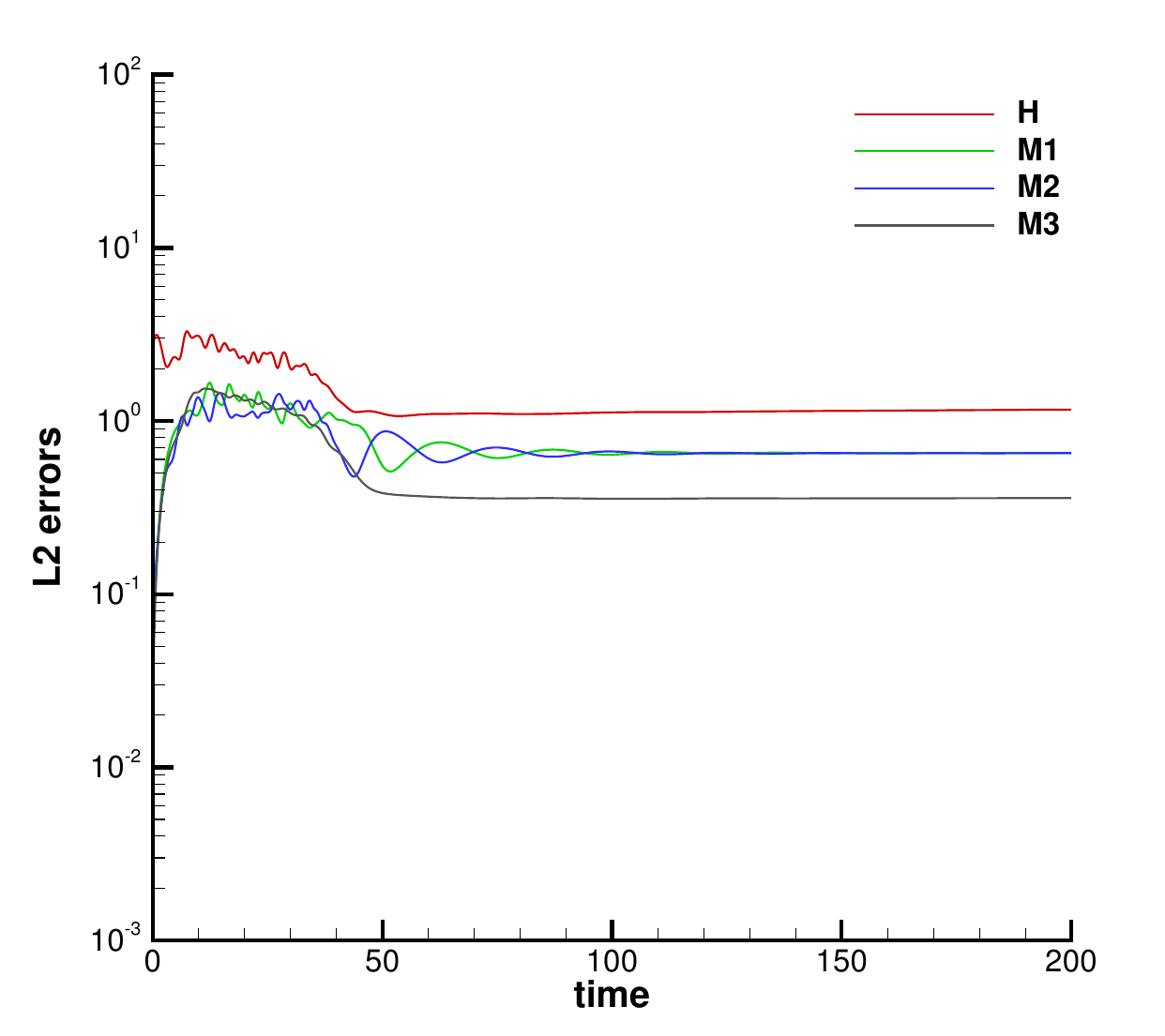} \\
    \end{tabular}
	\caption{Time evolution of the Einstein constraints.
	Left: Head on collision of Fig.~\ref{fig.HeadOn}.
	Right: Inspiralling merger of Fig.~\ref{fig.BinaryMerger}
	}
    \label{fig.binaryConstraints}
\end{center}
\end{figure}

The simulation is run until a final time of $t=200$. In Fig.~\ref{fig.BinaryMerger} we show the iso-contour surfaces of the scalar $\psi$  in the $z=0$ plane, together with the shift vector.  
The two black holes merge after a brief inspiral phase and at time $t=50$ a single black hole is already formed. 
The corresponding evolution of the Einstein constraints is instead visible in the right panel of 
Fig.~\ref{fig.binaryConstraints}. The ADM constraints show that the simulation is perfectly stable until the final simulation time and the obtained results are similar to those reported in \cite{CWENOBSSNOK}. 

%

\section{Conclusions}
\label{sec:conclusions}
In this work we have presented an ADER-DG scheme that solves the Einstein--Euler equations 
written as a single PDE system that is  based on  the first--order BSSNOK formulation recently proposed by \cite{CWENOBSSNOK}. Such a formulation is strongly hyperbolic, namely with all the eigenvalues reals and with a complete set of eigenvectors,
and this motivated us to consider Discontinuous Galerkin methods for its solution.
We recall, in fact, that, because of their very little numerical diffusivity, DG schemes are extremely sensitive to any minimal loss of strong hyperbolicity. Our implementation of DG schemes follows the ADER approach
for the time integration, which, by means of a local predictor \cite{DumbserEnauxToro}, allows to
advance the solution in time by a single step,  avoiding to go through the sequence of sub-steps typical of Runge--Kutta DG schemes. The resulting numerical scheme can reach an arbitrary order of accuracy both in space and in time,
being ultimately limited only by computational resources. In our DG scheme we take care of the Gibbs phenomenon by means of a subcell finite volume limiter. Numerical oscillations would instead be
produced either by true discontinuities that form in the matter part or by
steep gradients that are generated close to spacetime singularities. This approach was proposed in \cite{Dumbser2014} and it has already proved to perform quite well in the relativistic regime \cite{Zanotti2015d,Dumbser2017strongly}. Finally, our computational infrastructure allows for Adaptive Mesh Refinement (AMR) and Well Balancing (WB). The former
was originally incorporated within an ADER scheme by \cite{AMR3DCL} following a cell-by-cell refinement criterion, while the latter follows the same logic as that presented in \cite{DumbserZanottiGaburroPeshkov2023}, and it allows to preserve the stationarity of equilibrium solutions up to machine precision. 

With all these numerical tools at disposal, our high order ADER-DG schemes can successfully solve the classical tests of numerical relativity, including those that involve a matter contribution, that we have modeled through a perfect fluid with an ideal gas equation of state.  
A particular feature of our approach is the monolithic formulation of the fully coupled Einstein-Euler system via the use of one single first order strongly hyperbolic PDE system that allows the application of one and the same numerical scheme within one and the same computer code for the coupled simulation of both matter and spacetime. In terms of CPU-time and accuracy comparison, our ADER-DG becomes hundreds of time slower (at fifth order) than a finite difference CWENO scheme~\cite{CWENOBSSNOK}, if the
same number of grid elements is adopted.  
However, it correspondingly reduces the error by 4 orders of magnitude.
Conversely, if the comparison is done keeping the same total number of degrees of freedom, ADER-DG is as fast as its FD competitor, while still gaining one order of magnitude in accuracy. 

As an important achievement of this work, we can solve the inspiralling merger of two black holes, a result that, to the best of our knowledge, has never been obtained before with high order Discontinuous Galerkin schemes. For a specific test, namely the single puncture black hole, we have also performed a thorough comparison among three different approaches: (i) our new DG scheme applied to the FO-BSSNOK formulation, (ii) a simpler central WENO finite difference scheme, again with the FO-BSSNOK formulation as presented in \cite{CWENOBSSNOK}, and (iii) a traditional central finite difference scheme according to the widely adopted second--order formulation of BSSNOK. All of the them reproduce the same profile to a high accuracy, serving as a corroboration of the current DG scheme. The present work offers a number of natural extensions, the most attractive of which is  the computation of gravitational waveforms from astrophysical sources. This analysis will be the subject of a dedicated investigation.


\section{Acknowledgments}

This work was financially supported by the Italian
Ministry of Education, University and Research (MIUR)
in the framework of the PRIN 2022 project \textit{High order
structure-preserving semi-implicit schemes for hyperbolic
equations} , by the European Union—Next Generation EU,
Mission 4 Component 2-CUP E53D23005840006, and via
the Departments of Excellence Initiative 2018–2027 attrib-
uted to DICAM of the University of Trento (grant L. 232/
2016). M. D. was also funded by the European Union
NextGenerationEU project PNRR Spoke 7 CN HPC and by
the European Research Council (ERC) under the European
Union s Horizon 2020 research and innovation pro-
gramme, Grant agreement No. ERC-ADG-2021-
101052956-BEYOND. Views and opinions expressed are
however those of the author(s) only and do not necessarily
reflect those of the European Union or the European
Research Council. Neither the European Union nor the
granting authority can be held responsible for them. M. D.
and I. P. are members of the INdAM GNCS group in Italy.
We acknowledge the CINECA award under the ISCRA
initiative, for the availability of high–performance compu-
tational resources and support. The authors declare that not
a single line of this paper has been written with AI.



\bibliographystyle{apsrev4-1} 
\bibliography{referencesZ4}

\begin{thebibliography}{135}%
\makeatletter
\providecommand \@ifxundefined [1]{%
 \@ifx{#1\undefined}
}%
\providecommand \@ifnum [1]{%
 \ifnum #1\expandafter \@firstoftwo
 \else \expandafter \@secondoftwo
 \fi
}%
\providecommand \@ifx [1]{%
 \ifx #1\expandafter \@firstoftwo
 \else \expandafter \@secondoftwo
 \fi
}%
\providecommand \natexlab [1]{#1}%
\providecommand \enquote  [1]{``#1''}%
\providecommand \bibnamefont  [1]{#1}%
\providecommand \bibfnamefont [1]{#1}%
\providecommand \citenamefont [1]{#1}%
\providecommand \href@noop [0]{\@secondoftwo}%
\providecommand \href [0]{\begingroup \@sanitize@url \@href}%
\providecommand \@href[1]{\@@startlink{#1}\@@href}%
\providecommand \@@href[1]{\endgroup#1\@@endlink}%
\providecommand \@sanitize@url [0]{\catcode `\\12\catcode `\$12\catcode
  `\&12\catcode `\#12\catcode `\^12\catcode `\_12\catcode `\%12\relax}%
\providecommand \@@startlink[1]{}%
\providecommand \@@endlink[0]{}%
\providecommand \url  [0]{\begingroup\@sanitize@url \@url }%
\providecommand \@url [1]{\endgroup\@href {#1}{\urlprefix }}%
\providecommand \urlprefix  [0]{URL }%
\providecommand \Eprint [0]{\href }%
\providecommand \doibase [0]{http://dx.doi.org/}%
\providecommand \selectlanguage [0]{\@gobble}%
\providecommand \bibinfo  [0]{\@secondoftwo}%
\providecommand \bibfield  [0]{\@secondoftwo}%
\providecommand \translation [1]{[#1]}%
\providecommand \BibitemOpen [0]{}%
\providecommand \bibitemStop [0]{}%
\providecommand \bibitemNoStop [0]{.\EOS\space}%
\providecommand \EOS [0]{\spacefactor3000\relax}%
\providecommand \BibitemShut  [1]{\csname bibitem#1\endcsname}%
\let\auto@bib@innerbib\@empty
\bibitem [{\citenamefont {L{\"u}ck}\ \emph {et~al.}(2022)\citenamefont
  {L{\"u}ck}, \citenamefont {Smith},\ and\ \citenamefont {Punturo}}]{Luck2022}%
  \BibitemOpen
  \bibfield  {author} {\bibinfo {author} {\bibfnamefont {H.}~\bibnamefont
  {L{\"u}ck}}, \bibinfo {author} {\bibfnamefont {J.}~\bibnamefont {Smith}}, \
  and\ \bibinfo {author} {\bibfnamefont {M.}~\bibnamefont {Punturo}},\
  }\enquote {\bibinfo {title} {Third-generation gravitational-wave
  observatories},}\ in\ \href {\doibase 10.1007/978-981-16-4306-4_7} {\emph
  {\bibinfo {booktitle} {Handbook of Gravitational Wave Astronomy}}}\ (\bibinfo
   {publisher} {Springer Nature Singapore},\ \bibinfo {address} {Singapore},\
  \bibinfo {year} {2022})\ pp.\ \bibinfo {pages} {283--300}\BibitemShut
  {NoStop}%
\bibitem [{\citenamefont {{Thornburg}}(2007)}]{Thornburg2007LRR}%
  \BibitemOpen
  \bibfield  {author} {\bibinfo {author} {\bibfnamefont {J.}~\bibnamefont
  {{Thornburg}}},\ }\href {\doibase 10.12942/lrr-2007-3} {\bibfield  {journal}
  {\bibinfo  {journal} {Living Reviews in Relativity}\ }\textbf {\bibinfo
  {volume} {10}},\ \bibinfo {eid} {3} (\bibinfo {year} {2007})}\BibitemShut
  {NoStop}%
\bibitem [{\citenamefont {Alcubierre}(2008)}]{Alcubierre:2008}%
  \BibitemOpen
  \bibfield  {author} {\bibinfo {author} {\bibfnamefont {M.}~\bibnamefont
  {Alcubierre}},\ }\href@noop {} {\emph {\bibinfo {title} {{Introduction to 3+1
  numerical relativity}}}},\ Vol.\ \bibinfo {volume} {140}\ (\bibinfo
  {publisher} {Oxford University Press},\ \bibinfo {year} {2008})\BibitemShut
  {NoStop}%
\bibitem [{\citenamefont {Font}(2008)}]{Font08}%
  \BibitemOpen
  \bibfield  {author} {\bibinfo {author} {\bibfnamefont {J.~A.}\ \bibnamefont
  {Font}},\ }\href@noop {} {\bibfield  {journal} {\bibinfo  {journal} {Living
  Reviews in Relativity}\ }\textbf {\bibinfo {volume} {11}},\ \bibinfo {pages}
  {1} (\bibinfo {year} {2008})}\BibitemShut {NoStop}%
\bibitem [{\citenamefont {{Grandcl{\'e}ment}}\ and\ \citenamefont
  {{Novak}}(2009)}]{Grandclement2009}%
  \BibitemOpen
  \bibfield  {author} {\bibinfo {author} {\bibfnamefont {P.}~\bibnamefont
  {{Grandcl{\'e}ment}}}\ and\ \bibinfo {author} {\bibfnamefont
  {J.}~\bibnamefont {{Novak}}},\ }\href {\doibase 10.12942/lrr-2009-1}
  {\bibfield  {journal} {\bibinfo  {journal} {Living Reviews in Relativity}\
  }\textbf {\bibinfo {volume} {12}},\ \bibinfo {eid} {1} (\bibinfo {year}
  {2009})}\BibitemShut {NoStop}%
\bibitem [{\citenamefont {Bona}\ and\ \citenamefont
  {Palenzuela-Luque}(2005)}]{Bona-and-Palenzuela-Luque-2005:numrel-book}%
  \BibitemOpen
  \bibfield  {author} {\bibinfo {author} {\bibfnamefont {C.}~\bibnamefont
  {Bona}}\ and\ \bibinfo {author} {\bibfnamefont {C.}~\bibnamefont
  {Palenzuela-Luque}},\ }\href {\doibase 10.1007/b135928} {\emph {\bibinfo
  {title} {Elements of Numerical Relativity}}}\ (\bibinfo  {publisher}
  {Springer-Verlag},\ \bibinfo {address} {Berlin},\ \bibinfo {year}
  {2005})\BibitemShut {NoStop}%
\bibitem [{\citenamefont {Baumgarte}\ and\ \citenamefont
  {Shapiro}(2010)}]{Baumgarte2010}%
  \BibitemOpen
  \bibfield  {author} {\bibinfo {author} {\bibfnamefont {T.~W.}\ \bibnamefont
  {Baumgarte}}\ and\ \bibinfo {author} {\bibfnamefont {S.~L.}\ \bibnamefont
  {Shapiro}},\ }\href@noop {} {\emph {\bibinfo {title} {{Numerical relativity:
  solving Einstein's equations on the computer}}}}\ (\bibinfo  {publisher}
  {Cambridge University Press},\ \bibinfo {year} {2010})\BibitemShut {NoStop}%
\bibitem [{\citenamefont {{Gourgoulhon}}(2012)}]{Gourgoulhon2012}%
  \BibitemOpen
  \bibfield  {author} {\bibinfo {author} {\bibfnamefont {E.}~\bibnamefont
  {{Gourgoulhon}}},\ }\href {\doibase 10.1007/978-3-642-24525-1} {\emph
  {\bibinfo {title} {3+1 Formalism in General Relativity}}},\ Vol.\ \bibinfo
  {volume} {846}\ (\bibinfo {year} {2012})\BibitemShut {NoStop}%
\bibitem [{\citenamefont {{Faber}}\ and\ \citenamefont
  {{Rasio}}(2012)}]{Faber2012LRR}%
  \BibitemOpen
  \bibfield  {author} {\bibinfo {author} {\bibfnamefont {J.~A.}\ \bibnamefont
  {{Faber}}}\ and\ \bibinfo {author} {\bibfnamefont {F.~A.}\ \bibnamefont
  {{Rasio}}},\ }\href {\doibase 10.12942/lrr-2012-8} {\bibfield  {journal}
  {\bibinfo  {journal} {Living Reviews in Relativity}\ }\textbf {\bibinfo
  {volume} {15}},\ \bibinfo {eid} {8} (\bibinfo {year} {2012})},\ \Eprint
  {http://arxiv.org/abs/1204.3858} {arXiv:1204.3858 [gr-qc]} \BibitemShut
  {NoStop}%
\bibitem [{\citenamefont {Rezzolla}\ and\ \citenamefont
  {Zanotti}(2013)}]{Rezzolla_book:2013}%
  \BibitemOpen
  \bibfield  {author} {\bibinfo {author} {\bibfnamefont {L.}~\bibnamefont
  {Rezzolla}}\ and\ \bibinfo {author} {\bibfnamefont {O.}~\bibnamefont
  {Zanotti}},\ }\href@noop {} {\emph {\bibinfo {title} {{Relativistic
  hydrodynamics}}}}\ (\bibinfo  {publisher} {Oxford University Press},\
  \bibinfo {year} {2013})\BibitemShut {NoStop}%
\bibitem [{\citenamefont {{Bishop}}\ and\ \citenamefont
  {{Rezzolla}}(2016)}]{Bishop2016LRR}%
  \BibitemOpen
  \bibfield  {author} {\bibinfo {author} {\bibfnamefont {N.~T.}\ \bibnamefont
  {{Bishop}}}\ and\ \bibinfo {author} {\bibfnamefont {L.}~\bibnamefont
  {{Rezzolla}}},\ }\href {\doibase 10.1007/s41114-016-0001-9} {\bibfield
  {journal} {\bibinfo  {journal} {Living Reviews in Relativity}\ }\textbf
  {\bibinfo {volume} {19}},\ \bibinfo {eid} {2} (\bibinfo {year}
  {2016})}\BibitemShut {NoStop}%
\bibitem [{\citenamefont {{Kyutoku}}\ \emph {et~al.}(2021)\citenamefont
  {{Kyutoku}}, \citenamefont {{Shibata}},\ and\ \citenamefont
  {{Taniguchi}}}]{Kyutoku2021LRR}%
  \BibitemOpen
  \bibfield  {author} {\bibinfo {author} {\bibfnamefont {K.}~\bibnamefont
  {{Kyutoku}}}, \bibinfo {author} {\bibfnamefont {M.}~\bibnamefont
  {{Shibata}}}, \ and\ \bibinfo {author} {\bibfnamefont {K.}~\bibnamefont
  {{Taniguchi}}},\ }\href {\doibase 10.1007/s41114-021-00033-4} {\bibfield
  {journal} {\bibinfo  {journal} {Living Reviews in Relativity}\ }\textbf
  {\bibinfo {volume} {24}},\ \bibinfo {eid} {5} (\bibinfo {year} {2021})},\
  \Eprint {http://arxiv.org/abs/2110.06218} {arXiv:2110.06218 [astro-ph.HE]}
  \BibitemShut {NoStop}%
\bibitem [{\citenamefont {Pretorius}(2005)}]{Pretorius2005}%
  \BibitemOpen
  \bibfield  {author} {\bibinfo {author} {\bibfnamefont {F.}~\bibnamefont
  {Pretorius}},\ }\href {\doibase 10.1103/PhysRevLett.95.121101} {\bibfield
  {journal} {\bibinfo  {journal} {Phys. Rev. Lett.}\ }\textbf {\bibinfo
  {volume} {95}},\ \bibinfo {pages} {121101} (\bibinfo {year}
  {2005})}\BibitemShut {NoStop}%
\bibitem [{\citenamefont {{Lindblom}}\ \emph {et~al.}(2006)\citenamefont
  {{Lindblom}}, \citenamefont {{Scheel}}, \citenamefont {{Kidder}},
  \citenamefont {{Owen}},\ and\ \citenamefont {{Rinne}}}]{Lindblom2006}%
  \BibitemOpen
  \bibfield  {author} {\bibinfo {author} {\bibfnamefont {L.}~\bibnamefont
  {{Lindblom}}}, \bibinfo {author} {\bibfnamefont {M.~A.}\ \bibnamefont
  {{Scheel}}}, \bibinfo {author} {\bibfnamefont {L.~E.}\ \bibnamefont
  {{Kidder}}}, \bibinfo {author} {\bibfnamefont {R.}~\bibnamefont {{Owen}}}, \
  and\ \bibinfo {author} {\bibfnamefont {O.}~\bibnamefont {{Rinne}}},\ }\href
  {\doibase 10.1088/0264-9381/23/16/S09} {\bibfield  {journal} {\bibinfo
  {journal} {Classical and Quantum Gravity}\ }\textbf {\bibinfo {volume}
  {23}},\ \bibinfo {pages} {S447} (\bibinfo {year} {2006})},\ \Eprint
  {http://arxiv.org/abs/gr-qc/0512093} {arXiv:gr-qc/0512093 [gr-qc]}
  \BibitemShut {NoStop}%
\bibitem [{\citenamefont {{Tichy}}\ \emph {et~al.}(2023)\citenamefont
  {{Tichy}}, \citenamefont {{Ji}}, \citenamefont {{Adhikari}}, \citenamefont
  {{Rashti}},\ and\ \citenamefont {{Pirog}}}]{Tichy2023}%
  \BibitemOpen
  \bibfield  {author} {\bibinfo {author} {\bibfnamefont {W.}~\bibnamefont
  {{Tichy}}}, \bibinfo {author} {\bibfnamefont {L.}~\bibnamefont {{Ji}}},
  \bibinfo {author} {\bibfnamefont {A.}~\bibnamefont {{Adhikari}}}, \bibinfo
  {author} {\bibfnamefont {A.}~\bibnamefont {{Rashti}}}, \ and\ \bibinfo
  {author} {\bibfnamefont {M.}~\bibnamefont {{Pirog}}},\ }\href {\doibase
  10.1088/1361-6382/acaae7} {\bibfield  {journal} {\bibinfo  {journal}
  {Classical and Quantum Gravity}\ }\textbf {\bibinfo {volume} {40}},\ \bibinfo
  {eid} {025004} (\bibinfo {year} {2023})}\BibitemShut {NoStop}%
\bibitem [{\citenamefont {Nakamura}\ \emph {et~al.}(1987)\citenamefont
  {Nakamura}, \citenamefont {Oohara},\ and\ \citenamefont
  {Kojima}}]{Nakamura87}%
  \BibitemOpen
  \bibfield  {author} {\bibinfo {author} {\bibfnamefont {T.}~\bibnamefont
  {Nakamura}}, \bibinfo {author} {\bibfnamefont {K.}~\bibnamefont {Oohara}}, \
  and\ \bibinfo {author} {\bibfnamefont {Y.}~\bibnamefont {Kojima}},\
  }\href@noop {} {\bibfield  {journal} {\bibinfo  {journal} {Progress of
  Theoretical Physics Supplement}\ }\textbf {\bibinfo {volume} {90}},\ \bibinfo
  {pages} {1} (\bibinfo {year} {1987})}\BibitemShut {NoStop}%
\bibitem [{\citenamefont {Shibata}\ and\ \citenamefont
  {Nakamura}(1995)}]{Shibata95}%
  \BibitemOpen
  \bibfield  {author} {\bibinfo {author} {\bibfnamefont {M.}~\bibnamefont
  {Shibata}}\ and\ \bibinfo {author} {\bibfnamefont {T.}~\bibnamefont
  {Nakamura}},\ }\href@noop {} {\bibfield  {journal} {\bibinfo  {journal}
  {Physical Review D}\ }\textbf {\bibinfo {volume} {52}},\ \bibinfo {pages}
  {5428} (\bibinfo {year} {1995})}\BibitemShut {NoStop}%
\bibitem [{\citenamefont {{Baumgarte}}\ and\ \citenamefont
  {{Shapiro}}(1998)}]{Baumgarte1998}%
  \BibitemOpen
  \bibfield  {author} {\bibinfo {author} {\bibfnamefont {T.~W.}\ \bibnamefont
  {{Baumgarte}}}\ and\ \bibinfo {author} {\bibfnamefont {S.~L.}\ \bibnamefont
  {{Shapiro}}},\ }\href {\doibase 10.1103/PhysRevD.59.024007} {\bibfield
  {journal} {\bibinfo  {journal} {\prd}\ }\textbf {\bibinfo {volume} {59}},\
  \bibinfo {eid} {024007} (\bibinfo {year} {1998})}\BibitemShut {NoStop}%
\bibitem [{\citenamefont {Löffler}\ \emph {et~al.}(2012)\citenamefont
  {Löffler}, \citenamefont {Faber}, \citenamefont {Bentivegna}, \citenamefont
  {Bode}, \citenamefont {Diener}, \citenamefont {Haas}, \citenamefont {Hinder},
  \citenamefont {Mundim}, \citenamefont {Ott}, \citenamefont {Schnetter},
  \citenamefont {Allen}, \citenamefont {Campanelli},\ and\ \citenamefont
  {Laguna}}]{Loffler2012}%
  \BibitemOpen
  \bibfield  {author} {\bibinfo {author} {\bibfnamefont {F.}~\bibnamefont
  {Löffler}}, \bibinfo {author} {\bibfnamefont {J.}~\bibnamefont {Faber}},
  \bibinfo {author} {\bibfnamefont {E.}~\bibnamefont {Bentivegna}}, \bibinfo
  {author} {\bibfnamefont {T.}~\bibnamefont {Bode}}, \bibinfo {author}
  {\bibfnamefont {P.}~\bibnamefont {Diener}}, \bibinfo {author} {\bibfnamefont
  {R.}~\bibnamefont {Haas}}, \bibinfo {author} {\bibfnamefont {I.}~\bibnamefont
  {Hinder}}, \bibinfo {author} {\bibfnamefont {B.~C.}\ \bibnamefont {Mundim}},
  \bibinfo {author} {\bibfnamefont {C.~D.}\ \bibnamefont {Ott}}, \bibinfo
  {author} {\bibfnamefont {E.}~\bibnamefont {Schnetter}}, \bibinfo {author}
  {\bibfnamefont {G.}~\bibnamefont {Allen}}, \bibinfo {author} {\bibfnamefont
  {M.}~\bibnamefont {Campanelli}}, \ and\ \bibinfo {author} {\bibfnamefont
  {P.}~\bibnamefont {Laguna}},\ }\href {\doibase
  10.1088/0264-9381/29/11/115001} {\bibfield  {journal} {\bibinfo  {journal}
  {Classical and Quantum Gravity}\ }\textbf {\bibinfo {volume} {29}},\ \bibinfo
  {pages} {115001} (\bibinfo {year} {2012})}\BibitemShut {NoStop}%
\bibitem [{\citenamefont {Zlochower}\ \emph {et~al.}(2005)\citenamefont
  {Zlochower}, \citenamefont {Baker}, \citenamefont {Campanelli},\ and\
  \citenamefont {Lousto}}]{Zlochower2005}%
  \BibitemOpen
  \bibfield  {author} {\bibinfo {author} {\bibfnamefont {Y.}~\bibnamefont
  {Zlochower}}, \bibinfo {author} {\bibfnamefont {J.~G.}\ \bibnamefont
  {Baker}}, \bibinfo {author} {\bibfnamefont {M.}~\bibnamefont {Campanelli}}, \
  and\ \bibinfo {author} {\bibfnamefont {C.~O.}\ \bibnamefont {Lousto}},\
  }\href {\doibase 10.1103/PhysRevD.72.024021} {\bibfield  {journal} {\bibinfo
  {journal} {Phys. Rev. D}\ }\textbf {\bibinfo {volume} {72}},\ \bibinfo
  {pages} {024021} (\bibinfo {year} {2005})}\BibitemShut {NoStop}%
\bibitem [{\citenamefont {{Lousto}}\ and\ \citenamefont
  {{Healy}}(2023)}]{Lousto2023}%
  \BibitemOpen
  \bibfield  {author} {\bibinfo {author} {\bibfnamefont {C.~O.}\ \bibnamefont
  {{Lousto}}}\ and\ \bibinfo {author} {\bibfnamefont {J.}~\bibnamefont
  {{Healy}}},\ }\href {\doibase 10.1088/1361-6382/acc7ef} {\bibfield  {journal}
  {\bibinfo  {journal} {Classical and Quantum Gravity}\ }\textbf {\bibinfo
  {volume} {40}},\ \bibinfo {eid} {09LT01} (\bibinfo {year}
  {2023})}\BibitemShut {NoStop}%
\bibitem [{\citenamefont {Br\"ugmann}\ \emph {et~al.}(2008)\citenamefont
  {Br\"ugmann}, \citenamefont {Gonz\'alez}, \citenamefont {Hannam},
  \citenamefont {Husa}, \citenamefont {Sperhake},\ and\ \citenamefont
  {Tichy}}]{Bruegmann2008}%
  \BibitemOpen
  \bibfield  {author} {\bibinfo {author} {\bibfnamefont {B.}~\bibnamefont
  {Br\"ugmann}}, \bibinfo {author} {\bibfnamefont {J.~A.}\ \bibnamefont
  {Gonz\'alez}}, \bibinfo {author} {\bibfnamefont {M.}~\bibnamefont {Hannam}},
  \bibinfo {author} {\bibfnamefont {S.}~\bibnamefont {Husa}}, \bibinfo {author}
  {\bibfnamefont {U.}~\bibnamefont {Sperhake}}, \ and\ \bibinfo {author}
  {\bibfnamefont {W.}~\bibnamefont {Tichy}},\ }\href {\doibase
  10.1103/PhysRevD.77.024027} {\bibfield  {journal} {\bibinfo  {journal} {Phys.
  Rev. D}\ }\textbf {\bibinfo {volume} {77}},\ \bibinfo {pages} {024027}
  (\bibinfo {year} {2008})}\BibitemShut {NoStop}%
\bibitem [{\citenamefont {Thierfelder}\ \emph {et~al.}(2011)\citenamefont
  {Thierfelder}, \citenamefont {Bernuzzi},\ and\ \citenamefont
  {Br\"ugmann}}]{Thierfelder2011}%
  \BibitemOpen
  \bibfield  {author} {\bibinfo {author} {\bibfnamefont {M.}~\bibnamefont
  {Thierfelder}}, \bibinfo {author} {\bibfnamefont {S.}~\bibnamefont
  {Bernuzzi}}, \ and\ \bibinfo {author} {\bibfnamefont {B.}~\bibnamefont
  {Br\"ugmann}},\ }\href {\doibase 10.1103/PhysRevD.84.044012} {\bibfield
  {journal} {\bibinfo  {journal} {Phys. Rev. D}\ }\textbf {\bibinfo {volume}
  {84}},\ \bibinfo {pages} {044012} (\bibinfo {year} {2011})}\BibitemShut
  {NoStop}%
\bibitem [{McLachlan()}]{McLachlan:web}%
  \BibitemOpen
  McLachlan,\ \href {http://www.cct.lsu.edu/~eschnett/McLachlan/} {\enquote
  {\bibinfo {title} {{McLachlan}, a public {BSSN} code},}\ } (\bibinfo {year}
  {2008})\BibitemShut {NoStop}%
\bibitem [{\citenamefont {Clough}\ \emph {et~al.}(2015)\citenamefont {Clough},
  \citenamefont {Figueras}, \citenamefont {Finkel}, \citenamefont {Kunesch},
  \citenamefont {Lim},\ and\ \citenamefont {Tunyasuvunakool}}]{Clough_2015}%
  \BibitemOpen
  \bibfield  {author} {\bibinfo {author} {\bibfnamefont {K.}~\bibnamefont
  {Clough}}, \bibinfo {author} {\bibfnamefont {P.}~\bibnamefont {Figueras}},
  \bibinfo {author} {\bibfnamefont {H.}~\bibnamefont {Finkel}}, \bibinfo
  {author} {\bibfnamefont {M.}~\bibnamefont {Kunesch}}, \bibinfo {author}
  {\bibfnamefont {E.~A.}\ \bibnamefont {Lim}}, \ and\ \bibinfo {author}
  {\bibfnamefont {S.}~\bibnamefont {Tunyasuvunakool}},\ }\href {\doibase
  10.1088/0264-9381/32/24/245011} {\bibfield  {journal} {\bibinfo  {journal}
  {Classical and Quantum Gravity}\ }\textbf {\bibinfo {volume} {32}},\ \bibinfo
  {pages} {245011} (\bibinfo {year} {2015})}\BibitemShut {NoStop}%
\bibitem [{\citenamefont {Peterson}\ \emph {et~al.}(2023)\citenamefont
  {Peterson}, \citenamefont {Willcox},\ and\ \citenamefont
  {Mösta}}]{Peterson_2023}%
  \BibitemOpen
  \bibfield  {author} {\bibinfo {author} {\bibfnamefont {A.~J.}\ \bibnamefont
  {Peterson}}, \bibinfo {author} {\bibfnamefont {D.}~\bibnamefont {Willcox}}, \
  and\ \bibinfo {author} {\bibfnamefont {P.}~\bibnamefont {Mösta}},\ }\href
  {\doibase 10.1088/1361-6382/ad0b37} {\bibfield  {journal} {\bibinfo
  {journal} {Classical and Quantum Gravity}\ }\textbf {\bibinfo {volume}
  {40}},\ \bibinfo {pages} {245013} (\bibinfo {year} {2023})}\BibitemShut
  {NoStop}%
\bibitem [{\citenamefont {Yamamoto}\ \emph {et~al.}(2008)\citenamefont
  {Yamamoto}, \citenamefont {Shibata},\ and\ \citenamefont
  {Taniguchi}}]{Yamamoto2009}%
  \BibitemOpen
  \bibfield  {author} {\bibinfo {author} {\bibfnamefont {T.}~\bibnamefont
  {Yamamoto}}, \bibinfo {author} {\bibfnamefont {M.}~\bibnamefont {Shibata}}, \
  and\ \bibinfo {author} {\bibfnamefont {K.}~\bibnamefont {Taniguchi}},\ }\href
  {\doibase 10.1103/PhysRevD.78.064054} {\bibfield  {journal} {\bibinfo
  {journal} {Phys. Rev. D}\ }\textbf {\bibinfo {volume} {78}},\ \bibinfo
  {pages} {064054} (\bibinfo {year} {2008})}\BibitemShut {NoStop}%
\bibitem [{\citenamefont {Kiuchi}\ \emph {et~al.}(2017)\citenamefont {Kiuchi},
  \citenamefont {Kawaguchi}, \citenamefont {Kyutoku}, \citenamefont
  {Sekiguchi}, \citenamefont {Shibata},\ and\ \citenamefont
  {Taniguchi}}]{Kiuchi2017}%
  \BibitemOpen
  \bibfield  {author} {\bibinfo {author} {\bibfnamefont {K.}~\bibnamefont
  {Kiuchi}}, \bibinfo {author} {\bibfnamefont {K.}~\bibnamefont {Kawaguchi}},
  \bibinfo {author} {\bibfnamefont {K.}~\bibnamefont {Kyutoku}}, \bibinfo
  {author} {\bibfnamefont {Y.}~\bibnamefont {Sekiguchi}}, \bibinfo {author}
  {\bibfnamefont {M.}~\bibnamefont {Shibata}}, \ and\ \bibinfo {author}
  {\bibfnamefont {K.}~\bibnamefont {Taniguchi}},\ }\href {\doibase
  10.1103/PhysRevD.96.084060} {\bibfield  {journal} {\bibinfo  {journal} {Phys.
  Rev. D}\ }\textbf {\bibinfo {volume} {96}},\ \bibinfo {pages} {084060}
  (\bibinfo {year} {2017})}\BibitemShut {NoStop}%
\bibitem [{\citenamefont {Palenzuela}\ \emph {et~al.}(2018)\citenamefont
  {Palenzuela}, \citenamefont {Miñano}, \citenamefont {Viganò}, \citenamefont
  {Arbona}, \citenamefont {Bona-Casas}, \citenamefont {Rigo}, \citenamefont
  {Bezares}, \citenamefont {Bona},\ and\ \citenamefont
  {Massó}}]{Palenzuela2018}%
  \BibitemOpen
  \bibfield  {author} {\bibinfo {author} {\bibfnamefont {C.}~\bibnamefont
  {Palenzuela}}, \bibinfo {author} {\bibfnamefont {B.}~\bibnamefont {Miñano}},
  \bibinfo {author} {\bibfnamefont {D.}~\bibnamefont {Viganò}}, \bibinfo
  {author} {\bibfnamefont {A.}~\bibnamefont {Arbona}}, \bibinfo {author}
  {\bibfnamefont {C.}~\bibnamefont {Bona-Casas}}, \bibinfo {author}
  {\bibfnamefont {A.}~\bibnamefont {Rigo}}, \bibinfo {author} {\bibfnamefont
  {M.}~\bibnamefont {Bezares}}, \bibinfo {author} {\bibfnamefont
  {C.}~\bibnamefont {Bona}}, \ and\ \bibinfo {author} {\bibfnamefont
  {J.}~\bibnamefont {Massó}},\ }\href {\doibase 10.1088/1361-6382/aad7f6}
  {\bibfield  {journal} {\bibinfo  {journal} {Classical and Quantum Gravity}\
  }\textbf {\bibinfo {volume} {35}},\ \bibinfo {pages} {185007} (\bibinfo
  {year} {2018})}\BibitemShut {NoStop}%
\bibitem [{\citenamefont {{Zhang}}\ \emph {et~al.}(2024)\citenamefont
  {{Zhang}}, \citenamefont {{Li}}, \citenamefont {{Weinzierl}},\ and\
  \citenamefont {{Barrera-Hinojosa}}}]{Zhang2024}%
  \BibitemOpen
  \bibfield  {author} {\bibinfo {author} {\bibfnamefont {H.}~\bibnamefont
  {{Zhang}}}, \bibinfo {author} {\bibfnamefont {B.}~\bibnamefont {{Li}}},
  \bibinfo {author} {\bibfnamefont {T.}~\bibnamefont {{Weinzierl}}}, \ and\
  \bibinfo {author} {\bibfnamefont {C.}~\bibnamefont {{Barrera-Hinojosa}}},\
  }\href@noop {} {\bibfield  {journal} {\bibinfo  {journal} {arXiv e-prints}\ }
  (\bibinfo {year} {2024})},\ \Eprint {http://arxiv.org/abs/2406.11626}
  {arXiv:2406.11626 [gr-qc]} \BibitemShut {NoStop}%
\bibitem [{\citenamefont {Cockburn}\ and\ \citenamefont {Shu}(1989)}]{cbs1}%
  \BibitemOpen
  \bibfield  {author} {\bibinfo {author} {\bibfnamefont {B.}~\bibnamefont
  {Cockburn}}\ and\ \bibinfo {author} {\bibfnamefont {C.~W.}\ \bibnamefont
  {Shu}},\ }\href@noop {} {\bibfield  {journal} {\bibinfo  {journal}
  {Mathematics of Computation}\ }\textbf {\bibinfo {volume} {52}},\ \bibinfo
  {pages} {411} (\bibinfo {year} {1989})}\BibitemShut {NoStop}%
\bibitem [{\citenamefont {Cockburn}\ \emph {et~al.}(1989)\citenamefont
  {Cockburn}, \citenamefont {Lin},\ and\ \citenamefont {Shu}}]{cbs2}%
  \BibitemOpen
  \bibfield  {author} {\bibinfo {author} {\bibfnamefont {B.}~\bibnamefont
  {Cockburn}}, \bibinfo {author} {\bibfnamefont {S.~Y.}\ \bibnamefont {Lin}}, \
  and\ \bibinfo {author} {\bibfnamefont {C.}~\bibnamefont {Shu}},\ }\href@noop
  {} {\bibfield  {journal} {\bibinfo  {journal} {Journal of Computational
  Physics}\ }\textbf {\bibinfo {volume} {84}},\ \bibinfo {pages} {90} (\bibinfo
  {year} {1989})}\BibitemShut {NoStop}%
\bibitem [{\citenamefont {Cockburn}\ \emph {et~al.}(1990)\citenamefont
  {Cockburn}, \citenamefont {Hou},\ and\ \citenamefont {Shu}}]{cbs3}%
  \BibitemOpen
  \bibfield  {author} {\bibinfo {author} {\bibfnamefont {B.}~\bibnamefont
  {Cockburn}}, \bibinfo {author} {\bibfnamefont {S.}~\bibnamefont {Hou}}, \
  and\ \bibinfo {author} {\bibfnamefont {C.~W.}\ \bibnamefont {Shu}},\
  }\href@noop {} {\bibfield  {journal} {\bibinfo  {journal} {Mathematics of
  Computation}\ }\textbf {\bibinfo {volume} {54}},\ \bibinfo {pages} {545}
  (\bibinfo {year} {1990})}\BibitemShut {NoStop}%
\bibitem [{\citenamefont {Cockburn}\ and\ \citenamefont
  {Shu}(1998{\natexlab{a}})}]{cbs4}%
  \BibitemOpen
  \bibfield  {author} {\bibinfo {author} {\bibfnamefont {B.}~\bibnamefont
  {Cockburn}}\ and\ \bibinfo {author} {\bibfnamefont {C.~W.}\ \bibnamefont
  {Shu}},\ }\href@noop {} {\bibfield  {journal} {\bibinfo  {journal} {Journal
  of Computational Physics}\ }\textbf {\bibinfo {volume} {141}},\ \bibinfo
  {pages} {199} (\bibinfo {year} {1998}{\natexlab{a}})}\BibitemShut {NoStop}%
\bibitem [{\citenamefont {Dumbser}\ and\ \citenamefont
  {Zanotti}(2009)}]{DumbserZanotti}%
  \BibitemOpen
  \bibfield  {author} {\bibinfo {author} {\bibfnamefont {M.}~\bibnamefont
  {Dumbser}}\ and\ \bibinfo {author} {\bibfnamefont {O.}~\bibnamefont
  {Zanotti}},\ }\href@noop {} {\bibfield  {journal} {\bibinfo  {journal}
  {Journal of Computational Physics}\ }\textbf {\bibinfo {volume} {228}},\
  \bibinfo {pages} {6991} (\bibinfo {year} {2009})}\BibitemShut {NoStop}%
\bibitem [{\citenamefont {Radice}\ and\ \citenamefont
  {Rezzolla}(2011)}]{Radice2011}%
  \BibitemOpen
  \bibfield  {author} {\bibinfo {author} {\bibfnamefont {D.}~\bibnamefont
  {Radice}}\ and\ \bibinfo {author} {\bibfnamefont {L.}~\bibnamefont
  {Rezzolla}},\ }\href {\doibase 10.1103/PhysRevD.84.024010} {\bibfield
  {journal} {\bibinfo  {journal} {Phys. Rev. D}\ }\textbf {\bibinfo {volume}
  {84}},\ \bibinfo {pages} {024010} (\bibinfo {year} {2011})}\BibitemShut
  {NoStop}%
\bibitem [{\citenamefont {R.Hartmann}\ and\ \citenamefont
  {P.Houston}(2002)}]{Hartman_02}%
  \BibitemOpen
  \bibfield  {author} {\bibinfo {author} {\bibnamefont {R.Hartmann}}\ and\
  \bibinfo {author} {\bibnamefont {P.Houston}},\ }\href@noop {} {\bibfield
  {journal} {\bibinfo  {journal} {J. Comp. Phys.}\ }\textbf {\bibinfo {volume}
  {183}},\ \bibinfo {pages} {508} (\bibinfo {year} {2002})}\BibitemShut
  {NoStop}%
\bibitem [{\citenamefont {Persson}\ and\ \citenamefont
  {Peraire}(2006)}]{Persson_06}%
  \BibitemOpen
  \bibfield  {author} {\bibinfo {author} {\bibfnamefont {P.-O.}\ \bibnamefont
  {Persson}}\ and\ \bibinfo {author} {\bibfnamefont {J.}~\bibnamefont
  {Peraire}},\ }\href@noop {} {\bibfield  {journal} {\bibinfo  {journal} {AIAA
  Paper 2006-112}\ } (\bibinfo {year} {2006})}\BibitemShut {NoStop}%
\bibitem [{\citenamefont {Cesenek}\ \emph {et~al.}(2013)\citenamefont
  {Cesenek}, \citenamefont {Feistauer}, \citenamefont {Horacek}, \citenamefont
  {Kucera},\ and\ \citenamefont {Prokopova}}]{Feistauer4}%
  \BibitemOpen
  \bibfield  {author} {\bibinfo {author} {\bibfnamefont {J.}~\bibnamefont
  {Cesenek}}, \bibinfo {author} {\bibfnamefont {M.}~\bibnamefont {Feistauer}},
  \bibinfo {author} {\bibfnamefont {J.}~\bibnamefont {Horacek}}, \bibinfo
  {author} {\bibfnamefont {V.}~\bibnamefont {Kucera}}, \ and\ \bibinfo {author}
  {\bibfnamefont {J.}~\bibnamefont {Prokopova}},\ }\href@noop {} {\bibfield
  {journal} {\bibinfo  {journal} {Applied Mathematics and Computation}\
  }\textbf {\bibinfo {volume} {219}},\ \bibinfo {pages} {7139} (\bibinfo {year}
  {2013})}\BibitemShut {NoStop}%
\bibitem [{\citenamefont {Qiu}\ and\ \citenamefont {Shu}(2005)}]{QiuShu1}%
  \BibitemOpen
  \bibfield  {author} {\bibinfo {author} {\bibfnamefont {J.}~\bibnamefont
  {Qiu}}\ and\ \bibinfo {author} {\bibfnamefont {C.}~\bibnamefont {Shu}},\
  }\href@noop {} {\bibfield  {journal} {\bibinfo  {journal} {SIAM Journal on
  Scientific Computing}\ }\textbf {\bibinfo {volume} {26}},\ \bibinfo {pages}
  {907} (\bibinfo {year} {2005})}\BibitemShut {NoStop}%
\bibitem [{\citenamefont {J.Qiu}\ and\ \citenamefont
  {C-W.Shu}(2004)}]{Qiu_2004}%
  \BibitemOpen
  \bibfield  {author} {\bibinfo {author} {\bibnamefont {J.Qiu}}\ and\ \bibinfo
  {author} {\bibnamefont {C-W.Shu}},\ }\href {\doibase
  10.1016/j.jcp.2003.07.026} {\bibfield  {journal} {\bibinfo  {journal} {J.
  Comput. Phys.}\ }\textbf {\bibinfo {volume} {193}},\ \bibinfo {pages} {115}
  (\bibinfo {year} {2004})}\BibitemShut {NoStop}%
\bibitem [{\citenamefont {Balsara}\ \emph {et~al.}(2007)\citenamefont
  {Balsara}, \citenamefont {Altmann}, \citenamefont {Munz},\ and\ \citenamefont
  {Dumbser}}]{balsara2007}%
  \BibitemOpen
  \bibfield  {author} {\bibinfo {author} {\bibfnamefont {D.}~\bibnamefont
  {Balsara}}, \bibinfo {author} {\bibfnamefont {C.}~\bibnamefont {Altmann}},
  \bibinfo {author} {\bibfnamefont {C.}~\bibnamefont {Munz}}, \ and\ \bibinfo
  {author} {\bibfnamefont {M.}~\bibnamefont {Dumbser}},\ }\href@noop {}
  {\bibfield  {journal} {\bibinfo  {journal} {Journal of Computational
  Physics}\ }\textbf {\bibinfo {volume} {226}},\ \bibinfo {pages} {586}
  (\bibinfo {year} {2007})}\BibitemShut {NoStop}%
\bibitem [{\citenamefont {J.Zhu}\ \emph {et~al.}(2008)\citenamefont {J.Zhu},
  \citenamefont {Qiu}, \citenamefont {C.-W.Shu},\ and\ \citenamefont
  {M.Dumbser}}]{Zhu_2008}%
  \BibitemOpen
  \bibfield  {author} {\bibinfo {author} {\bibnamefont {J.Zhu}}, \bibinfo
  {author} {\bibfnamefont {J.}~\bibnamefont {Qiu}}, \bibinfo {author}
  {\bibnamefont {C.-W.Shu}}, \ and\ \bibinfo {author} {\bibnamefont
  {M.Dumbser}},\ }\href {\doibase 10.1016/j.jcp.2007.12.024} {\bibfield
  {journal} {\bibinfo  {journal} {J. Comput. Phys.}\ }\textbf {\bibinfo
  {volume} {227}},\ \bibinfo {pages} {4330} (\bibinfo {year}
  {2008})}\BibitemShut {NoStop}%
\bibitem [{\citenamefont {J.Zhu}\ and\ \citenamefont {Qiu}(2013)}]{Zhu_13}%
  \BibitemOpen
  \bibfield  {author} {\bibinfo {author} {\bibfnamefont {C.~S.}\ \bibnamefont
  {J.Zhu}, \bibfnamefont {X.Zhong}}\ and\ \bibinfo {author} {\bibfnamefont
  {J.}~\bibnamefont {Qiu}},\ }\href@noop {} {\bibfield  {journal} {\bibinfo
  {journal} {J. Comp. Phys.}\ }\textbf {\bibinfo {volume} {248}},\ \bibinfo
  {pages} {200} (\bibinfo {year} {2013})}\BibitemShut {NoStop}%
\bibitem [{\citenamefont {H.Luo}\ \emph {et~al.}(2007)\citenamefont {H.Luo},
  \citenamefont {J.D.Baum},\ and\ \citenamefont {R.L\"{o}hner}}]{Luo_2007}%
  \BibitemOpen
  \bibfield  {author} {\bibinfo {author} {\bibnamefont {H.Luo}}, \bibinfo
  {author} {\bibnamefont {J.D.Baum}}, \ and\ \bibinfo {author} {\bibnamefont
  {R.L\"{o}hner}},\ }\href {\doibase 10.1016/j.jcp.2006.12.017} {\bibfield
  {journal} {\bibinfo  {journal} {J. Comput. Phys.}\ }\textbf {\bibinfo
  {volume} {225}},\ \bibinfo {pages} {686} (\bibinfo {year}
  {2007})}\BibitemShut {NoStop}%
\bibitem [{\citenamefont {L.{Krivodonova}}(2007)}]{Krivodonova_2007}%
  \BibitemOpen
  \bibfield  {author} {\bibinfo {author} {\bibnamefont {L.{Krivodonova}}},\
  }\href {\doibase 10.1016/j.jcp.2007.05.011} {\bibfield  {journal} {\bibinfo
  {journal} {Journal of Computational Physics}\ }\textbf {\bibinfo {volume}
  {226}},\ \bibinfo {pages} {879} (\bibinfo {year} {2007})}\BibitemShut
  {NoStop}%
\bibitem [{\citenamefont {{Dumbser}}\ \emph {et~al.}(2014)\citenamefont
  {{Dumbser}}, \citenamefont {{Zanotti}}, \citenamefont {{Loub{\`e}re}},\ and\
  \citenamefont {{Diot}}}]{Dumbser2014}%
  \BibitemOpen
  \bibfield  {author} {\bibinfo {author} {\bibfnamefont {M.}~\bibnamefont
  {{Dumbser}}}, \bibinfo {author} {\bibfnamefont {O.}~\bibnamefont
  {{Zanotti}}}, \bibinfo {author} {\bibfnamefont {R.}~\bibnamefont
  {{Loub{\`e}re}}}, \ and\ \bibinfo {author} {\bibfnamefont {S.}~\bibnamefont
  {{Diot}}},\ }\href {\doibase 10.1016/j.jcp.2014.08.009} {\bibfield  {journal}
  {\bibinfo  {journal} {Journal of Computational Physics}\ }\textbf {\bibinfo
  {volume} {278}},\ \bibinfo {pages} {47} (\bibinfo {year} {2014})}\BibitemShut
  {NoStop}%
\bibitem [{\citenamefont {{Boscheri}}\ and\ \citenamefont
  {{Dumbser}}(2017)}]{Boscheri2017}%
  \BibitemOpen
  \bibfield  {author} {\bibinfo {author} {\bibfnamefont {W.}~\bibnamefont
  {{Boscheri}}}\ and\ \bibinfo {author} {\bibfnamefont {M.}~\bibnamefont
  {{Dumbser}}},\ }\href {\doibase 10.1016/j.jcp.2017.06.022} {\bibfield
  {journal} {\bibinfo  {journal} {Journal of Computational Physics}\ }\textbf
  {\bibinfo {volume} {346}},\ \bibinfo {pages} {449} (\bibinfo {year}
  {2017})},\ \Eprint {http://arxiv.org/abs/1612.04068} {arXiv:1612.04068
  [math.NA]} \BibitemShut {NoStop}%
\bibitem [{\citenamefont {{Zanotti}}\ and\ \citenamefont
  {{Dumbser}}(2011)}]{Zanotti2011}%
  \BibitemOpen
  \bibfield  {author} {\bibinfo {author} {\bibfnamefont {O.}~\bibnamefont
  {{Zanotti}}}\ and\ \bibinfo {author} {\bibfnamefont {M.}~\bibnamefont
  {{Dumbser}}},\ }\href {\doibase 10.1111/j.1365-2966.2011.19551.x} {\bibfield
  {journal} {\bibinfo  {journal} {Monthly Notices of the Royal Astronomical
  Society}\ }\textbf {\bibinfo {volume} {418}},\ \bibinfo {pages} {1004}
  (\bibinfo {year} {2011})}\BibitemShut {NoStop}%
\bibitem [{\citenamefont {{Zanotti}}\ \emph
  {et~al.}(2015{\natexlab{a}})\citenamefont {{Zanotti}}, \citenamefont
  {{Fambri}},\ and\ \citenamefont {{Dumbser}}}]{Zanotti2015d}%
  \BibitemOpen
  \bibfield  {author} {\bibinfo {author} {\bibfnamefont {O.}~\bibnamefont
  {{Zanotti}}}, \bibinfo {author} {\bibfnamefont {F.}~\bibnamefont {{Fambri}}},
  \ and\ \bibinfo {author} {\bibfnamefont {M.}~\bibnamefont {{Dumbser}}},\
  }\href {\doibase 10.1093/mnras/stv1510} {\bibfield  {journal} {\bibinfo
  {journal} {Mon. Not. R. Astron. Soc.}\ }\textbf {\bibinfo {volume} {452}},\
  \bibinfo {pages} {3010} (\bibinfo {year} {2015}{\natexlab{a}})}\BibitemShut
  {NoStop}%
\bibitem [{\citenamefont {{Teukolsky}}(2016)}]{Teukolsky2015}%
  \BibitemOpen
  \bibfield  {author} {\bibinfo {author} {\bibfnamefont {S.~A.}\ \bibnamefont
  {{Teukolsky}}},\ }\href {\doibase 10.1016/j.jcp.2016.02.031} {\bibfield
  {journal} {\bibinfo  {journal} {Journal of Computational Physics}\ }\textbf
  {\bibinfo {volume} {312}},\ \bibinfo {pages} {333} (\bibinfo {year}
  {2016})}\BibitemShut {NoStop}%
\bibitem [{\citenamefont {{Bugner}}\ \emph {et~al.}(2016)\citenamefont
  {{Bugner}}, \citenamefont {{Dietrich}}, \citenamefont {{Bernuzzi}},
  \citenamefont {{Weyhausen}},\ and\ \citenamefont
  {{Br{\"u}gmann}}}]{Bugner2016}%
  \BibitemOpen
  \bibfield  {author} {\bibinfo {author} {\bibfnamefont {M.}~\bibnamefont
  {{Bugner}}}, \bibinfo {author} {\bibfnamefont {T.}~\bibnamefont
  {{Dietrich}}}, \bibinfo {author} {\bibfnamefont {S.}~\bibnamefont
  {{Bernuzzi}}}, \bibinfo {author} {\bibfnamefont {A.}~\bibnamefont
  {{Weyhausen}}}, \ and\ \bibinfo {author} {\bibfnamefont {B.}~\bibnamefont
  {{Br{\"u}gmann}}},\ }\href {\doibase 10.1103/PhysRevD.94.084004} {\bibfield
  {journal} {\bibinfo  {journal} {\prd}\ }\textbf {\bibinfo {volume} {94}},\
  \bibinfo {eid} {084004} (\bibinfo {year} {2016})},\ \Eprint
  {http://arxiv.org/abs/1508.07147} {arXiv:1508.07147 [gr-qc]} \BibitemShut
  {NoStop}%
\bibitem [{\citenamefont {Zhao}\ and\ \citenamefont {Tang}(2017)}]{Zhao2017}%
  \BibitemOpen
  \bibfield  {author} {\bibinfo {author} {\bibfnamefont {J.}~\bibnamefont
  {Zhao}}\ and\ \bibinfo {author} {\bibfnamefont {H.}~\bibnamefont {Tang}},\
  }\href {\doibase https://doi.org/10.1016/j.jcp.2017.04.027} {\bibfield
  {journal} {\bibinfo  {journal} {Journal of Computational Physics}\ }\textbf
  {\bibinfo {volume} {343}},\ \bibinfo {pages} {33} (\bibinfo {year}
  {2017})}\BibitemShut {NoStop}%
\bibitem [{\citenamefont {Fambri}\ \emph {et~al.}(2018)\citenamefont {Fambri},
  \citenamefont {Dumbser}, \citenamefont {K{\"o}ppel}, \citenamefont
  {Rezzolla},\ and\ \citenamefont {Zanotti}}]{ADERGRMHD}%
  \BibitemOpen
  \bibfield  {author} {\bibinfo {author} {\bibfnamefont {F.}~\bibnamefont
  {Fambri}}, \bibinfo {author} {\bibfnamefont {M.}~\bibnamefont {Dumbser}},
  \bibinfo {author} {\bibfnamefont {S.}~\bibnamefont {K{\"o}ppel}}, \bibinfo
  {author} {\bibfnamefont {L.}~\bibnamefont {Rezzolla}}, \ and\ \bibinfo
  {author} {\bibfnamefont {O.}~\bibnamefont {Zanotti}},\ }\href@noop {}
  {\bibfield  {journal} {\bibinfo  {journal} {Monthly Notices of the Royal
  Astronomical Society}\ }\textbf {\bibinfo {volume} {477}},\ \bibinfo {pages}
  {4543} (\bibinfo {year} {2018})}\BibitemShut {NoStop}%
\bibitem [{\citenamefont {{H{\'e}bert}}\ \emph {et~al.}(2018)\citenamefont
  {{H{\'e}bert}}, \citenamefont {{Kidder}},\ and\ \citenamefont
  {{Teukolsky}}}]{Hebert2018}%
  \BibitemOpen
  \bibfield  {author} {\bibinfo {author} {\bibfnamefont {F.}~\bibnamefont
  {{H{\'e}bert}}}, \bibinfo {author} {\bibfnamefont {L.~E.}\ \bibnamefont
  {{Kidder}}}, \ and\ \bibinfo {author} {\bibfnamefont {S.~A.}\ \bibnamefont
  {{Teukolsky}}},\ }\href {\doibase 10.1103/PhysRevD.98.044041} {\bibfield
  {journal} {\bibinfo  {journal} {\prd}\ }\textbf {\bibinfo {volume} {98}},\
  \bibinfo {eid} {044041} (\bibinfo {year} {2018})}\BibitemShut {NoStop}%
\bibitem [{\citenamefont {{Duan}}\ and\ \citenamefont
  {{Tang}}(2020)}]{Junming2020}%
  \BibitemOpen
  \bibfield  {author} {\bibinfo {author} {\bibfnamefont {J.}~\bibnamefont
  {{Duan}}}\ and\ \bibinfo {author} {\bibfnamefont {H.}~\bibnamefont
  {{Tang}}},\ }\href {\doibase 10.1016/j.jcp.2020.109731} {\bibfield  {journal}
  {\bibinfo  {journal} {Journal of Computational Physics}\ }\textbf {\bibinfo
  {volume} {421}},\ \bibinfo {eid} {109731} (\bibinfo {year} {2020})},\ \Eprint
  {http://arxiv.org/abs/1911.03825} {arXiv:1911.03825 [math.NA]} \BibitemShut
  {NoStop}%
\bibitem [{\citenamefont {{Deppe}}\ \emph
  {et~al.}(2022{\natexlab{a}})\citenamefont {{Deppe}}, \citenamefont
  {{H{\'e}bert}}, \citenamefont {{Kidder}},\ and\ \citenamefont
  {{Teukolsky}}}]{Deppe2022}%
  \BibitemOpen
  \bibfield  {author} {\bibinfo {author} {\bibfnamefont {N.}~\bibnamefont
  {{Deppe}}}, \bibinfo {author} {\bibfnamefont {F.}~\bibnamefont
  {{H{\'e}bert}}}, \bibinfo {author} {\bibfnamefont {L.~E.}\ \bibnamefont
  {{Kidder}}}, \ and\ \bibinfo {author} {\bibfnamefont {S.~A.}\ \bibnamefont
  {{Teukolsky}}},\ }\href {\doibase 10.1088/1361-6382/ac8864} {\bibfield
  {journal} {\bibinfo  {journal} {Classical and Quantum Gravity}\ }\textbf
  {\bibinfo {volume} {39}},\ \bibinfo {eid} {195001} (\bibinfo {year}
  {2022}{\natexlab{a}})}\BibitemShut {NoStop}%
\bibitem [{\citenamefont {{Bhoriya}}\ \emph {et~al.}(2023)\citenamefont
  {{Bhoriya}}, \citenamefont {{Biswas}}, \citenamefont {{Kumar}},\ and\
  \citenamefont {{Chandrashekhar}}}]{Deepak2023}%
  \BibitemOpen
  \bibfield  {author} {\bibinfo {author} {\bibfnamefont {D.}~\bibnamefont
  {{Bhoriya}}}, \bibinfo {author} {\bibfnamefont {B.}~\bibnamefont {{Biswas}}},
  \bibinfo {author} {\bibfnamefont {H.}~\bibnamefont {{Kumar}}}, \ and\
  \bibinfo {author} {\bibfnamefont {P.}~\bibnamefont {{Chandrashekhar}}},\
  }\href {\doibase 10.48550/arXiv.2310.09566} {\bibfield  {journal} {\bibinfo
  {journal} {arXiv e-prints}\ ,\ \bibinfo {eid} {arXiv:2310.09566}} (\bibinfo
  {year} {2023})},\ \Eprint {http://arxiv.org/abs/2310.09566} {arXiv:2310.09566
  [math.NA]} \BibitemShut {NoStop}%
\bibitem [{\citenamefont {Cockburn}\ and\ \citenamefont
  {Shu}(2001)}]{CBS-convection-dominated}%
  \BibitemOpen
  \bibfield  {author} {\bibinfo {author} {\bibfnamefont {B.}~\bibnamefont
  {Cockburn}}\ and\ \bibinfo {author} {\bibfnamefont {C.~W.}\ \bibnamefont
  {Shu}},\ }\href@noop {} {\bibfield  {journal} {\bibinfo  {journal} {J. Sci.
  Comput.}\ }\textbf {\bibinfo {volume} {16}},\ \bibinfo {pages} {173}
  (\bibinfo {year} {2001})}\BibitemShut {NoStop}%
\bibitem [{\citenamefont {Cockburn}\ and\ \citenamefont
  {Shu}(1998{\natexlab{b}})}]{CBS-convection-diffusion}%
  \BibitemOpen
  \bibfield  {author} {\bibinfo {author} {\bibfnamefont {B.}~\bibnamefont
  {Cockburn}}\ and\ \bibinfo {author} {\bibfnamefont {C.~W.}\ \bibnamefont
  {Shu}},\ }\href@noop {} {\bibfield  {journal} {\bibinfo  {journal} {SIAM J.
  Numer. Anal.}\ }\textbf {\bibinfo {volume} {35}},\ \bibinfo {pages} {2440}
  (\bibinfo {year} {1998}{\natexlab{b}})}\BibitemShut {NoStop}%
\bibitem [{\citenamefont {Dumbser}\ and\ \citenamefont
  {Munz}(2006)}]{dumbser_jsc}%
  \BibitemOpen
  \bibfield  {author} {\bibinfo {author} {\bibfnamefont {M.}~\bibnamefont
  {Dumbser}}\ and\ \bibinfo {author} {\bibfnamefont {C.}~\bibnamefont {Munz}},\
  }\href@noop {} {\bibfield  {journal} {\bibinfo  {journal} {Journal of
  Scientific Computing}\ }\textbf {\bibinfo {volume} {27}},\ \bibinfo {pages}
  {215} (\bibinfo {year} {2006})}\BibitemShut {NoStop}%
\bibitem [{\citenamefont {Qiu}\ \emph {et~al.}(2005)\citenamefont {Qiu},
  \citenamefont {Dumbser},\ and\ \citenamefont {Shu}}]{QiuDumbserShu}%
  \BibitemOpen
  \bibfield  {author} {\bibinfo {author} {\bibfnamefont {J.}~\bibnamefont
  {Qiu}}, \bibinfo {author} {\bibfnamefont {M.}~\bibnamefont {Dumbser}}, \ and\
  \bibinfo {author} {\bibfnamefont {C.}~\bibnamefont {Shu}},\ }\href@noop {}
  {\bibfield  {journal} {\bibinfo  {journal} {Computer Methods in Applied
  Mechanics and Engineering}\ }\textbf {\bibinfo {volume} {194}},\ \bibinfo
  {pages} {4528} (\bibinfo {year} {2005})}\BibitemShut {NoStop}%
\bibitem [{\citenamefont {Dumbser}\ \emph
  {et~al.}(2008{\natexlab{a}})\citenamefont {Dumbser}, \citenamefont {Balsara},
  \citenamefont {Toro},\ and\ \citenamefont {Munz}}]{dumbser2008unified}%
  \BibitemOpen
  \bibfield  {author} {\bibinfo {author} {\bibfnamefont {M.}~\bibnamefont
  {Dumbser}}, \bibinfo {author} {\bibfnamefont {D.~S.}\ \bibnamefont
  {Balsara}}, \bibinfo {author} {\bibfnamefont {E.~F.}\ \bibnamefont {Toro}}, \
  and\ \bibinfo {author} {\bibfnamefont {C.-D.}\ \bibnamefont {Munz}},\
  }\href@noop {} {\bibfield  {journal} {\bibinfo  {journal} {Journal of
  Computational Physics}\ }\textbf {\bibinfo {volume} {227}},\ \bibinfo {pages}
  {8209} (\bibinfo {year} {2008}{\natexlab{a}})}\BibitemShut {NoStop}%
\bibitem [{\citenamefont {Miller}\ and\ \citenamefont
  {Schnetter}(2016)}]{Miller2017}%
  \BibitemOpen
  \bibfield  {author} {\bibinfo {author} {\bibfnamefont {J.~M.}\ \bibnamefont
  {Miller}}\ and\ \bibinfo {author} {\bibfnamefont {E.}~\bibnamefont
  {Schnetter}},\ }\href {\doibase 10.1088/1361-6382/34/1/015003} {\bibfield
  {journal} {\bibinfo  {journal} {Classical and Quantum Gravity}\ }\textbf
  {\bibinfo {volume} {34}},\ \bibinfo {pages} {015003} (\bibinfo {year}
  {2016})}\BibitemShut {NoStop}%
\bibitem [{\citenamefont {{Deppe}}\ \emph {et~al.}(2024)\citenamefont
  {{Deppe}}, \citenamefont {{Foucart}}, \citenamefont {{Bonilla}},
  \citenamefont {{Boyle}}, \citenamefont {{Corso}}, \citenamefont {{Duez}},
  \citenamefont {{Giesler}}, \citenamefont {{H{\'e}bert}}, \citenamefont
  {{Kidder}}, \citenamefont {{Kim}}, \citenamefont {{Kumar}}, \citenamefont
  {{Legred}}, \citenamefont {{Lovelace}}, \citenamefont {{Most}}, \citenamefont
  {{Moxon}}, \citenamefont {{Nelli}}, \citenamefont {{Pfeiffer}}, \citenamefont
  {{Scheel}}, \citenamefont {{Teukolsky}}, \citenamefont {{Throwe}},\ and\
  \citenamefont {{Vu}}}]{Deppe2024}%
  \BibitemOpen
  \bibfield  {author} {\bibinfo {author} {\bibfnamefont {N.}~\bibnamefont
  {{Deppe}}}, \bibinfo {author} {\bibfnamefont {F.}~\bibnamefont {{Foucart}}},
  \bibinfo {author} {\bibfnamefont {M.~S.}\ \bibnamefont {{Bonilla}}}, \bibinfo
  {author} {\bibfnamefont {M.}~\bibnamefont {{Boyle}}}, \bibinfo {author}
  {\bibfnamefont {N.~J.}\ \bibnamefont {{Corso}}}, \bibinfo {author}
  {\bibfnamefont {M.~D.}\ \bibnamefont {{Duez}}}, \bibinfo {author}
  {\bibfnamefont {M.}~\bibnamefont {{Giesler}}}, \bibinfo {author}
  {\bibfnamefont {F.}~\bibnamefont {{H{\'e}bert}}}, \bibinfo {author}
  {\bibfnamefont {L.~E.}\ \bibnamefont {{Kidder}}}, \bibinfo {author}
  {\bibfnamefont {Y.}~\bibnamefont {{Kim}}}, \bibinfo {author} {\bibfnamefont
  {P.}~\bibnamefont {{Kumar}}}, \bibinfo {author} {\bibfnamefont
  {I.}~\bibnamefont {{Legred}}}, \bibinfo {author} {\bibfnamefont
  {G.}~\bibnamefont {{Lovelace}}}, \bibinfo {author} {\bibfnamefont {E.~R.}\
  \bibnamefont {{Most}}}, \bibinfo {author} {\bibfnamefont {J.}~\bibnamefont
  {{Moxon}}}, \bibinfo {author} {\bibfnamefont {K.~C.}\ \bibnamefont
  {{Nelli}}}, \bibinfo {author} {\bibfnamefont {H.~P.}\ \bibnamefont
  {{Pfeiffer}}}, \bibinfo {author} {\bibfnamefont {M.~A.}\ \bibnamefont
  {{Scheel}}}, \bibinfo {author} {\bibfnamefont {S.~A.}\ \bibnamefont
  {{Teukolsky}}}, \bibinfo {author} {\bibfnamefont {W.}~\bibnamefont
  {{Throwe}}}, \ and\ \bibinfo {author} {\bibfnamefont {N.~L.}\ \bibnamefont
  {{Vu}}},\ }\href {\doibase 10.48550/arXiv.2406.19038} {\bibfield  {journal}
  {\bibinfo  {journal} {arXiv e-prints}\ ,\ \bibinfo {eid} {arXiv:2406.19038}}
  (\bibinfo {year} {2024})}\BibitemShut {NoStop}%
\bibitem [{\citenamefont {Kidder}\ \emph {et~al.}(2000)\citenamefont {Kidder},
  \citenamefont {Scheel}, \citenamefont {Teukolsky}, \citenamefont {Carlson},\
  and\ \citenamefont {Cook}}]{Kidder2000}%
  \BibitemOpen
  \bibfield  {author} {\bibinfo {author} {\bibfnamefont {L.~E.}\ \bibnamefont
  {Kidder}}, \bibinfo {author} {\bibfnamefont {M.~A.}\ \bibnamefont {Scheel}},
  \bibinfo {author} {\bibfnamefont {S.~A.}\ \bibnamefont {Teukolsky}}, \bibinfo
  {author} {\bibfnamefont {E.~D.}\ \bibnamefont {Carlson}}, \ and\ \bibinfo
  {author} {\bibfnamefont {G.~B.}\ \bibnamefont {Cook}},\ }\href {\doibase
  10.1103/PhysRevD.62.084032} {\bibfield  {journal} {\bibinfo  {journal} {Phys.
  Rev. D}\ }\textbf {\bibinfo {volume} {62}},\ \bibinfo {pages} {084032}
  (\bibinfo {year} {2000})}\BibitemShut {NoStop}%
\bibitem [{\citenamefont {Duez}\ \emph {et~al.}(2008)\citenamefont {Duez},
  \citenamefont {Foucart}, \citenamefont {Kidder}, \citenamefont {Pfeiffer},
  \citenamefont {Scheel},\ and\ \citenamefont {Teukolsky}}]{Duez2008}%
  \BibitemOpen
  \bibfield  {author} {\bibinfo {author} {\bibfnamefont {M.~D.}\ \bibnamefont
  {Duez}}, \bibinfo {author} {\bibfnamefont {F.}~\bibnamefont {Foucart}},
  \bibinfo {author} {\bibfnamefont {L.~E.}\ \bibnamefont {Kidder}}, \bibinfo
  {author} {\bibfnamefont {H.~P.}\ \bibnamefont {Pfeiffer}}, \bibinfo {author}
  {\bibfnamefont {M.~A.}\ \bibnamefont {Scheel}}, \ and\ \bibinfo {author}
  {\bibfnamefont {S.~A.}\ \bibnamefont {Teukolsky}},\ }\href {\doibase
  10.1103/PhysRevD.78.104015} {\bibfield  {journal} {\bibinfo  {journal} {Phys.
  Rev. D}\ }\textbf {\bibinfo {volume} {78}},\ \bibinfo {pages} {104015}
  (\bibinfo {year} {2008})}\BibitemShut {NoStop}%
\bibitem [{\citenamefont {Haas}\ \emph {et~al.}(2016)\citenamefont {Haas},
  \citenamefont {Ott}, \citenamefont {Szilagyi}, \citenamefont {Kaplan},
  \citenamefont {Lippuner}, \citenamefont {Scheel}, \citenamefont {Barkett},
  \citenamefont {Muhlberger}, \citenamefont {Dietrich}, \citenamefont {Duez},
  \citenamefont {Foucart}, \citenamefont {Pfeiffer}, \citenamefont {Kidder},\
  and\ \citenamefont {Teukolsky}}]{Haas2016}%
  \BibitemOpen
  \bibfield  {author} {\bibinfo {author} {\bibfnamefont {R.}~\bibnamefont
  {Haas}}, \bibinfo {author} {\bibfnamefont {C.~D.}\ \bibnamefont {Ott}},
  \bibinfo {author} {\bibfnamefont {B.}~\bibnamefont {Szilagyi}}, \bibinfo
  {author} {\bibfnamefont {J.~D.}\ \bibnamefont {Kaplan}}, \bibinfo {author}
  {\bibfnamefont {J.}~\bibnamefont {Lippuner}}, \bibinfo {author}
  {\bibfnamefont {M.~A.}\ \bibnamefont {Scheel}}, \bibinfo {author}
  {\bibfnamefont {K.}~\bibnamefont {Barkett}}, \bibinfo {author} {\bibfnamefont
  {C.~D.}\ \bibnamefont {Muhlberger}}, \bibinfo {author} {\bibfnamefont
  {T.}~\bibnamefont {Dietrich}}, \bibinfo {author} {\bibfnamefont {M.~D.}\
  \bibnamefont {Duez}}, \bibinfo {author} {\bibfnamefont {F.}~\bibnamefont
  {Foucart}}, \bibinfo {author} {\bibfnamefont {H.~P.}\ \bibnamefont
  {Pfeiffer}}, \bibinfo {author} {\bibfnamefont {L.~E.}\ \bibnamefont
  {Kidder}}, \ and\ \bibinfo {author} {\bibfnamefont {S.~A.}\ \bibnamefont
  {Teukolsky}},\ }\href {\doibase 10.1103/PhysRevD.93.124062} {\bibfield
  {journal} {\bibinfo  {journal} {Phys. Rev. D}\ }\textbf {\bibinfo {volume}
  {93}},\ \bibinfo {pages} {124062} (\bibinfo {year} {2016})}\BibitemShut
  {NoStop}%
\bibitem [{\citenamefont {{Brown}}\ \emph {et~al.}(2012)\citenamefont
  {{Brown}}, \citenamefont {{Diener}}, \citenamefont {{Field}}, \citenamefont
  {{Hesthaven}}, \citenamefont {{Herrmann}}, \citenamefont {{Mrou{\'e}}},
  \citenamefont {{Sarbach}}, \citenamefont {{Schnetter}}, \citenamefont
  {{Tiglio}},\ and\ \citenamefont {{Wagman}}}]{Brown2012}%
  \BibitemOpen
  \bibfield  {author} {\bibinfo {author} {\bibfnamefont {J.~D.}\ \bibnamefont
  {{Brown}}}, \bibinfo {author} {\bibfnamefont {P.}~\bibnamefont {{Diener}}},
  \bibinfo {author} {\bibfnamefont {S.~E.}\ \bibnamefont {{Field}}}, \bibinfo
  {author} {\bibfnamefont {J.~S.}\ \bibnamefont {{Hesthaven}}}, \bibinfo
  {author} {\bibfnamefont {F.}~\bibnamefont {{Herrmann}}}, \bibinfo {author}
  {\bibfnamefont {A.~H.}\ \bibnamefont {{Mrou{\'e}}}}, \bibinfo {author}
  {\bibfnamefont {O.}~\bibnamefont {{Sarbach}}}, \bibinfo {author}
  {\bibfnamefont {E.}~\bibnamefont {{Schnetter}}}, \bibinfo {author}
  {\bibfnamefont {M.}~\bibnamefont {{Tiglio}}}, \ and\ \bibinfo {author}
  {\bibfnamefont {M.}~\bibnamefont {{Wagman}}},\ }\href {\doibase
  10.1103/PhysRevD.85.084004} {\bibfield  {journal} {\bibinfo  {journal} {Phys.
  Rev. D}\ }\textbf {\bibinfo {volume} {85}},\ \bibinfo {eid} {084004}
  (\bibinfo {year} {2012})}\BibitemShut {NoStop}%
\bibitem [{\citenamefont {Bona}\ \emph {et~al.}(2003)\citenamefont {Bona},
  \citenamefont {Ledvinka}, \citenamefont {Palenzuela},\ and\ \citenamefont
  {Z{\'{a}}cek}}]{Bona:2003fj}%
  \BibitemOpen
  \bibfield  {author} {\bibinfo {author} {\bibfnamefont {C.}~\bibnamefont
  {Bona}}, \bibinfo {author} {\bibfnamefont {T.}~\bibnamefont {Ledvinka}},
  \bibinfo {author} {\bibfnamefont {C.}~\bibnamefont {Palenzuela}}, \ and\
  \bibinfo {author} {\bibfnamefont {M.}~\bibnamefont {Z{\'{a}}cek}},\
  }\href@noop {} {\bibfield  {journal} {\bibinfo  {journal} {Phys. Rev. D}\
  }\textbf {\bibinfo {volume} {67}},\ \bibinfo {pages} {104005} (\bibinfo
  {year} {2003})}\BibitemShut {NoStop}%
\bibitem [{\citenamefont {Bona}\ \emph {et~al.}(2004)\citenamefont {Bona},
  \citenamefont {Ledvinka}, \citenamefont {Palenzuela},\ and\ \citenamefont
  {\ifmmode \check{Z}\else \v{Z}\fi{}\'a\ifmmode~\check{c}\else
  \v{c}\fi{}ek}}]{Bona:2004}%
  \BibitemOpen
  \bibfield  {author} {\bibinfo {author} {\bibfnamefont {C.}~\bibnamefont
  {Bona}}, \bibinfo {author} {\bibfnamefont {T.}~\bibnamefont {Ledvinka}},
  \bibinfo {author} {\bibfnamefont {C.}~\bibnamefont {Palenzuela}}, \ and\
  \bibinfo {author} {\bibfnamefont {M.}~\bibnamefont {\ifmmode \check{Z}\else
  \v{Z}\fi{}\'a\ifmmode~\check{c}\else \v{c}\fi{}ek}},\ }\href {\doibase
  10.1103/PhysRevD.69.064036} {\bibfield  {journal} {\bibinfo  {journal} {Phys.
  Rev. D}\ }\textbf {\bibinfo {volume} {69}},\ \bibinfo {pages} {064036}
  (\bibinfo {year} {2004})}\BibitemShut {NoStop}%
\bibitem [{\citenamefont {Dumbser}\ \emph {et~al.}(2018)\citenamefont
  {Dumbser}, \citenamefont {Guercilena}, \citenamefont {K{\"o}ppel},
  \citenamefont {Rezzolla},\ and\ \citenamefont
  {Zanotti}}]{Dumbser2017strongly}%
  \BibitemOpen
  \bibfield  {author} {\bibinfo {author} {\bibfnamefont {M.}~\bibnamefont
  {Dumbser}}, \bibinfo {author} {\bibfnamefont {F.}~\bibnamefont {Guercilena}},
  \bibinfo {author} {\bibfnamefont {S.}~\bibnamefont {K{\"o}ppel}}, \bibinfo
  {author} {\bibfnamefont {L.}~\bibnamefont {Rezzolla}}, \ and\ \bibinfo
  {author} {\bibfnamefont {O.}~\bibnamefont {Zanotti}},\ }\href@noop {}
  {\bibfield  {journal} {\bibinfo  {journal} {Physical Review D}\ }\textbf
  {\bibinfo {volume} {97}},\ \bibinfo {pages} {084053} (\bibinfo {year}
  {2018})}\BibitemShut {NoStop}%
\bibitem [{\citenamefont {Dumbser}\ \emph {et~al.}(2020)\citenamefont
  {Dumbser}, \citenamefont {Fambri}, \citenamefont {Gaburro},\ and\
  \citenamefont {Reinarz}}]{Dumbser2020GLM}%
  \BibitemOpen
  \bibfield  {author} {\bibinfo {author} {\bibfnamefont {M.}~\bibnamefont
  {Dumbser}}, \bibinfo {author} {\bibfnamefont {F.}~\bibnamefont {Fambri}},
  \bibinfo {author} {\bibfnamefont {E.}~\bibnamefont {Gaburro}}, \ and\
  \bibinfo {author} {\bibfnamefont {A.}~\bibnamefont {Reinarz}},\ }\href@noop
  {} {\bibfield  {journal} {\bibinfo  {journal} {Journal of Computational
  Physics}\ }\textbf {\bibinfo {volume} {404}},\ \bibinfo {pages} {109088}
  (\bibinfo {year} {2020})}\BibitemShut {NoStop}%
\bibitem [{\citenamefont {Alic}\ \emph {et~al.}(2012)\citenamefont {Alic},
  \citenamefont {Bona-Casas}, \citenamefont {Bona}, \citenamefont {Rezzolla},\
  and\ \citenamefont {Palenzuela}}]{Alic:2011a}%
  \BibitemOpen
  \bibfield  {author} {\bibinfo {author} {\bibfnamefont {D.}~\bibnamefont
  {Alic}}, \bibinfo {author} {\bibfnamefont {C.}~\bibnamefont {Bona-Casas}},
  \bibinfo {author} {\bibfnamefont {C.}~\bibnamefont {Bona}}, \bibinfo {author}
  {\bibfnamefont {L.}~\bibnamefont {Rezzolla}}, \ and\ \bibinfo {author}
  {\bibfnamefont {C.}~\bibnamefont {Palenzuela}},\ }\href@noop {} {\bibfield
  {journal} {\bibinfo  {journal} {Physical Review D}\ }\textbf {\bibinfo
  {volume} {85}},\ \bibinfo {pages} {064040} (\bibinfo {year}
  {2012})}\BibitemShut {NoStop}%
\bibitem [{\citenamefont {{Gaburro}}\ \emph {et~al.}(2021)\citenamefont
  {{Gaburro}}, \citenamefont {{Castro}},\ and\ \citenamefont
  {{Dumbser}}}]{Gaburro2021WBGR1D}%
  \BibitemOpen
  \bibfield  {author} {\bibinfo {author} {\bibfnamefont {E.}~\bibnamefont
  {{Gaburro}}}, \bibinfo {author} {\bibfnamefont {M.~J.}\ \bibnamefont
  {{Castro}}}, \ and\ \bibinfo {author} {\bibfnamefont {M.}~\bibnamefont
  {{Dumbser}}},\ }\href@noop {} {\bibfield  {journal} {\bibinfo  {journal}
  {SIAM Journal on Scientific Computing}\ }\textbf {\bibinfo {volume} {43}},\
  \bibinfo {pages} {B1226} (\bibinfo {year} {2021})}\BibitemShut {NoStop}%
\bibitem [{\citenamefont {Dumbser}\ \emph {et~al.}(2024)\citenamefont
  {Dumbser}, \citenamefont {Zanotti}, \citenamefont {Gaburro},\ and\
  \citenamefont {Peshkov}}]{DumbserZanottiGaburroPeshkov2023}%
  \BibitemOpen
  \bibfield  {author} {\bibinfo {author} {\bibfnamefont {M.}~\bibnamefont
  {Dumbser}}, \bibinfo {author} {\bibfnamefont {O.}~\bibnamefont {Zanotti}},
  \bibinfo {author} {\bibfnamefont {E.}~\bibnamefont {Gaburro}}, \ and\
  \bibinfo {author} {\bibfnamefont {I.}~\bibnamefont {Peshkov}},\ }\href
  {\doibase 10.1016/j.jcp.2024.112875} {\bibfield  {journal} {\bibinfo
  {journal} {J. Comput. Phys.}\ }\textbf {\bibinfo {volume} {504}},\ \bibinfo
  {pages} {112875} (\bibinfo {year} {2024})}\BibitemShut {NoStop}%
\bibitem [{\citenamefont {{Dumbser}}\ \emph {et~al.}(2024)\citenamefont
  {{Dumbser}}, \citenamefont {{Zanotti}},\ and\ \citenamefont
  {{Puppo}}}]{CWENOBSSNOK}%
  \BibitemOpen
  \bibfield  {author} {\bibinfo {author} {\bibfnamefont {M.}~\bibnamefont
  {{Dumbser}}}, \bibinfo {author} {\bibfnamefont {O.}~\bibnamefont
  {{Zanotti}}}, \ and\ \bibinfo {author} {\bibfnamefont {G.}~\bibnamefont
  {{Puppo}}},\ }\href@noop {} {\bibfield  {journal} {\bibinfo  {journal} {arXiv
  e-prints}\ } (\bibinfo {year} {2024})},\ \Eprint
  {http://arxiv.org/abs/2406.12055} {arXiv:2406.12055 [gr-qc]} \BibitemShut
  {NoStop}%
\bibitem [{\citenamefont {Levy}\ \emph {et~al.}(2000)\citenamefont {Levy},
  \citenamefont {Puppo},\ and\ \citenamefont {Russo}}]{Levy2000}%
  \BibitemOpen
  \bibfield  {author} {\bibinfo {author} {\bibfnamefont {D.}~\bibnamefont
  {Levy}}, \bibinfo {author} {\bibfnamefont {G.}~\bibnamefont {Puppo}}, \ and\
  \bibinfo {author} {\bibfnamefont {G.}~\bibnamefont {Russo}},\ }\href
  {\doibase 10.1016/S0168-9274(99)00108-7} {\bibfield  {journal} {\bibinfo
  {journal} {Applied Numerical Mathematics}\ }\textbf {\bibinfo {volume}
  {33}},\ \bibinfo {pages} {415 – 421} (\bibinfo {year} {2000})}\BibitemShut
  {NoStop}%
\bibitem [{\citenamefont {Levy}\ \emph {et~al.}(2001)\citenamefont {Levy},
  \citenamefont {Puppo},\ and\ \citenamefont {Russo}}]{Levy2001}%
  \BibitemOpen
  \bibfield  {author} {\bibinfo {author} {\bibfnamefont {D.}~\bibnamefont
  {Levy}}, \bibinfo {author} {\bibfnamefont {G.}~\bibnamefont {Puppo}}, \ and\
  \bibinfo {author} {\bibfnamefont {G.}~\bibnamefont {Russo}},\ }\href
  {\doibase 10.1137/S1064827599359461} {\bibfield  {journal} {\bibinfo
  {journal} {SIAM Journal on Scientific Computing}\ }\textbf {\bibinfo {volume}
  {22}},\ \bibinfo {pages} {656 – 672} (\bibinfo {year} {2001})}\BibitemShut
  {NoStop}%
\bibitem [{\citenamefont {Balsara}\ \emph {et~al.}(2024)\citenamefont
  {Balsara}, \citenamefont {Deepak}, \citenamefont {Zanotti},\ and\
  \citenamefont {Dumbser}}]{Balsara2024c}%
  \BibitemOpen
  \bibfield  {author} {\bibinfo {author} {\bibfnamefont {D.}~\bibnamefont
  {Balsara}}, \bibinfo {author} {\bibfnamefont {B.}~\bibnamefont {Deepak}},
  \bibinfo {author} {\bibfnamefont {O.}~\bibnamefont {Zanotti}}, \ and\
  \bibinfo {author} {\bibfnamefont {M.}~\bibnamefont {Dumbser}},\ }\href@noop
  {} {\  (\bibinfo {year} {2024})},\ \Eprint {http://arxiv.org/abs/2406.05450}
  {arXiv:2406.05450 [gr-qc]} \BibitemShut {NoStop}%
\bibitem [{\citenamefont {{Alcubierre}}\ \emph {et~al.}(2004)\citenamefont
  {{Alcubierre}}, \citenamefont {{Allen}},\ and\ \citenamefont {{Bona et
  al.}}}]{Alcubierre2004}%
  \BibitemOpen
  \bibfield  {author} {\bibinfo {author} {\bibfnamefont {M.}~\bibnamefont
  {{Alcubierre}}}, \bibinfo {author} {\bibfnamefont {G.}~\bibnamefont
  {{Allen}}}, \ and\ \bibinfo {author} {\bibfnamefont {C.}~\bibnamefont {{Bona
  et al.}}},\ }\href {\doibase 10.1088/0264-9381/21/2/019} {\bibfield
  {journal} {\bibinfo  {journal} {Classical and Quantum Gravity}\ }\textbf
  {\bibinfo {volume} {21}},\ \bibinfo {pages} {589} (\bibinfo {year}
  {2004})}\BibitemShut {NoStop}%
\bibitem [{\citenamefont {Rhebergen}\ \emph {et~al.}(2008)\citenamefont
  {Rhebergen}, \citenamefont {Bokhove},\ and\ \citenamefont {van~der
  Vegt}}]{Rhebergen2008}%
  \BibitemOpen
  \bibfield  {author} {\bibinfo {author} {\bibfnamefont {S.}~\bibnamefont
  {Rhebergen}}, \bibinfo {author} {\bibfnamefont {O.}~\bibnamefont {Bokhove}},
  \ and\ \bibinfo {author} {\bibfnamefont {J.}~\bibnamefont {van~der Vegt}},\
  }\href@noop {} {\bibfield  {journal} {\bibinfo  {journal} {Journal of
  Computational Physics}\ }\textbf {\bibinfo {volume} {227}},\ \bibinfo {pages}
  {1887} (\bibinfo {year} {2008})}\BibitemShut {NoStop}%
\bibitem [{\citenamefont {Dumbser}\ \emph {et~al.}(2009)\citenamefont
  {Dumbser}, \citenamefont {Castro}, \citenamefont {{Par\'es}},\ and\
  \citenamefont {Toro}}]{Dumbser2009a}%
  \BibitemOpen
  \bibfield  {author} {\bibinfo {author} {\bibfnamefont {M.}~\bibnamefont
  {Dumbser}}, \bibinfo {author} {\bibfnamefont {M.}~\bibnamefont {Castro}},
  \bibinfo {author} {\bibfnamefont {C.}~\bibnamefont {{Par\'es}}}, \ and\
  \bibinfo {author} {\bibfnamefont {E.}~\bibnamefont {Toro}},\ }\href@noop {}
  {\bibfield  {journal} {\bibinfo  {journal} {Computers and Fluids}\ }\textbf
  {\bibinfo {volume} {38}},\ \bibinfo {pages} {1731} (\bibinfo {year}
  {2009})}\BibitemShut {NoStop}%
\bibitem [{\citenamefont {Dumbser}\ \emph {et~al.}(2010)\citenamefont
  {Dumbser}, \citenamefont {Hidalgo}, \citenamefont {Castro}, \citenamefont
  {Parés},\ and\ \citenamefont {Toro}}]{Dumbser2010}%
  \BibitemOpen
  \bibfield  {author} {\bibinfo {author} {\bibfnamefont {M.}~\bibnamefont
  {Dumbser}}, \bibinfo {author} {\bibfnamefont {A.}~\bibnamefont {Hidalgo}},
  \bibinfo {author} {\bibfnamefont {M.}~\bibnamefont {Castro}}, \bibinfo
  {author} {\bibfnamefont {C.}~\bibnamefont {Parés}}, \ and\ \bibinfo {author}
  {\bibfnamefont {E.~F.}\ \bibnamefont {Toro}},\ }\href {\doibase
  https://doi.org/10.1016/j.cma.2009.10.016} {\bibfield  {journal} {\bibinfo
  {journal} {Computer Methods in Applied Mechanics and Engineering}\ }\textbf
  {\bibinfo {volume} {199}},\ \bibinfo {pages} {625} (\bibinfo {year}
  {2010})}\BibitemShut {NoStop}%
\bibitem [{\citenamefont {Castro}\ \emph {et~al.}(2006)\citenamefont {Castro},
  \citenamefont {Gallardo},\ and\ \citenamefont {Par{\'e}s}}]{Castro2006}%
  \BibitemOpen
  \bibfield  {author} {\bibinfo {author} {\bibfnamefont {M.}~\bibnamefont
  {Castro}}, \bibinfo {author} {\bibfnamefont {J.}~\bibnamefont {Gallardo}}, \
  and\ \bibinfo {author} {\bibfnamefont {C.}~\bibnamefont {Par{\'e}s}},\
  }\href@noop {} {\bibfield  {journal} {\bibinfo  {journal} {Mathematics of
  computation}\ }\textbf {\bibinfo {volume} {75}},\ \bibinfo {pages} {1103}
  (\bibinfo {year} {2006})}\BibitemShut {NoStop}%
\bibitem [{\citenamefont {Par{\'e}s}(2006)}]{Pares2006}%
  \BibitemOpen
  \bibfield  {author} {\bibinfo {author} {\bibfnamefont {C.}~\bibnamefont
  {Par{\'e}s}},\ }\href@noop {} {\bibfield  {journal} {\bibinfo  {journal}
  {SIAM Journal on Numerical Analysis}\ }\textbf {\bibinfo {volume} {44}},\
  \bibinfo {pages} {300} (\bibinfo {year} {2006})}\BibitemShut {NoStop}%
\bibitem [{\citenamefont {Dal~Maso}\ \emph {et~al.}(1995)\citenamefont
  {Dal~Maso}, \citenamefont {Lefloch},\ and\ \citenamefont
  {Murat}}]{DalMaso1995}%
  \BibitemOpen
  \bibfield  {author} {\bibinfo {author} {\bibfnamefont {G.}~\bibnamefont
  {Dal~Maso}}, \bibinfo {author} {\bibfnamefont {P.~G.}\ \bibnamefont
  {Lefloch}}, \ and\ \bibinfo {author} {\bibfnamefont {F.}~\bibnamefont
  {Murat}},\ }\href@noop {} {\bibfield  {journal} {\bibinfo  {journal} {Journal
  de math{\'e}matiques pures et appliqu{\'e}es}\ }\textbf {\bibinfo {volume}
  {74}},\ \bibinfo {pages} {483} (\bibinfo {year} {1995})}\BibitemShut
  {NoStop}%
\bibitem [{\citenamefont {Toro}(1999)}]{toro-book}%
  \BibitemOpen
  \bibfield  {author} {\bibinfo {author} {\bibfnamefont {E.~F.}\ \bibnamefont
  {Toro}},\ }\href@noop {} {\emph {\bibinfo {title} {{{Riemann} Solvers and
  Numerical Methods for Fluid Dynamics}}}},\ \bibinfo {edition} {2nd}\ ed.\
  (\bibinfo  {publisher} {Springer},\ \bibinfo {year} {1999})\BibitemShut
  {NoStop}%
\bibitem [{\citenamefont {Dumbser}\ \emph
  {et~al.}(2008{\natexlab{b}})\citenamefont {Dumbser}, \citenamefont {Enaux},\
  and\ \citenamefont {Toro}}]{DumbserEnauxToro}%
  \BibitemOpen
  \bibfield  {author} {\bibinfo {author} {\bibfnamefont {M.}~\bibnamefont
  {Dumbser}}, \bibinfo {author} {\bibfnamefont {C.}~\bibnamefont {Enaux}}, \
  and\ \bibinfo {author} {\bibfnamefont {E.}~\bibnamefont {Toro}},\ }\href@noop
  {} {\bibfield  {journal} {\bibinfo  {journal} {Journal of Computational
  Physics}\ }\textbf {\bibinfo {volume} {227}},\ \bibinfo {pages} {3971}
  (\bibinfo {year} {2008}{\natexlab{b}})}\BibitemShut {NoStop}%
\bibitem [{\citenamefont {{Titarev}}\ and\ \citenamefont
  {{Toro}}(2002)}]{Titarev2002}%
  \BibitemOpen
  \bibfield  {author} {\bibinfo {author} {\bibfnamefont {V.~A.}\ \bibnamefont
  {{Titarev}}}\ and\ \bibinfo {author} {\bibfnamefont {E.~F.}\ \bibnamefont
  {{Toro}}},\ }\href@noop {} {\bibfield  {journal} {\bibinfo  {journal}
  {Journal of Scientific Computing}\ }\textbf {\bibinfo {volume} {17}},\
  \bibinfo {pages} {609} (\bibinfo {year} {2002})}\BibitemShut {NoStop}%
\bibitem [{\citenamefont {{Titarev}}\ and\ \citenamefont
  {{Toro}}(2005)}]{Titarev2005}%
  \BibitemOpen
  \bibfield  {author} {\bibinfo {author} {\bibfnamefont {V.~A.}\ \bibnamefont
  {{Titarev}}}\ and\ \bibinfo {author} {\bibfnamefont {E.~F.}\ \bibnamefont
  {{Toro}}},\ }\href {\doibase 10.1016/j.jcp.2004.10.028} {\bibfield  {journal}
  {\bibinfo  {journal} {Journal of Computational Physics}\ }\textbf {\bibinfo
  {volume} {204}},\ \bibinfo {pages} {715} (\bibinfo {year}
  {2005})}\BibitemShut {NoStop}%
\bibitem [{\citenamefont {Toro}\ and\ \citenamefont
  {Titarev}(2006)}]{Toro2006}%
  \BibitemOpen
  \bibfield  {author} {\bibinfo {author} {\bibfnamefont {E.~F.}\ \bibnamefont
  {Toro}}\ and\ \bibinfo {author} {\bibfnamefont {V.~A.}\ \bibnamefont
  {Titarev}},\ }\href@noop {} {\bibfield  {journal} {\bibinfo  {journal}
  {Journal of Computational Physics}\ }\textbf {\bibinfo {volume} {212}},\
  \bibinfo {pages} {150} (\bibinfo {year} {2006})}\BibitemShut {NoStop}%
\bibitem [{\citenamefont {Dumbser}\ and\ \citenamefont
  {Toro}(2011)}]{Dumbser2011}%
  \BibitemOpen
  \bibfield  {author} {\bibinfo {author} {\bibfnamefont {M.}~\bibnamefont
  {Dumbser}}\ and\ \bibinfo {author} {\bibfnamefont {E.~F.}\ \bibnamefont
  {Toro}},\ }\href@noop {} {\bibfield  {journal} {\bibinfo  {journal} {Journal
  of Scientific Computing}\ }\textbf {\bibinfo {volume} {48}},\ \bibinfo
  {pages} {70} (\bibinfo {year} {2011})}\BibitemShut {NoStop}%
\bibitem [{\citenamefont {Busto}\ \emph {et~al.}(2020)\citenamefont {Busto},
  \citenamefont {Chiocchetti}, \citenamefont {Dumbser}, \citenamefont
  {Gaburro},\ and\ \citenamefont {Peshkov}}]{BCDGP20}%
  \BibitemOpen
  \bibfield  {author} {\bibinfo {author} {\bibfnamefont {S.}~\bibnamefont
  {Busto}}, \bibinfo {author} {\bibfnamefont {S.}~\bibnamefont {Chiocchetti}},
  \bibinfo {author} {\bibfnamefont {M.}~\bibnamefont {Dumbser}}, \bibinfo
  {author} {\bibfnamefont {E.}~\bibnamefont {Gaburro}}, \ and\ \bibinfo
  {author} {\bibfnamefont {I.}~\bibnamefont {Peshkov}},\ }\href@noop {}
  {\bibfield  {journal} {\bibinfo  {journal} {Frontiers in Physics}\ }\textbf
  {\bibinfo {volume} {8}},\ \bibinfo {pages} {32} (\bibinfo {year}
  {2020})}\BibitemShut {NoStop}%
\bibitem [{\citenamefont {Godunov}(1959)}]{godunov}%
  \BibitemOpen
  \bibfield  {author} {\bibinfo {author} {\bibfnamefont {S.}~\bibnamefont
  {Godunov}},\ }\href@noop {} {\bibfield  {journal} {\bibinfo  {journal}
  {Mathematics of the USSR: Sbornik}\ }\textbf {\bibinfo {volume} {47}},\
  \bibinfo {pages} {271} (\bibinfo {year} {1959})}\BibitemShut {NoStop}%
\bibitem [{\citenamefont {{Kidder}}\ \emph {et~al.}(2017)\citenamefont
  {{Kidder}}, \citenamefont {{Field}}, \citenamefont {{Foucart}}, \citenamefont
  {{Schnetter}}, \citenamefont {{Teukolsky}}, \citenamefont {{Bohn}},
  \citenamefont {{Deppe}}, \citenamefont {{Diener}}, \citenamefont
  {{H{\'e}bert}}, \citenamefont {{Lippuner}}, \citenamefont {{Miller}},
  \citenamefont {{Ott}}, \citenamefont {{Scheel}},\ and\ \citenamefont
  {{Vincent}}}]{Kidder2017}%
  \BibitemOpen
  \bibfield  {author} {\bibinfo {author} {\bibfnamefont {L.~E.}\ \bibnamefont
  {{Kidder}}}, \bibinfo {author} {\bibfnamefont {S.~E.}\ \bibnamefont
  {{Field}}}, \bibinfo {author} {\bibfnamefont {F.}~\bibnamefont {{Foucart}}},
  \bibinfo {author} {\bibfnamefont {E.}~\bibnamefont {{Schnetter}}}, \bibinfo
  {author} {\bibfnamefont {S.~A.}\ \bibnamefont {{Teukolsky}}}, \bibinfo
  {author} {\bibfnamefont {A.}~\bibnamefont {{Bohn}}}, \bibinfo {author}
  {\bibfnamefont {N.}~\bibnamefont {{Deppe}}}, \bibinfo {author} {\bibfnamefont
  {P.}~\bibnamefont {{Diener}}}, \bibinfo {author} {\bibfnamefont
  {F.}~\bibnamefont {{H{\'e}bert}}}, \bibinfo {author} {\bibfnamefont
  {J.}~\bibnamefont {{Lippuner}}}, \bibinfo {author} {\bibfnamefont
  {J.}~\bibnamefont {{Miller}}}, \bibinfo {author} {\bibfnamefont {C.~D.}\
  \bibnamefont {{Ott}}}, \bibinfo {author} {\bibfnamefont {M.~A.}\ \bibnamefont
  {{Scheel}}}, \ and\ \bibinfo {author} {\bibfnamefont {T.}~\bibnamefont
  {{Vincent}}},\ }\href {\doibase 10.1016/j.jcp.2016.12.059} {\bibfield
  {journal} {\bibinfo  {journal} {Journal of Computational Physics}\ }\textbf
  {\bibinfo {volume} {335}},\ \bibinfo {pages} {84} (\bibinfo {year}
  {2017})}\BibitemShut {NoStop}%
\bibitem [{\citenamefont {{Deppe}}\ \emph
  {et~al.}(2022{\natexlab{b}})\citenamefont {{Deppe}}, \citenamefont
  {{H{\'e}bert}}, \citenamefont {{Kidder}}, \citenamefont {{Throwe}},
  \citenamefont {{Anantpurkar}}, \citenamefont {{Armaza}}, \citenamefont
  {{Bonilla}}, \citenamefont {{Boyle}}, \citenamefont {{Chaudhary}},
  \citenamefont {{Duez}}, \citenamefont {{Vu}}, \citenamefont {{Foucart}},
  \citenamefont {{Giesler}}, \citenamefont {{Guo}}, \citenamefont {{Kim}},
  \citenamefont {{Kumar}}, \citenamefont {{Legred}}, \citenamefont {{Li}},
  \citenamefont {{Lovelace}}, \citenamefont {{Ma}}, \citenamefont {{Macedo}},
  \citenamefont {{Melchor}}, \citenamefont {{Morales}}, \citenamefont
  {{Moxon}}, \citenamefont {{Nelli}}, \citenamefont {{O'Shea}}, \citenamefont
  {{Pfeiffer}}, \citenamefont {{Ramirez}}, \citenamefont {{R{\"u}ter}},
  \citenamefont {{Sanchez}}, \citenamefont {{Scheel}}, \citenamefont
  {{Thomas}}, \citenamefont {{Vieira}}, \citenamefont {{Wittek}}, \citenamefont
  {{Wlodarczyk}},\ and\ \citenamefont {{Teukolsky}}}]{Deppe2022b}%
  \BibitemOpen
  \bibfield  {author} {\bibinfo {author} {\bibfnamefont {N.}~\bibnamefont
  {{Deppe}}}, \bibinfo {author} {\bibfnamefont {F.}~\bibnamefont
  {{H{\'e}bert}}}, \bibinfo {author} {\bibfnamefont {L.~E.}\ \bibnamefont
  {{Kidder}}}, \bibinfo {author} {\bibfnamefont {W.}~\bibnamefont {{Throwe}}},
  \bibinfo {author} {\bibfnamefont {I.}~\bibnamefont {{Anantpurkar}}}, \bibinfo
  {author} {\bibfnamefont {C.}~\bibnamefont {{Armaza}}}, \bibinfo {author}
  {\bibfnamefont {G.~S.}\ \bibnamefont {{Bonilla}}}, \bibinfo {author}
  {\bibfnamefont {M.}~\bibnamefont {{Boyle}}}, \bibinfo {author} {\bibfnamefont
  {H.}~\bibnamefont {{Chaudhary}}}, \bibinfo {author} {\bibfnamefont {M.~D.}\
  \bibnamefont {{Duez}}}, \bibinfo {author} {\bibfnamefont {N.~L.}\
  \bibnamefont {{Vu}}}, \bibinfo {author} {\bibfnamefont {F.}~\bibnamefont
  {{Foucart}}}, \bibinfo {author} {\bibfnamefont {M.}~\bibnamefont
  {{Giesler}}}, \bibinfo {author} {\bibfnamefont {J.~S.}\ \bibnamefont
  {{Guo}}}, \bibinfo {author} {\bibfnamefont {Y.}~\bibnamefont {{Kim}}},
  \bibinfo {author} {\bibfnamefont {P.}~\bibnamefont {{Kumar}}}, \bibinfo
  {author} {\bibfnamefont {I.}~\bibnamefont {{Legred}}}, \bibinfo {author}
  {\bibfnamefont {D.}~\bibnamefont {{Li}}}, \bibinfo {author} {\bibfnamefont
  {G.}~\bibnamefont {{Lovelace}}}, \bibinfo {author} {\bibfnamefont
  {S.}~\bibnamefont {{Ma}}}, \bibinfo {author} {\bibfnamefont {A.}~\bibnamefont
  {{Macedo}}}, \bibinfo {author} {\bibfnamefont {D.}~\bibnamefont {{Melchor}}},
  \bibinfo {author} {\bibfnamefont {M.}~\bibnamefont {{Morales}}}, \bibinfo
  {author} {\bibfnamefont {J.}~\bibnamefont {{Moxon}}}, \bibinfo {author}
  {\bibfnamefont {K.~C.}\ \bibnamefont {{Nelli}}}, \bibinfo {author}
  {\bibfnamefont {E.}~\bibnamefont {{O'Shea}}}, \bibinfo {author}
  {\bibfnamefont {H.~P.}\ \bibnamefont {{Pfeiffer}}}, \bibinfo {author}
  {\bibfnamefont {T.}~\bibnamefont {{Ramirez}}}, \bibinfo {author}
  {\bibfnamefont {H.~R.}\ \bibnamefont {{R{\"u}ter}}}, \bibinfo {author}
  {\bibfnamefont {J.}~\bibnamefont {{Sanchez}}}, \bibinfo {author}
  {\bibfnamefont {M.~A.}\ \bibnamefont {{Scheel}}}, \bibinfo {author}
  {\bibfnamefont {S.}~\bibnamefont {{Thomas}}}, \bibinfo {author}
  {\bibfnamefont {D.}~\bibnamefont {{Vieira}}}, \bibinfo {author}
  {\bibfnamefont {N.~A.}\ \bibnamefont {{Wittek}}}, \bibinfo {author}
  {\bibfnamefont {T.}~\bibnamefont {{Wlodarczyk}}}, \ and\ \bibinfo {author}
  {\bibfnamefont {S.~A.}\ \bibnamefont {{Teukolsky}}},\ }\href {\doibase
  10.1103/PhysRevD.105.123031} {\bibfield  {journal} {\bibinfo  {journal}
  {\prd}\ }\textbf {\bibinfo {volume} {105}},\ \bibinfo {eid} {123031}
  (\bibinfo {year} {2022}{\natexlab{b}})},\ \Eprint
  {http://arxiv.org/abs/2109.12033} {arXiv:2109.12033 [gr-qc]} \BibitemShut
  {NoStop}%
\bibitem [{\citenamefont {Wu}\ and\ \citenamefont {Tang}(2016)}]{Kailiang2017}%
  \BibitemOpen
  \bibfield  {author} {\bibinfo {author} {\bibfnamefont {K.}~\bibnamefont
  {Wu}}\ and\ \bibinfo {author} {\bibfnamefont {H.}~\bibnamefont {Tang}},\
  }\href {\doibase 10.3847/1538-4365/228/1/3} {\bibfield  {journal} {\bibinfo
  {journal} {The Astrophysical Journal Supplement Series}\ }\textbf {\bibinfo
  {volume} {228}},\ \bibinfo {pages} {3} (\bibinfo {year} {2016})}\BibitemShut
  {NoStop}%
\bibitem [{\citenamefont {{Zanotti}}\ \emph
  {et~al.}(2015{\natexlab{b}})\citenamefont {{Zanotti}}, \citenamefont
  {{Fambri}}, \citenamefont {{Dumbser}},\ and\ \citenamefont
  {{Hidalgo}}}]{Zanotti2015c}%
  \BibitemOpen
  \bibfield  {author} {\bibinfo {author} {\bibfnamefont {O.}~\bibnamefont
  {{Zanotti}}}, \bibinfo {author} {\bibfnamefont {F.}~\bibnamefont {{Fambri}}},
  \bibinfo {author} {\bibfnamefont {M.}~\bibnamefont {{Dumbser}}}, \ and\
  \bibinfo {author} {\bibfnamefont {A.}~\bibnamefont {{Hidalgo}}},\ }\href@noop
  {} {\bibfield  {journal} {\bibinfo  {journal} {Computers and Fluids}\
  }\textbf {\bibinfo {volume} {118}},\ \bibinfo {pages} {204 } (\bibinfo {year}
  {2015}{\natexlab{b}})}\BibitemShut {NoStop}%
\bibitem [{\citenamefont {Sonntag}\ and\ \citenamefont
  {Munz}(2017)}]{Sonntag2}%
  \BibitemOpen
  \bibfield  {author} {\bibinfo {author} {\bibfnamefont {M.}~\bibnamefont
  {Sonntag}}\ and\ \bibinfo {author} {\bibfnamefont {C.}~\bibnamefont {Munz}},\
  }\href@noop {} {\bibfield  {journal} {\bibinfo  {journal} {Journal of
  Scientific Computing}\ }\textbf {\bibinfo {volume} {70}},\ \bibinfo {pages}
  {1262} (\bibinfo {year} {2017})}\BibitemShut {NoStop}%
\bibitem [{\citenamefont {Titarev}\ and\ \citenamefont
  {Toro}(2004)}]{Titarev2004}%
  \BibitemOpen
  \bibfield  {author} {\bibinfo {author} {\bibfnamefont {V.}~\bibnamefont
  {Titarev}}\ and\ \bibinfo {author} {\bibfnamefont {E.}~\bibnamefont {Toro}},\
  }\href {\doibase https://doi.org/10.1016/j.jcp.2004.05.015} {\bibfield
  {journal} {\bibinfo  {journal} {Journal of Computational Physics}\ }\textbf
  {\bibinfo {volume} {201}},\ \bibinfo {pages} {238} (\bibinfo {year}
  {2004})}\BibitemShut {NoStop}%
\bibitem [{\citenamefont {Bermudez}\ and\ \citenamefont
  {V\'azquez-Cend\'on}(1994)}]{Bermudez1994}%
  \BibitemOpen
  \bibfield  {author} {\bibinfo {author} {\bibfnamefont {A.}~\bibnamefont
  {Bermudez}}\ and\ \bibinfo {author} {\bibfnamefont {M.}~\bibnamefont
  {V\'azquez-Cend\'on}},\ }\href@noop {} {\bibfield  {journal} {\bibinfo
  {journal} {Computers \& Fluids}\ }\textbf {\bibinfo {volume} {23}},\ \bibinfo
  {pages} {1049} (\bibinfo {year} {1994})}\BibitemShut {NoStop}%
\bibitem [{\citenamefont {LeVeque}(1998)}]{leveque1998balancing}%
  \BibitemOpen
  \bibfield  {author} {\bibinfo {author} {\bibfnamefont {R.}~\bibnamefont
  {LeVeque}},\ }\href@noop {} {\bibfield  {journal} {\bibinfo  {journal}
  {Journal of Computational Physics}\ }\textbf {\bibinfo {volume} {146}},\
  \bibinfo {pages} {346} (\bibinfo {year} {1998})}\BibitemShut {NoStop}%
\bibitem [{\citenamefont {Gosse}(2001)}]{gosse2001well}%
  \BibitemOpen
  \bibfield  {author} {\bibinfo {author} {\bibfnamefont {L.}~\bibnamefont
  {Gosse}},\ }\href@noop {} {\bibfield  {journal} {\bibinfo  {journal}
  {Mathematical Models and Methods in Applied Sciences}\ }\textbf {\bibinfo
  {volume} {11}},\ \bibinfo {pages} {339} (\bibinfo {year} {2001})}\BibitemShut
  {NoStop}%
\bibitem [{\citenamefont {Audusse}\ \emph {et~al.}(2004)\citenamefont
  {Audusse}, \citenamefont {Bouchut}, \citenamefont {Bristeau}, \citenamefont
  {Klein},\ and\ \citenamefont {Perthame}}]{audusse2004fast}%
  \BibitemOpen
  \bibfield  {author} {\bibinfo {author} {\bibfnamefont {E.}~\bibnamefont
  {Audusse}}, \bibinfo {author} {\bibfnamefont {F.}~\bibnamefont {Bouchut}},
  \bibinfo {author} {\bibfnamefont {M.}~\bibnamefont {Bristeau}}, \bibinfo
  {author} {\bibfnamefont {R.}~\bibnamefont {Klein}}, \ and\ \bibinfo {author}
  {\bibfnamefont {B.}~\bibnamefont {Perthame}},\ }\href@noop {} {\bibfield
  {journal} {\bibinfo  {journal} {SIAM Journal on Scientific Computing}\
  }\textbf {\bibinfo {volume} {25}},\ \bibinfo {pages} {2050} (\bibinfo {year}
  {2004})}\BibitemShut {NoStop}%
\bibitem [{\citenamefont {Botta}\ \emph {et~al.}(2004)\citenamefont {Botta},
  \citenamefont {Klein}, \citenamefont {Langenberg},\ and\ \citenamefont
  {L{\"u}tzenkirchen}}]{BottaKlein}%
  \BibitemOpen
  \bibfield  {author} {\bibinfo {author} {\bibfnamefont {N.}~\bibnamefont
  {Botta}}, \bibinfo {author} {\bibfnamefont {R.}~\bibnamefont {Klein}},
  \bibinfo {author} {\bibfnamefont {S.}~\bibnamefont {Langenberg}}, \ and\
  \bibinfo {author} {\bibfnamefont {S.}~\bibnamefont {L{\"u}tzenkirchen}},\
  }\href@noop {} {\bibfield  {journal} {\bibinfo  {journal} {Journal of
  Computational Physics}\ }\textbf {\bibinfo {volume} {196}},\ \bibinfo {pages}
  {539} (\bibinfo {year} {2004})}\BibitemShut {NoStop}%
\bibitem [{\citenamefont {Castro}\ \emph {et~al.}(2017)\citenamefont {Castro},
  \citenamefont {de~Luna},\ and\ \citenamefont {Par{\'e}s}}]{Castro2017Book}%
  \BibitemOpen
  \bibfield  {author} {\bibinfo {author} {\bibfnamefont {M.~J.}\ \bibnamefont
  {Castro}}, \bibinfo {author} {\bibfnamefont {T.~M.}\ \bibnamefont {de~Luna}},
  \ and\ \bibinfo {author} {\bibfnamefont {C.}~\bibnamefont {Par{\'e}s}},\ }in\
  \href@noop {} {\emph {\bibinfo {booktitle} {Handbook of Numerical
  Analysis}}},\ Vol.~\bibinfo {volume} {18}\ (\bibinfo  {publisher}
  {Elsevier},\ \bibinfo {year} {2017})\ pp.\ \bibinfo {pages}
  {131--175}\BibitemShut {NoStop}%
\bibitem [{\citenamefont {Castro}\ and\ \citenamefont
  {Par{\'e}s}(2020)}]{castro2020well}%
  \BibitemOpen
  \bibfield  {author} {\bibinfo {author} {\bibfnamefont {M.~J.}\ \bibnamefont
  {Castro}}\ and\ \bibinfo {author} {\bibfnamefont {C.}~\bibnamefont
  {Par{\'e}s}},\ }\href@noop {} {\bibfield  {journal} {\bibinfo  {journal}
  {Journal of Scientific Computing}\ }\textbf {\bibinfo {volume} {82}},\
  \bibinfo {pages} {1} (\bibinfo {year} {2020})}\BibitemShut {NoStop}%
\bibitem [{\citenamefont {Xu}\ and\ \citenamefont {Shu}(2024)}]{xu2024}%
  \BibitemOpen
  \bibfield  {author} {\bibinfo {author} {\bibfnamefont {Z.}~\bibnamefont
  {Xu}}\ and\ \bibinfo {author} {\bibfnamefont {C.-W.}\ \bibnamefont {Shu}},\
  }\href@noop {} {\  (\bibinfo {year} {2024})}\BibitemShut {NoStop}%
\bibitem [{\citenamefont {Pareschi}\ and\ \citenamefont
  {Rey}(2017)}]{PareschiRey}%
  \BibitemOpen
  \bibfield  {author} {\bibinfo {author} {\bibfnamefont {L.}~\bibnamefont
  {Pareschi}}\ and\ \bibinfo {author} {\bibfnamefont {T.}~\bibnamefont {Rey}},\
  }\href@noop {} {\bibfield  {journal} {\bibinfo  {journal} {Computers \&
  Fluids}\ }\textbf {\bibinfo {volume} {156}},\ \bibinfo {pages} {329}
  (\bibinfo {year} {2017})}\BibitemShut {NoStop}%
\bibitem [{\citenamefont {Berberich}\ \emph {et~al.}(2021)\citenamefont
  {Berberich}, \citenamefont {Chandrashekar},\ and\ \citenamefont
  {Klingenberg}}]{berberich2021high}%
  \BibitemOpen
  \bibfield  {author} {\bibinfo {author} {\bibfnamefont {J.}~\bibnamefont
  {Berberich}}, \bibinfo {author} {\bibfnamefont {P.}~\bibnamefont
  {Chandrashekar}}, \ and\ \bibinfo {author} {\bibfnamefont {C.}~\bibnamefont
  {Klingenberg}},\ }\href@noop {} {\bibfield  {journal} {\bibinfo  {journal}
  {Computers \& Fluids}\ ,\ \bibinfo {pages} {104858}} (\bibinfo {year}
  {2021})}\BibitemShut {NoStop}%
\bibitem [{\citenamefont {Dumbser}\ \emph {et~al.}(2013)\citenamefont
  {Dumbser}, \citenamefont {Zanotti}, \citenamefont {Hidalgo},\ and\
  \citenamefont {Balsara}}]{AMR3DCL}%
  \BibitemOpen
  \bibfield  {author} {\bibinfo {author} {\bibfnamefont {M.}~\bibnamefont
  {Dumbser}}, \bibinfo {author} {\bibfnamefont {O.}~\bibnamefont {Zanotti}},
  \bibinfo {author} {\bibfnamefont {A.}~\bibnamefont {Hidalgo}}, \ and\
  \bibinfo {author} {\bibfnamefont {D.}~\bibnamefont {Balsara}},\ }\href@noop
  {} {\bibfield  {journal} {\bibinfo  {journal} {J. Comput. Phys.}\ }\textbf
  {\bibinfo {volume} {248}},\ \bibinfo {pages} {257} (\bibinfo {year}
  {2013})}\BibitemShut {NoStop}%
\bibitem [{\citenamefont {Zanotti}\ and\ \citenamefont
  {Dumbser}(2015)}]{Zanotti2015}%
  \BibitemOpen
  \bibfield  {author} {\bibinfo {author} {\bibfnamefont {O.}~\bibnamefont
  {Zanotti}}\ and\ \bibinfo {author} {\bibfnamefont {M.}~\bibnamefont
  {Dumbser}},\ }\href
  {http://www.scopus.com/inward/record.url?eid=2-s2.0-84920650018&partnerID=40&md5=3b1f2cc6325d5e96abf5b76c36a6b886}
  {\bibfield  {journal} {\bibinfo  {journal} {Computer Physics Communications}\
  }\textbf {\bibinfo {volume} {188}},\ \bibinfo {pages} {110} (\bibinfo {year}
  {2015})}\BibitemShut {NoStop}%
\bibitem [{\citenamefont {{Reinarz}}\ \emph {et~al.}(2020)\citenamefont
  {{Reinarz}}, \citenamefont {{Charrier}}, \citenamefont {{Bader}},
  \citenamefont {{Bovard}}, \citenamefont {{Dumbser}}, \citenamefont {{Duru}},
  \citenamefont {{Fambri}}, \citenamefont {{Gabriel}}, \citenamefont
  {{Gallard}}, \citenamefont {{K{\"o}ppel}}, \citenamefont {{Krenz}},
  \citenamefont {{Rannabauer}}, \citenamefont {{Rezzolla}}, \citenamefont
  {{Samfass}}, \citenamefont {{Tavelli}},\ and\ \citenamefont
  {{Weinzierl}}}]{ExaHype2020}%
  \BibitemOpen
  \bibfield  {author} {\bibinfo {author} {\bibfnamefont {A.}~\bibnamefont
  {{Reinarz}}}, \bibinfo {author} {\bibfnamefont {D.~E.}\ \bibnamefont
  {{Charrier}}}, \bibinfo {author} {\bibfnamefont {M.}~\bibnamefont {{Bader}}},
  \bibinfo {author} {\bibfnamefont {L.}~\bibnamefont {{Bovard}}}, \bibinfo
  {author} {\bibfnamefont {M.}~\bibnamefont {{Dumbser}}}, \bibinfo {author}
  {\bibfnamefont {K.}~\bibnamefont {{Duru}}}, \bibinfo {author} {\bibfnamefont
  {F.}~\bibnamefont {{Fambri}}}, \bibinfo {author} {\bibfnamefont {A.-A.}\
  \bibnamefont {{Gabriel}}}, \bibinfo {author} {\bibfnamefont {J.-M.}\
  \bibnamefont {{Gallard}}}, \bibinfo {author} {\bibfnamefont {S.}~\bibnamefont
  {{K{\"o}ppel}}}, \bibinfo {author} {\bibfnamefont {L.}~\bibnamefont
  {{Krenz}}}, \bibinfo {author} {\bibfnamefont {L.}~\bibnamefont
  {{Rannabauer}}}, \bibinfo {author} {\bibfnamefont {L.}~\bibnamefont
  {{Rezzolla}}}, \bibinfo {author} {\bibfnamefont {P.}~\bibnamefont
  {{Samfass}}}, \bibinfo {author} {\bibfnamefont {M.}~\bibnamefont
  {{Tavelli}}}, \ and\ \bibinfo {author} {\bibfnamefont {T.}~\bibnamefont
  {{Weinzierl}}},\ }\href {\doibase 10.1016/j.cpc.2020.107251} {\bibfield
  {journal} {\bibinfo  {journal} {Computer Physics Communications}\ }\textbf
  {\bibinfo {volume} {254}},\ \bibinfo {eid} {107251} (\bibinfo {year}
  {2020})}\BibitemShut {NoStop}%
\bibitem [{\citenamefont {Khokhlov}(1998)}]{Khokhlov1998}%
  \BibitemOpen
  \bibfield  {author} {\bibinfo {author} {\bibfnamefont {A.}~\bibnamefont
  {Khokhlov}},\ }\href {\doibase https://doi.org/10.1006/jcph.1998.9998}
  {\bibfield  {journal} {\bibinfo  {journal} {Journal of Computational
  Physics}\ }\textbf {\bibinfo {volume} {143}},\ \bibinfo {pages} {519}
  (\bibinfo {year} {1998})}\BibitemShut {NoStop}%
\bibitem [{\citenamefont {Babiuc}\ \emph {et~al.}(2008)\citenamefont {Babiuc},
  \citenamefont {Husa}, \citenamefont {Alic}, \citenamefont {Hinder},
  \citenamefont {Lechner}, \citenamefont {Schnetter}, \citenamefont
  {Szilágyi}, \citenamefont {Zlochower}, \citenamefont {Dorband},
  \citenamefont {Pollney},\ and\ \citenamefont {Winicour}}]{Babiuc_2008}%
  \BibitemOpen
  \bibfield  {author} {\bibinfo {author} {\bibfnamefont {M.~C.}\ \bibnamefont
  {Babiuc}}, \bibinfo {author} {\bibfnamefont {S.}~\bibnamefont {Husa}},
  \bibinfo {author} {\bibfnamefont {D.}~\bibnamefont {Alic}}, \bibinfo {author}
  {\bibfnamefont {I.}~\bibnamefont {Hinder}}, \bibinfo {author} {\bibfnamefont
  {C.}~\bibnamefont {Lechner}}, \bibinfo {author} {\bibfnamefont
  {E.}~\bibnamefont {Schnetter}}, \bibinfo {author} {\bibfnamefont
  {B.}~\bibnamefont {Szilágyi}}, \bibinfo {author} {\bibfnamefont
  {Y.}~\bibnamefont {Zlochower}}, \bibinfo {author} {\bibfnamefont
  {N.}~\bibnamefont {Dorband}}, \bibinfo {author} {\bibfnamefont
  {D.}~\bibnamefont {Pollney}}, \ and\ \bibinfo {author} {\bibfnamefont
  {J.}~\bibnamefont {Winicour}},\ }\href {\doibase
  10.1088/0264-9381/25/12/125012} {\bibfield  {journal} {\bibinfo  {journal}
  {Classical and Quantum Gravity}\ }\textbf {\bibinfo {volume} {25}},\ \bibinfo
  {pages} {125012} (\bibinfo {year} {2008})}\BibitemShut {NoStop}%
\bibitem [{\citenamefont {{Tichy}}(2009)}]{Tichy2009}%
  \BibitemOpen
  \bibfield  {author} {\bibinfo {author} {\bibfnamefont {W.}~\bibnamefont
  {{Tichy}}},\ }\href {\doibase 10.1103/PhysRevD.80.104034} {\bibfield
  {journal} {\bibinfo  {journal} {Phys. Rev. D}\ }\textbf {\bibinfo {volume}
  {80}},\ \bibinfo {eid} {104034} (\bibinfo {year} {2009})},\ \Eprint
  {http://arxiv.org/abs/0911.0973} {arXiv:0911.0973 [gr-qc]} \BibitemShut
  {NoStop}%
\bibitem [{\citenamefont {{Mignone}}\ and\ \citenamefont
  {{Bodo}}(2005)}]{Mignone2005}%
  \BibitemOpen
  \bibfield  {author} {\bibinfo {author} {\bibfnamefont {A.}~\bibnamefont
  {{Mignone}}}\ and\ \bibinfo {author} {\bibfnamefont {G.}~\bibnamefont
  {{Bodo}}},\ }\href {\doibase 10.1111/j.1365-2966.2005.09546.x} {\bibfield
  {journal} {\bibinfo  {journal} {Mon. Not. R. Astron. Soc.}\ }\textbf
  {\bibinfo {volume} {364}},\ \bibinfo {pages} {126} (\bibinfo {year}
  {2005})}\BibitemShut {NoStop}%
\bibitem [{\citenamefont {Radice}\ and\ \citenamefont
  {Rezzolla}(2012)}]{Radice2012a}%
  \BibitemOpen
  \bibfield  {author} {\bibinfo {author} {\bibfnamefont {D.}~\bibnamefont
  {Radice}}\ and\ \bibinfo {author} {\bibfnamefont {L.}~\bibnamefont
  {Rezzolla}},\ }\href@noop {} {\bibfield  {journal} {\bibinfo  {journal}
  {Astronomy \& Astrophysics}\ }\textbf {\bibinfo {volume} {547}},\ \bibinfo
  {pages} {A26} (\bibinfo {year} {2012})}\BibitemShut {NoStop}%
\bibitem [{\citenamefont {Zhang}\ and\ \citenamefont
  {MacFadyen}(2006)}]{Zhang2006}%
  \BibitemOpen
  \bibfield  {author} {\bibinfo {author} {\bibfnamefont {W.}~\bibnamefont
  {Zhang}}\ and\ \bibinfo {author} {\bibfnamefont {A.}~\bibnamefont
  {MacFadyen}},\ }\href {http://iopscience.iop.org/0067-0049/164/1/255}
  {\bibfield  {journal} {\bibinfo  {journal} {The Astrophysical Journal
  Supplement Series}\ }\textbf {\bibinfo {volume} {164}},\ \bibinfo {pages}
  {255} (\bibinfo {year} {2006})}\BibitemShut {NoStop}%
\bibitem [{\citenamefont {Mart{\'\i}}\ and\ \citenamefont
  {M{\"{u}}ller}(1994)}]{Marti94}%
  \BibitemOpen
  \bibfield  {author} {\bibinfo {author} {\bibfnamefont {J.~M.}\ \bibnamefont
  {Mart{\'\i}}}\ and\ \bibinfo {author} {\bibfnamefont {E.}~\bibnamefont
  {M{\"{u}}ller}},\ }\href@noop {} {\bibfield  {journal} {\bibinfo  {journal}
  {J. Fluid Mech.}\ }\textbf {\bibinfo {volume} {258}},\ \bibinfo {pages} {317}
  (\bibinfo {year} {1994})}\BibitemShut {NoStop}%
\bibitem [{\citenamefont {{Rezzolla}}\ and\ \citenamefont
  {{Zanotti}}(2001)}]{Rezzolla2001}%
  \BibitemOpen
  \bibfield  {author} {\bibinfo {author} {\bibfnamefont {L.}~\bibnamefont
  {{Rezzolla}}}\ and\ \bibinfo {author} {\bibfnamefont {O.}~\bibnamefont
  {{Zanotti}}},\ }\href {\doibase 10.1017/S0022112001006450} {\bibfield
  {journal} {\bibinfo  {journal} {Journal of Fluid Mechanics}\ }\textbf
  {\bibinfo {volume} {449}},\ \bibinfo {pages} {395} (\bibinfo {year}
  {2001})},\ \Eprint {http://arxiv.org/abs/gr-qc/0103005} {arXiv:gr-qc/0103005
  [gr-qc]} \BibitemShut {NoStop}%
\bibitem [{\citenamefont {Ant{\'o}n}\ \emph {et~al.}(2006)\citenamefont
  {Ant{\'o}n}, \citenamefont {Zanotti}, \citenamefont {Miralles}, \citenamefont
  {Mart{\'\i}}, \citenamefont {Ib{\'a}{\~n}ez}, \citenamefont {Font},\ and\
  \citenamefont {Pons}}]{Anton06}%
  \BibitemOpen
  \bibfield  {author} {\bibinfo {author} {\bibfnamefont {L.}~\bibnamefont
  {Ant{\'o}n}}, \bibinfo {author} {\bibfnamefont {O.}~\bibnamefont {Zanotti}},
  \bibinfo {author} {\bibfnamefont {J.~A.}\ \bibnamefont {Miralles}}, \bibinfo
  {author} {\bibfnamefont {J.~M.}\ \bibnamefont {Mart{\'\i}}}, \bibinfo
  {author} {\bibfnamefont {J.~M.}\ \bibnamefont {Ib{\'a}{\~n}ez}}, \bibinfo
  {author} {\bibfnamefont {J.~A.}\ \bibnamefont {Font}}, \ and\ \bibinfo
  {author} {\bibfnamefont {J.~A.}\ \bibnamefont {Pons}},\ }\href@noop {}
  {\bibfield  {journal} {\bibinfo  {journal} {The Astrophysical Journal}\
  }\textbf {\bibinfo {volume} {637}},\ \bibinfo {pages} {296} (\bibinfo {year}
  {2006})}\BibitemShut {NoStop}%
\bibitem [{\citenamefont {Tolman}(1939)}]{Tolman}%
  \BibitemOpen
  \bibfield  {author} {\bibinfo {author} {\bibfnamefont {R.~C.}\ \bibnamefont
  {Tolman}},\ }\href@noop {} {\bibfield  {journal} {\bibinfo  {journal}
  {Physical Review}\ }\textbf {\bibinfo {volume} {55}},\ \bibinfo {pages} {364}
  (\bibinfo {year} {1939})}\BibitemShut {NoStop}%
\bibitem [{\citenamefont {Oppenheimer}\ and\ \citenamefont
  {Volkoff}(1939)}]{Oppenheimer39b}%
  \BibitemOpen
  \bibfield  {author} {\bibinfo {author} {\bibfnamefont {J.~R.}\ \bibnamefont
  {Oppenheimer}}\ and\ \bibinfo {author} {\bibfnamefont {G.~M.}\ \bibnamefont
  {Volkoff}},\ }\href@noop {} {\bibfield  {journal} {\bibinfo  {journal}
  {Physical Review}\ }\textbf {\bibinfo {volume} {55}},\ \bibinfo {pages} {374}
  (\bibinfo {year} {1939})}\BibitemShut {NoStop}%
\bibitem [{\citenamefont {{Camenzind}}(2007)}]{Camenzind2007}%
  \BibitemOpen
  \bibfield  {author} {\bibinfo {author} {\bibfnamefont {M.}~\bibnamefont
  {{Camenzind}}},\ }\href {\doibase 10.1007/978-3-540-49912-1} {\emph {\bibinfo
  {title} {{Compact objects in astrophysics : white dwarfs, neutron stars, and
  black holes}}}}\ (\bibinfo {year} {2007})\BibitemShut {NoStop}%
\bibitem [{\citenamefont {{Font}}\ \emph {et~al.}(2002)\citenamefont {{Font}},
  \citenamefont {{Goodale}}, \citenamefont {{Iyer}}, \citenamefont {{Miller}},
  \citenamefont {{Rezzolla}}, \citenamefont {{Seidel}}, \citenamefont
  {{Stergioulas}}, \citenamefont {{Suen}},\ and\ \citenamefont
  {{Tobias}}}]{Font2002}%
  \BibitemOpen
  \bibfield  {author} {\bibinfo {author} {\bibfnamefont {J.~A.}\ \bibnamefont
  {{Font}}}, \bibinfo {author} {\bibfnamefont {T.}~\bibnamefont {{Goodale}}},
  \bibinfo {author} {\bibfnamefont {S.}~\bibnamefont {{Iyer}}}, \bibinfo
  {author} {\bibfnamefont {M.}~\bibnamefont {{Miller}}}, \bibinfo {author}
  {\bibfnamefont {L.}~\bibnamefont {{Rezzolla}}}, \bibinfo {author}
  {\bibfnamefont {E.}~\bibnamefont {{Seidel}}}, \bibinfo {author}
  {\bibfnamefont {N.}~\bibnamefont {{Stergioulas}}}, \bibinfo {author}
  {\bibfnamefont {W.-M.}\ \bibnamefont {{Suen}}}, \ and\ \bibinfo {author}
  {\bibfnamefont {M.}~\bibnamefont {{Tobias}}},\ }\href {\doibase
  10.1103/PhysRevD.65.084024} {\bibfield  {journal} {\bibinfo  {journal}
  {Physical Review D}\ }\textbf {\bibinfo {volume} {65}},\ \bibinfo {eid}
  {084024} (\bibinfo {year} {2002})}\BibitemShut {NoStop}%
\bibitem [{\citenamefont {Dumbser}(2010)}]{ADERNSE}%
  \BibitemOpen
  \bibfield  {author} {\bibinfo {author} {\bibfnamefont {M.}~\bibnamefont
  {Dumbser}},\ }\href@noop {} {\bibfield  {journal} {\bibinfo  {journal}
  {Computers \& Fluids}\ }\textbf {\bibinfo {volume} {39}},\ \bibinfo {pages}
  {60} (\bibinfo {year} {2010})}\BibitemShut {NoStop}%
\bibitem [{\citenamefont {Brandt}\ and\ \citenamefont
  {Br\"ugmann}(1997)}]{Brandt1997}%
  \BibitemOpen
  \bibfield  {author} {\bibinfo {author} {\bibfnamefont {S.}~\bibnamefont
  {Brandt}}\ and\ \bibinfo {author} {\bibfnamefont {B.}~\bibnamefont
  {Br\"ugmann}},\ }\href {\doibase 10.1103/PhysRevLett.78.3606} {\bibfield
  {journal} {\bibinfo  {journal} {Phys. Rev. Lett.}\ }\textbf {\bibinfo
  {volume} {78}},\ \bibinfo {pages} {3606} (\bibinfo {year}
  {1997})}\BibitemShut {NoStop}%
\bibitem [{\citenamefont {Alcubierre}\ \emph {et~al.}(2003)\citenamefont
  {Alcubierre}, \citenamefont {Br\"ugmann}, \citenamefont {Diener},
  \citenamefont {Koppitz}, \citenamefont {Pollney}, \citenamefont {Seidel},\
  and\ \citenamefont {Takahashi}}]{Alcubierre2003}%
  \BibitemOpen
  \bibfield  {author} {\bibinfo {author} {\bibfnamefont {M.}~\bibnamefont
  {Alcubierre}}, \bibinfo {author} {\bibfnamefont {B.}~\bibnamefont
  {Br\"ugmann}}, \bibinfo {author} {\bibfnamefont {P.}~\bibnamefont {Diener}},
  \bibinfo {author} {\bibfnamefont {M.}~\bibnamefont {Koppitz}}, \bibinfo
  {author} {\bibfnamefont {D.}~\bibnamefont {Pollney}}, \bibinfo {author}
  {\bibfnamefont {E.}~\bibnamefont {Seidel}}, \ and\ \bibinfo {author}
  {\bibfnamefont {R.}~\bibnamefont {Takahashi}},\ }\href {\doibase
  10.1103/PhysRevD.67.084023} {\bibfield  {journal} {\bibinfo  {journal} {Phys.
  Rev. D}\ }\textbf {\bibinfo {volume} {67}},\ \bibinfo {pages} {084023}
  (\bibinfo {year} {2003})}\BibitemShut {NoStop}%
\bibitem [{\citenamefont {Campanelli}\ \emph {et~al.}(2006)\citenamefont
  {Campanelli}, \citenamefont {Lousto}, \citenamefont {Marronetti},\ and\
  \citenamefont {Zlochower}}]{Campanelli2006}%
  \BibitemOpen
  \bibfield  {author} {\bibinfo {author} {\bibfnamefont {M.}~\bibnamefont
  {Campanelli}}, \bibinfo {author} {\bibfnamefont {C.~O.}\ \bibnamefont
  {Lousto}}, \bibinfo {author} {\bibfnamefont {P.}~\bibnamefont {Marronetti}},
  \ and\ \bibinfo {author} {\bibfnamefont {Y.}~\bibnamefont {Zlochower}},\
  }\href {\doibase 10.1103/PhysRevLett.96.111101} {\bibfield  {journal}
  {\bibinfo  {journal} {Phys. Rev. Lett.}\ }\textbf {\bibinfo {volume} {96}},\
  \bibinfo {pages} {111101} (\bibinfo {year} {2006})}\BibitemShut {NoStop}%
\bibitem [{\citenamefont {Baker}\ \emph {et~al.}(2006)\citenamefont {Baker},
  \citenamefont {Centrella}, \citenamefont {Choi}, \citenamefont {Koppitz},\
  and\ \citenamefont {van Meter}}]{Baker2006}%
  \BibitemOpen
  \bibfield  {author} {\bibinfo {author} {\bibfnamefont {J.~G.}\ \bibnamefont
  {Baker}}, \bibinfo {author} {\bibfnamefont {J.}~\bibnamefont {Centrella}},
  \bibinfo {author} {\bibfnamefont {D.-I.}\ \bibnamefont {Choi}}, \bibinfo
  {author} {\bibfnamefont {M.}~\bibnamefont {Koppitz}}, \ and\ \bibinfo
  {author} {\bibfnamefont {J.}~\bibnamefont {van Meter}},\ }\href {\doibase
  10.1103/PhysRevLett.96.111102} {\bibfield  {journal} {\bibinfo  {journal}
  {Phys. Rev. Lett.}\ }\textbf {\bibinfo {volume} {96}},\ \bibinfo {pages}
  {111102} (\bibinfo {year} {2006})}\BibitemShut {NoStop}%
\bibitem [{\citenamefont {Reisswig}\ \emph {et~al.}(2009)\citenamefont
  {Reisswig}, \citenamefont {Bishop}, \citenamefont {Pollney},\ and\
  \citenamefont {Szil\'agyi}}]{Reisswig2009}%
  \BibitemOpen
  \bibfield  {author} {\bibinfo {author} {\bibfnamefont {C.}~\bibnamefont
  {Reisswig}}, \bibinfo {author} {\bibfnamefont {N.~T.}\ \bibnamefont
  {Bishop}}, \bibinfo {author} {\bibfnamefont {D.}~\bibnamefont {Pollney}}, \
  and\ \bibinfo {author} {\bibfnamefont {B.}~\bibnamefont {Szil\'agyi}},\
  }\href {\doibase 10.1103/PhysRevLett.103.221101} {\bibfield  {journal}
  {\bibinfo  {journal} {Phys. Rev. Lett.}\ }\textbf {\bibinfo {volume} {103}},\
  \bibinfo {pages} {221101} (\bibinfo {year} {2009})}\BibitemShut {NoStop}%
\bibitem [{\citenamefont {Hannam}\ \emph {et~al.}(2010)\citenamefont {Hannam},
  \citenamefont {Husa}, \citenamefont {Ohme}, \citenamefont {M\"uller},\ and\
  \citenamefont {Br\"ugmann}}]{Hannam2010}%
  \BibitemOpen
  \bibfield  {author} {\bibinfo {author} {\bibfnamefont {M.}~\bibnamefont
  {Hannam}}, \bibinfo {author} {\bibfnamefont {S.}~\bibnamefont {Husa}},
  \bibinfo {author} {\bibfnamefont {F.}~\bibnamefont {Ohme}}, \bibinfo {author}
  {\bibfnamefont {D.}~\bibnamefont {M\"uller}}, \ and\ \bibinfo {author}
  {\bibfnamefont {B.}~\bibnamefont {Br\"ugmann}},\ }\href {\doibase
  10.1103/PhysRevD.82.124008} {\bibfield  {journal} {\bibinfo  {journal} {Phys.
  Rev. D}\ }\textbf {\bibinfo {volume} {82}},\ \bibinfo {pages} {124008}
  (\bibinfo {year} {2010})}\BibitemShut {NoStop}%
\bibitem [{\citenamefont {Ansorg}\ \emph {et~al.}(2004)\citenamefont {Ansorg},
  \citenamefont {Br{\"u}gmann},\ and\ \citenamefont {Tichy}}]{Ansorg:2004ds}%
  \BibitemOpen
  \bibfield  {author} {\bibinfo {author} {\bibfnamefont {M.}~\bibnamefont
  {Ansorg}}, \bibinfo {author} {\bibfnamefont {B.}~\bibnamefont
  {Br{\"u}gmann}}, \ and\ \bibinfo {author} {\bibfnamefont {W.}~\bibnamefont
  {Tichy}},\ }\href@noop {} {\bibfield  {journal} {\bibinfo  {journal} {Phys.
  Rev. D}\ }\textbf {\bibinfo {volume} {70}},\ \bibinfo {pages} {064011}
  (\bibinfo {year} {2004})}\BibitemShut {NoStop}%
\end{thebibliography}%

\end{document}